\documentclass{aa}

\usepackage{xcolor}
\usepackage{graphicx}
\usepackage{rotating}
\usepackage{siunitx}
\usepackage{txfonts}
\usepackage{graphicx}
\usepackage{subfig}
\usepackage{hyperref}
\usepackage{tablefootnote}
\usepackage{threeparttable}
\hypersetup{
    colorlinks=true,
    citecolor=blue,
    }
\usepackage[normalem]{ulem}    
\usepackage{xcolor}
\usepackage{placeins}    
\usepackage[capitalise]{cleveref}
\usepackage{placeins}
\usepackage{makecell}
\usepackage{multirow}
\usepackage{appendix}

\begin{document}

   \title{X-Shooting ULLYSES: Massive stars at low metallicity}

   \subtitle{\MakeUppercase{\romannumeral 13}. Putting the bi-stability jump to the test in the LMC }

   \author{
          T.\ Alkousa\inst{\ref{inst:sheff}}
          \and
          P.A.\ Crowther\inst{\ref{inst:sheff}}
          \and
          J.M.\ Bestenlehner\inst{\ref{inst:sheff},\ref{inst:sheff2}}
          \and
          H.\ Sana\inst{\ref{inst:KUL}}
          \and
          F.\ Tramper\inst{\ref{inst:madrid}}
          \and
          J.S.\ Vink\inst{\ref{inst:armagh}}
          \and
          D.\ Pauli\inst{\ref{inst:KUL}}
          \and 
          J.Th.\ van Loon\inst{\ref{inst:keele}}
          \and
          \newline
          F.\ Najarro\inst{\ref{inst:madrid}}          
          \and
          R.\ Kuiper\inst{\ref{inst:essen}}
          \and
          A.A.C.\ Sander\inst{\ref{inst:ari}}
          \and
          M.\ Bernini-Peron\inst{\ref{inst:ari}}
          \and
          The XShootU collaboration
          }

   \institute{
            {Astrophysics Research Cluster, School of Mathematical and Physical Sciences, University of Sheffield, Hicks Building, Hounsfield Road, Sheffield S3 7RH, United Kingdom \label{inst:sheff}}
              \email{Talkousa1@sheffield.ac.uk}
            \and
            {School of Chemical, Materials and Biological Engineering, University of Sheffield, Sir Robert Hadfield Building, Mappin Street, Sheffield S1 3JD, United Kingdom \label{inst:sheff2}}
            \and
            {Institute of Astronomy, KU Leuven, Celestijnenlaan 200D, B-3001, Leuven, Belgium \label{inst:KUL}}
            \and
            {Departamento de Astrofísica, Centro de Astrobiología, (CSIC-INTA), Ctra. Torrejón a Ajalvir, km 4, 28850 Torrejón de Ardoz, Madrid, Spain \label{inst:madrid}}
            \and   
            {Armagh Observatory and Planetarium, College Hill, Armagh BT61 9DG, United Kingdom \label{inst:armagh}}      
            \and
            {Lennard-Jones Laboratories, Keele University, ST5 5BG, United Kingdom \label{inst:keele}}   
            \and
            {Faculty of Physics, University of Duisburg-Essen, Lotharstra{\ss}e 1, D-47057 Duisburg, Germany \label{inst:essen}} 
            \and
            {Zentrum f{\"u}r Astronomie der Universit{\"a}t Heidelberg, Astronomisches Rechen-Institut, M{\"o}nchhofstr. 12-14, 69120 Heidelberg \label{inst:ari}}
            }

   \date{Recieived 21 January 2025, Accepted 01 June 2025}

  \abstract 
   {Due to the important role massive stars ($>~8\,M_{\odot}$) play in galactic evolution across cosmic ages, it is important to obtain a deeper understanding of the behaviour of mass-loss in 
   low metallicity environments, which largely determines the path a massive star takes throughout its life and its final fate. This would allow us to better predict the evolution of massive stars in 
   the early universe.}
   {We aim to investigate the theoretical bi-stability jump, which predicts an increase in mass-loss rates below $T_{\rm eff} \approx 25{\rm -}21~{\rm kK}$. We further aim to constrain the 
   photospheric and wind parameters of a sample of LMC late-O and B-supergiant.}
   {We utilise the 1D, non-LTE radiative transfer model \textsc{CMFGEN} in a grid-based approach, and subsequent fine-tuned spectroscopic fitting procedure that allows us to determine the 
   stellar and wind parameters of each star. We apply this method to ultra-violet data from the ULLYSES programme and complementary optical data from the XShootU collaboration. We also utilise 
   evolutionary models to obtain the evolutionary masses and compare them to our derived spectroscopic masses.}
   {We derive physical parameters and wind properties of 16 late O- and B-supergiants that span a wide temperature range $T_{\rm eff} \approx 12{\rm -}30~{\rm kK}$, surface gravity range 
   $\log{(g/{\rm cm\,s^{-2}})}\approx 1.8{\rm -}3.1$, and a mass-loss rate range $\dot{M} \approx 10^{-7.6}{\rm -}10^{-5.7}~M_{\odot}\,{\rm yr}^{-1}$. We also compare our results 
   to previous studies that attempted to investigate the metallicity dependence of wind properties.}
   {We find that our derived photospheric and wind properties are consistent with multiple previous studies. For most of our sample, we find that the evolutionary masses and the spectroscopic 
   masses are consistent within the uncertainties. Our results do not reproduce a bi-stability jump in any temperature range, but rather a monotonic decrease in mass-loss rate at lower 
   temperatures. We obtain a wind terminal velocity-effective temperature relation for LMC supergiants $\varv_{\infty}/{\rm km\,s^{-1}} = 0.076(\pm0.011)T_{\rm eff}/{\rm K} - 884(\pm260)$. We find that our derived mass-loss rates do not agree with 
   the mass-loss rates predicted by any of the numerical recipes. This is also the case for the ratio of the terminal wind velocity to the escape velocity $\varv_{\infty}/\varv_{\rm esc}$, and we derive the relation 
   $\varv_{\infty}/\varv_{\rm esc} = 4.1 (\pm 0.8)\log{(T_{\rm eff}/{\rm K})} - 16.3 (\pm 3.5)$. We find that there is a metallicity dependence of wind parameters from a comparison with a previous SMC study, 
   and we obtain the modified wind momentum-luminosity relation $\log{D_{\rm mom}^{\rm LMC}} = 1.39 (\pm 0.54) \log{(L_{\rm bol}/L_{\odot})} + 20.4 (\pm 3.0)$.
   }


   \maketitle

\section{Introduction}
Massive stars ($> 8M_{\odot}$) are hot and luminous stars that possess powerful winds that provide significant radiative, chemical and mechanical feedback to their 
surroundings at every evolutionary stage .
Due to significant mass-loss via stellar winds, the evolution of massive stars cannot be predicted solely by determining the initial mass, so the mass that is lost
throughout the life of a massive star could be the difference between its life ending in core-collapse supernova (ccSNe II/Ib/Ic) and leaving behind a black hole (BH)
or a neutron star (NS) or directly collapse to a BH without a ccSN \citep{smartt2009}. 

Massive stars are rare by absolute numbers, but their high temperatures, and subsequently, their extreme ultra-violet (UV) fluxes are thought to have played an essential role in re-ionizing the 
Universe \citep{1997ApJ...483...21H}. Their ionizing radiation may also drive star formation in their host galaxies \citep{crowther2019}

Massive stars eject mass during all evolutionary stages via stellar winds. In the advanced evolutionary stages (supergiants), their stellar winds become more powerful than on the zero age 
main sequence (ZAMS), leading to copious amounts of mechanical feedback to their surroundings \citep[for a general review on massive star feedback see e.g.][]{geen2023}. 

Due to efficient internal mixing processes, these stellar winds become enriched in elements synthesized in the interior layers of the star \citep{langer2012}. This process plays an 
important role in the chemical enrichment of the ISM, which in turn has a significant impact on the chemical evolution of the parent galaxy. The explosive nucleosynthesis in ccSNe 
yields elements heavier than iron, which also drives the chemical evolution and metallicity $Z$ of the host galaxy \citep{smith2014}.

Empirically derived stellar parameters accompanied by evolutionary \citep{yoon2006, brott2011} and population synthesis models \citep{leitherer1999, BPASS2018} can be used to peer into the 
collective evolutionary paths of massive stars. Thus, bridging the gap between the empirical and theoretically predicted properties of massive stars \citep{vink2001, krticka2021, bjorklund2023} 
is a fundamental pillar for an overall better understanding of the galactic evolution on a cosmic timeline. 

Blue supergiants are visually the brightest stars in external galaxies \citep{bresolin2001}. They have successfully been used as extragalactic distance indicators and diagnostics of heavy-metal 
metallicities \citep{kudritzki2003, urbaneja2005, przybilla2006, kudritzki2024}.

The principal motivation for the present study is to investigate the behaviour of the winds of late O and early B-supergiants  
in the LMC, more specifically in the temperature range associated with the ``bi-stability jump". The term ``bi-stability jump" describes a phenomenon of a steep ``jump" in stellar wind density
around $T_{\rm eff}\approx25{\rm -}21~{\rm kK}$, which was first coined by \citet{pauldrachandpuls1990} from a spectroscopic analysis of P Cygni, where two solutions were possible. The first 
solution has a high temperature (the ``hot" side of the jump) involved higher ionization levels which would produce a faster and relatively low density wind. The second
solution had a lower temperature (the ``cool" side of the jump), where the wind recombines to lower ionization stages, which results in a significant drop in the terminal velocity and a denser wind.
Later, \citet{lamers1995}  observed a jump in the ratio $\varv_{\infty}/\varv_{{\rm esc}}$  from $\approx2.6$ for supergiants 
earlier than B1 ($\approx25~{\rm kK}$) to $\varv_{\infty}/\varv_{{\rm esc}}\approx1.3$ for supergiants of types later than B1 \citep{kudritzkiandpuls2000}.
The reason for this jump is attributed to the recombination of $\ion{Fe}{IV}$ to $\ion{Fe}{III}$ as explained in \citet{vink2000}, since these lines dominate the acceleration in
the subsonic part of the wind \citep{vink1999}. $\ion{Fe}{III}$ has far more lines in the UV region close to the peak of the spectral energy distribution (SED) compared to $\ion{Fe}{IV}$. Thus, 
more momentum is transmitted to the material and a much higher mass loss is produced. 

\citet{vink2001} provides a numerically derived, mass-loss prescription, where the  temperature of the jump which divides the range into ``cool" and ``hot" depends on the Eddington parameter $\Gamma_{\rm e}$
and the metallicity. The introduction of the bi-stability jump could potentially increase the mass-loss rate by a factor of seven for the ``cool" solution compared to the ``hot" for stars that
are located roughly around the bi-stability jump.

On the other hand, newer mass-loss prescriptions using a different approach to calculate the radiative acceleration, such as \citet{Bjorklund2021} and \citet{krticka2021} do not predict such a 
steep increase in mass-loss rates. There have been multiple efforts to explore the behaviour of the wind of blue supergiants around the bi-stability jump in the
Milky Way \citep{Crowther2006, benaglia2007, deBurgos2024}, the LMC \citep{verhamme2024}, and the SMC \citep{bernini2024}. 

This investigation is facilitated by the advent of the Ultra-violet Legacy Library of Young Stars as Essential Standard \citep[ULLYSES,][]{roman-duval2025}, to which 1000 orbits of
{\it Hubble Space Telescope} (HST) were dedicated, making this the largest HST Director's Discretionary program ever conducted. ULLYSES compiled an 
ultra-violet (UV) spectroscopic Legacy Atlas of about 250 OB-stars in low-$Z$ regions, spanning the upper Hertzsprung-Russell diagram. The XShooting ULLYSES (XShootU) collaboration \citep{xshootU1} 
also compiled a complementary optical spectral library of the same stars using the medium resolution spectrograph X-shooter \citep{vernet2011} mounted on the Very Large Telescope (VLT). 
This complimentary dataset is referred to as Xshooting ULLYSES (XshootU). =The ULLYSES and XShootU observations were not conducted simultaneously. Consequently, the impact of the time-dependent 
nature of stellar winds on non-photospheric spectral lines cannot be explored. Such an analysis can only be conducted with extensive time-series of optical and UV spectra 
\citep[see e.g,][]{markova2005, massa2024}.

Although several studies investigate the properties of OB-supergiants in the Milky Way \citep[MW, e.g.,][]{herrero2002, repolust2004, deBurgos2023}, and the Magellanic Clouds 
\citep[MCs, e.g.,][]{crowther2002, bestenlehner2020, Brands2022}, using various analysis techniques, none of these studies had the unique ULLYSES/XshootU dataset. Unlike optical-only studies, 
the UV spectra of OB-supergiants provide a deeper insight into the properties of the wind, allowing the direct measurement of wind velocities and the 
breaking of the mass-loss rate-clumping degeneracy via saturated and unsaturated P Cygni profiles \citep{xshootU1}. This degeneracy has been extensively discussed in literature. Simply spoken, 
optical wind diagnostics such as H$\alpha$ or $\ion{He}{II}~\lambda4686$ (in the case of O-supergiants) tend to overestimate the mass-loss rates 
if wind clumping is neglected \citep{puls2008}. On the other hand, mass-loss rates that are estimated using only unsaturated P Cygni profiles tend to have very large 
uncertainties due to intrinsic wind-line variability \citep{massa2024} and are subject to degeneracies due to the presence of X-rays generated via shocks in the wind 
\citep{puls2008}.

In Section \ref{observations}, we present a detailed account of the observational data that was used in the study. In Section \ref{methods}, we give a detailed description 
of the methods and techniques that were used in the analysis. In Section \ref{results}, we present our results, including the values of the physical and wind parameters 
that were obtained through our pipeline. In Section \ref{discussion}, we compare our results to previous empirical studies and numerical predictions and 
we discuss the implications of our findings on the bi-stability jump and the effect of metallicity on wind parameters. In Section \ref{summary}, we summarize the 
interpretations of our results and discuss our plan for follow up studies.

\section{Observations}
\label{observations}
\begin{figure*}
    \centering
     \includegraphics[scale = 0.95]{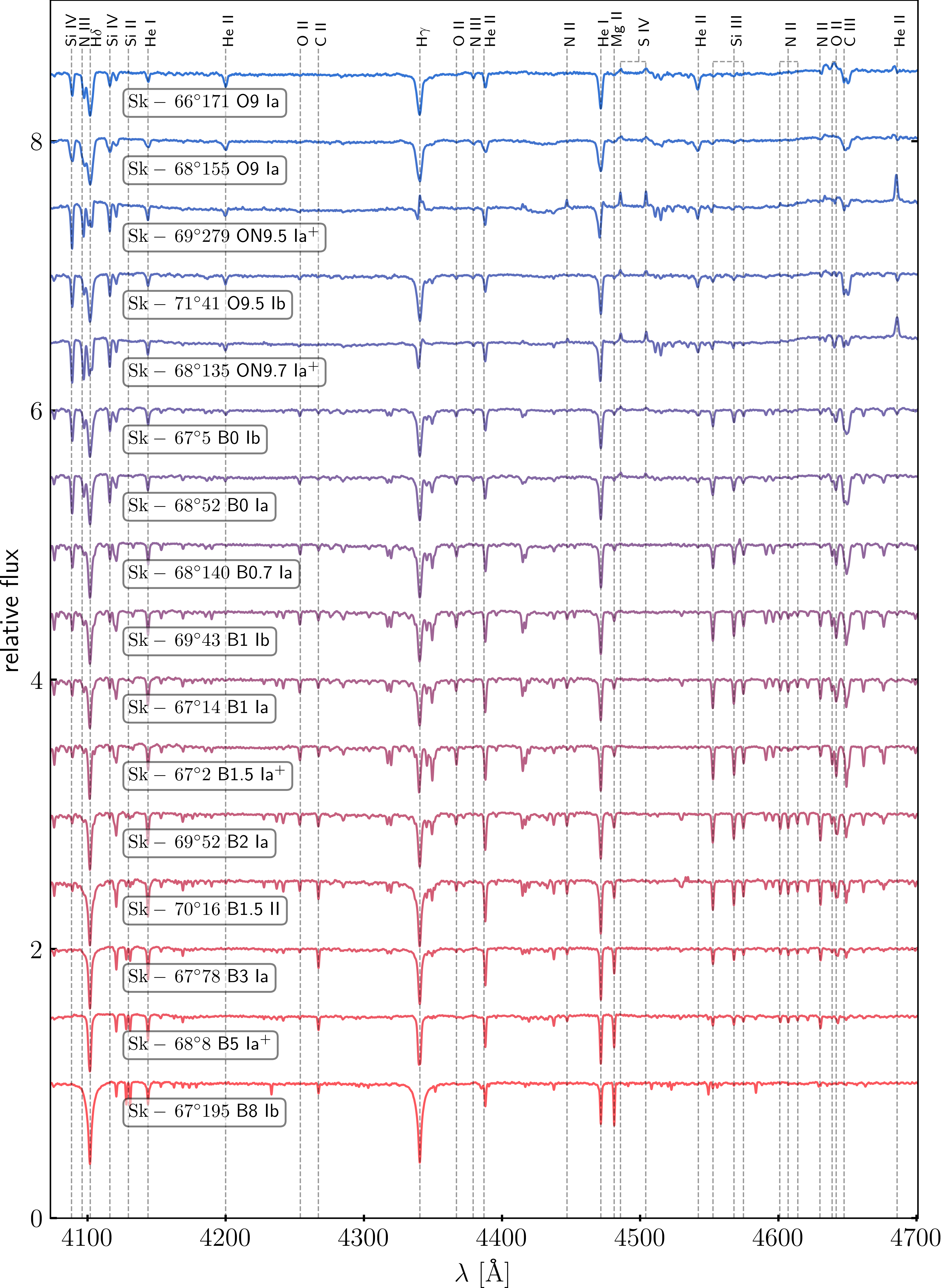}
         \caption{Normalised XshootU spectra \citep[DR1]{Sana2024} in the blue optical range of the sample analysed in this work with identification for a subset of the optical lines used in the analysis. 
         For illustration purposes an arbitrary offset of $0.5$ for each spectrum was added. Fits to the violet, yellow, and red spectra for all stars are included in Appendix~\ref{overall_app}.}
         \label{all_opt}
 \end{figure*}
\subsection{Sample}
\label{sample}
We initially chose a sample of LMC supergiants in the spectral type range O7-B9, the classification of which was obtained from various sources and compiled in \citet{xshootU1}. In Table~\ref{table:1},  we provide the updated
spectral type taken from \citet{Bestenlehner2025}, in which the classification was determined using O-star templates from \citet{sota2011} and B-star templates from \citet{negueruela2024}.
In the revised classification, the luminosity class of Sk\,$-$70$^{\circ}$~16 (Fig.~\ref{SK-70D16}) was changed from a low luminosity supergiant (Ib) to a bright giant (II).  
During this study, we had to exclude some objects due to signs of binarity or odd features in the morphology of optical wind-lines (H$\alpha$ and $\ion{He}{II}~\lambda4686$) that could hint toward a circumstellar disks, 
leaving the sample with objects in the spectral range O9-B8, covering a wide temperature range that includes the bi-stability jump (theoretically predicted to be around B1 spectral type). We briefly discuss the omitted objects in 
Appendix~\ref{Omitted}.

\subsection{UV data (ULLYSES)} 
The UV data used in this work is a subset of the {\it HST} ULLYSES sample \citep{roman-duval2025}, which obtained moderate-resolution spectra of OB-stars with selected 
wavelength settings of the Cosmic Origins Spectrograph \citep[COS, ][]{COS} G130M/1291, G160M/1611 and G185M/1953 with resolutions $R\approx12,000-16,000$, $R\approx13,000-20,000$, and $R\approx16,000-20,000$
respectively, and the Space Telescope Imaging Spectrograph \citep[STIS, ][]{STIS} E140M/1425 $(R\approx45,800)$, and E230M/1978 $(R\approx30,000)$ gratings in the far- and near-UV during HST cycles 27 - 29. 
Those new spectra were combined with suitable existing spectra (previously obtained with {\it FUSE} and/or {\it HST}). Far Ultra-violet Spectroscopic Explorer (FUSE) spectra covers the wavelength 
$\approx 900-1160 {\rm \AA}$ through the $4'' \times 20''$ (MDRS) or $30'' \times 30''$ (LWRS) appertures, with a resolving power of $\approx 15,000$ \citep{FUSE}. 

\subsection{Optical data (XShootU)} 
To complement the UV spectra, high quality optical/NIR spectroscopy was carried out with the X-shooter instrument \citep{vernet2011}, which is mounted on the 
Very Large Telescope (VLT). This slit-fed ($11 \arcsec$ slit length) 
spectrograph provides simultaneous coverage of the wavelength region between $3000 - 10200$ nm, across two arms; UVB ($300 \le \lambda \le 560 $ nm), VIS ($560 \le \lambda \le 1000$ nm). 
X-shooter’s wide wavelength coverage made it the instrument of choice to build an optical legacy data-set \citep{xshootU1}. The XshootU dataset 
was observed with the following settings: $0.8\arcsec$ slit width for the UBV arm achieving spectra resolution $R\approx6700$, and $0.7\arcsec$ for the VIS arm ($R \approx 11400$).
Fig.~\ref{all_opt} show the optical spectra used in this work, which were combined, flux calibrated, corrected for Telluric contamination and normalised by \citet[DR]{Sana2024}.

\subsection{Auxiliary data (MIKE)} 
Another complimentary spectroscopic optical dataset was collected using the Magellan Inamori Kyocera Echelle (MIKE) spectrograph which is mounted on the Magellan Clay Telescope, for known slow 
rotating ULLYSES stars. The higher spectral resolution $(R\approx 35000-40000)$ and the wide wavelength coverage ($3350-5000\AA$ blue arm) and ($4900-9500\AA$ red arm) is needed 
to resolve spectral features and determine their rotation rates using metal lines \citep{mike}. Four of the stars in our analysed sample are also included in the Magellan/MIKE sample 
Sk\,$-$67$^{\circ}$~78, Sk\,$-$70$^{\circ}$~16, Sk\,$-$68$^{\circ}$~8, and Sk\,$-$67$^{\circ}$~195.

    \begin{table*}
        \caption{List of the targets in our sample. The last three columns are the wavelengths (in $\AA$) of the UV ranges and the corresponding instruments
        that were used for observations. The literature spectral types are adopted from \citet{xshootU1}. The revised classifications are adopted from \citet{Bestenlehner2025}.}
        \label{table:1}    
        \centering            
        \small
        \begin{tabular}{c c c c c c c c}        
            \hline
            Sk\,$-$   &HDE        &SpT (revised) &SpT (literature) &$(900-1160)$ &$(1150-1700)$ &$(1700-2370)$ \\
            \hline 
            66$^{\circ}$~171 &269889   &O9 Ia         &O9 Ia        &FUSE &STIS E140M &STIS E230M   \\
            68$^{\circ}$~155 &-        &O9 Ia         &B0.5 I       &FUSE &COS  G130M+G160M &STIS E230M   \\
            69$^{\circ}$~279 &-        &ON9.5 Ia$^+$  &O9.2 Iaf     &FUSE &STIS E140M &STIS E230M   \\
	          71$^{\circ}$~41  &-        &O9.5 Ib	      &O9.7 Iab	    &FUSE &STIS E140M &-            \\
	          68$^{\circ}$~135 &-      	 &ON9.7 Ia$^+$  &ON9.7 Ia$^+$ &FUSE &STIS E140M &STIS E230M   \\
            67$^{\circ}$~5   &268605   &B0 Ib         &O9.7 Ib      &FUSE &STIS E140M &STIS E230M   \\
            68$^{\circ}$~52  &269050   &B0 Ia         &B0 Ia        &FUSE &STIS E140M &STIS E230M   \\
            69$^{\circ}$~43  &268809   &B1 Ib         &B0.5 Ia      &FUSE &STIS E140M &STIS E230M   \\
            68$^{\circ}$~140 &-        &B0.7 Ia       &B0.7 Iab     &FUSE &COS  G130M+G160M &STIS E230M   \\
            67$^{\circ}$~2   &270754   &B1.5 Ia$^+$   &B1 Ia$^+$    &FUSE &STIS E140M &STIS E230M   \\
            67$^{\circ}$~14, &268685   &B1 Ib         &B1.5 Ia      &FUSE &COS  G130M+G160M &STIS E230M   \\
            69$^{\circ}$~52, &268867   &B2 Ia         &B2 Ia        &FUSE &COS  G130M+G160M &STIS E230M   \\
            67$^{\circ}$~78, &269371   &B3 Ia         &B3 Ia        &FUSE &COS  G130M+G160M &STIS E230M   \\
            70$^{\circ}$~16  &-        &B1.5 II       &B4 I         &-    &COS  G130M+G160M &COS  G185M   \\
            68$^{\circ}$~8   &268729   &B5 Ia$^+$     &B5 Ia$^+$    &-    &COS  G130M+G160M &STIS E230M   \\
            67$^{\circ}$~195 &-        &B8 Ib         &B6 I         &-    &COS  G130M+G160M &STIS E230M   \\ 
            \noalign{\smallskip}
            \hline
        \end{tabular}
    \end{table*}
\subsection{Photometry}   
The photometric magnitudes utilised in the SED fitting to obtain the bolometric luminosities of the targets were taken from various sources and compiled in \citet{xshootU1}. The $U B V$ 
photometry are drawn from \citet{ardeberg1972, Schmidt-Kaler1999, massey2002}.

The infrared $J K_{S}$ photometry are preferably taken from the \textit{VISTA near-infrared $Y J K_{S}$ survey of the Magellanic System} \citep[VMC,][]{VMC}, with $H$-band photometry taken from
the \textit{Two Micron All Sky Survey} \citep[2MASS,][]{cutri2003, skrutskie2006}. For very bright sources that are saturated in VMC \citep{VMC} we use the $J$ and $K_{S}$ photometry provided in 2MASS 
\citep{skrutskie2006}.

\subsection{UV normalization}  
\begin{sidewaysfigure*}
    \centering
     \includegraphics[scale = 0.95]{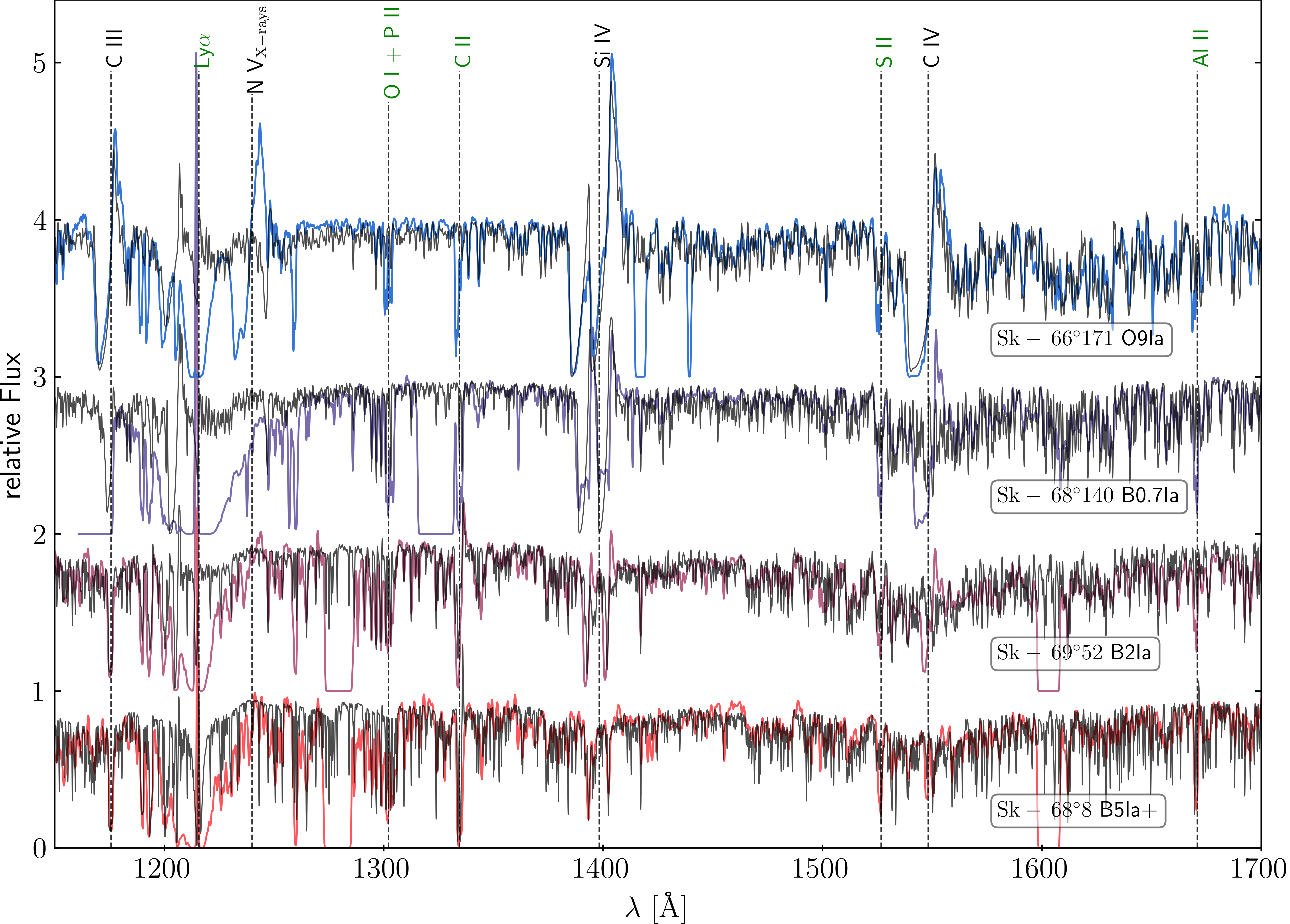}
         \caption{Normalised UV spectra in the range of the sample analysed in this work with line identification for the UV lines used in the analysis. 
         For illustration purposes an offset of $1$ was added. Coloured lines: observations. Black lines: model fits for each of the respective stars. 
         Stellar features are indicated in black line labels, whereas green line labels indicate interstellar features. The HST gap at $\approx 1300$ 
         in the spectra of Sk\, 69$^{\circ}$~52, Sk\, 68$^{\circ}$~8 is due to appending observations from COS G130M and COS G160M. The gap at $1600~\AA$ is due 
         to appending observations from COS G160M and COS G185M. Other UV fits are compiled in Appendix~\ref{overall_app}}
         \label{all_UV}
 \end{sidewaysfigure*}

The UV spectra require normalization, which is challenging due to the iron forest that heavily contaminates the continuum. Fig.~\ref{all_UV} shows normalised UV spectra of a subset of our sample.
The quality of the fits to the UV lines is highly dependent on the quality of the normalization of the observed spectrum. On average, we obtained very 
good normalised UV and far UV spectra for the targets by using the SED fitting explained later in Section \ref{luminosity} to obtain the bolometric luminosities of the 
targets. In essence, we apply the extinction from \citet{Gordon2003} to the normalised synthetic spectrum of the best-fitting model and scale it to the flux levels of the 
observed spectrum using the $K_{S}$ magnitude, then simply divide the observed spectrum by the extinct and scaled continuum of the model to obtain a normalised spectrum 
with relative flux values. 

   \begin{figure}
   \centering
	\includegraphics[width=\hsize]{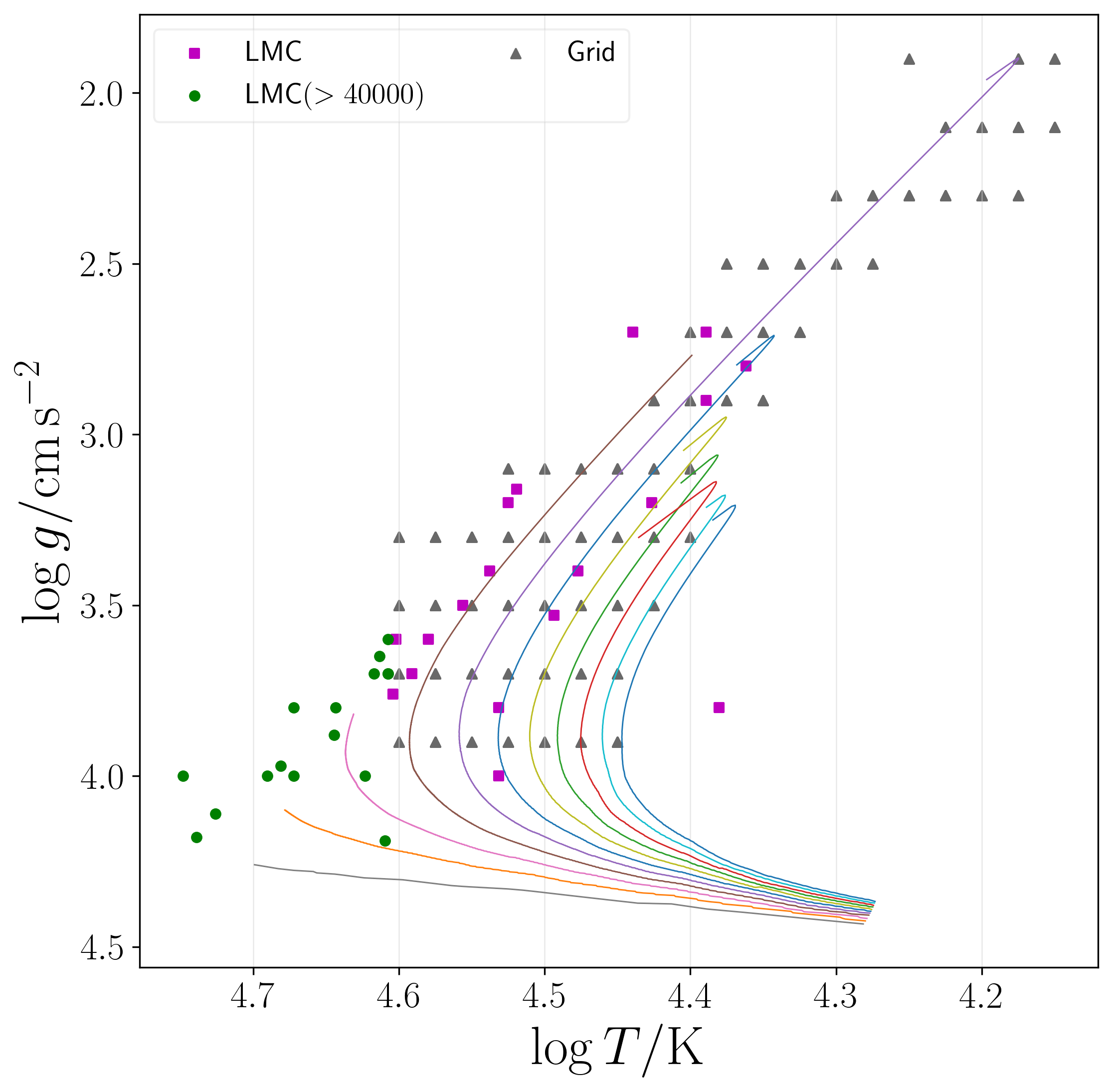}
        \caption{Grey triangles: model grid, purple squares: LMC targets with temperatures $\le 40~{\rm kK}$ (from literature), green points: 
        LMC targets with temperatures $\ge 40~{\rm kK}$. The overlaid lines are isochrones taken from \citet{brott2011}.}
        \label{t-g}
   \end{figure}

\section{Method}
\label{methods}
Line blanketed, plane parallel model atmospheres such as TLUSTY \citep[non-LTE, ][]{hubeny1995} and ATLAS \citep[LTE, ][]{kurucz1979} coupled to SURFACE/DETAIL \citep{butler1985newsletter} 
have been widely applied to early-type stars with weak winds, to derive physical parameters and accurate elemental abundances \citep[e.g., ][]{przybilla2006, hunter2007, przybilla2008}. 
This combined non-LTE method has been also been recently employed in studies of B-supergiants \citep[e.g., ][]{wessmayer2022, wessmayer2023}.
For stars with strong stellar winds, it is necessary to employ line blanketed model atmospheres with spherical geometry. Examples include \textsc{FASTWIND} \citep{Puls2005, rivero2012}, 
\textsc{PoWR} \citep{grafener2002, hamann2003} and \textsc{CMFGEN} \citep{Hillier1990, hillier1998}. These codes have the advantage of incorporating stellar winds, albeit at the expense of computational 
resources, such that the determination of physical parameters and elemental abundances are more costly than the plane parallel case.

\subsection{Model atmosphere code and grid}
In this work, we use \textsc{CMFGEN} \citep{Hillier1990, hillier1998} which solves the radiative transfer equations in a 1-D non-LTE, spherical geometry with a radial outflow of material 
(that can also be optically thick in the continuum) in the co-moving frame. We choose to utilise \textsc{CMFGEN} because it accounts for the influence of extreme UV line blanketing on the wind
populations and ionization structure, due to thousands of overlapping lines. It does that with a detailed treatment using ``super-levels" which was pioneered by \citet{anderson1985, anderson1989},
in which several levels with similar energies and properties are treated as a single or super level. The idea of super-levels is of tremendous importance for the iron-group elements.
An individual ionization stage can have hundreds of levels and line transitions (cf. the iron forest in Fig.~\ref{all_UV}) that need to be considered in the full model atom, but applying the super-levels method reduces significantly the number of statistical 
equilibrium equations to be considered in the NLTE treatment. There is also a time-dependent variant of \textsc{CMFGEN} for the simulation of supernovae spectra \citep[e.g.,][]{Dessart2010}, but the scheme we employ in this work assumes stationary outflows.

For the initial setup of the (quasi-)hydrostatic layers, a previous \textsc{CMFGEN} model can be used. \textsc{CMFGEN} does not calculate the velocity field stratification with the radiative 
acceleration. Instead, the density (velocity) structure of the extended atmosphere is set through the continuity equation by a parametrized velocity law $\varv(r)$ on top of a solution of the 
hydrostatic equation from a connection velocity to the inner photosphere \citep[e.g.,][]{Martins2012}. 

In \textsc{CMFGEN}, clumping is treated in the ``optically-thin clumping" approximation, also known as “micro-clumping”, assuming a void inter-clump medium and a volume filling factor 
$f_{\rm vol}$. To solve the radiative transfer equation with the micro-clumping assumption, the size of the clumps is assumed to be smaller than the mean free path of the photons \citep{hillier1996}.
The treatment of wind inhomogeneities in \textsc{CMFGEN} has been extensively discussed in the literature \citep{hillier1997, hillier1998, HillierandMiller1999}. In the grid, we adopt 
a a velocity-dependent clumping law that was introduced for O-stars in \citet{Hillier2003}:
\begin{equation}
  \label{eq:clumping} 
    f(r) = f_{\rm vol,\infty} + (1-f_{\rm vol,\infty})\exp{(-\frac{\varv (r)}{\varv_{\rm cl}})},
\end{equation}
where $f_{\rm vol,\infty}$ is the terminal volume-filling factor, $\varv_{\rm cl}$ is the onset clumping velocity, and $\varv(r)$ is the velocity of the wind at a given radius $r$. In its essence, 
$f_{\rm vol,\infty}$ determines the degree of clumping in the wind, where smaller values of $f_{\rm vol,\infty}$ indicate a more highly clumped wind at $r \to \infty$. $\varv_{\rm cl}$ 
determines the location (depth) at which clumping starts. This means that Equation~\ref{eq:clumping}, describes winds that become more smooth as $\varv(r)$ becomes lower approaching the photosphere, 
and become rapidly more clumped at larger radii. In our grid, we fix $f_{\rm vol,\infty}$ to a value of $0.1$. Later in the fine-tuning procedure, $f_{\rm vol,\infty}$ is treated as a free 
parameter. This is essential for obtaining satisfactory fits for the electron scattering wings of H$\alpha$ and unsaturated P Cygni resonance lines.

Fig.~\ref{t-g} shows a slice of the grid in $T_{\rm eff}$-$\log{g}$ space. We fixed the luminosity in our models to $\log{(L/ L_{\odot})} = 5.8$ and varied $\log{(T_{\rm eff}/{\rm K})}$ in the 
range $\approx\left[4.15, 4.60\right]$ and $\log{(g/{\rm cm\,s^{-2}})}$ in the range $\approx \left[1.7, 3.9\right]$ depending on the temperature, in steps of $0.025$~dex for 
$\log{(T_{\rm eff}/{\rm K})}$ and $0.2$~dex for $\log{(g/{\rm cm\,s^{-2}})}$. We employed the empirical temperature and metallicity dependent terminal wind velocity recipe from \citet{xshootu3}: 
\begin{equation}
  \label{eq:vinf_callum} 
    \varv_{\infty}~({\rm km\,s^{-1}}) = \left[0.092(\pm 0.003)T_{\rm eff}~({\rm K})-1040(\pm 100)\right]Z/Z_{\odot}^{(0.22\pm 0.03)},
\end{equation}
where $Z \approx 0.43 Z_{\odot}$ for the LMC \citep{choudhury2016}. In \citet{xshootu3}, Sobolev with Exact Integration (SEI) modelling was employed to measure the terminal wind velocities for objects 
no later than B1.5 spectral types, using the $\ion{C}{iv}~\lambda\lambda1548-1551$ resonance doublet, with radial velocities adopted from the UV.

In our grid, we utilised a modified $\beta$ velocity law that was introduced in \citet{Hillier2003}:
\begin{equation}
  \label{eq:vel_law} 
    \varv(r) = \varv_{0}+(\varv_{\infty}-\varv_{0})\left(1-\frac{R_{\ast}}{r}\right)^{\beta},
\end{equation}
where $\varv_{\infty}$ is the terminal wind velocity, $\varv_{0}$ is the connection velocity, which is estimated as two thirds the speed of sound $\approx10~{\rm km\,s^{-1}}$, and $R_{\tau = 100}$ is 
the radius of the star, which is defined at optical depth $\tau = 100$. For our grid we adopt adopt a fixed value of $\beta = 1$. This procedure yielded a 2-D grid in the temperature-gravity 
parameter space (grid$_{base}$).

\begin{table}
  \caption{Metal abundances ($\epsilon_{\rm X} = \log{{\rm X}/{\rm H}}+12$ by number) adopted in the model grid. Baseline LMC abundances $\epsilon_{\rm X}^{\rm LMC}$ are drawn from the compilation 
  of \citet{xshootU1}. $\Delta \epsilon_{\rm X}$ represents the difference between the LMC baseline values and the values adopted in our grid.} 
  \label{table:abundances}     
  \centering                                     
  \small                                
  \begin{tabular}{c c c c}          
      \hline\hline{\smallskip}
      Element     &$\epsilon_{\rm X}^{\rm LMC}$  &$\Delta \epsilon_{\rm X}$    &Reference\\    
      \hline      
      C           &$7.03$                        &$-0.52$                      &\citep{hunter2008}\\              
      N           &$8.01$                        &$+0.47$                      &\citep{hunter2008}\\      
      O           &$8.40$                        &$-0.07$                      &\citep{hunter2008}\\      
      Ne          &$7.70$                        &-                            &-\\        
      Mg          &$7.13$                        &-                            &-\\      
      Al          &$6.20$                        &-                            &-\\      
      Si          &$7.06$                        &-                            &-\\      
      P           &$5.11$                        &-                            &-\\    
      S           &$6.88$                        &-                            &-\\    
      Ca          &$6.10$                        &-                            &-\\      
      Fe          &$7.23$                        &-                            &-\\      
      Ni          &$5.90$                        &-                            &-\\      
     \noalign{\smallskip}
     \hline
  \end{tabular}
\end{table}
In Table~\ref{table:abundances}, we present the values of metal abundances $\log{X/H}+12$ (by number) adopted in the model grid. Henceforth, we substitute $\log{{\rm X}/{\rm H}}+12$ with 
the notation $\epsilon_{\rm X}$. Rather than adopting baseline LMC CNO abundance values for metals such as the one compiled in \citet{xshootU1}, we chose to adopt processed abundances 
correlating to low luminosity B-stars in the LMC from \citet{hunter2008}, that show nitrogen enhancement of $\Delta \epsilon_{\rm N}\approx+0.47$~dex, compared to the LMC baseline 
in \citet{xshootU1}, and at the expense of carbon and oxygen deficiencies of $\Delta \epsilon_{\rm C}\approx-0.5$~dex and $\Delta \epsilon_{\rm O}\approx-0.07$~dex. 
Since \citet{hunter2008} excludes B-supergiants, we checked the validity of these values for luminous B-supergiants (which might be expected to show the greatest N enhancements) by resorting to 
the work of \citet{McEvoy2015}, in which \textsc{TLUSTY} non-LTE model atmosphere calculations have been used to determine atmospheric parameters and nitrogen abundances for 34 single and 18 
binary supergiants. Their analysis shows a nitrogen enrichment value of $\approx 0.5$~dex relative to \citet{xshootU1}, which is very close to our grid's nitrogen abundance. For other key 
elements in our grid like silicon, magnesium, and iron, we implemented baseline LMC abundances from \citet{xshootU1}. Abundances of metals heavier than oxygen are not modified during the 
fitting procedure. 

Lastly, we covered the temperature-gravity-wind density parameter space. We did that by iterating over the mass-loss rates $\log{(\dot{M}/M_{\odot}\,{\rm yr}^{-1})}$ in the range 
$\approx\left[-5.5, -7.3\right]$ in steps of $0.3$~dex for each of the points from grid$_{base}$. The final grid is a 3-D grid in the temperature-gravity-wind density space. 

Since the strength of emission features scales not only with the mass-loss rate but also with the volume-filling factor ($f_{\rm vol}$ ), terminal velocity, and radius of the star, it is convenient 
to compress these parameters into one parameter when compartmentalizing the model grids. To spectroscopically quantify mass-loss rates of hot massive stars using scaling relations, we chose to 
adopt the transformed radius \citep{schmutz1989} originally used for optically thick winds of Wolf-Rayet stars (WR), where the line equivalent width is preserved:
\begin{equation}
  \label{eq:R_t} 
    R_t = R_{\ast}\left[\frac{v_{\infty}}{2500~{\rm km\,s^{-1}}}\Big/\frac{\dot{M}}{10^{-4}~M_{\odot}{\rm yr}^{-1} \sqrt{f_{\rm vol}}}\right] ^{2/3},
\end{equation}
The grid covers a wide range of $\log{R_t}$ from $1.6$ to $3.1$, with lower values relating to denser winds. This $\log{R_t}$ range corresponds to an optical depth-invariant 
wind-strength parameter $\log{Q} = \log{\dot{M}/(R_{\ast} \varv_{\infty})^{3/2}}$ \citep{puls1996} range of $-13.9$ to $-11.5$, where larger values of $\log{Q}$ correspond to denser winds. We employ $R_t$ to scale
$\dot{M}$ to the derived bolometric luminosity of the star.

\subsection{Atomic data and X-rays} 
In our models we include 14 species and 50 different ions. We exclude higher ionization stages for the cooler ($<25~{\rm kK}$) models  and include the lower ionization stages. 
This is of special importance to the iron lines which dominate the UV in the B-star regime. We include our detailed underlying model atom structure in the appendix in Table~\ref{table_atomic}. 

X-rays can be included in \textsc{CMFGEN} using X-ray emissivities from collisional plasma models \citep{smith2001}, where the source of this emission is assumed to be shocks forming in the winds
\citep{pauldrach1994}. The detailed approach used in \textsc{CMFGEN} is described in \citet{hillier1998}. The general spectral appearance is not affected by X-rays, 
although the highest ionization UV lines can be enhanced at the expense of lower ionization lines \citep{baum1992} at relatively low stellar temperatures due to Auger processes.  
We elect to exclude X-rays from our analysis, because the ad hoc addition of X-rays in \textsc{CMFGEN} does not provide phenomenological description of the physical shock parameters and 
reduces the number of unconstrained parameters in the fitting procedure. 

The main drawbacks of excluding X-rays in our models is that, in some cases, it is difficult to obtain satisfactory fits for high ionization UV P Cygni lines. This could potentially lead to 
over estimating the mass-loss rates. \citet{bernini2023}, who included X-rays in their CMFGEN models, found that their mass-loss rates for Galactic B-supergiants are lower by a factor of two 
compared to the mass-loss rates obtained by \citet{Crowther2006} and \citet{searle2008}, who exclude X-rays. In Fig.~\ref{all_UV}, the discrepancy between the predicted and observed 
P Cygni $\ion{N}{V}~\lambda\lambda 1238-1242$ line in Sk\,$-$66$^{\circ}$~171 is due to the lack of X-rays in the model. This is also the case for the P Cygni 
$\ion{C}{iv}~\lambda\lambda1548-1551$ profile in Sk\,$-$68$^{\circ}$~140 and Sk\,$-$69$^{\circ}$~52.

\subsection{Summary of the fitting procedure}
Since \textsc{CMFGEN} is time and resource-intensive, we are limited by a small number of models relative to the number of parameters that we have to extract from the stellar spectra. 
Therefore, as a first-order approximation (initial pinpoint), we use the results of the analysis done using the pipeline that was introduced in \citet{bestenlehner2022}, which is a grid-based
$\chi^2$- minimization algorithm, that utilises the entire optical spectrum rather than selected diagnostic lines allowing a wider range of temperature from B to early O-stars to be analysed. This 
grid of synthetic spectra was computed with the non-LTE stellar atmosphere and radiative transfer code \textsc{FASTWIND} \citep{Puls2005}, which is efficient and quick but lacks a 
detailed treatment of the iron forest in the UV range. This pipeline provides a first approximation of $T_{\rm eff}$, $\log{g}$, $\varv_{\rm rot}\sin{i}$ using only the optical XShootU data, 
which we then can refine using the UV range provided by our \textsc{CMFGEN} grid, and then produce a fine-tuned model for the entire spectrum based on the best fitting grid model. 
To summarize, our entire procedure consists of the following steps:
\begin{itemize}
    \item 1$^{\rm st}$ step: Approximation of $T_{\rm eff}$ and $\log{g}$ from the optical spectrum using the model de-idealization pipeline introduced in \citet{bestenlehner2022}, 
          which we use to pin-point the closest fitting model from the grid in $T_{\rm eff}$-$\log{g}$-$\dot{M}$ parameter space.
    \item 2$^{\rm nd}$ step: Fine-tune the values of $T_{\rm eff}$, $\log{g}$, $\varv_{\infty}$ and the helium abundance of the model from the previous step. 
    \item 3$^{\rm rd}$ step: Fine-tune wind parameters $\dot{M}$, $\beta$-law, $f_{\rm vol}$ and clumping on-set velocity $\varv_{\rm cl}$.
    \item 4$^{\rm th}$ step: Refine CNO- surface abundances.
\end{itemize}
The second, third and fourth steps take on average ten tailored models in total to obtain a satisfactory fit.

\subsection{Diagnostics}
\label{diag_intro}
Underlying systematic errors arise from our grid and fitting procedure. The most notable are the inclusion of a limited number of species and ions, fixing certain parameters like the micro- and 
macro-turbulent velocities, and fixing the abundances of elements heavier than oxygen. An important part of our analysis is wind clumping, which can vary greatly depending on the treatment 
utilised in the model \citep[e.g.,][]{Brands2022}. Finally, the finite spectral resolution of the observation adds another layer of uncertainty to the overall analysis. We estimate the model 
uncertainties (parameters with superscript notation 'm') the derived physical parameters and wind properties. By 'model uncertainty' we simply mean the smallest variation in a given input 
parameter that would produce a noticeable change in the quality of the fit. 

\subsubsection{Effective temperature and helium abundance}
After we derived a rough estimate of the stellar and wind parameters from our grid, we start fine-tuning the model parameters to reproduce the observed spectrum. In order to do that, we first vary the temperature, 
which has a large impact on the morphology of the spectrum for different spectral classes. Consequently, obtaining an accurate temperature determines the overall quality of the fit 
and of the other estimated stellar and wind parameters. In our fitting procedure, $T_{\rm eff}$ is derived using the ionization balance of helium (He) and silicon (Si). In practice, lines from 
successive ions, of the same elements must be observed. 

The most reliable lines for O-stars are $\ion{He}{ii}$ and $\ion{He}{i}$, and historically $\ion{He}{ii}~\lambda4542$ and $\ion{He}{i}~\lambda4471$ have been used \citep{martins2011}, which is 
what we used for the O-stars in our sample as primary diagnostics. Additionally we use $\ion{He}{II}~\lambda 5411$ and $\ion{He}{I}~\lambda4922$ as a sanity check\\

\begin{figure}
  \centering
   \includegraphics[width=\hsize]{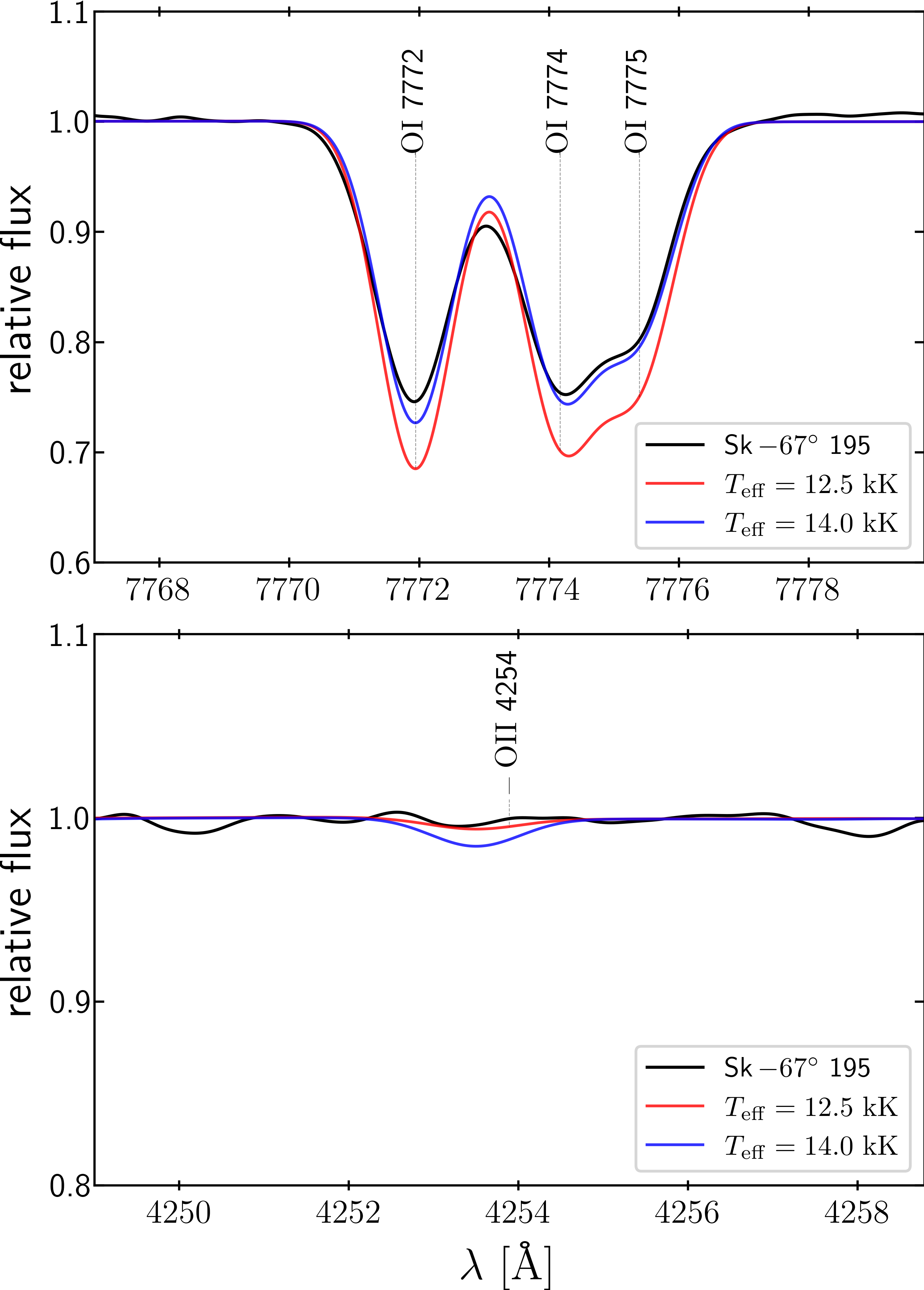}
       \caption{$\ion{O}{I}~\lambda\lambda\lambda 7772-7774-7775$ and $\ion{O}{II}~\lambda 4254$ lines of the B8 supergiant Sk\,$-$67$^{\circ}$ 195 (black line). Red line: the preferred model 
        the temperature of which ($T_{\rm eff} = 12.5$~kK) was obtained from fitting $\ion{He}{i}~\lambda4471$ and $\ion{Mg}{ii}~\lambda4481$, and a comparison with higher $T_{\rm eff}= 14.0$~kK.}
       \label{OI}
  \end{figure}

For B-stars the main diagnostic is the silicon ionization balance \citep{mcerlean1999, trundle2004I, trundle2004II}. For the earliest B-stars (B0 - B2) spanning a temperature range 
$T_{\rm eff}\approx28{\rm -}21~{\rm kK}$ we use $\ion{Si}{iv}~\lambda\lambda 4089-4116$ \footnote{The $\ion{Si}{iv}$ absorption line at $\lambda = 4089~{\rm \AA}$ is actually a blend of 
$\ion{Si}{iv}$ and $\ion{O}{II}$ \citep{deBurgos2024b}. Considering models with extended $\ion{O}{II}$ atomic levels, we find that the $\ion{O}{II}$ component contributes 
$\approx 9\%,\,~10\%,\,~62\%$ to the equivalent width of the combined line at $T_{\rm eff}=30,\,~25,\,~20$~kK, respectively, with both lines being very weak in the low $T_{\rm eff}$ case} and 
$\ion{Si}{iii}~\lambda\lambda\lambda 4553-4568-4575$ as our main diagnostics. For mid B-stars (B3-B5) in the temperature range 
$T_{\rm eff}\approx20{\rm -}14~{\rm kK}$ $\ion{Si}{iii}~\lambda\lambda\lambda 4553-4568-4575$ and $\ion{Si}{ii}~\lambda\lambda 4128-4131$ are used.
For late B-stars (B6-B8) we resort to fitting $\ion{He}{i}~\lambda4471$ and $\ion{Mg}{ii}~\lambda4481$, which is not as reliable as fitting consecutive ions of the same species.

An alternative indicator of $T_{\rm eff}$ in late B-stars is the ratio of $\ion{O}{II}$ to $\ion{O}{I}$ line. By way of example, in Fig.~\ref{OI} we show the triplet 
$\ion{O}{I}~\lambda\lambda\lambda 7772-7774-7775$ (upper panel) and $\ion{O}{II}~\lambda 4254$ (lower panel). The preferred model (red solid line) for the B8 supergiant Sk\,$-$67$^{\circ}$ 195, 
the temperature of which was obtained from fitting $\ion{He}{i}~\lambda4471$ and $\ion{Mg}{ii}~\lambda4481$, reproduces the $\ion{O}{I}~\lambda\lambda\lambda 7772-7774-7775$ and 
$\ion{O}{II}~\lambda 4254$ lines relatively well. For comparison, we display another model that differs from the preferred model only in its temperature with $\Delta T_{\rm eff} = 1500$~K 
(blue solid line). We select this model as it is the nearest model on the grid to our preferred model in the temperature parameter-space that presents $\ion{O}{II}$ lines. This fit implies that 
decreasing the oxygen abundance in the comparison model could produce a better match to both lines. This could be used either as a primary diagnostic for $T_{\rm eff}$ or as a sanity check in late B-supergiants.  

By way of a sanity check, we have compared the predicted Balmer jump strengths to observations \citep{kudritzki2008, urbaneja2017}. We find that, for the most part, the strength of the predicted 
Balmer jumps -- the $T_{\rm eff}$ of which was obtained from line diagnostics -- are in good agreement with observations. Nevertheless, a few hundred K higher temperatures produce a better match 
for a subset of mid to late B-supergiants, albeit within the quoted uncertainties.

The earliest B-stars (B0-B0.7) in the temperature range $T_{\rm eff}\approx25{\rm -}22~{\rm kK}$ show weak $\ion{He}{ii}$ lines, so we use $\ion{He}{i}~\lambda 4471$ and $\ion{He}{ii}~\lambda 4542$ 
as a sanity check when fine-tuning the temperature for such objects. There is also a temperature range $T_{\rm eff}\approx20{\rm -}17~{\rm kK}$ at which silicon lines of all three ionization stages 
($\ion{Si}{IV}$, $\ion{Si}{III}$, and $\ion{Si}{II}$) are present. In those cases, we attempted to fit all lines but focused mainly on the stronger consecutive pair. 

Since silicon is an alpha-process element, its abundance is primarily determined by the environment, so is not heavily influenced by the evolution of the star in the supergiant phase. Also, 
\citet{korn2005} obtains a value of $\epsilon_{\rm Si} = 7.07\pm 0.3$ from fast rotating B-stars in the LMC, and \citet{hunter2007} finds a silicon abundance $\epsilon_{\rm Si} = 7.19\pm 0.07$ 
from narrow-lined late O and B-stars in the LMC, whereas \citet{dopita2019} derives a value $\epsilon_{\rm Si}= 7.11\pm 0.04$ from supernova remnants in the LMC. The low variance between those 
values is what led us to fixing the silicon abundance. For those reasons, we did not attempt to fit the silicon lines' strength by changing the abundance of silicon at the fine tuning stage. We 
also do not change the abundance of magnesium during our fine tuning stage. Similar to silicon, this is motivated by the low variance between the magnesium abundance values obtained via different 
methods in the literature. $\epsilon_{\rm Mg}$ values from the literature are as follows: $\epsilon_{\rm Mg}= 7.15\pm 0.3$ from \citet{korn2005}, and $\epsilon_{\rm Mg}= 7.06\pm 0.09$ from 
\citet{hunter2007}, and $\epsilon_{\rm Mg}= 7.19\pm 0.09$ from \citet{dopita2019}

For O-stars, the helium abundance is essential for an accurate temperature estimate. Changing the helium abundance at the expense of hydrogen modifies the 
ionization structure in the atmosphere which, in turn, affects the shape and strength of H$\alpha$ especially when it is in emission. This is the case for the majority of 
the targets in our sample. Hence it is important to also obtain an accurate helium abundance for both O-stars and B-stars. In our fitting procedure, as we are fitting
temperature line diagnostics, we simultaneously fine-tune abundances of helium and hydrogen before attempting to fit wind lines such as H$\alpha$. 

\begin{figure}
    \centering
      \includegraphics[width= \columnwidth]{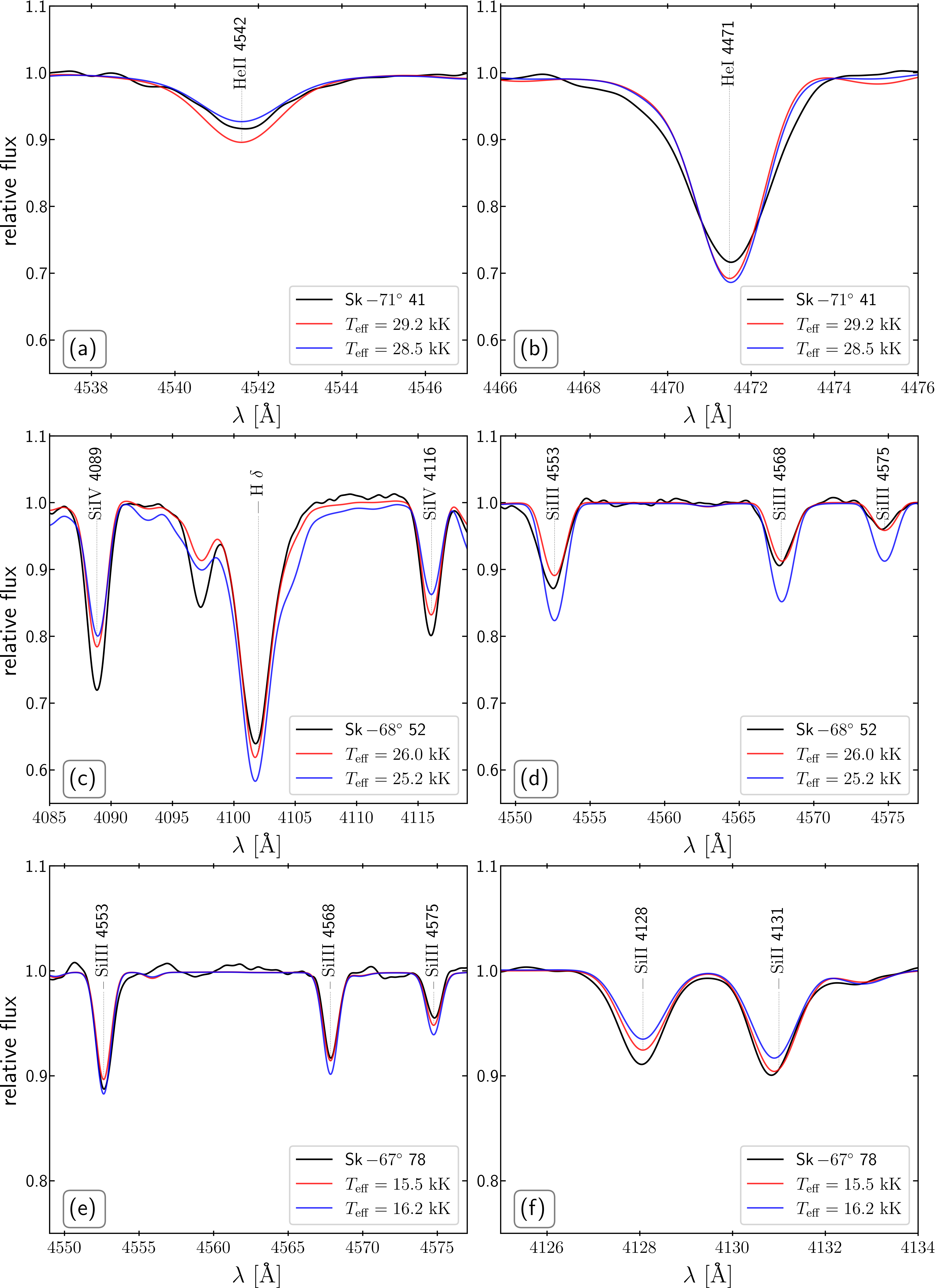}
    \caption{Examples of the effect of varying $T_{\rm eff}$ on the quality of the fit. Best fitting model: red solid line. Comparison model: blue solid line ($\Delta T_{\rm eff} = 700~{\rm kK}$). 
    Observed XShootU spectrum: black solid line. a-b: Sk\,$-$71$^{\circ}$~41 (O9.5 Ib). c-d: Sk\,$-$68$^{\circ}$~52 (B0 Ia). e-f: Sk\,$-$67$^{\circ}$~78 (B3 Ia).}
    \label{temp_sense}
\end{figure}

It is important to understand that varying $T_{\rm eff}$ by small amounts can have a different effect on the quality of the fit depending on the temperature range of the model. To illustrate this, 
we show in Fig.~\ref{temp_sense} the effect of varying the temperature of the best fitting model (red solid line) by $\Delta T_{\rm eff} = -700$~K with all other parameters fixed. The top panels 
shows the fits to $\ion{He}{ii}~\lambda 4542$ and $\ion{He}{i}~\lambda 4471$ (panels $a$ and $b$ respectively in Fig.~\ref{temp_sense}) for the O9.5 Ib supergiant Sk\,$-$71$^{\circ}$~41 
(Fig.~\ref{SK-71D41}), where the best fitting model has a $T_{\rm eff} = 29.2~{\rm kK}$ and we can see that while the model with lower temperature does not yield a noticeably different fit for 
$\ion{He}{i}~\lambda 4471$, $\ion{He}{ii}~\lambda 4542$ is significantly changed. The equivalent widths (EW) ratio of $\ion{He}{i}~\lambda 4471$ to $\ion{He}{ii}~\lambda 4542$ in the observations 
is $\approx 3.20\pm0.05$. our preferred model yield an EW ratio of $\approx 3.15\pm0.05$ compared to an EW ratio $\approx 3.45\pm0.05$ for the comparison model with the higher $T_{\rm eff}$.

Panels $c$ and $d$ in Fig.~\ref{temp_sense} show the comparison of the fits of $\ion{Si}{iv}~\lambda\lambda 4088-4116$ and $\ion{Si}{iii}~\lambda\lambda\lambda 4553-4568-4575$, respectively,  
for the B0 Ia supergiant Sk\,$-$68$^{\circ}$~52 (best fitting model $T_{\rm eff} = 26.0~{\rm kK}$). In this case lowering the temperature by $\Delta T_{\rm eff} = 700~{\rm K}$ slightly 
worsens the quality of the fits for both sets of lines. The EW ratio of $\ion{Si}{iii}~\lambda 4552$ to $\ion{Si}{iv}~\lambda 4088$ in the observed spectrum of Sk\,$-$68$^{\circ}$~52 
is $\approx 0.55\pm0.05$. Our preferred model yields an EW ratio of $\approx 0.51\pm0.05$ compared to an EW ratio of $0.45\pm0.05$ for the comparison model with the lower $T_{\rm eff}$.

In panels $e$ and $f$ of Fig.~\ref{temp_sense}, we show the fits of the B3 Ia supergiant Sk\,$-$67$^{\circ}$~78 for $\ion{Si}{iii}~\lambda\lambda\lambda 4553-4568-4575$  and 
$\ion{Si}{ii}~\lambda\lambda 4128-4131$, respectively, where the best fitting model has $T_{\rm eff} = 15.5~{\rm kK}$. We can see that raising the temperature of the model by 
$\Delta T_{\rm eff} = 700~{\rm K}$ noticeably deteriorates the quality of the fit. The EW ratio of $\ion{Si}{ii}~\lambda 4128$ to $\ion{Si}{iv}~\lambda 4552$ in the observed spectrum of 
Sk\,$-$67$^{\circ}$~78 is $\approx 0.60\pm 0.05$. Our preferred model yields an EW ratio of $\approx 0.62\pm0.05$ compared to an EW ratio $0.53\pm0.05$ for the comparison model with the 
higher $T_{\rm eff}$.

We estimate the model uncertainty from the spectral fitting as $\Delta T_{\rm eff}^{m}\approx\pm500{\rm -}1000~{\rm K}$ depending on the quality of the fit, and we estimate a more conservative and 
realistic uncertainty as double the model uncertainty in attempt to take into account limitations of the model atmosphere code, giving us $\Delta T_{\rm eff}\approx\pm1000{\rm -}2000~{\rm K}$.

\subsubsection{Effective surface gravity} 
The surface gravity ($\log{g}$) is obtained via fitting the wings of Balmer lines. Balmer lines are broadened by collisional processes and are most 
sensitive to Stark broadening (linear Stark broadening affects the wings of the H and, to a lesser extent, $\ion{He}{ii}$ lines; indeed, this effect is mainly used to constrain 
the surface gravity), which is related to gas pressure, and in turn, gas pressure is intimately connected to electron pressure ($P_{\rm e} = N_{\rm e}KT$) in hot stars.
This means that Balmer lines are broader in higher gravity stars. Degeneracies that occur when fitting Balmer lines are alleviated once temperature diagnostic lines are included (fitting $\log{g}$
and $T_{\rm eff}$ simultaneously and iteratively).

We use H$\gamma$ as the main indicator since it is usually in absorption and its wings are mostly not contaminated by wind emission or by blending with other lines. H$\eta$ and H$\zeta$ are 
used as a sanity check. These Balmer lines are usually strong and well resolved  \citep{diaz2020, martins2011}, but in the case of the O hypergiants Sk\,$-$68$^{\circ}$~135 and Sk\,$-$69$^{\circ}$~279 
we resorted to fitting H$\eta$ as our primary surface gravity diagnostic due to wind contamination in all the lower Balmer lines.

\begin{figure}
    \centering{
      \includegraphics[width = \columnwidth]{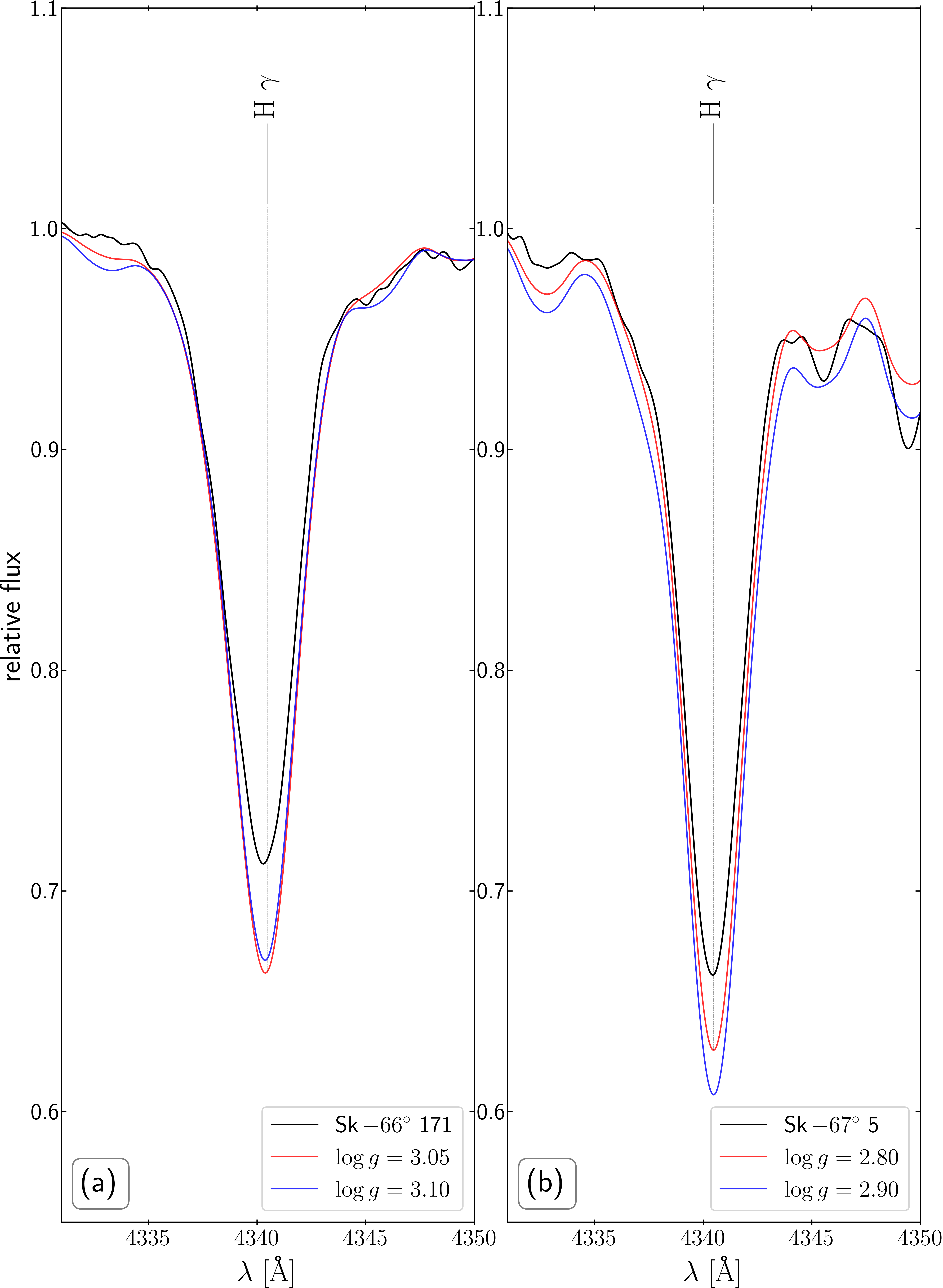}
    }
    \caption{Examples of the effect of varying $\log{g}$ on the quality of the fit. Best fitting model: red solid line. Comparison model: blue solid line. Observed XShootU spectrum: black solid 
    line. a: Sk\,$-$66$^{\circ}$~171 (O9 Ia) ($\Delta\log{g} = 0.05$). b: Sk\,$-$67$^{\circ}$~5 (B0 Ib) ($\Delta\log{g} = 0.1$).}
    \label{grav_sense}
\end{figure}

To show the level of sensitivity of the model to modest changes in $\log{g}$, we present two cases of fitting H$\gamma$. For the O9 Ia supergiant Sk\,$-$66$^{\circ}$~171
(panel (a) in Fig.~\ref{grav_sense}), the best fitting model (red solid line) was computed with $\log{(g/{\rm cm\,s^{-2}})}=3.05$~dex and the comparison model (blue solid line) with 
$\log{(g/{\rm cm\,s^{-2}})}=3.10$~dex. Both models were computed with the same $T_{\rm eff}= 29.9~{\rm kK}$ and with all other parameters kept the same, and as shown the two fits are very similar.
The second case (panel (b) in Fig.~\ref{grav_sense}) is for the B0 Ib supergiant Sk\,$-$67$^{\circ}$~5, where the best fitting model and the comparison model were computed for 
$\log{(g/{\rm cm\,s^{-2}})}=2.8$~dex and $2.9$, respectively, and the same $T_{\rm eff} = 25.6~{\rm kK}$. In contrast to the previous case, a $\Delta\log{(g/{\rm cm\,s^{-2}})}=0.1$~dex dramatically changes the quality of the fit. 

We estimate the model uncertainty at $\Delta \log{g^m}= \pm0.05-0.10$~dex and adopt a conservative uncertainty $\Delta \log{g_c}= \pm0.2-0.25$~dex which takes into account the 
additional uncertainty of $\varv_{\rm rot}\sin{i}$ and $R_{*}$ and the quality of the normalization (a $1\%$ change in the wings due to normalization can lead to a $\log{g}$ difference 
of $0.1$~dex).

\subsubsection{Luminosity}
\label{luminosity}

\begin{table}
  \caption{Properties of the \textsc{PYPHOT} filter functions \citep{zenodopyphot} employed in this study.}            
  \label{table:pyphot}      
  \centering                                      
  \small                                
  \begin{tabular}{c c c c c}          
      \hline\hline{\smallskip}
      filter        &$\lambda_{central}$           &$\lambda_{\rm pivot}$       &$\lambda_{\rm effective}$    &Vega\\
                    &$\AA$                         &$\AA$                       &$\AA$                        &\\ 
      \hline
      $B$           &4347.526                      &4296.702                    &4316.312                     &20.512\\ 
      $V$           &5504.666                      &5469.853                    &5438.689                     &21.099\\
      $K_{S}$       &21655.864                     &21638.169                   &21521.413                    &25.921\\
     \noalign{\smallskip}
     \hline
  \end{tabular}
\end{table}
For the determination of the bolometric luminosities $L_{\rm bol}$ of our stars, we use the flux from the model computations, which was computed for a fixed luminosity of 
$\log{(L_{\rm model}/L_{\odot})}=5.8$. We select Vega magnitude systems photometric zero-point. First, we apply suitable filter functions from \textsc{pyphot} \citep{zenodopyphot} using the effective 
wavelengths of the filters and we obtain the $B V K_{S}$ fluxes of the model using the function $get\_flux$, and from the fluxes we calculate magnitudes of the model 
($B^{\rm m}$, $V^{\rm m}$ and $K_{S}^{\rm m}$) as $-2.5\log{F_{B V K_{S}}} - m^{\rm vega}_{B V K_{S}}$. The filter function properties and Vega magnitudes are 
presented in Table~\ref{table:pyphot}. We then apply the extinction law from \citet{Gordon2003} and fit the relative extinction $R_{\rm V}$ and other parameters that determine the shape of the UV 
extinction curve \citep{FM1990} to match the shape of the observed SED. The model SED is then scaled to the observed SED using a factor equal to $K_{S}^{\rm m}/K_{S}$. Using the model magnitudes, 
the color excess $E(B-V)$ is calculated as $(B-V)-(B^{\rm m}-V^{\rm m})$. Assuming the distance modulus $DM$ for the LMC is $18.50\pm0.02$~mag \citep{alves2004}, we calculate the bolometric 
correction of the visual magnitude of the model $BC_{\rm V}^{\rm m}$ with the equation $BC_{V}^{\rm m} = -2.5\log{L_{\rm model}}+4.74-V^{\rm m}$. Finally, the bolometric luminosity is 
calculated via $BC_{V}^{\rm m}$ and the apparent visual magnitude of star $m_{V}$ as $\log{L_{\rm bol}} = (DM+A_{V}-BC_{V}^{\rm m}-m_{V}+4.74)/2.5$, where $A_{V}$ is the 
total extinction and is equal to $E(B-V)\,R_{V}$. The absolute visual magnitude is calculated as $M_{V} = m_{V}-DM-A_{V}$.
 
With this procedure we simultaneously obtain the extinction parameter $R_{V}$ and the color excess $E(B-V)$. This method relies on the relation of the $B$ and $V$ bands to the $K_{S}$ band, and 
since the extinction is minimal in the $K_{S}$, the obtained bolometric luminosities are highly reliable. The uncertainty of the derived bolometric luminosity is dominated by the uncertainty in the 
distance to the target, for which we adopt a random error of $0.1$~dex. Moreover, the SED fitting method (by eye) also result in an uncertainty in the $A_{V}$, which is 
inherited by the bolometric luminosity. Finally, the 2MASS photometry formal errors range from $0.01$ to $0.1$, averaging $\approx 0.05$. Therefore, we estimate the uncertainty to be 
$\Delta L_{\rm bol}\approx \pm 0.1{\rm -}0.2$~dex. In Appendix~\ref{SED_app} we present the SED fits for the stars in our sample.

\subsubsection{Line-broadening parameters} 
\label{lbp}
Out of all parameters that cause spectral line-broadening (macro-turbulent velocity $\varv_{\rm mac}$, micro-turbulent velocity $\varv_{\rm mic}$, and projected rotational velocity 
$\varv_{\rm rot}\sin{i}$) we elected to include only $\varv_{\rm rot}\sin{i}$ as part of the fitting procedure. When calculating the formal integral using \textsc{CMFFLUX}, we fix the 
$\varv_{\rm mic}$ in the photosphere to $10~{\rm km\,s^{-1}}$ and the maximum $\varv_{\rm mic}$ in the wind to $100{\rm km\,s^{-1}}$. 

Changing $\varv_{\rm mic}$ in the photosphere has little effect on wind lines, but the changes in line opacities do affect photospheric lines (primarily metal but also $\ion{He}{I}$ lines). 
This produces a degeneracy between $\varv_{\rm mic}$ and chemical abundances. Therefore, increasing $\varv_{\rm mic}$ could lead to underestimating the abundances and temperatures 
\citep{Brands2022}, hence our decision to exclude $\varv_{\rm mic}$ from our fitting procedure. This degeneracy can be alleviated by simultaneously fitting $\varv_{\rm mic}$ and the abundances 
using multiplet lines. A common diagnostic of $\varv_{\rm mic}$ in B-stars is the depth of the components of $\ion{Si}{iii}~\lambda\lambda\lambda 4553-4568-4575$, which are sensitive and react 
differently to changes in $\varv_{\rm mic}$ \citep{mcerlean1998}. 

The values of macro-turbulent velocities on OB-stars range from a few ${\rm km\,s^{-1}}$ for dwarfs up to a few tens of ${\rm km\,s^{-1}}$ for supergiants. We fix $\varv_{\rm mac}$ at 
$20~{\rm km\,s^{-1}}$ in our fitting procedure. Fixing $\varv_{\rm mac}$ is appropriate for our sample, especially when taking into account the velocity resolution of the UBV and VIS arms of 
X-shooter that $\approx 45$ and $25~{\rm km\,s^{-1}}$, respectively. The main diagnostic line for O-stars is $\ion{O}{III}~5592$. For early B-stars $\ion{O}{III}~5592$ is also used and as a 
sanity check we fit $\ion{C}{III}~\lambda 4267$ $\ion{Si}{iv}~\lambda\lambda 4089-4116$, $\ion{Si}{iii}~\lambda\lambda\lambda 4553-4568-4575$. for mid to late B-stars we fit the
$\ion{Mg}{II}~4481$ line and as a sanity check we use $\ion{C}{II}$ and $\ion{Si}{II}$.

The uncertainty in $\varv_{\rm rot}\sin{i}$  measurements is dominated by the velocity resolution of the UBV arm of X-shooter \citep{vernet2011} $\Delta\varv \approx 45~{\rm km\,s^{-1}}$. Later in 
Section \ref{LBP}, we investigate the line-broadening characteristics of a subsample of our stars that have MIKE data by using the \textsc{IACOB-BROAD} tool \citep{simondiaz2014}. 

\subsubsection{Terminal wind velocity ($\varv_{\infty}$)}  
The ``black velocity" $\varv_{\rm black}$ is the velocity measured at the bluest extent of fully saturated P Cygni absorption \citep{prinja1990}. This is thought to provide a more robust estimate of the wind terminal 
velocity than the ``edge velocity" $\varv_{\rm edge}$, which is measured at the point where the blue trough of the P Cygni profile intersects the local continuum \citep{abbot1985}. 
The difference $\varv_{\rm edge} - \varv_{\rm black}$ arises from the turbulence in the velocity field of the wind.

In this study, we obtained $\varv_{\rm edge}$ and $\varv_{\rm black}$ using direct measurements of all the viable P Cygni resonance lines in the range 
$\approx \left[1200, 2900\right]~{\rm \AA}$, with the advantage of accurate radial velocity estimates for each object from optical metal lines. 
$\varv_{\rm edge}$ and $\varv_{\rm black}$ were determined as the mean value of all velocities obtained from the individual resonance lines. The resonance lines used 
are $\ion{Si}{iv}~\lambda\lambda1394-1403$, $\ion{C}{iv}~\lambda\lambda1548-1551$, $\ion{Al}{iii}~\lambda\lambda1855-1863$, 
$\ion{Mg}{ii}~\lambda\lambda2796-2803$.

Stars at low metallicity are well known for having absorption profiles that may not reach zero intensity, consequently it is sometimes not possible to obtain an accurate measurement
of $\varv_{\rm black}$, and one has to calculate $\varv_{\infty}$ as a fraction of $\varv_{\rm edge}$. The ratio $\varv_{\rm black}\big/\varv_{\rm edge}$ is obtained from Tables 
\ref{table_app_1} and \ref{table_app_2}.

\subsubsection{Wind density parameters} 
The primary optical diagnostic line used to constrain the wind mass-loss rate is H$\alpha$, and in the case of O-supergiants, $\ion{He}{II}~4686$ is used to a lesser extent. 
H$\alpha$ is formed relatively close to the photosphere ($\approx 1-2~R_{*}$) and since it is a recombination line it is very sensitive to wind density ($\sim \rho^{2}$). The wind density itself is a 
function of mass-loss rate and velocity field structure. Therefore the parameters we try to fit that affect the strength and the morphology of H$\alpha$ within the framework of \textsc{CMFGEN} 
are wind terminal velocity $\varv_{\infty}$, mass-loss rate $\dot{M}$, $\beta$ and clumping parameters (volume filling factor $f_{\rm vol}$ and the onset clumping velocity $\varv_{\rm cl}$). 
UV lines are used to further constrain the density and to break degeneracies of clumping and mass-loss. The most consistently available lines in our sample which we use to constrain the 
mass-loss rate and clumping are the unsaturated P Cygni sulphur doublet $\ion{S}{IV}~1063-1073$, plus $\ion{C}{III}~\lambda 1176$ and $\ion{Si}{IV}~\lambda\lambda 1394-1403$ if 
it is not fully saturated. We also use $\ion{C}{iv}~\lambda\lambda1548-1551$ for O-stars as a sanity check and for late B-stars we use $\ion{Al}{III}~\lambda 1856-1862$ and 
$\ion{Mg}{II}~\lambda2796-2803$.

In our fine-tuned model, we adopt a value for the on-set clumping velocity $\varv_{\rm cl}$ equal to double the sound speed of the model, which comes to a value in the range 
$\approx25{\rm -}35~{\rm km\,s^{-1}}$ depending mainly on the temperature and helium abundance. This is a common procedure in the analysis of OB stars \citep{marcolino2009, puebla2016}. 
Since the connection velocity in our models is set by default to $10~{\rm km\,s^{-1}}$, the adopted values of $\varv_{\rm cl}$ indicate that the clumping becomes significant in the base of the 
super-sonic winds, rather than in the subsonic layers or the photosphere. Although, recent 2-D global simulations of O-stars indicate that wind inhomogeneities could originate from ``photospheric 
turbulence'' arising in the iron opacity peak zone due to the unstable nature of convection \citep{debnath2024}. 

Having acquired $R_{*}$ (from the derived $L_{\rm bol}$ and $T_{\rm eff}$), $\varv_{\infty}$ and model mass-loss rate ($\dot{M}^{m}$), we derive the luminosity-adjusted mass-loss rate ($\dot{M}$) by 
scaling $\dot{M}^{m}$ to the transformed radius $R_{\rm t}$ via Eq.~\ref{eq:R_t}.

We estimate uncertainties of the wind parameters in a way that is suitable for our method of analysis. For $\varv_{\infty}$ that is determined as $\varv_{\rm black}$ we calculate the uncertainty by quadrature of 
the systematic and stochastic errors. The systematic errors stem from the resolution of the instrument, and the stochastic errors are the combination of the standard deviation of the mean 
$\varv_{\infty}$ and mean radial velocity $\varv_{\rm rad}$. For $\varv_{\infty}$ that is calculated as a fraction of $\varv_{\rm edge}$, other than the resolution and the dispersion of the mean $\varv_{\rm edge}$ and $\varv_{\rm rad}$, we 
additionally take into account in quadrature the standard deviation of $\varv_{\infty}\big/\varv_{\rm edge}$ of the mean over our sample, which yields relatively higher $\varv_{\infty}$ errors for stars that 
do not show saturated P Cygni profiles in their UV spectra.

It is quite difficult to quantify the uncertainty of $\beta$, but by varying beta in our model and adjusting the mass-loss rate accordingly we were able to get a general idea on the range 
of beta that would reproduce similar quality fit for H$\alpha$ which is defined by the uncertainty $\Delta \beta= \pm0.2$. The model uncertainty of the mass-loss rate 
$\Delta\log{\dot{M^m}}= \pm0.05$ to $0.1$~dex depending on the quality of the fit. Using Eq.~\ref{eq:R_t} we calculate the uncertainty in the scaled mass-loss rates as:
\begin{equation}
\begin{array}{l}
\Delta \log{\dot{M}} =  \sqrt{\left(\frac{4}{3}\frac{ \Delta T_{\rm eff}}{T_{\rm eff}}\right)^{2}+}\ldots \\
\\
~~~~~~~~~~~~~\sqrt{\left(\frac{2}{\ln{10}}\cdot \frac{\Delta\dot{M}^{m}}{\dot{M}^{m}}\right)^{2} + \left(\cdot \frac{\Delta \varv_{\infty}}{ \varv_{\infty}}\right)^{2}}.
\end{array}
\end{equation}

\subsubsection{He and CNO abundances}
\label{abundances_method}
The accurate determination of both the effective temperature and surface gravity depends on the helium abundance in the model, especially for O-stars because of the 
way the temperature is gauged by $\ion{He}{I}$ to $\ion{He}{II}$ equivalent widths ratio. To restrict the helium mass fraction (Y) we fix the mass fraction of all 
included elements except for hydrogen, which means that increasing or decreasing the mass fraction of helium in the model would respectively deplete or enrich the 
model with hydrogen.  We use the following Helium line diagnostics:\\
$\ion{He}{I}~\lambda4026$, $ \ion{He}{I}~\lambda4471$,$ \ion{He}{I}~\lambda4922$, ($\ion{He}{I}~\lambda6678$), ($\ion{He}{I}~\lambda7065$), ($\ion{He}{I}~\lambda7281$), $\ion{He}{II}~\lambda4542$, 
$ \ion{He}{II}~\lambda5411$.

Recalling Section 3, plane parallel model atmosphere have been very successful at determining metal abundances of early type stars with weak winds \citep[e.g., ][]{hunter2007, przybilla2008}, 
which have included late B-supergiants \citep{przybilla2006}. Models employing spherical geometry generally require significantly higher resources, hindering abundance determinations, although 
FASTWIND has been successfully used to obtain CNO abundances in early B-supergiants \citep{urbaneja2005b}. Since CMFGEN is used for the present study, we acknowledge larger uncertainties in 
derived CNO abundances with respect to other analyses. Nevertheless, after obtaining the stellar and wind parameters we try to fit multiple lines that adhere to different ionization levels of CNO elements. The lines depend on the 
spectral type of the star, but the primary lines that we use to refine the CNO abundances are:\\
\begin{itemize}
    \item Carbon:\\
                O-stars: $\ion{C}{IV}\,\lambda\lambda5801-5811$, $\ion{C}{III}\,\lambda\lambda4647-4650$\\
                B-stars: $\ion{C}{III}\,\lambda\lambda4647-4650$, $\ion{C}{III}\,\lambda5696$, $\ion{C}{II}\,\lambda4070$, $\ion{C}{II}\,\lambda4267$, $\ion{C}{II}\,\lambda\lambda6578-6582$, $\ion{C}{II}\,\lambda\lambda\lambda7231-7236-7237$
    \item Nitrogen:\\
                O-stars: $\ion{N}{III}\,\lambda4097$, $\ion{N}{III}\,\lambda\lambda4510-4515$, $\ion{N}{III}\,\lambda\lambda4634-4641$  
                B-stars: $\ion{N}{III}\,\lambda4097$, $\ion{N}{II}\,\lambda3995$, $\ion{N}{II}\,\lambda4447$, $\ion{N}{II}\,\lambda\lambda\lambda4601-4607-4614$, $\ion{N}{II}\,\lambda4630$ 
    \item Oxygen:\\
                O-stars: $\ion{O}{III}\,\lambda\lambda3261-3265$, $\ion{O}{III}\,\lambda3760$, $\ion{O}{III}\,\lambda5592$\\                
                B-stars: $\ion{O}{II}\,\lambda4254$, $\ion{O}{II}\,\lambda4367$, $\ion{O}{II}\,\lambda\lambda4415-4417$, $\ion{O}{II}\,\lambda\lambda4638-4641$, ($\ion{O}{I}\,\lambda\lambda\lambda7772-7774-7775$).\\

\end{itemize}
The triplet $\ion{N}{III}\,\lambda4634-4641$ is usually in emission and is notoriously difficult to fit \citep{rivero2011}. In some objects $\ion{C}{III}\,\lambda5696$ is also in emission.
We leave fitting CNO abundances as the last step in our fitting procedure since it does not effect other diagnostic lines that are used to gauge other parameters.
$\ion{He}{I}$ and $\ion{O}{I}$ lines quoted above in parentheses were not part of the analysis and are only used as supplementary diagnostics.

The CNO abundances we present in Section \ref{abund_res} are subject to large uncertainties, for which we adopt a value of $\approx \pm 0.3$~dex. This is due to the high sensitivity of metal 
lines to changes in $T_{\rm eff}$ and $\log{g}$, in addition to changes in $\varv_{\rm mic}$. We are more confident in our helium abundances, and we determine our uncertainties based on the way we 
varied the helium mass fraction in our models (in steps of $10\%$ of the baseline grid abundance), so we adopt a $20\%$ uncertainty in the mass-fraction with translates to $0.09$~dex for the 
helium abundance by number $\epsilon_{\rm He}=\log{\frac{\rm He}{\rm H}} + 12$.

\section{Results}
\label{results}
In this section we present an overview of the results and compare them to other theoretical and empirical results that utilise the UV and optical or the optical range exclusively. The quality of 
our analysis is described for each star individually in the Appendix \ref{individ_com}. Best fitting physical parameters are presented in Table~\ref{table:2}, including inferred evolutionary 
masses and ages from {\sc Bonnsai} \citep{bonnsai2014} applied to \citet{brott2011} rotating single-star evolutionary models for LMC metallicity.

\begin{sidewaystable*}
        \caption{Derived stellar parameters based on the best fitting \textsc{CMFGEN} model. $M_{init}$, $M_{\rm evo}$, and ages of the stars are derived using an updated Bayesian inference method 
        (Bronner et al. in prep) that is similar to {\sc Bonnsai} \citep{bonnsai2014} applied to \citet{brott2011} evolutioanry tracks. $\Gamma_e$ is calculated using $M_{\rm evo}$.}        
        \label{table:2}      
        \def\arraystretch{1.5}
        \centering                                     
        \small
        \addtolength{\tabcolsep}{-0.1em}
        \begin{tabular}{c c c c c | c c c c c c | c c c | c c c c}         
            \hline\hline{\smallskip}
          Sk\,$-$        &$T_{\rm eff}$  &Diag.                          &$\log{g_c}$         &$R_{*}$      &$\log{L_{\rm bol}}$  &$R_{\rm V}$ &$E(B-V)$ &$M_{\rm V}$ &$BC_{\rm V}$ &$v_{\rm rad}$          &$v_{\rm rot}\sin{i}$   &$M_{\rm spec}$   &$\Gamma_{\rm e}^{\rm spec}$  &$M_{init}$                 &$M_{\rm evo}$              &Age                       &$\Gamma_{\rm e}^{\rm evo}$\\
                         &kK             &                               &${\rm cm\,s^{-2}}$  &$R_{\odot}$  &$L_{\odot}$          &            &mag      &mag         &mag          &${\rm km\,s^{-1}}$     &${\rm km\,s^{-1}}$     &$M_{\odot}$      &                             &$M_{\odot}$                &$M_{\odot}$                &My                        &\\
            \hline
	      66$^{\circ}$~171 &29.9$\pm1.0$   &$\ion{He}{I}-\ion{He}{II}$     &3.11$\pm0.20$       & 25          &5.67$\pm0.11$        &1.9         &0.16   &$-6.6$        &$-2.83$        &410                    &75                     &31 $\pm 7$     &0.73                         &$34.61^{+ 9.03}_{-4.34}$   &$30.04^{+ 8.14}_{-2.26}$   &$ 5.27^{+0.67}_{-1.03}$   &$0.74^{+0.27}_{-0.08}$\\
	      68$^{\circ}$~155 &29.0$\pm1.0$   &$\ion{He}{I}-\ion{He}{II}$     &3.06$\pm0.20$       & 26          &5.64$\pm0.11$        &2.9         &0.29   &$-6.6$        &$-2.76$        &240                    &80                     &29 $\pm 6$     &0.72                         &$34.21^{+ 8.41}_{-7.98}$   &$29.89^{+ 8.49}_{-4.59}$   &$ 5.47^{+0.22}_{-0.95}$   &$0.70^{+0.29}_{-0.16}$\\
	      69$^{\circ}$~279 &28.5$\pm2.0$   &$\ion{He}{I}-\ion{He}{II}$     &2.95$\pm0.25$       & 27          &5.63$\pm0.11$        &2.6         &0.34   &$-6.6$        &$-2.76$        &230                    &40                     &24 $\pm 6$     &0.90                         &$33.79^{+ 7.11}_{-3.21}$   &$30.25^{+ 7.71}_{-1.37}$   &$ 4.23^{+0.60}_{-0.44}$   &$0.70^{+0.26}_{-0.08}$\\
	      71$^{\circ}$~41  &29.2$\pm1.0$   &$\ion{He}{I}-\ion{He}{II}$     &3.12$\pm0.20$       & 23          &5.53$\pm0.11$        &2.8         &0.25   &$-6.3$        &$-2.77$        &260                    &45                     &25 $\pm 5$     &0.69                         &$32.80^{+ 4.82}_{-3.48}$   &$30.30^{+ 5.30}_{-2.09}$   &$ 4.26^{+0.60}_{-0.31}$   &$0.57^{+0.18}_{-0.08}$\\
	      68$^{\circ}$~135 &26.9$\pm1.5$   &$\ion{He}{I}-\ion{He}{II}$     &2.82$\pm0.24$       & 51          &6.10$\pm0.12$        &2.8         &0.28   &$-7.9$        &$-2.60$        &270                    &45                     &65 $\pm16$     &0.96                         &$60.74^{+16.54}_{-6.69}$   &$52.35^{+14.90}_{-4.63}$   &$ 3.25^{+0.74}_{-0.71}$   &{\color{gray} $0.89_{-0.00}^{+0.2}$}\\
	      67$^{\circ}$~5   &25.6$\pm1.0$   &$\ion{He}{I}-\ion{He}{II}$     &2.81$\pm0.20$       & 45          &5.89$\pm0.12$        &2.2         &0.17   &$-7.5$        &$-2.46$        &295                    &65                     &47 $\pm11$     &0.76                         &$44.19^{+ 9.10}_{-4.57}$   &$38.12^{+ 6.09}_{-3.63}$   &$ 4.79^{+0.28}_{-0.95}$   &$0.95^{+0.16}_{-0.10}$\\
	      68$^{\circ}$~52  &26.0$\pm1.0$   &$\ion{Si}{III}-\ion{Si}{IV}$   &2.85$\pm0.20$       & 42          &5.87$\pm0.12$        &2.9         &0.22   &$-7.4$        &$-2.51$        &255                    &50                     &46 $\pm10$     &0.83                         &$49.52^{+ 7.30}_{-8.41}$   &$44.97^{+ 6.72}_{-6.46}$   &$ 3.40^{+0.45}_{-0.34}$   &$0.85^{+0.15}_{-0.15}$\\
        69$^{\circ}$~43  &22.4$\pm1.0$   &$\ion{Si}{III}-\ion{Si}{IV}$   &2.71$\pm0.21$       & 40          &5.55$\pm0.11$        &2.5         &0.19   &$-7.0$        &$-2.14$        &255                    &50                     &29 $\pm 6$     &0.63                         &$31.94^{+ 4.11}_{-3.55}$   &$29.47^{+ 4.06}_{-2.30}$   &$ 5.15^{+0.43}_{-0.69}$   &$0.62^{+0.14}_{-0.09}$\\
        68$^{\circ}$~140 &24.1$\pm1.0$   &$\ion{Si}{III}-\ion{Si}{IV}$   &2.81$\pm0.21$       & 33          &5.52$\pm0.11$        &3.1         &0.34   &$-6.7$        &$-2.32$        &260                    &50                     &26 $\pm 7$     &0.67                         &$30.21^{+ 4.47}_{-3.23}$   &$29.06^{+ 3.81}_{-3.08}$   &$ 5.21^{+0.61}_{-0.72}$   &$0.59^{+0.14}_{-0.11}$\\
        67$^{\circ}$~2   &18.8$\pm1.0$   &$\ion{Si}{III}-\ion{Si}{IV}$   &2.31$\pm0.21$       & 71          &5.76$\pm0.12$        &2.7         &0.27   &$-7.9$        &$-1.73$        &320                    &45                     &38 $\pm 6$     &0.76                         &$41.53^{+ 7.90}_{-3.04}$   &$36.64^{+ 6.16}_{-1.93}$   &$ 4.24^{+0.37}_{-0.49}$   &$0.75^{+0.18}_{-0.07}$\\
        67$^{\circ}$~14  &21.1$\pm1.0$   &$\ion{Si}{III}-\ion{Si}{IV}$   &2.51$\pm0.21$       & 46          &5.58$\pm0.11$        &1.6         &0.16   &$-7.2$        &$-2.00$        &300                    &50                     &25 $\pm10$     &0.73                         &$39.98^{+ 2.50}_{-6.18}$   &$35.50^{+ 2.09}_{-4.06}$   &$ 4.75^{+0.40}_{-0.35}$   &$0.53^{+0.08}_{-0.12}$\\
	      69$^{\circ}$~52  &18.8$\pm1.0$   &$\ion{Si}{III}-\ion{Si}{IV}$   &2.31$\pm0.21$       & 59          &5.60$\pm0.11$        &1.9         &0.28   &$-7.5$        &$-1.73$        &260                    &50                     &26 $\pm 7$     &0.72                         &$40.57^{+ 5.77}_{-1.92}$   &$35.56^{+ 4.48}_{-1.05}$   &$ 4.44^{+0.26}_{-0.24}$   &$0.53^{+0.14}_{-0.06}$\\
	      67$^{\circ}$~78  &15.5$\pm1.0$   &$\ion{Si}{II}-\ion{Si}{III}$   &2.10$\pm0.21$       & 75          &5.47$\pm0.11$        &2.2         &0.17   &$-7.6$        &$-1.33$        &310                    &30                     &26 $\pm 7$     &0.58                         &$28.56^{+ 3.61}_{-3.80}$   &$27.72^{+ 2.61}_{-4.02}$   &$ 5.49^{+1.03}_{-0.60}$   &$0.55^{+0.11}_{-0.16}$\\
	      70$^{\circ}$~16  &18.4$\pm1.0$   &$\ion{Si}{II}-\ion{Si}{III}$   &2.61$\pm0.21$       & 27          &4.88$\pm0.10$        &1.9         &0.19   &$-5.8$        &$-1.70$        &265                    &30                     &11 $\pm 3$     &0.36                         &$16.21^{+ 1.45}_{-1.41}$   &$16.02^{+ 1.45}_{-1.31}$   &$10.40^{+1.57}_{-1.12}$   &$0.24^{+0.11}_{-0.10}$\\
	      68$^{\circ}$~8   &14.1$\pm1.0$   &$\ion{Si}{II}-\ion{Si}{III}$   &1.81$\pm0.21$       &102          &5.57$\pm0.11$        &2.4         &0.25   &$-8.1$        &$-1.13$        &250                    &35                     &24 $\pm 7$     &0.74                         &$31.59^{+ 4.56}_{-4.25}$   &$29.59^{+ 5.08}_{-4.93}$   &$ 5.20^{+0.74}_{-0.77}$   &$0.61^{+0.19}_{-0.18}$\\
	      67$^{\circ}$~195 &12.6$\pm1.0$   &$\ion{He}{I}-\ion{Mg}{II}$     &2.11$\pm0.22$       & 44          &4.65$\pm0.10$        &2.9         &0.14   &$-6.1$        &$-0.83$        &300                    &25                     & 9 $\pm 3$     &0.24                         &$13.23^{+ 1.24}_{-0.98}$   &$13.20^{+ 1.17}_{-0.10}$   &$13.94^{+1.70}_{-1.87}$   &$0.17^{+0.12}_{-0.08}$\\
            \noalign{\smallskip}
            \hline
        \end{tabular}
  \footnotetext{Presented uncertainties for effective temperatures $T_{\rm eff}$ are double the model uncertainty (see Section~\ref{diag_intro}). In the uncertainties of centrifugal 
  force-corrected surface gravity $\log{g_c}$, we take into account the model uncertainty of $\log{g}$ and the uncertainty of the projected rotational velocity $v_{\rm rot}\sin{i}$ and the radius $R_{*}$. 
  The uncertainty in radial velocity $v_{\rm rad}$ measurements is dominated by the velocity resolution of the UBV part of the spectrum which is $\Delta\varv\approx45~{\rm km\,s^{-1}}$.
  The relative extinction $R_{\rm V}$ ($\Delta R_{\rm V} \approx 0.3$~mag), color excess $E(B-V)$ ($\Delta E(B-V) \approx 0.05$~mag), absolute V-band magnitude $M_{\rm V}$, and bolometric correction $BC_{\rm V}$ are produced from the SED fits that are shown in 
  Appendix \ref{SED_app}. $\Gamma_{\rm e}^{\rm spec}$ is subject to uncertainties of $\Delta\Gamma_{\rm e}^{\rm spec} \approx 0.18$ which takes into account the uncertainties of $M_{\rm spec}$ and $\log{L_{\rm bol}}$.}
\end{sidewaystable*}

\subsection{Hertzsprung-Russell diagram}
\begin{figure}
    \centering
     \includegraphics[width=\hsize]{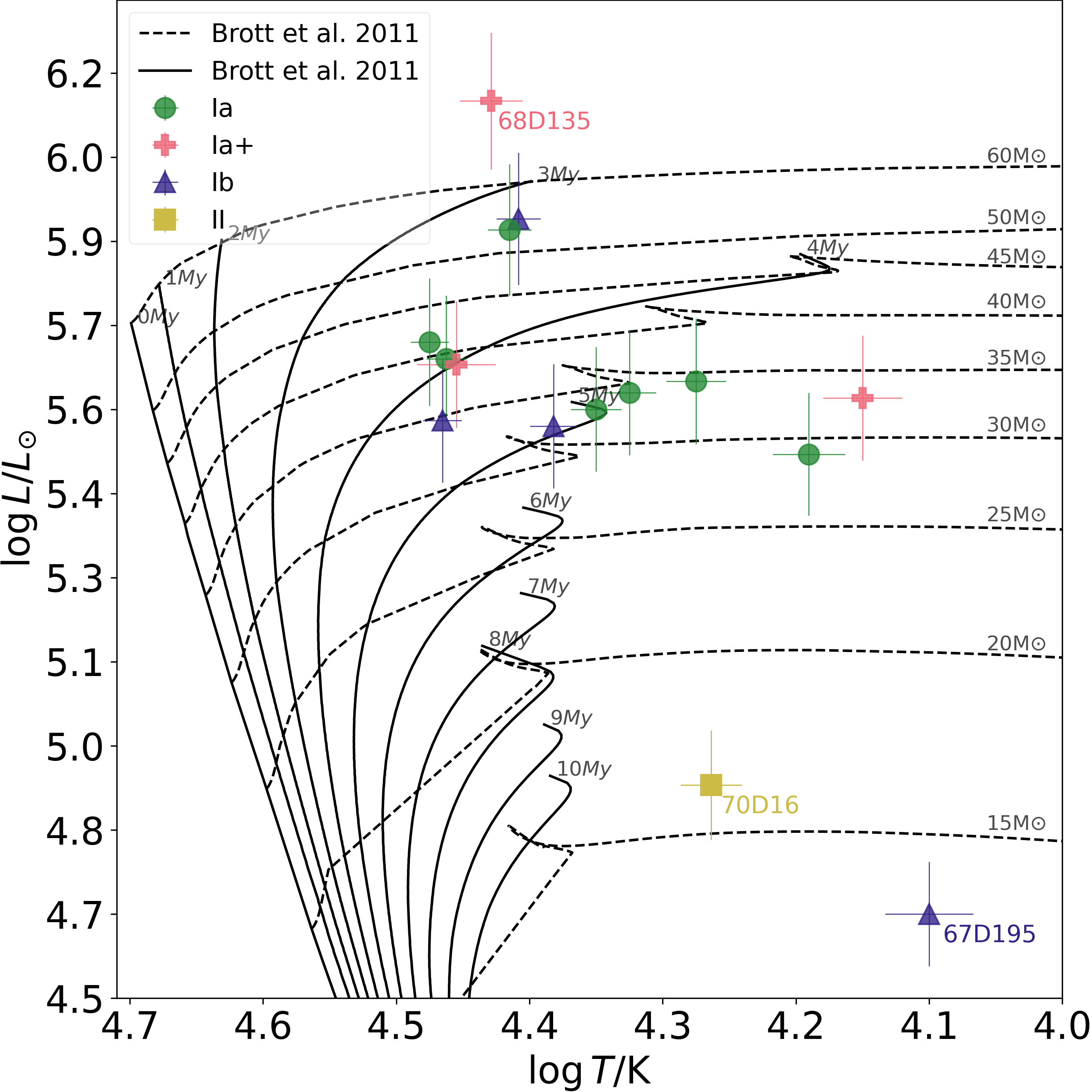}
         \caption{The Hertzsprung-Russell diagram for our sample. Overlaid in solid black lines are non-rotating isochrones for different ages ($\approx[0-10]$ Myr) and dashed black lines are the 
         evolutionary tracks for stellar masses in the range $\approx[10-60]~M_{\odot}$ with a rotational velocity of $50~{\rm km\,s^{-1}}$. Both the isochrones and evolutionary tracks are adopted 
         from \citet{brott2011}.}
         \label{HRD}
    \end{figure}

Fig.~\ref{HRD} shows the location of our targets on a Hertzsprung-Russell diagram (HRD). Our sample spans a range of temperatures $\log{(T_{\rm eff}/{\rm K})}\approx4.1{\rm -}4.5$, while 
luminosities cover a broad range of $\log{(L_{\rm bol}/L_{\odot})}\approx4.50{\rm -}6.10$ (see Table~\ref{table:2}). The extremey luminous star is the hypergiant Sk\,$-$68$^{\circ}$~135 
(Fig.~\ref{SK-68D135}) with $\log{(L_{\rm bol}/L_{\odot})}\approx 6.1 $.

The majority of our sample excdeed $\log{(L_{\rm bol}/L_{\odot})}\approx5.4$ aside from Sk\,$-$70$^{\circ}$~16, which is a bright giant (II), and Sk\,$-$67$^{\circ}$~195, which is a B8 Ib supergiant 
according to \citet{Bestenlehner2025}. Overall, the majority of the sample lies between the zero age main sequence (ZAMS) and the terminal age main sequence 
(TAMS), according to the evolutionary models of \citet{brott2011}, but several targets, assuming a single-star evolutionary scenario, are located beyond the TAMS and would therefore be 
identified as post-main sequence stars.

\begin{figure}
  \centering
   \includegraphics[width=\hsize]{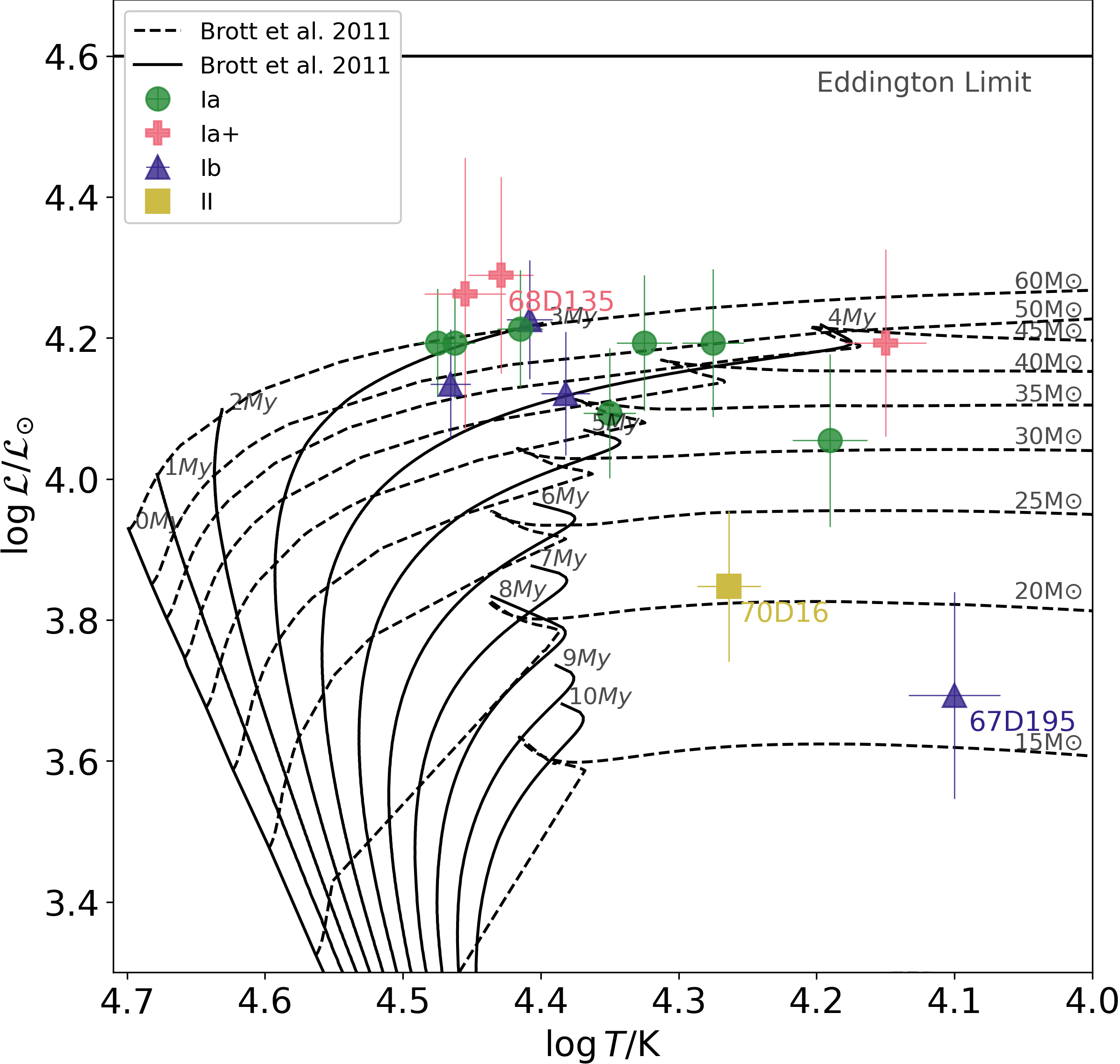}
       \caption{The spectroscopic Hertzsprung-Russell diagram for our sample. The evolutionary tracks (solid black lines) and the isochrones (dashed black lines) are the same as in Fig.~\ref{HRD}, 
       which are taken from \citet{brott2011}. The Encoding of the luminosity classes is the same as in Fig.~\ref{HRD}}
       \label{SHRD}
  \end{figure}
Fig.~\ref{SHRD} shows the location of our targets on the spectroscopic Hertzsprung-Russell diagram (sHRD) \citep{langerandkudritzki2014}, where $\mathcal{L} = T_{\rm eff}^{4}/g$ and 
$\mathcal{L_{\odot}}$ is calculated with solar values $T_{\rm eff}=5777$~kK and $\log{(g/{\rm cm\,s^{-1}})}=4.44$. The majority of our stars fall in the $\log{\mathcal{L}/\mathcal{L}_{\odot}}$ 
range of $4.0{\rm -}4.3$. We find that the hypergiants are the closest to the Eddington limit of $\log{\mathcal{L}/\mathcal{L}_{\odot}}\approx4.6$ (assuming a hydrogen mass fraction of $0.73$).
Also, as expected, the low luminosity supergiant Sk\,$-$67$^{\circ}$~195 and the bright giant Sk\,$-$70$^{\circ}$~16 are the furthest away from the Eddington limit with 
$\log{\mathcal{L}/\mathcal{L}_{\odot}}$ equal to $3.68$ and $3.82$, respectively.

We also compare a calibration of the bolometric correction in the visual band $BC_{\rm V}$ versus $T_{\rm eff}$ to the calibration obtained by \citet{Lanz2007} from \textsc{TLUSTY} 
non-LTE plane-parallel models of B-stars. A simple fit to our dataset reveals:
\begin{equation}
    \label{eq:BC} 
    BC_{\rm V}/{\rm mag} = 21.00 - 5.33 \log{(T_{\rm eff}/{\rm K})},
\end{equation}
with a standard deviation of $\approx 0.1~{\rm mag}$, which support the results of \citet{Lanz2007} for the LMC, who finds: \\
\\$BC_{\rm V}/{\rm mag}=21.08-5.36\log{(T_{\rm eff}/{\rm K})}$\\

As to be expected for a UV-bright sample, extinctions are relatively low with an average color excess $E(B-V) = 0.23\pm0.06~{\rm mag}$, and an average total extinction in the V-band 
$A^{\rm average}_{\rm V} = 0.6 \pm0.4~{\rm mag}$. This agrees with the findings of \citet{Gordon2003}.

\subsection{Stellar masses}
\label{mass}
Evolutionary masses ($M_{\rm evo}$) presented in Table~\ref{table:2} are derived via a Bayesian inference method that is similar to {\sc Bonnsai} \citep{bonnsai2014} with updated 
techniques (Bronner et al. in prep) applied to LMC tracks from \citet{brott2011} with $L_{\rm bol}$, $T_{\rm eff}$, $\log{g}$, $\varv_{\rm rot}\sin{i}$, and $\log{\frac{\rm He}{\rm H}} + 12$ as input parameters. 
Our sample spans a wide range range of ($M_{\rm evo}$) of $13.20^{+ 1.17}_{-0.10}~M_{\odot}$ (Sk\,$-$67$^{\circ}$~195) to $52.35^{+14.90}_{-4.63}~M_{\odot}$ (Sk\,$-$68$^{\circ}$~135). 
Table~\ref{table:2} also includes stellar ages that range from $3.40^{+0.45}_{-0.34}$~Myr (Sk\,$-$68$^{\circ}$~135) to $13.94^{+1.70}_{-1.87}$~Myr (Sk\,$-$67$^{\circ}$~195).

In Table~\ref{table:2} we present our values of the true surface effective gravity $\log{(g_c/{\rm cm\,s^{-2}})} = \log{(g_{\rm model} + \varv_{\rm rot}\sin{i}^{2}/R_{*})}$, where $R_{*}$ is 
obtained from Stefan–Boltzmann law. $g_c$ takes into account the centrifugal force due to stellar rotation \citep{herrero1992}. Our values range from $\log{g_c}\approx 1.8$ to $3.1$ due to the 
stars possessing large radii ($R_{*} > 20~R_{\odot}$), which is expected for a sample that consists of evolved OB-stars. 

\begin{figure}
    \centering
     \includegraphics[width=\hsize]{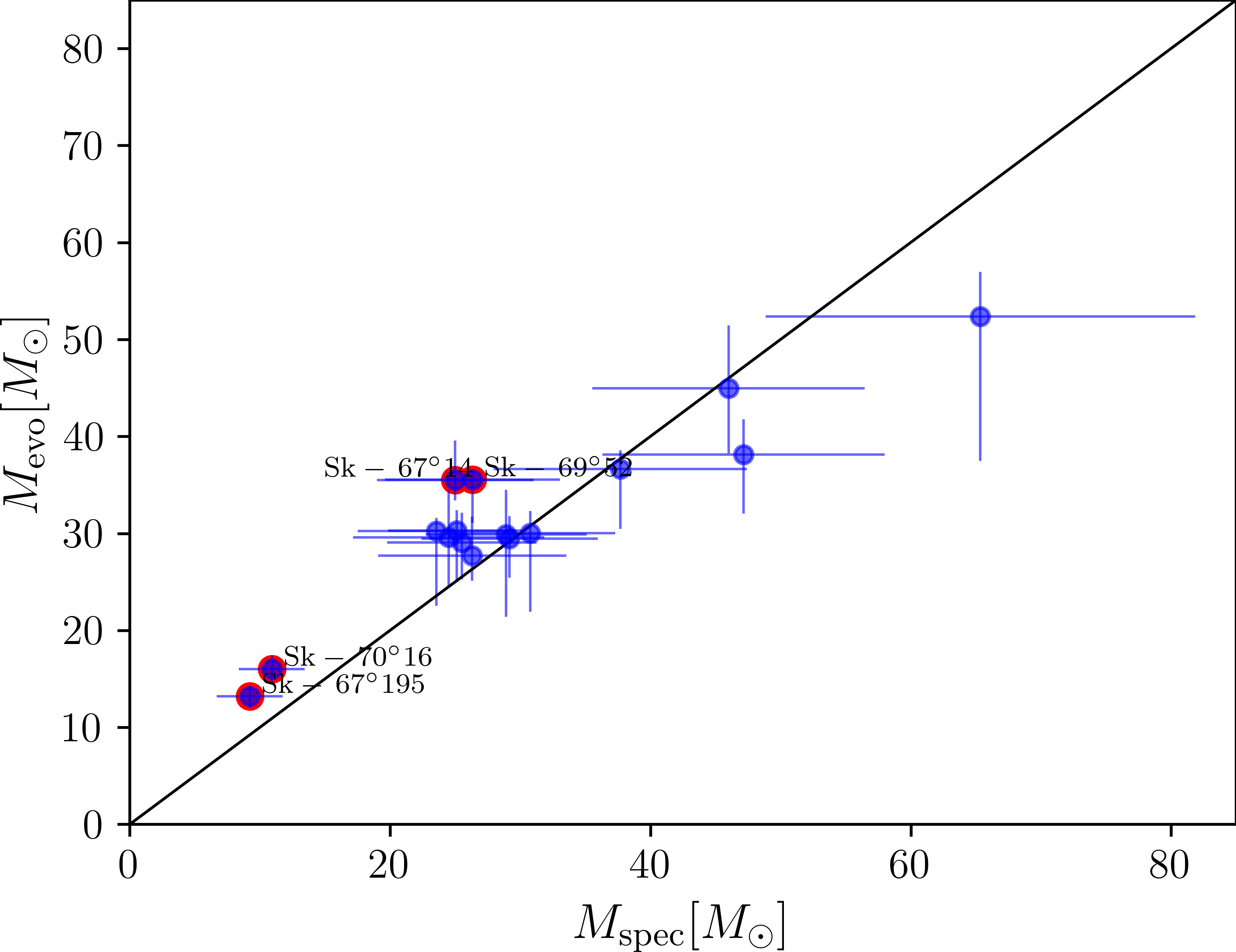}
         \caption{Blue circles: spectroscopically obtained mass from our analysis $M_{\rm spec}$ versus the mass produced by {\sc Bonnsai} \citep{bonnsai2014} applied to evolutionary tracks from \citet{brott2011} $M_{\rm evo}$. 
         Red circles overlaid onto the blue points depicts stars that have a sginificant discrepancy between their $M_{\rm spec}$ and $M_{\rm evo}$.}
         \label{Mevo_Mspec}
\end{figure}
We derive the spectroscopic masses ($M_{\rm spec}$) from the gravities via $g = G\,M/R_{*}^{2}$. Therefore, we derive relative uncertainties of $~20\%-30\%$, which mainly depend on the 
uncertainties of $\log{g}$, which is largely affected by quality of normalization. To put this into perspective, a difference in the wings of the Balmer lines of $1\%$ can potentailly lead to a 
$\log{g}$ difference up to $\approx 0.1$~dex. Such a difference leads to a relative uncertainty of $\Delta M_{\rm spec}\approx20\%$. 

Spectroscopic masses are compared to evolutionary masses in Fig.~\ref{Mevo_Mspec}. This shows that, for most of the stars, $M_{\rm spec}$ and $M_{\rm evo}$ are consistent within 
the uncertainties, similar to what \citet{schneider2018} finds in a large sample of OB-stars. For some of the stars (red circles in Fig.~\ref{Mevo_Mspec}), the mass discrepancy that was established 
for Galactic O-supergiants by \citet{herrero1992}, and later expanded to SMC O-stars by \citet{trundle2004I} and further to Galactic B-supergiants in \citet{Crowther2006} persists. We find that the 
spectroscopic masses for these stars are significantly lower than the masses produced by evolutionary models. 

The Eddington parameter $\Gamma_{\rm e}^{\rm evo}$ presented in Table~\ref{table:2} is derived using the evolutionary masses and it ranges from $\Gamma_{\rm e} = 0.18_{-0.03}^{+0.03}$ (Sk\,$-$67$^{\circ}$~195) to $0.95_{-0.12}^{+0.02}$
(Sk\,$-$67$^{\circ}$~5). We note that for the hypergiant Sk\,$-$68$^{\circ}$~135 we obtain $\Gamma_{\rm e} = 1.04_{-0.15}^{+0.05}$, which is unphysical, therefore we adopt the value corresponding 
to the lower limit of $\Gamma_{\rm e} = 0.89$.

\subsection{Wind properties}
\label{sec:wind}
\begin{table}
    \caption{Wind parameters for our targets. The mass-loss rates of Sk\,$-$70$^{\circ}$~16 and Sk\,$-$67$^{\circ}$~195 are upper limits. Escape velocities are adjusted for $\Gamma_{\rm e}$.}   
    \def\arraystretch{1.5}
    \label{table:4}      
    \centering                                     
    \small                                
    \addtolength{\tabcolsep}{-0.3em}
    \begin{tabular}{c c c c c c c c}        
        \hline\hline{\smallskip}
        Target        &$L_{\rm bol}$  &$\log{\dot{M}}$            &$\varv_{\infty}$   &$\varv_{{\rm esc}, (1-\Gamma_{\rm e})}$  &$\beta$  &$f_{\rm vol}$  &$\varv_{\rm cl}$  \\
        Sk\,$-$       &$L_{\odot}$    &$M_{\odot}\,{\rm yr}^{-1}$ &${\rm km\,s^{-1}}$ &${\rm km\,s^{-1}}$                       &         &               &${\rm km\,s^{-1}}$\\ 
    66$^{\circ}$~171  &5.67  &$-6.07$ $\pm0.33$                   &1775 $\pm63$       &680                                      &1.7      &0.03           &30\\
    68$^{\circ}$~155  &5.64  &$-6.19$ $\pm0.25$                   &1520 $\pm61$       &650                                      &1.5      &0.03           &30\\
    69$^{\circ}$~279  &5.63  &$-5.70$ $\pm0.39$                   &630  $\pm62$       &580                                      &2.7      &0.1            &30\\
    71$^{\circ}$~41   &5.53  &$-6.03$ $\pm0.25$                   &1390 $\pm112$      &650                                      &1.2      &0.1            &35\\
    68$^{\circ}$~135  &6.10  &$-5.70$ $\pm0.31$                   &880  $\pm62$       &690                                      &2.3      &0.1            &35\\
    67$^{\circ}$~5    &5.89  &$-6.05$ $\pm0.35$                   &1230 $\pm61$       &630                                      &1.3      &0.1            &30\\
    68$^{\circ}$~52   &5.87  &$-6.28$ $\pm0.25$                   &1140 $\pm62$       &640                                      &2.0      &0.03           &35\\
    69$^{\circ}$~43   &5.55  &$-6.49$ $\pm0.25$                   &825  $\pm64$       &530                                      &2.0      &0.1            &30\\
    68$^{\circ}$~140  &5.52  &$-6.46$ $\pm0.25$                   &1000 $\pm62$       &540                                      &2.2      &0.1            &30\\
    67$^{\circ}$~2    &5.76  &$-6.21$ $\pm0.26$                   &435  $\pm61$       &450                                      &3.0      &0.14           &30\\
    67$^{\circ}$~14   &5.58  &$-6.33$ $\pm0.29$                   &810  $\pm67$       &460                                      &2.0      &0.1            &30\\
    69$^{\circ}$~52   &5.60  &$-6.62$ $\pm0.38$                   &465  $\pm61$       &410                                      &2.5      &0.14           &25\\
    67$^{\circ}$~78   &5.47  &$-6.70$ $\pm0.40$                   &380  $\pm75$       &360                                      &3.0      &0.2            &30\\
    70$^{\circ}$~16   &4.88  &$-7.60$ $\pm0.34$                   &235  $\pm70$       &390                                      &1.0      &0.1            &25\\
    68$^{\circ}$~8    &5.57  &$-6.50$ $\pm0.26$                   &210  $\pm65$       &300                                      &1.0      &0.1            &25\\
    67$^{\circ}$~195  &4.65  &$-7.50$ $\pm0.35$                   &210  $\pm62$       &280                                      &1.0      &0.1            &25\\
       \noalign{\smallskip}
       \hline
    \end{tabular}
\end{table}

\subsubsection{Terminal wind velocity $\varv_{\infty}$} 
\begin{figure}
	\includegraphics[width=\hsize]{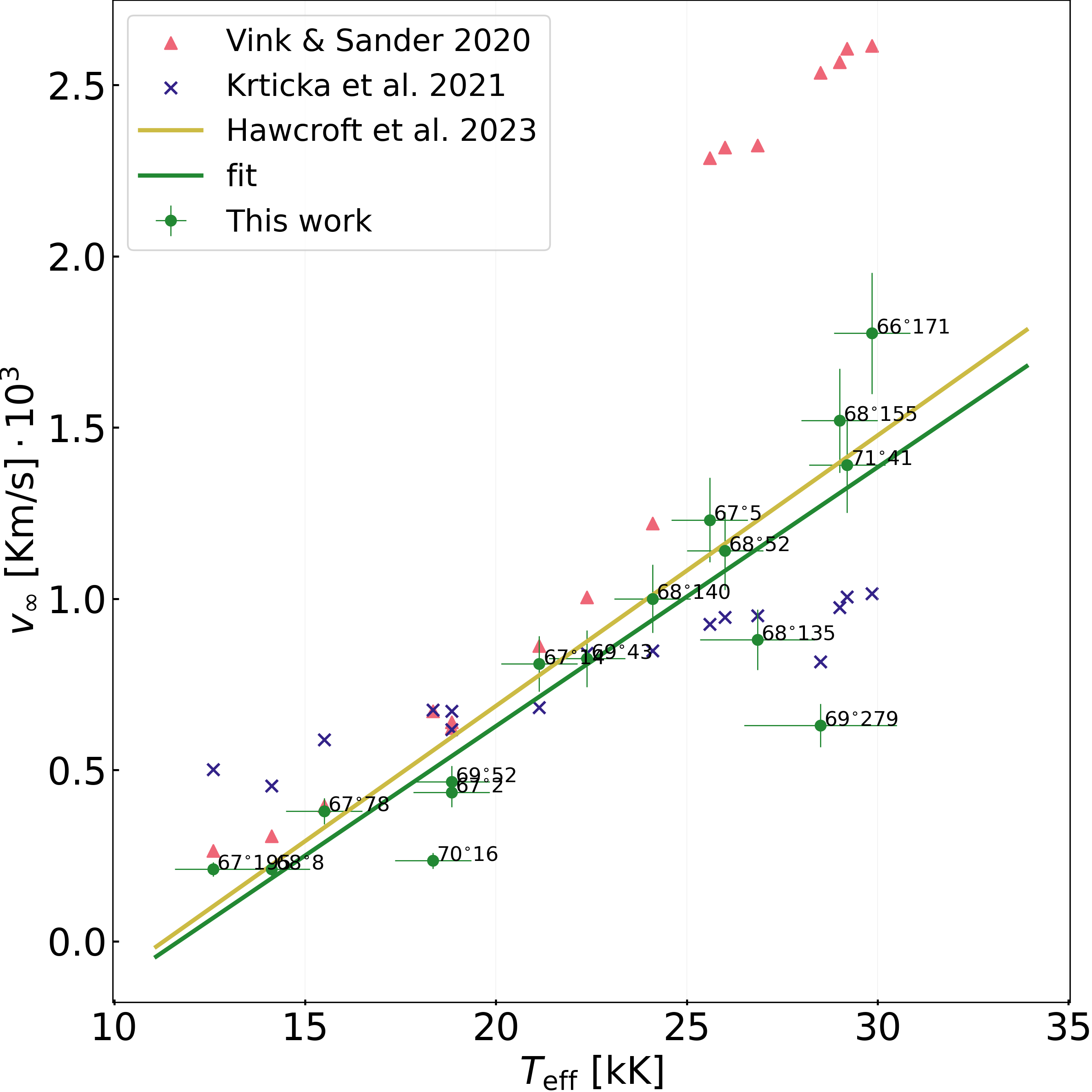}
    \caption{$v_{\infty}$ versus $T_{\rm eff}$ for LMC supergiants. Green dots: results from this work, green line: linear fit of our results, yellow line: $\varv_{\infty}$-$T_{\rm eff}$
    relation from \citet{xshootu3}, pink triangles: $\varv_{\infty}$ calculated velocities from \citet{vinksander2021} recipe, violet diagonal cross:  $\varv_{\infty}$ calculated velocities from 
    \citep{krticka2021} recipe.}
    \label{v-t}
\end{figure}

measured wind velocities are presented in Table~\ref{table:4}. In Fig.~\ref{v-t} we compare our $\varv_{\infty}-(T_{\rm eff})$ to the empirical recipe of \citet{xshootu3},
adopting a simple linear fit of the form: 
\begin{equation}
  \label{eq:v-t} 
   \varv_{\infty} [{\rm km\,s^{-1}}] = aT_{\rm eff} [{\rm K}]~-~b.
\end{equation}
\begin{table}
  \caption{Slopes $a$, and offsets, $b$, of the linear fits to the terminal wind velocity equation (Equation~\ref{eq:v-t}) of this study, and \citet{xshootu3}}       
  \def\arraystretch{1.5}
  \label{table:vinft}     
  \centering      
  \small                               
  \addtolength{\tabcolsep}{+1.0em}
  \begin{tabular}{c c c}          
      \hline\hline{\smallskip}
      LMC           &$a$                   &$b$\\
                    &K                     &${\rm km\,s^{-1}}$ \\
\hline
This study          &$0.076\pm0.011$        &$884\pm260$    \\                          
\citet{xshootu3}    &$0.085\pm0.050$        &$1150\pm170$    \\                                                      
    \noalign{\smallskip}
      \hline
  \end{tabular}
\end{table}
In Table~\ref{table:vinft} we present the paramteres of the linear fit in Equation~\ref{eq:v-t}. We find that our $\varv_{\infty}$-$T_{\rm eff}$ relation agrees with the relation obtained by 
\citet{xshootu3} within the uncertainties. We note that the two outliers are the hypergiants Sk\,-69$^{\circ}$ 279 (Fig.~\ref{SK-69D279}) and Sk\,-68$^{\circ}$ 135 (Fig.~\ref{SK-68D135}), which have 
abnormally low $\varv_{\infty}$ compared to stars of similar temperatures.

We also compare our results to numerical predictions of wind velocity from \citet{vinksander2021} and \citet{krticka2024}. The velocities calculated using the recipe provided in 
\citet{vinksander2021}, which was obtained using a locally consistent Monte Carlo radiative transfer model \citep{mullerandvink2008}, agree with our measurements for stars located at the 'cool' side of the bi-stability jump predicted in \citet{vink2001} ($<25$~kK), where as we find large discrepancies in 
the wind velocities of the objects that are located above the 'hot' edge of the bi-stability jump. As for the velocities calculated by the recipe presented in \citet{krticka2024}, we find that 
there is a discrepancy in the general trend, but just as in our results, there is a lack of a downward jump in the temperature range $T_{\rm eff}\approx25{\rm -}21$~kK.

We also calculate the values of the escape velocity of each star (see Table~\ref{table:4}) using the evolutionary mass obtained via {\sc Bonnsai} as explained in Section~\ref{mass}. We discuss 
the implications of our findings later in Section \ref{dowespot}.

\subsubsection{Mass-loss and clumping}
\begin{figure}
    \centering
     \includegraphics[width=\hsize]{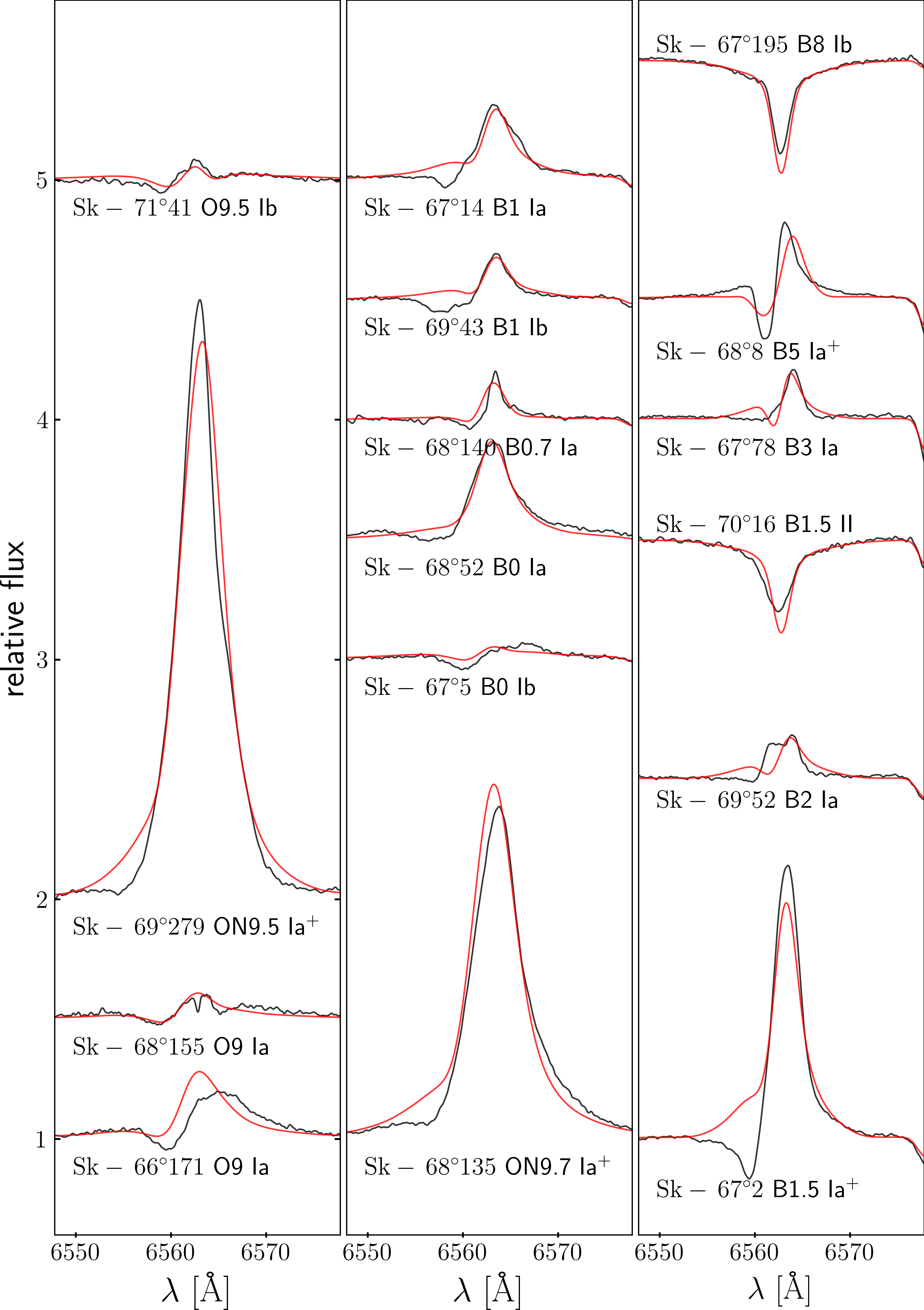}
         \caption{Spectral fits  to H$\alpha$ for our sample of OB-supergiants. Black line: observations, red line: model fit}
         \label{all_halpha}
 \end{figure}
\begin{figure}
	\includegraphics[width=\hsize]{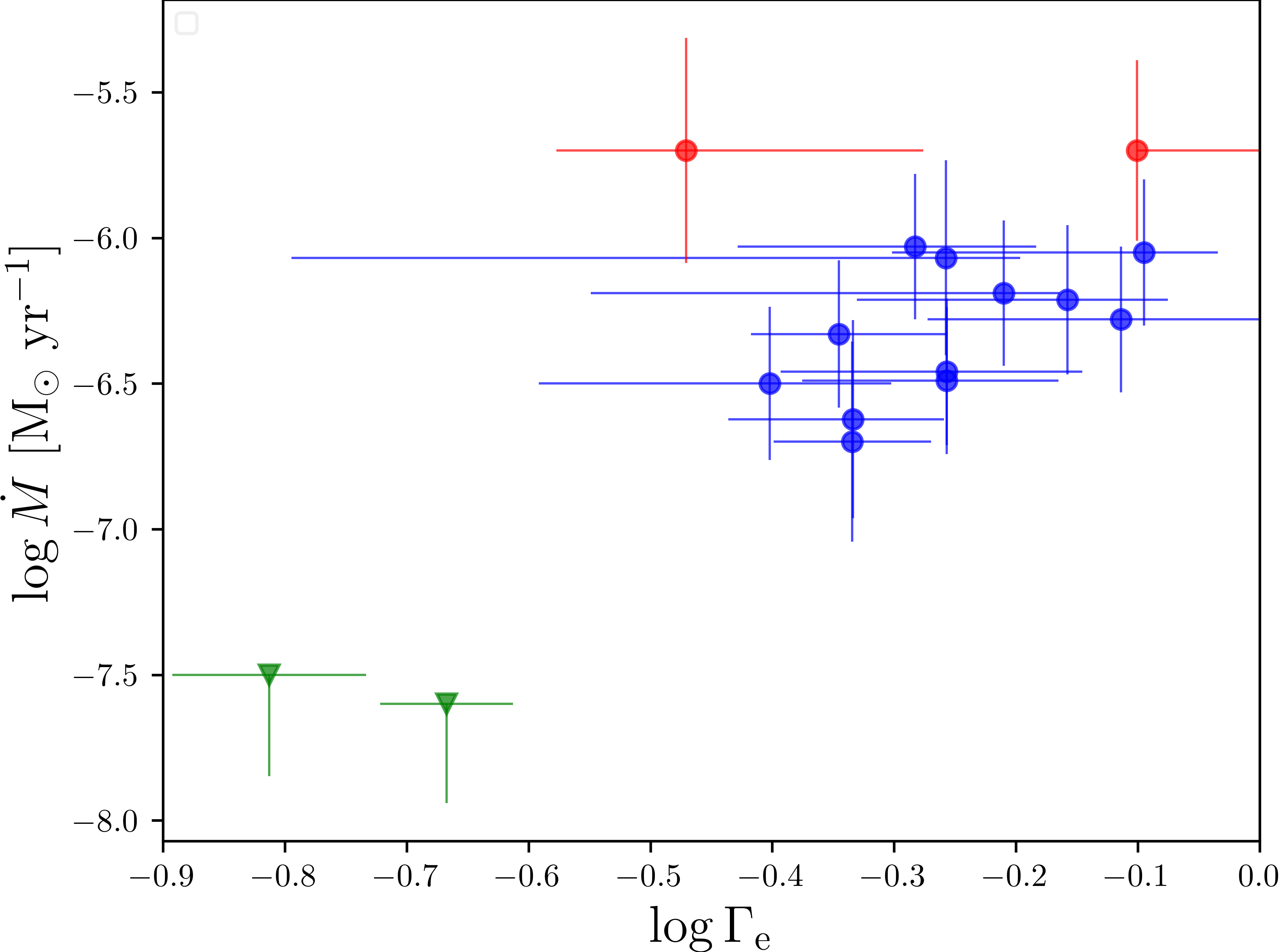}
    \caption{$\dot{M}$ versus $\Gamma_{\rm e}$. The green triangles indicate the objects Sk\,$-$67$^{\circ}$~195 and Sk\,$-$70$^{\circ}$~16, the mass-loss rates of which are 
    considered upper limits. The red points correspond to the hypergiants Sk\,$-$67$^{\circ}$~279 and Sk\,$-$68$^{\circ}$~135.}
    \label{edding}
\end{figure}
Fig.~\ref{all_halpha} presents the spectral fits to H$\alpha$ for all the stars in our sample. Aside from a few peculiar cases, our models well reproduce the overall shape and intensity of the 
emission feature in H$\alpha$ as shown in. The results for the true mass-loss rates that are scaled to the transformed radius and volume-filling factor are presented in Table~\ref{table:4}. 

A good indicator of wind strength is the Eddington ratio ($\Gamma_{\rm e} = L_{\rm bol}/L_{\rm edd}$), and it is expected that objects with higher $\Gamma_{\rm e}$ will have a greater mass-loss rate 
\citep{vink2011, bestenlehner2014}. In Fig.~\ref{edding} a correlation between $\Gamma_{\rm e}$ and $\dot{M}$ is revealed. 

\begin{figure}
  \centering
   \includegraphics[width=\hsize]{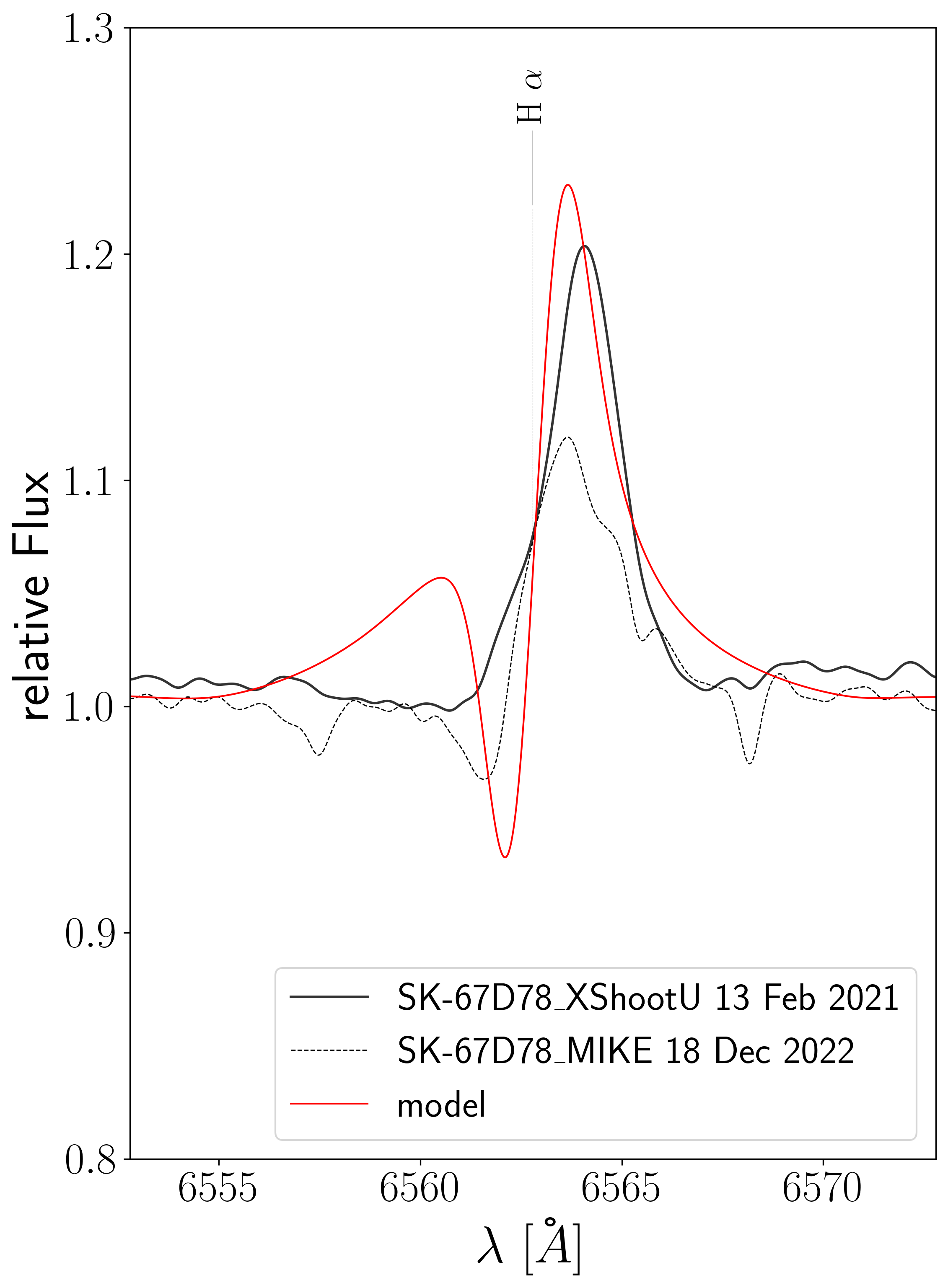}
       \caption{H$\alpha$ of Sk\,$-$67$^{\circ}$~78 from MIKE observations (black dashed line) overlaid onto H$\alpha$ from XshootU (black solid line) with our best fitting model (red solid line).}
       \label{Mike}
\end{figure}

Spectral fits in this study were applied to a single XshootU observation, so we present the Magellan/MIKE \citep{mike} H$\alpha$ spectrum of Sk\,$-$67$^{\circ}$~78 in Fig.~\ref{Mike}. 
H$\alpha$ is significantly weaker in MIKE observations with respect to XshootU. This would lead to different $\dot{M}$, $\beta$ and $f_{\rm vol}$ if we were to apply our fitting to MIKE data. 
Therefore, we include a factor of 2 in the uncertainties of the model mass-loss rates to take into account the possible variation in H$\alpha$ and the additional error from having non-simultaneous UV and optical 
observation. This puts the uncertainty for the derived mass loss rates in the range $\Delta \log{\dot{M}}\approx\pm0.2{\rm -}0.4$~dex.  

For Sk\,$-$70$^{\circ}$~16 (Fig.~\ref{SK-70D16}) and Sk\,$-$67$^{\circ}$~195 (Fig.~\ref{SK-67D195}) we adopt $\beta = 1.0$ due to H$\alpha$ being fully in absorption. For the rest of our 
sample, we find that higher $\beta=2.0\pm0.6$ are preferred to achieve a satisfactory fit for H$\alpha$. This is in agreement with \citet{Crowther2006} in which \textsc{CMFGEN} was used to model 
cool Galactic B-supergiants and they find an average value of $2.0$ for $\beta$ is necessary to achieve a good fit. This is also in agreement with \citet{haucke2018}, in which they analysed Galactic 
B-supergiants using \textsc{FASTWIND} and find that the suitable average value for $\beta$ is also $\approx 2.0$. We present our $\beta$ values in Table~\ref{table:4}.

\begin{figure}
  \centering
\includegraphics[width=\hsize]{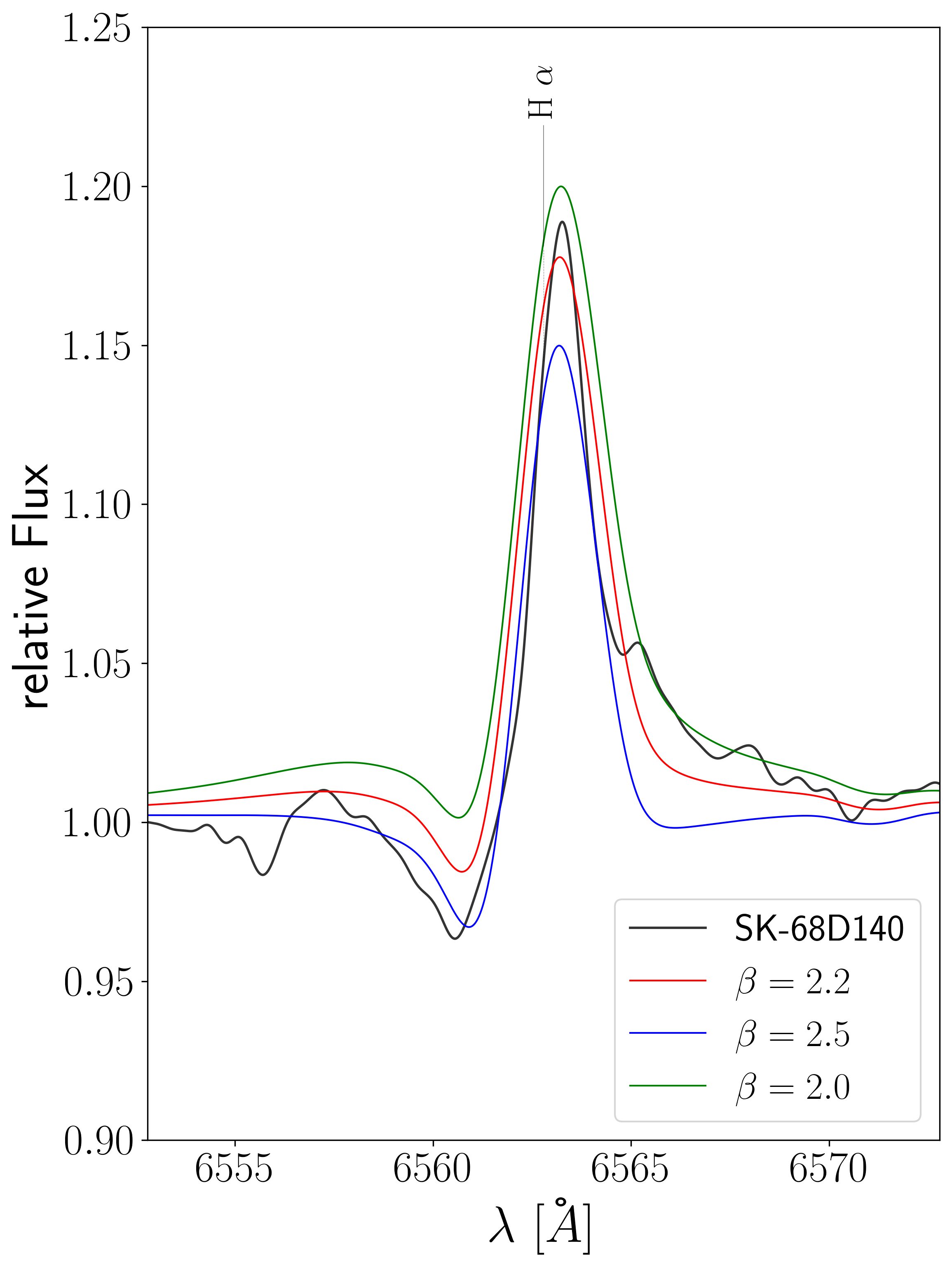}
      \caption{Best fitting model $\beta = 2.2$ (red solid line) versus two other models with $\beta=2.5$ (blue solid line) and $\beta=2.0$ (green solid line) for H$\alpha$ of Sk\,$-$68$^{\circ}$~140 (black solid line)}
      \label{beta_effect}
 \end{figure}

 \begin{figure*}
  \centering
  \includegraphics[scale=0.75]{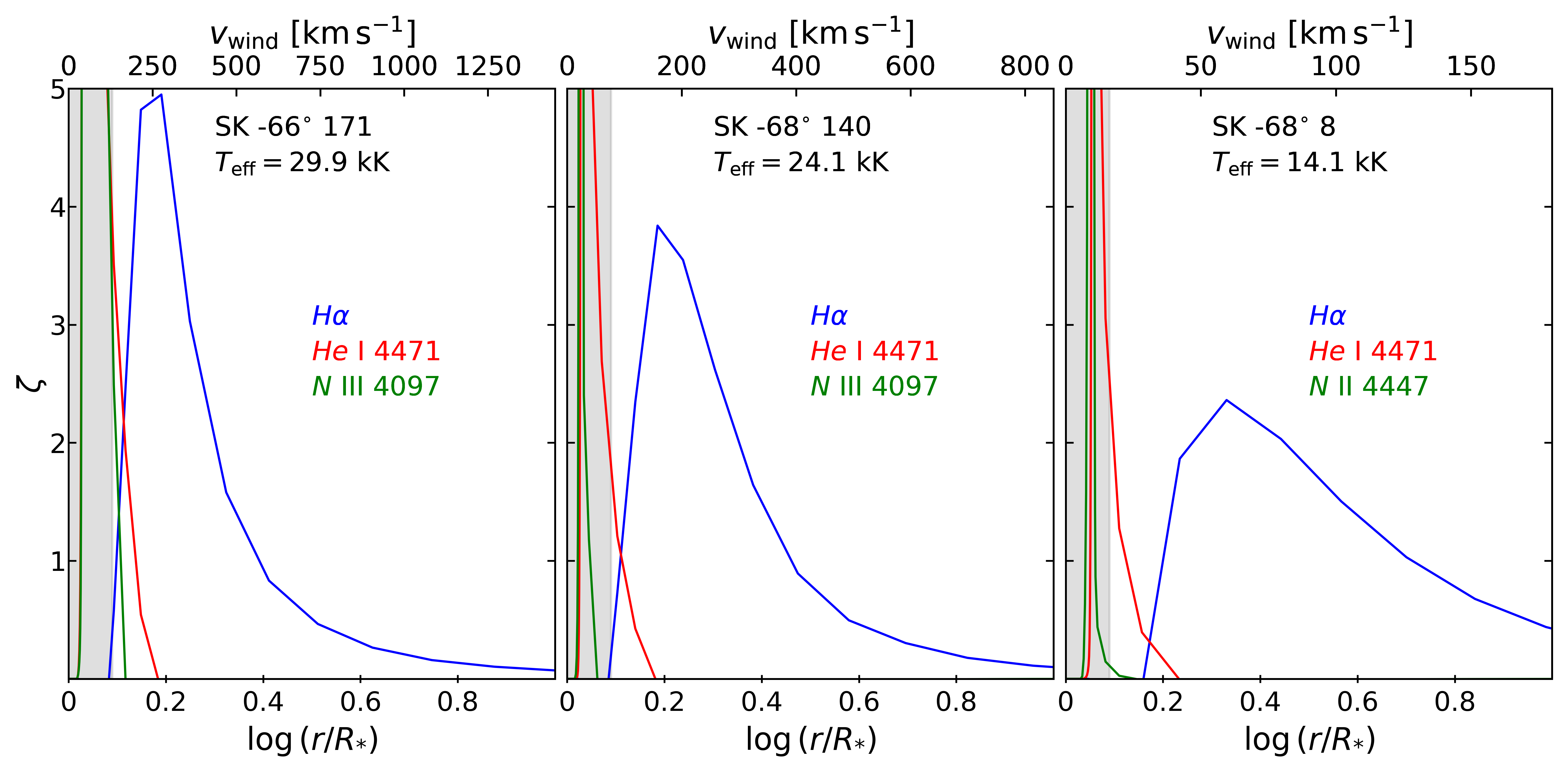}   
  \caption{Line formation region ($\zeta$ versus $\log{(r/R_{*})}$) of H$\alpha$ (blue solid line), $\ion{He}{I}~\lambda4471$ (red solid line), and a line from the dominating nitrogen ion 
  (green solid line), for Sk\,$-$66$^{\circ}$~171 ($T_{\rm eff} = 29.9~{\rm kK}$, left panel), Sk\,$-$68$^{\circ}$~140 ($T_{\rm eff} = 24.1~{\rm kK}$, middle panel), and Sk\,$-$68$^{\circ}$~8 ($T_{\rm eff} = 14.1~{\rm kK}$, right panel). 
  The grey shaded region indicates the photosphere.}
  \label{line_formation}
\end{figure*}

Fig.~\ref{beta_effect} demonstrates the effect of varying $\beta$ for Sk\,$-$68$^{\circ}$~140. In the middle panel of Fig.~\ref{line_formation}, we present the line-formation region of H$\alpha$ 
(blue solid line) for Sk\,$-$68$^{\circ}$~140. The peak is at $\log{(r/R_{*})}\approx0.2-0.4$, which is well into the sonic regime of the atmosphere, and is averaged over a wide range 
of $\varv_{\rm wind}$ ($\log{(r/R_{*})}\approx0.1-0.8$). From Equation~\ref{eq:vel_law} it is clear that increasing $\beta$ lowers the acceleration of the wind. This leads to lower wind velocities 
at H$\alpha$ line-formation region, where the wind has not yet reached it's terminal velocity, resulting in higher densities at this region. This yields narrower red-shifted emission and 
blue-shifted absorption in H$\alpha$ with less extended wings

In Fig.~\ref{beta_effect}, although the model with $\beta = 2.5$ (the blue solid line on Fig.~\ref{beta_effect}) fits the blue shifted absorption better its emission is weaker than the observation, 
and the model with $\beta = 2.0$ (green solid line) fits the extended red wing of the emission but has an overly extended blue wing and is stronger than the observed emission. Therefore we select 
the model with $\beta = 2.2$ (red solid line), which matches the observed emission and fits the overall morphology of the line. 

\subsection{Line-broadening parameters}
\label{LBP}
In Table~\ref{table:2}, we include $\varv_{\rm rot}\sin{i}$, the values of which are within reasonable agreement with the spectroscopic pipeline results from \citet{Bestenlehner2025}. 
The values of $\varv_{\rm rot}\sin{i}$ we obtain for our stars are in the range $\approx25{\rm -}80~{\rm km\,s^{-1}}$.

\begin{table}
  \caption{Comparison of the broadening paramaters adopted in our analysis to those obtained via \textsc{IACOB-BROAD} \citep{simondiaz2014} applied to high resolution MIKE data.}   
  \def\arraystretch{1.5}
  \label{table:rot}      
  \centering      
  \small                              
  \addtolength{\tabcolsep}{-0.0em}
  \begin{tabular}{c c c | c c c}      
      \hline\hline{\smallskip}
Sk\,$-$                           &\multicolumn{2}{c}{XShootU}                                                      &\multicolumn{3}{c}{\textsc{IACOB-BROAD}} MIKE\\
      \hline{\smallskip}
                    &$\varv_{\rm rot}\sin{i}$                 &$\varv_{\rm mac}$                &$\varv_{\rm rot}\sin{i}$                     &$\varv_{\rm mac}$                    &line\\
                    &${\rm km\,s^{-1}}$                       &${\rm km\,s^{-1}}$               &${\rm km\,s^{-1}}$                           &${\rm km\,s^{-1}}$                  &\\ 
\hline
67$^{\circ}$~2      &45                                       &20                               &$46^{+10}_{-15}$                             &$40^{+17}_{-20}$                    &$\ion{Si}{III}~\lambda4552$\\
67$^{\circ}$~78     &30                                       &20                               &$39^{+10}_{-29}$                             &$23^{+33}_{-22}$                    &$\ion{Si}{III}~\lambda4552$\\
70$^{\circ}$~16     &35                                       &20                               &$40^{+10}_{-25}$                             &$35^{+13}_{-17}$                    &$\ion{Si}{III}~\lambda4552$\\
68$^{\circ}$~8      &40                                       &20                               &$38^{+7}_{-15}$                              &$22^{+22}_{-20}$                    &$\ion{Si}{III}~\lambda4552$\\
67$^{\circ}$~195    &30                                       &20                               &$28^{+8}_{-14}$                              &$28^{+14}_{-16}$                    &$\ion{Si}{II}~\lambda6347$\\
    \noalign{\smallskip}
      \hline
  \end{tabular}
\end{table}
In Table~\ref{table:rot}, we compare our derived $\varv_{\rm rot}\sin{i}$, which were obtained with the assumption of $\varv_{\rm mac}=20~{\rm km\,s^{-1}}$, to the $\varv_{\rm rot}\sin{i}$ and 
$\varv_{\rm mac}$ that were obtained by applying the \textsc{IACOB-BROAD} \citep{simondiaz2014} tool to a subsample of the stars analysed in this study that has high resolution MIKE data. 
\textsc{IACOB-BROAD} is a procedure based on a combined fourier transform and goodness of fit approach that allows for the extraction of line-broadening parameters from a single snapshot of OB-type 
star spectra. 

We utilise $\ion{Si}{III}~\lambda4552$ for all stars in the subsample except the for the late supergiant Sk\,$-$67$^{\circ}$~195, for which we use $\ion{Si}{II}~\lambda6347$. We 
find that our derived $\varv_{\rm rot}\sin{i}$ are in good agreement with those obtained from \textsc{IACOB-BROAD}. On the other hand, we systematically underestimate $\varv_{\rm mac}$ by 
$\approx~10~{\rm km\,s^{-1}}$. This implies that we could potentially be underestimating the CNO abundances in our sample, albeit this understimation is well within our adopted uncertainties 
of $0.3$~dex.

\subsection{He and CNO abundances}
\label{abund_res}
\begin{table}
    \caption{Best fitting photospheric abundances $\epsilon_{\rm X}=\log{{\rm X}/{\rm H}}+12$. The final two columns are the cumulative CNO abundances relative to LMC baseline \citep{xshootU1} and 
    solar \citep{Asplund2005}, respectively. The adopted uncertainty on CNO abundances is $\pm 0.3$}          
    \def\arraystretch{1.2}
    \label{table:3}   
    \centering      
    \small                        
    \begin{tabular}{c c c c c c c}   
        \hline\hline{\smallskip}
    Sk\,$-$             &$Y$                &$\epsilon_{\rm C}$            &$\epsilon_{\rm N}$           &$\epsilon_{\rm O}$          &$\frac{\Sigma {\rm CNO}}{\Sigma {\rm CNO}_{\rm LMC}}$  &$\frac{\Sigma {\rm CNO}}{\Sigma {\rm CNO}_{\odot}}$\\
\hline
66$^{\circ}$~171        &0.43 $\pm$ 0.085   &8.05                          &8.19                         &8.42                        &1.2                                                    &0.5\\
68$^{\circ}$~155        &0.43 $\pm$ 0.085   &7.53                          &8.25                         &8.42                        &1.1                                                    &0.5\\
69$^{\circ}$~279        &0.36 $\pm$ 0.073   &7.53                          &8.02                         &8.07                        &0.6                                                    &0.3\\
71$^{\circ}$~41         &0.30 $\pm$ 0.061   &7.79                          &7.50                         &8.70                        &1.7                                                    &0.8\\
68$^{\circ}$~135        &0.36 $\pm$ 0.073   &7.53                          &8.24                         &8.17                        &0.9                                                    &0.4\\
67$^{\circ}$~5          &0.47 $\pm$ 0.094   &7.61                          &7.62                         &8.45                        &0.8                                                    &0.4\\
68$^{\circ}$~52         &0.30 $\pm$ 0.061   &7.49                          &7.50                         &8.33                        &0.8                                                    &0.4\\
69$^{\circ}$~43         &0.30 $\pm$ 0.061   &7.49                          &7.50                         &8.33                        &0.8                                                    &0.4\\
68$^{\circ}$~140        &0.30 $\pm$ 0.061   &7.19                          &8.10                         &8.33                        &1.0                                                    &0.5\\
67$^{\circ}$~2          &0.43 $\pm$ 0.085   &7.32                          &8.40                         &8.34                        &1.1                                                    &0.5\\
67$^{\circ}$~14         &0.36 $\pm$ 0.073   &7.53                          &7.72                         &8.67                        &1.4                                                    &0.7\\
69$^{\circ}$~52         &0.43 $\pm$ 0.085   &7.53                          &8.36                         &8.42                        &1.2                                                    &0.5\\
70$^{\circ}$~16         &0.30 $\pm$ 0.061   &7.39                          &8.24                         &8.33                        &1.1                                                    &0.5\\
67$^{\circ}$~78         &0.30 $\pm$ 0.061   &7.65                          &7.98                         &8.33                        &1.0                                                    &0.5\\
68$^{\circ}$~8          &0.40 $\pm$ 0.079   &7.85                          &8.47                         &8.39                        &1.4                                                    &0.7\\
67$^{\circ}$~195        &0.36 $\pm$ 0.073   &7.71                          &8.02                         &8.37                        &1.0                                                    &0.5\\
      \noalign{\smallskip}
        \hline
    \end{tabular}
  \end{table}

\begin{figure}
  \centering
   \includegraphics[width=\hsize]{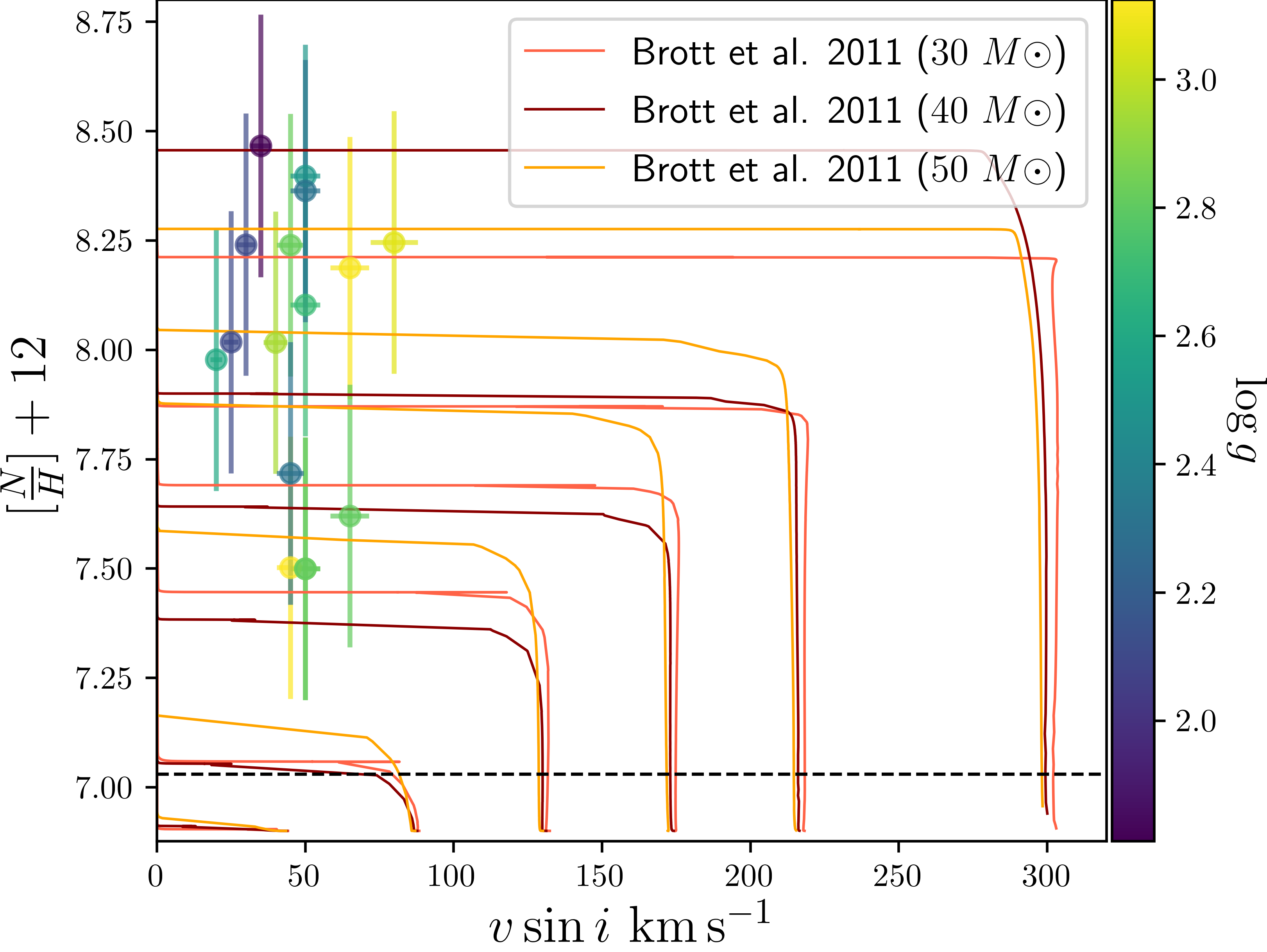}
       \caption{Nitrogen abundance (by number) versus $\varv_{\rm rot}\sin{i}$. The colour scheme corresponds to the value of $\log{g}$, and the evolutionary tracks are computed for 
       initial masses of $30~M_{\odot}$, $40~M_{\odot}$ and $50~M_{\odot}$ (red, brown, and orange solid lines respectively) for initial rotational velocities of $50,~110,~170,~220,~270~380~{\rm km\,s^{-1}}$ 
       \citep{brott2011}, which we multiplied by a factor of $\pi/4$ to take into account the inclination of the rotation axis \citep{hunter2008}. The black dashed line represents the LMC 
       baseline nitrogen abundance.}
       \label{N_vsini}
  \end{figure}
Table~\ref{table:3} presents the results for the chemical abundances of He and CNO elements. The analysis shows an overall boost in helium at the expense of hydrogen relative to 
the LMC baseline helium mass fraction of $Y = 0.25$ \citep{brott2011}. This is to be expected for a sample of evolved stars. We would also expect nitrogen enrichment at the expense of oxygen and 
carbon which get depleted due to CNO-cycle processing. Indeed, nitrogen abundances are significantly higher than the LMC baseline (Fig.~\ref{N_vsini}), with a mean enhancement of $1$~dex and a 
spread of $0.5$~dex. We also note that our sample is mostly carbon and oxygen depleted with the exception of Sk\,$-$71$^{\circ}$~41 (Fig.~\ref{SK-71D41}) and Sk\,$-$67$^{\circ}$~2 
(Fig.~\ref{SK-67D2}) where we find a $0.2$ dex enhancement in oxygen. 
  
\begin{figure*}
  \centering
  \includegraphics[width=\hsize]{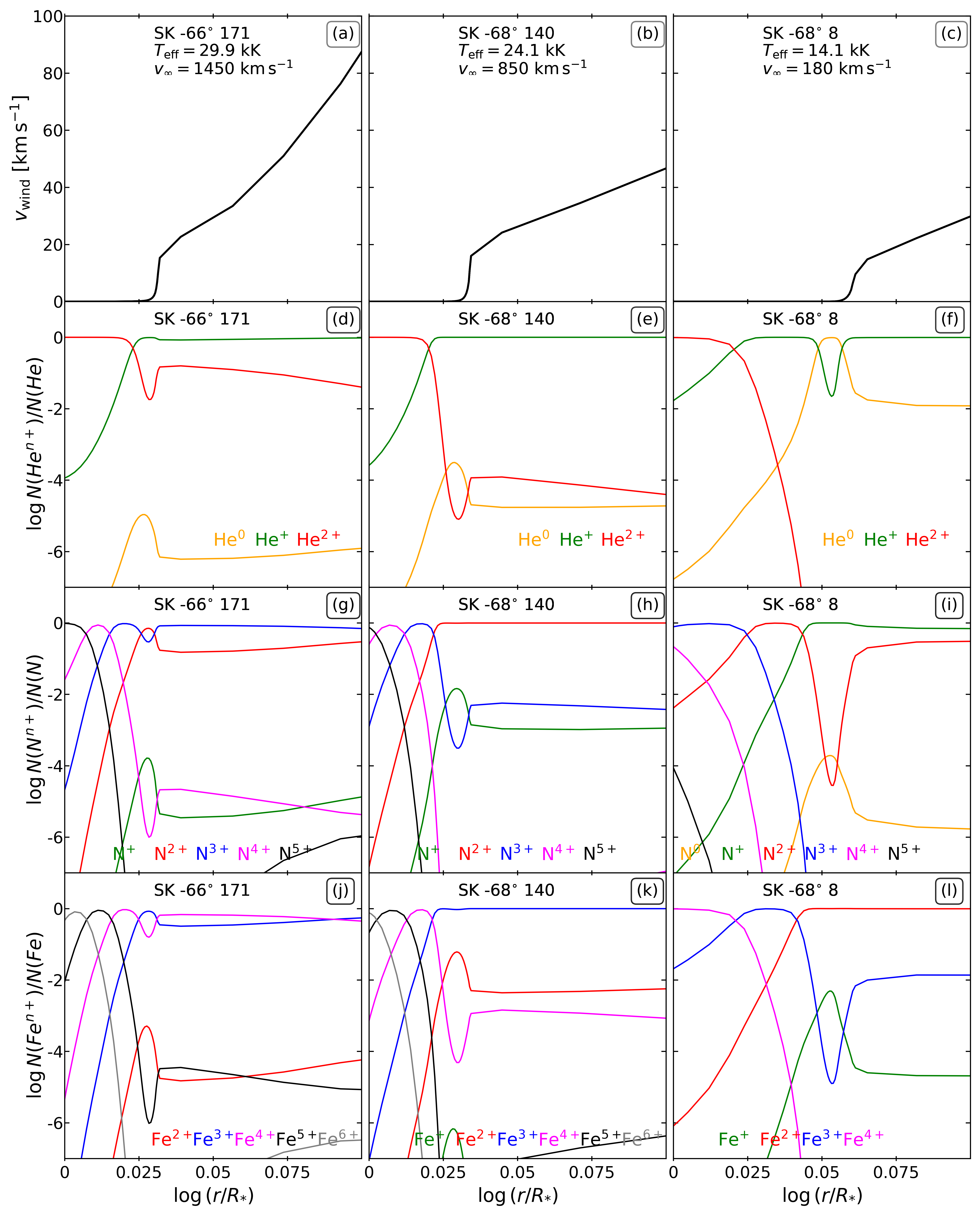}   
  \caption{Each column, from left to right, adheres to Sk\,$-$66$^{\circ}$~171 ($T_{\rm eff} = 29.9~{\rm kK}$), Sk\,$-$68$^{\circ}$~140 ($T_{\rm eff} = 24.1~{\rm kK}$), and Sk\,$-$68$^{\circ}$~8 ($T_{\rm eff} = 14.1~{\rm kK}$), 
          respectively.
          $a$, $b$, and $c$: Wind verlocity as a fucntion of radius in stellar radii units $\log{(r/R_{*})}$. 
          $d$, $e$, and $f$:ionization structure (relative ionic density $\log{(N(i^{n+})/N(i_{\rm tot}))}$ versus $\log{(r/R_{*})}$) of helium. 
          $g$, $h$, $i$: Ionization structure of nitrogen.
          $j$, $k$, and $l$: Ionization structure of iron. 
          $i^{0}$ (orange), $i^{+}$ (green), $i^{2+}$ (red), $i^{3+}$ (blue), $i^{4+}$ (magenta), $i^{5+}$ (black), $i^{6+}$ (grey) where i is the element.}
  \label{ion_struct}
\end{figure*}
Recalling Section \ref{abundances_method}, it is necessary to use lines of lower ionization stages for B-stars that O-stars. In panels $d$-$l$ of Fig.~\ref{ion_struct}, we use the best fitting 
models for Sk\,$-$66$^{\circ}$~171, Sk\,$-$68$^{\circ}$~140, and Sk\,$-$68$^{\circ}$~8 (left to right, respectively), which corresponds to temperatures of $29.9$, $24.1$, and $14.1~{\rm kK}$, 
respectively, to plot the radial ionization structure of helium, nitrogen and iron. This shows that the atmosphere is dominated by lines of different ionization levels depending on the temperature, 
which illustrates the reason behind fitting lines of lower ionization stages with lower $T_{\rm eff}$.

In Fig.~\ref{line_formation}, we show the line formation regions of different lines versus radius $\varv_{\rm wind}$, and we can see that photospheric lines like 
$\ion{He}{I}~\lambda4471$ (red solid lines), $\ion{N}{III}~\lambda4097$, and $\ion{N}{II}~\lambda4447$ (green solid lines) form around $\varv_{\rm wind}\approx10-15~{\rm km\,s^{-1}}$, which 
coincides with the transition point between the subsonic and super-sonic regimes, connecting the photosphere to the inner region of the stellar winds. On the topmost panels of 
Fig.~\ref{ion_struct}, the transition point is where the velocity jumps from $0~{\rm km\,s^{-1}}$ to $\approx15~{\rm km\,s^{-1}}$.

From Fig.~\ref{N_vsini}, we see that one could reproduce the overall distribution of our sample's observed nitrogen enhancement via evolutionary tracks that include rotational
mixing \citep{brott2011}. However, extremely high initial rotational velocities do not agree with what is observed for O-stars and the findings of \citet{ramirez2013}, who obtained the 
rotational properties of a large sample of LMC O-stars (216 stars), and concluded that the distribution of $\varv_{\rm rot}\sin{i}$ peaks at $\approx 80~{\rm km\,s^{-1}}$.
We also do not find a clear positive correlation between rotation rates and nitrogen enhancement which is similar to the findings of \citet{hunter2008}.

\begin{figure}
  \centering
   \includegraphics[width=\hsize]{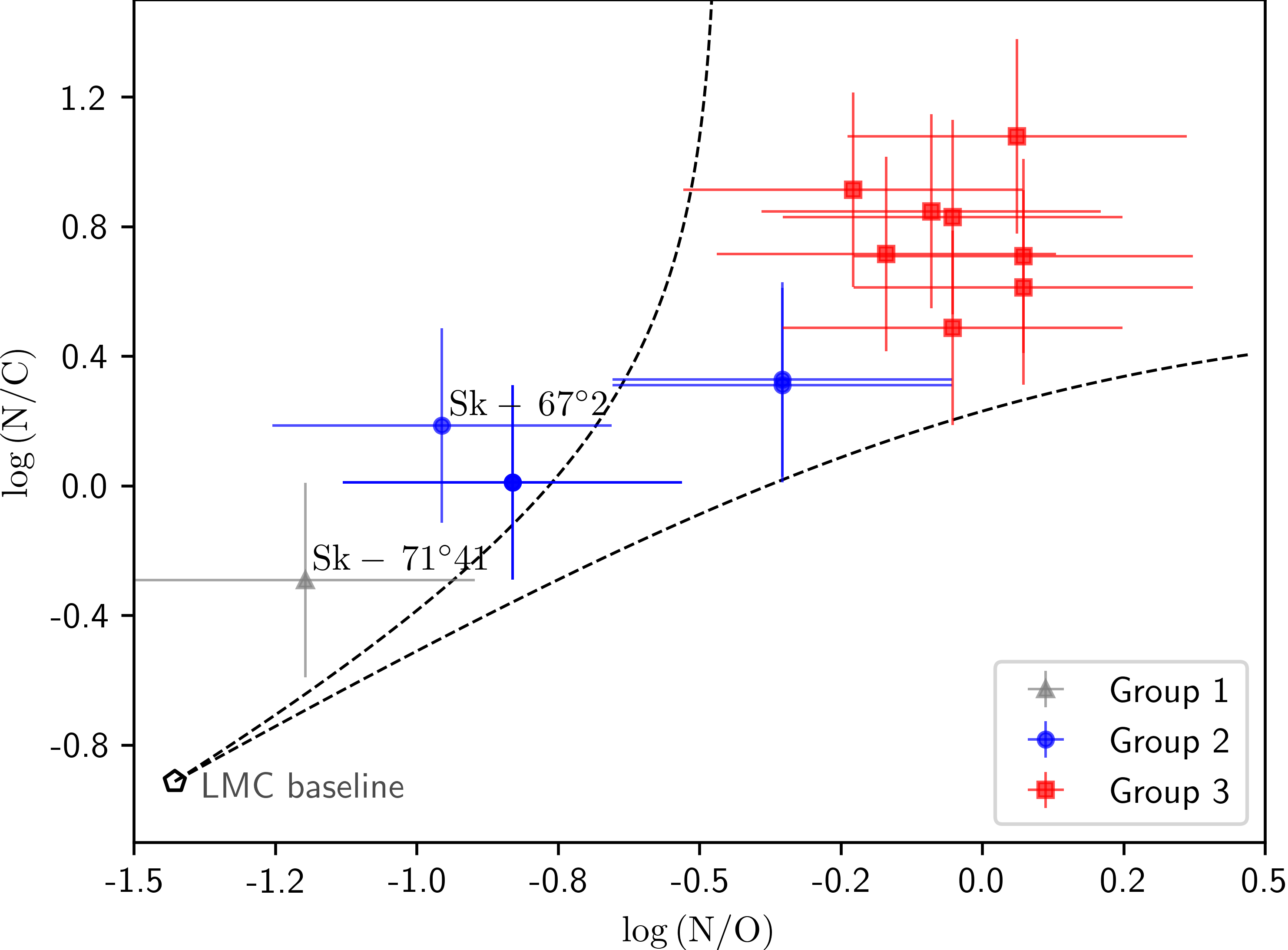}
       \caption{$\log{{\rm N}/{\rm C}}$ versus $\log{{\rm N}/{\rm O}}$ of our sample (grey triangles, blue circles, and red squares). The upper and lower boundry (dashed black lines) are adopted 
       from \citet{maeder2014}. The sample is divided into three groups depending on their N/C and N/O values similar to \citet{menon2024}.}
       \label{NO_NC}
  \end{figure}
In Fig.~\ref{NO_NC}, we present the logarithm number ratios N/C and N/O of our sample (blue circles). \citet{maeder2014} analytically obtains the upper and lower bounds (black dashed lines) 
from nuclear constaints ($^{12}$C is turned immediately into $^{14}$N and the number of carbon atoms is constant throughout the CNO cycle) on the the changes of N/C versus N/O ratios during the CNO cycle. 
We find that the majority of our stars are within those bounds. The exception are the stars Sk\,$-$71$^{\circ}$~41, Sk\,$-$67$^{\circ}$~5, Sk\,$-$68$^{\circ}$~52, Sk\,$-$68$^{\circ}$~140, which 
fall outside of the boundaries, but are within the uncertainties. In Fig.~\ref{NO_NC}, we divide the stars into groups following the criteria introduced by \citet{menon2024}, who obtains 
the CNO abundances of 59 early B-supergiants in the LMC using \textsc{FASTWIND}. We find that only Sk\,$-$71$^{\circ}$~41 is part of \textit{Group 1}, which is the group whose number ratios can 
be explained by a single star evolutionary scenario. \textit{Group 2}, whose number ratios can be obtained via blue loop models, fast rotating ($\varv_{\rm init, rot}=220{\rm km\,s^{-1}}$) TAMS 
models, or merger models. The rest of the sample falls within \textit{Group 3}, the number ratios of which can only be explained by merger scenarios.

Although it is expected for the combined mass fraction of CNO elements of any given star in our sample to stay constant throughout its evolution and to be similar to that of the LMC, we do find 
that, in some cases, the aggregate CNO mass fraction required to obtain a satisfactory fit to the designated CNO lines significantly diverges from the baseline CNO mass fraction of the LMC 
($\Sigma {\rm CNO}/\Sigma {\rm CNO}_{\rm LMC}$ in Table~\ref{table:3}). The most extreme examples of this are Sk\,$-$71$^{\circ}$~41 and Sk\,$-$69$^{\circ}$~279, with ratios of summed CNO mass 
fraction to LMC CNO mass fraction of $\approx 1.7$ and $0.6$, respectively, indicating that these objects have a respectively higher and lower cumulative CNO mass fraction. This likely arises 
from the sensitivity of abundances to other parameters, mainly $T_{\rm eff}$, $\log{g}$. Fixing $\varv_{\rm mic}$, as explained in Section~\ref{lbp}, could be a contributing factor 
to this discrepancy.

\section{Discussion}
\label{discussion}
In Section~\ref{dowespot} we discuss the existence of the bi-stability jump in our sample. In Section~\ref{emp_comp} we compare our results to previous empirical studies. 
In Section~\ref{theor_comp}, we compare our results to various numerical predictions. Finally, in Section~\ref{met_eff} we explore the dependence of wind properties on metallicity.

\subsection{Do we spot a bi-stability jump?}
\label{dowespot}
\begin{figure}
    \centering
	\includegraphics[width=\hsize]{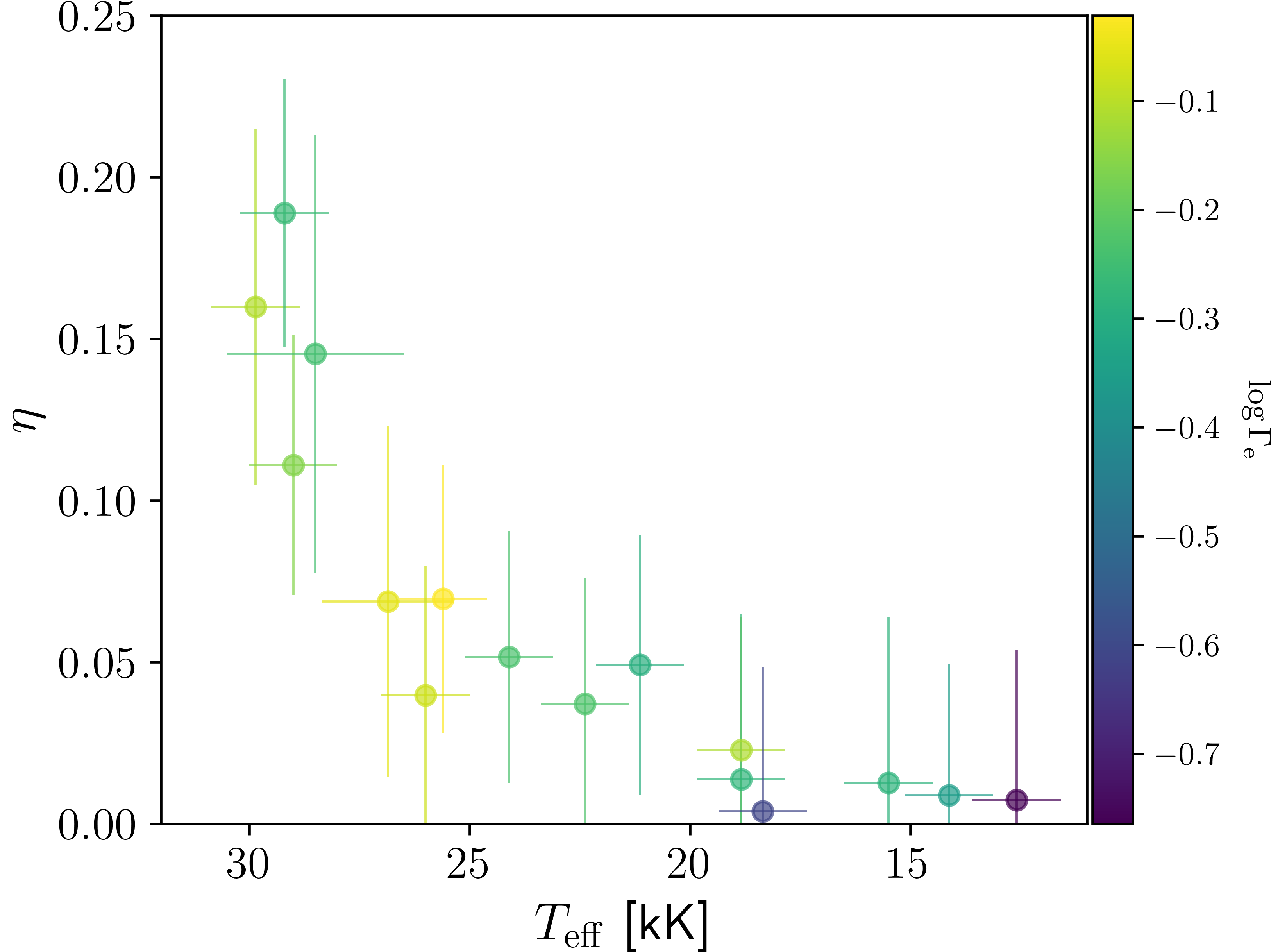}
        \caption{wind efficiency ($\eta = \frac{\dot{M}\varv_{\infty}}{L_{\rm bol}/c}$) in terms of effective temperature.}
        \label{wind_efficiency}
   \end{figure}
   \begin{figure}
    \centering
  \includegraphics[width=\hsize]{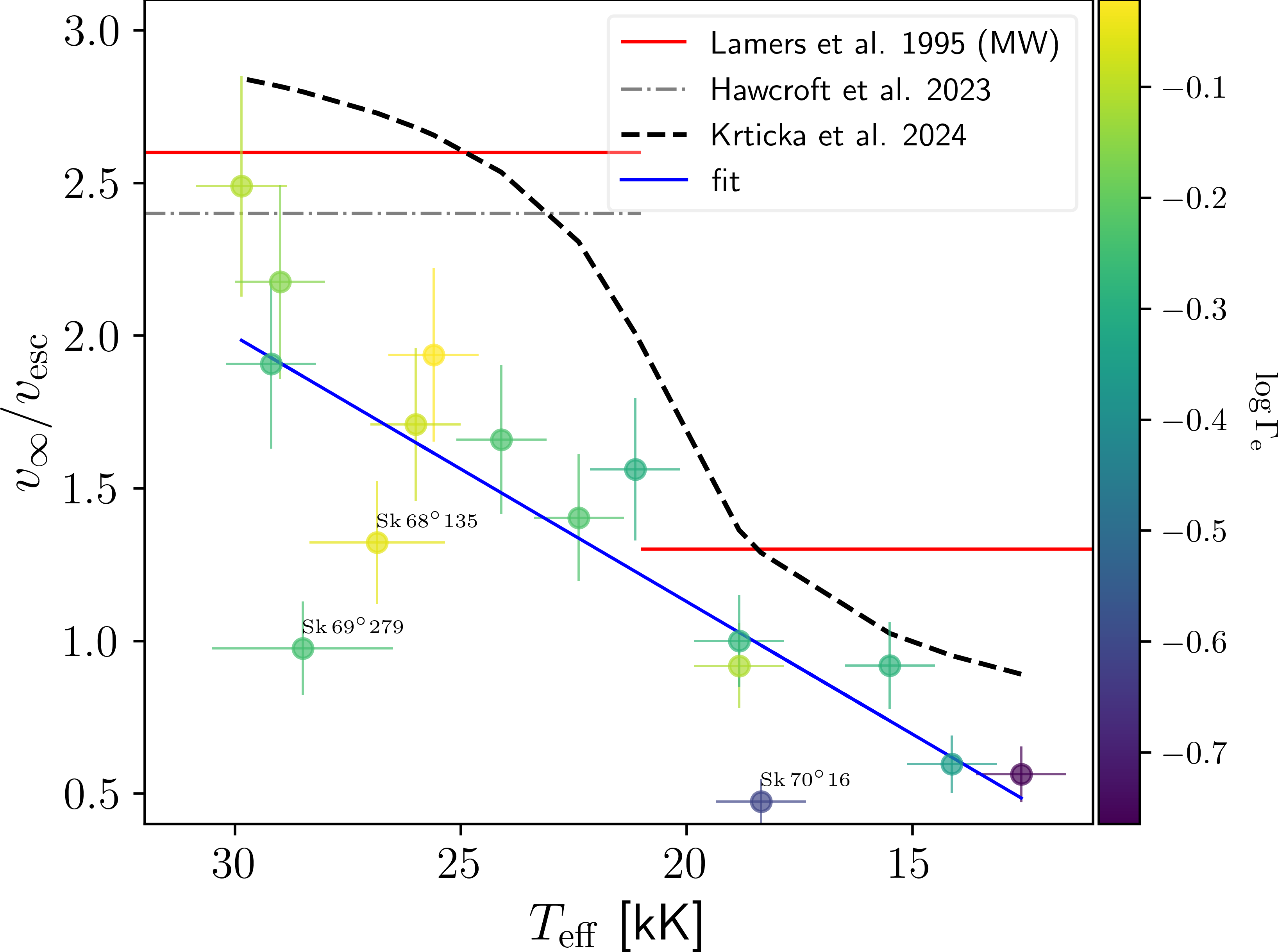}
        \caption{The ratio ($\varv_{\infty}/\varv_{{\rm esc}}$) as a function of temperature. The yellow dashed line is the relation presented in \citet{krticka2021}. The filled red lines represent the 
        ratios from \citet{lamers1995}. The green filled line is the average relation for stars of spectral type earlier than B1 from \citet{xshootu3}. The color gradient correlates to the
        value of $\Gamma_e$.}
        \label{vinfvesc}
   \end{figure}  

The main goal of this study is to explore the existence of the numerical mass-loss rate jump-like increase when going from the 'hot' regime to the 'cool' regime around $T_{\rm eff} = 25~{\rm kK}$.  
We attempted to alleviate the degeneracy of mass-loss and clumping that plagues the recombination line emission in H$\alpha$ by utilising the appropriate unsaturated P Cygni lines in the UV as 
previously mentioned. Fig.~\ref{wind_efficiency} shows our derived wind efficiency parameter ($\eta = \frac{\dot{M}\varv_{\infty}}{L_{\rm bol}/c}$) vs $T_{\rm eff}$. We see a clear downward trend 
in the overall sample. This is in agreement with \citet{benaglia2007}, who derive the mass-loss rates for a sample of Galactic OB-stars from thermal radio emission and find a similar monotonic downward 
trend in wind efficiency with lower temperatures. 

In Fig.~\ref{vinfvesc}, we compare the distribution of $\varv_{\infty}/\varv_{\rm esc}$ versus $T_{\rm eff}$ to empirical predictions from \citet{lamers1995} (MW) and \citet{xshootu3} 
(LMC), plus the $\varv_{\rm esc}$-$T_{\rm eff}$ prediction from \citet{krticka2021} which does not assume a bi-stability jump, but rather a smooth transition from high to low ratios. 
For our sample, we see a smooth decline in $\varv_{\infty}/\varv_{\rm esc}$ with lower temperatures and do not see any signs of a bi-stability jump around the proposed $T_{\rm eff}\approx25{\rm -}21~{\rm kK}$. 
Although the \citet{krticka2021} recipe produces a smooth decrease in $\varv_{\infty}/\varv_{\rm esc}$, our results do not match their prediction. The outliers are the hypergiants 
Sk\,$-$68$^{\circ}$~135 and Sk\,$-$69$^{\circ}$~279, which have low $\varv_{\infty}\big/\varv_{\rm esc}$. This is due to abnormally slow winds relative to other objects with similar temperatures. 

The vertical error bars in Fig.~\ref{vinfvesc} include the uncertainties of $\varv_{\infty}$ and the $\varv_{{\rm esc}}$. With a simple linear fit we obtain the relation:
\begin{equation}
  \label{eq:vinf_vesc_t} 
  \varv_{\infty}/\varv_{\rm esc}=4.1(\pm 0.8)\log{(T_{\rm eff}/{\rm K})}-16.3(\pm 3.5).
\end{equation}

\subsection{Comparison to previous empirical studies}
\label{emp_comp}
    \begin{figure}
    \centering
      \includegraphics[width=\hsize]{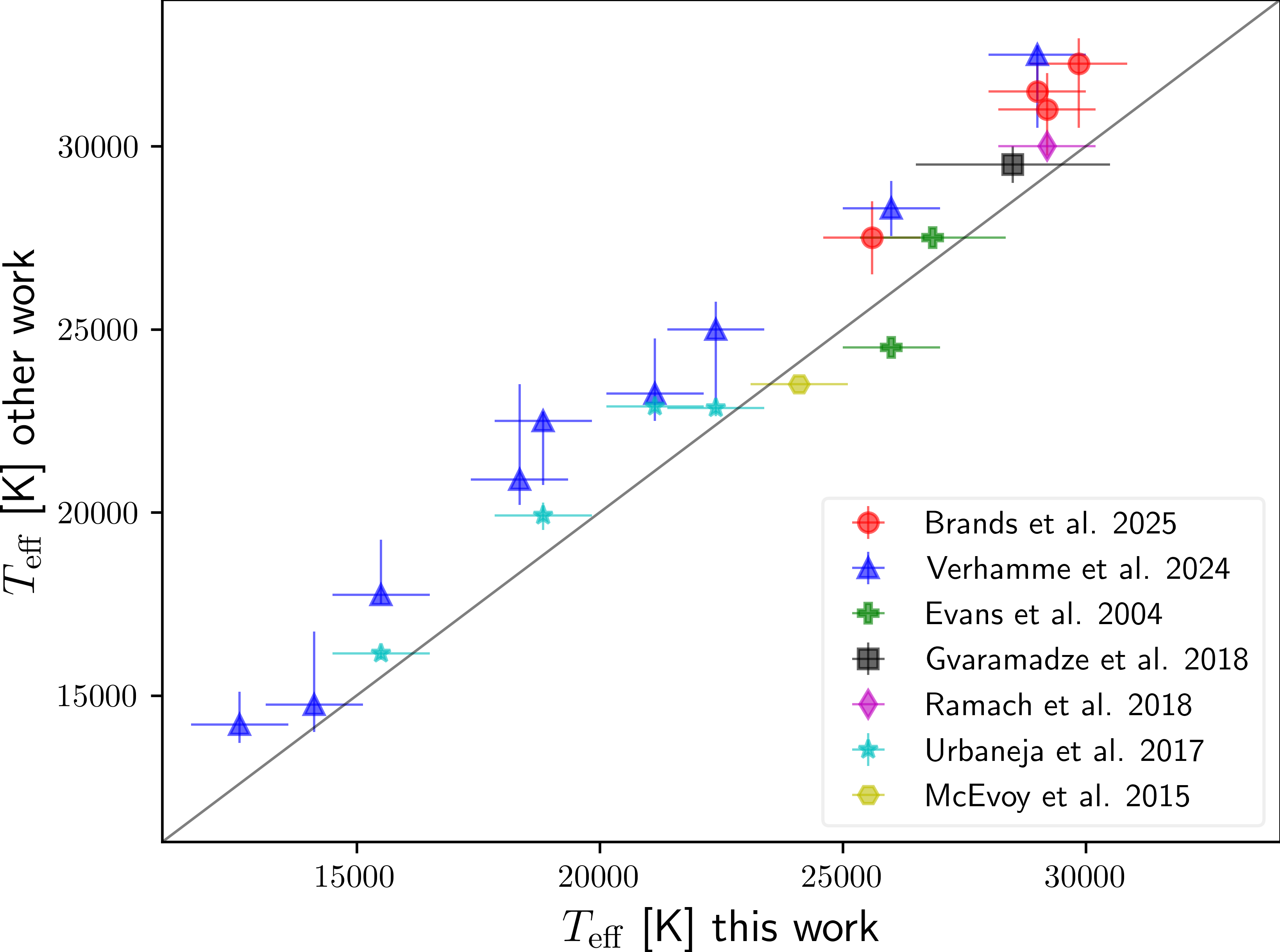}
    \caption{Comparison of our obtained $T_{\rm eff}$ to previous studies. on the x-axis we present our effective temperature for each star and on the y-axis is the the comparison $T_{\rm eff}$ 
    from previous analysis. Blue triangles: Comparison to values obtained by \citet{verhamme2024}, Red circles: Comparison to values obtained by \citet{brands2025}. 
    Green filled crosses: Comparison to values obtained by \citet{evans2004}. Black square: Comparison to values obtained by \citet{gvaramadze2018}. Magenta diamond: Comparison to values 
    obtained by \citet{ramachandran2018}. Cyan star: Our $T_{\rm eff}$ to the effective temperatures from \citet{urbaneja2017}. Yellow hexagon: Comparison to values obtained by \citet{McEvoy2015}.}
    \label{teff_compare}
    \end{figure}
    
    \begin{figure}
    \centering
      \includegraphics[width=\hsize]{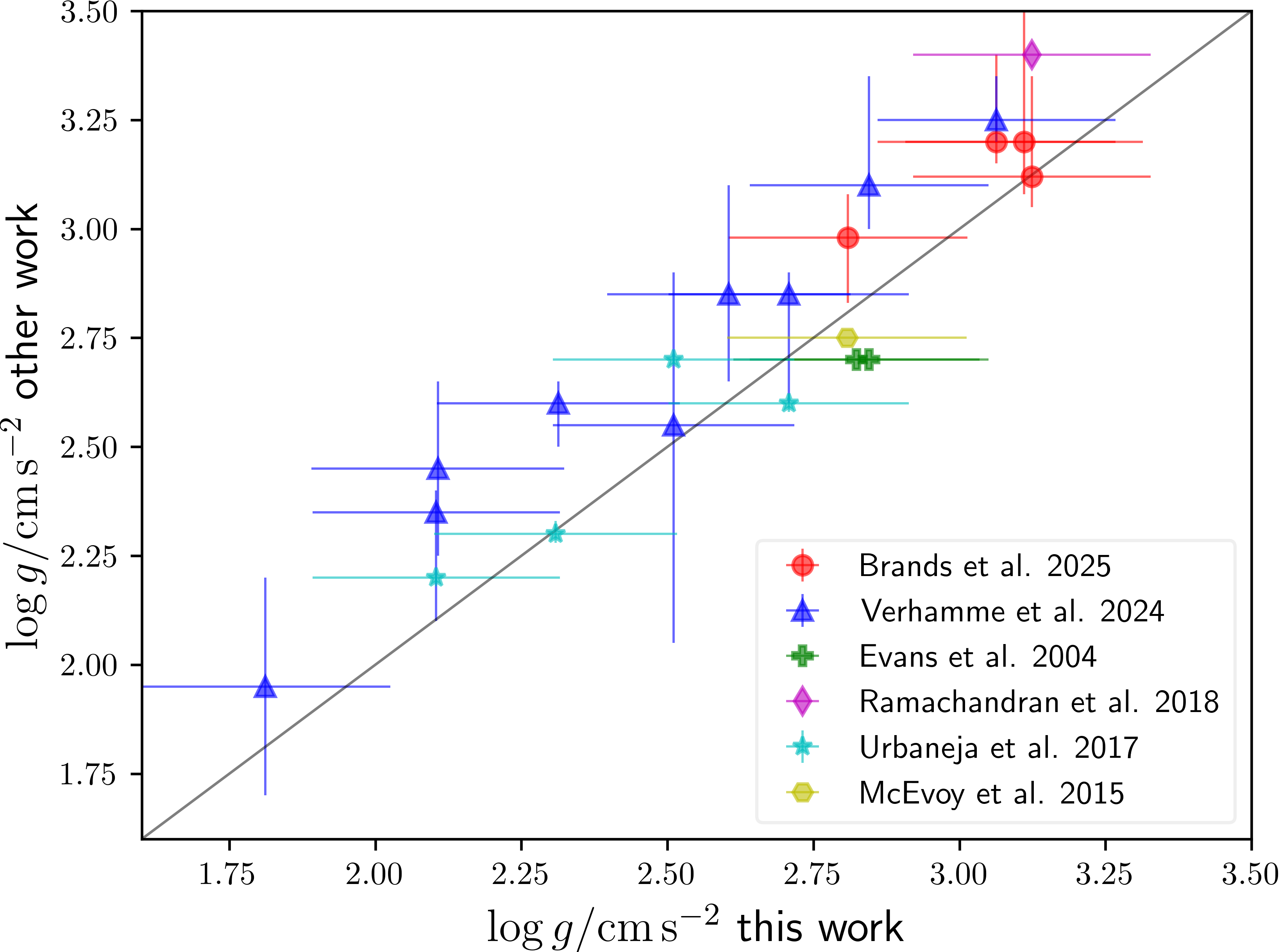}
    \caption{Comparison of our obtained $\log{g}$ to previous studies. The symbols follow the same encoding as in Fig.~\ref{teff_compare}.}
    \label{logg_compare}
    \end{figure}
    
    \begin{figure}
    \centering
      \includegraphics[width=\hsize]{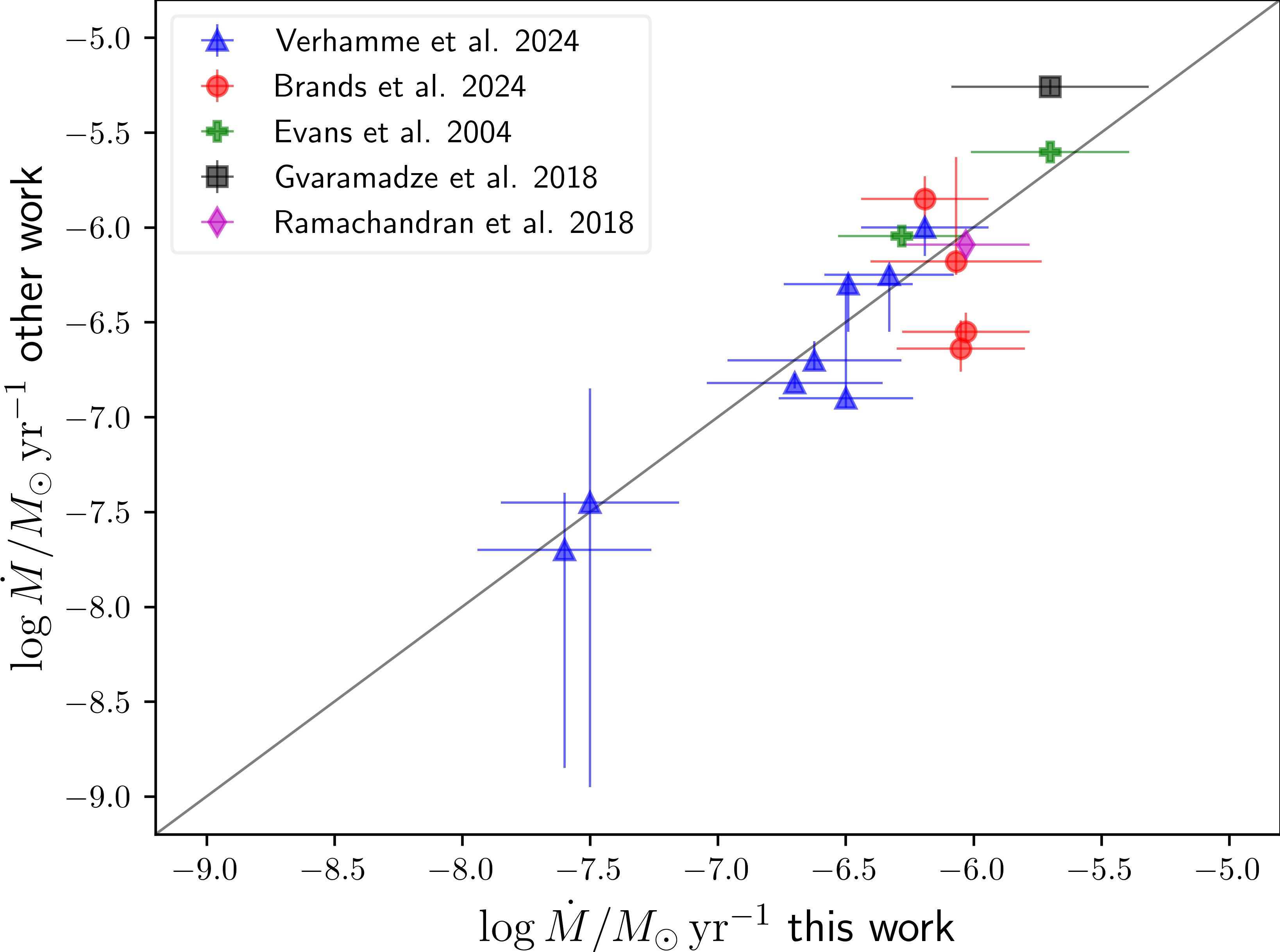}
    \caption{Comparison of our derived $\dot{M}$ to previous studies. The symbols follow the same encoding as in Fig.~\ref{teff_compare}.}
    \label{mdot_compare}
    \end{figure}  
        
    \begin{figure}
      \centering
      \includegraphics[width=\hsize]{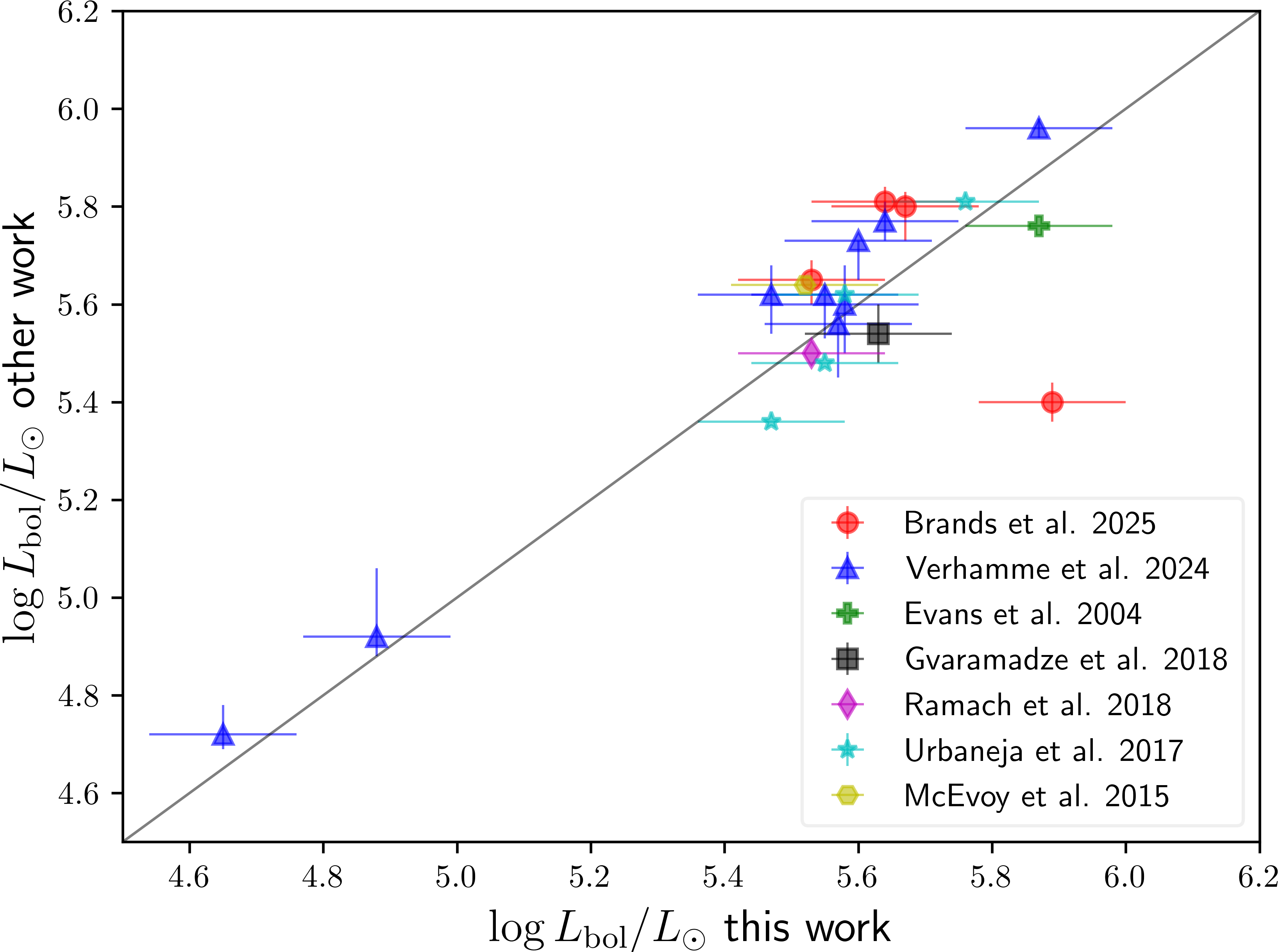}
    \caption{Comparison of our derived $\log{L_{\rm bol}}$ to previous studies. The symbols follow the same encoding as in Fig.~\ref{teff_compare}.}
    \label{logL_compare}
    \end{figure}

Due to differences in radiative transfer codes, in analysis methods and in observational datasets, and above all the lack of a clear unified methodology for calculating model and true 
uncertainties, we are not comparing like-for-like. The reason for the comparison is to check if the collection of those comparison studies produce an observable bi-stability jump.

\citet{sander2024} concludes that an average maximum spread in $T_{\rm eff}$, $\log{g}$, and $\log{\dot{M}}$ obtained with different methods applied to ULYSSES/XshootU data is $\Delta T_{\rm eff} \approx 3000~{\rm kK}$, 
$\Delta \log{(g/{\rm cm\,s^{-2}})} \approx 0.2$ and $\Delta \log{(\dot{M}/{\rm M_{\odot}\,{\rm yr}^{-1}})} \approx 0.4$. While the analysis conducted by \citet{sander2024} is very revealing 
of the discrepancies between the different radiative transfer codes and a great attempt at parametrizing the spread that could be expected when comparing results obtained by these codes, the sample 
was limited to two LMC and one SMC O-type stars. Therefore, the spread of parameters obtained in that study is not significant from a statistical point of view. 

In this section, we focus on comparing $T_{\rm eff}$, $\log{g}$, and $\log{\dot{M}}$. in Sections~\ref{UV+optical} and \ref{optical}, when comparing the mass-loss rates, we do not adjust the values
to the equivalent ``smoothed wind'' values (for $f_{\rm vol}=1$).

Fig.~\ref{teff_compare}-\ref{logL_compare} compare our derived values of $T_{\rm eff}$, $\log{g}$, $\dot{M}$, and $\log{L_{\rm bol}}$ respectively, versus values derived in previous studies. We 
compile our derived $T_{\rm eff}$, $\log{g}$, $\log{\dot{M}}$, $\log{L_{\rm bol}}$, $\beta$, and $f_{\rm vol}$ alongside the results of various previous empirical studies in 
Table~\ref{comparison_table} in Appendix~\ref{app:comp}.

\subsubsection{UV + Optical studies} 
\label{UV+optical} 
\citet{evans2004} has two targets in common with the present study, also employed \textsc{CMFGEN}. Their $T_{\rm eff}$ for the hypergiant Sk\,$-$68$^{\circ}$~135 is $1500$ K
lower than ours and a $T_{\rm eff}$ for Sk\,$-$68$^{\circ}$~52 is $800$ K higher than our estimate, whereas their $\log{g}$ is lower than ours by $0.15$~dex and $0.12$~dex respectively.
They provide two mass-loss rates for each object, one with the assumption of a smooth wind, and the other with the assumption of a generic volume filling factor of $0.1$, and we
compare our mass-loss rates to the latter and we find that their mass-loss rates are higher by $0.23$~dex and $0.10$~dex for  Sk\,$-$68$^{\circ}$~52 and Sk\,$-$68$^{\circ}$~135 respectively.
This can be attributed to the difference in photospheric parameters, the difference in $\beta$-law, which they assume is 2.75 and 3.50 respectively for Sk\,$-$68$^{\circ}$~52 and 
Sk\,$-$68$^{\circ}$~135, versus our $\beta$ that we estimate at $2$ and $2.3$, and most significantly the large difference in the terminal wind velocity estimates, as we determine the wind
velocities at $1140~{\rm km\,s^{-1}}$ and $880~{\rm km\,s^{-1}}$, versus their $\varv_{\infty}$ of $1400~{\rm km\,s^{-1}}$ and $1350~{\rm km\,s^{-1}}$ for Sk\,$-$68$^{\circ}$~52 and Sk\,$-$68$^{\circ}$~135, 
respectively, which they measure from the $\ion{N}{III }~\lambda991$ doublet in the FUV FUSE range. The luminosities obtained by \citet{evans2004} are consistent within $0.1$ dex of our luminosities.

We share Sk\,$-$71$^{\circ}$~41 in common with \citet{ramachandran2018}, who analysed a large sample of OB-stars in the LMC N206 superbubble using the Potsdam Wolf–Rayet (\textsc{PoWR}) code 
\citep{PoWR2002, PoWR2003, PoWR2015}. 
Their $T_{\rm eff}$ is slightly higher than ours (see Table~\ref{comparison_table}), but we do find a large discrepancy in $\log{g}$, where they estimate it to be $\approx 0.3$~dex higher than 
ours. The source of this observed discrepancy is unclear but could be the due to the difference in datasets. The mass-loss rate obtained by them is consistent with our mass-loss rates, albeit 
slightly lower.

We have seven targets in common with \citet{verhamme2024} and four targets with \citet{brands2025}, who conducted a UV+optical analysis for B-supergiants and 
hot O-supergiants, respectively, using the genetic algorithm \textsc{Kiwi-GA}, which utilises \textsc{FASTWIND} non-LTE radiative transfer models. Both of these studies used the same spectroscopic 
dataset as this study (ULLYSES/XShootU dataset). We find that our effective temperatures is systematically lower than those of \citet{verhamme2024} and 
\citet{brands2025} by $1000-4000$~K (see Fig.~\ref{teff_compare}). This could be attributed to the differences between \textsc{CMFGEN} and \textsc{FASTWIND} in 
the treatment of helium and silicon lines \citep{massey2013}, and the different analysis methods (genetic algorithm vs manual by-eye fitting).

For the coolest star Sk\,$-$67$^{\circ}$~195, \citet{verhamme2024} obtained the temperature through the $\ion{He}{I}$ and $\ion{Si}{II}$ lines, whereas we obtain our temperature from fitting the 
$\ion{He}{I}$ and $\ion{Mg}{II}$ lines. \textsc{FASTWIND} suffferce from known convergence problems below $15~{\rm kK}$ \citep{lorenzo2025}, therefore \textsc{CMFGEN} is more reliable.
The higher temperatures of \citet{verhamme2024} and \citet{brands2025} lead to higher $\log{g}$ in comparison to our work. We find that that our effective 
surface gravities are $0.1-0.25$~dex lower than theirs (see Fig.~\ref{logg_compare}). The luminosities obtained by \citet{verhamme2024} and \citet{brands2025} are systematically higher 
than ours by $\Delta L_{\rm bol} = 0.02-0.12$~dex, albeit they are consistent within the uncertainties. The sole outlier is Sk\,$-$67$^{\circ}$~5, for which the lumiosity derived by 
\citet{brands2025} is lower than ours by approximately $\Delta \log{L_{\rm bol}} = 0.5$~dex. This arises from adopting incorrect $UBV$ photometry by \citet{brands2025}. The use of the correct 
photometry yields $\log{L_{\rm bol}} = 6.00^{+0.04}_{-0.04}$ for Sk\,$-$67$^{\circ}$~5 \citep[][priv. comm.]{brands2025}, which is in better agreement with our derived luminosity of 
$\log{L_{\rm bol}}= 5.89$. We do not notice any trend in the difference between our mass-loss rates and those obtained by \citet{verhamme2024} and \citet{brands2025} despite the 
different clumping parametrization. We do note that mass-loss rates from \citet{verhamme2024} and \citet{brands2025} are consistent with our mass-loss rates within the uncertainties.

The advantages of our analysis methods compared to \citet{verhamme2024} are the detailed treatment of line blanketing in the UV provided in \textsc{CMFGEN} (see Fig.~\ref{all_UV} and 
Appendix~\ref{overall_app}), the robust velocity measurements from P Cygni profiles in the UV, and the employment of $\ion{Mg}{II}~4481$ when measuring the temperature of the coolest B-supergiants 
where $\ion{Si}{III}$ lines are no longer present. The main disadvantages of our method are the lack of robust calculations of statistical model errors. Also, unlike \textsc{FASTWIND}, the 
treatment of clumping in \textsc{CMFGEN} is limited to micro-clumping, which, in some cases, yields poor fits to the red emission in P Cygni profiles, and requires unrealialisticly low 
volume-filling factors (or high clumping factors) to obtain good fits for the blue absorption. \citet{muijres2011} describes how optically-thin clumps increase the predicted mass-loss rates, 
while the optically-thick clumps can display an opposite effect. This might have a differential effect on object at either side of the bi-stability jump. This effect was recently explored 
by \citet{deBurgos2024} in a large sample of Galactic OB-stars, and they report a lack of evidence for a bi-stability jump. In this study, as in \citet{verhamme2024} and \citet{brands2025}, we do not find a 
correlation between the mass-loss rate and the clumping parameters. 

\subsubsection{Optical only studies} 
\label{optical}
Sk\,$-$68$^{\circ}$~140 is in common with \citet{McEvoy2015}, who obtained the atmospheric parameters and nitrogen abundances for a B-supergiants in the LMC using \textsc{TLUSTY} and 
\textsc{SYNSPEC} codes \citep{TLUSTY1995}. We find that our $T_{\rm eff}$ and $\log{g}$ are slightly higher compared to theirs ($500$ K and $0.05$~dex respectively), which falls 
well within our uncertainties. The model grids they use are plane-parallel and do not provide a measure of the mass-loss rate.

We share 4 targets in common with \citet{urbaneja2017}, which is a photometric and spectroscopic (optical only) study for LMC OB-supergiants using \textsc{FASTWIND}. We find that, the effective temperatures are systematically larger than ours by $500-1500$ K, whereas for $\log{g}$ we do not identify any systematic trend. 
Since \citet{urbaneja2017} conducted their study to specifically explore the gravity–luminosity relationship, mass-loss rates are not investigated. We find that our $T_{\rm eff}$ and $\log{g}$ 
are consistent with theirs within the uncertainties. We calculate $\log{L_{\rm bol}}$ from the total extinction values, bolometric corrections and apparent magnitudes provided in 
\citet{urbaneja2017}. We find that they are in agreement with our derived $\log{L_{\rm bol}}$ within the uncertainties.

\citet{gvaramadze2018} performed a detailed optical only spectroscopic analysis using \textsc{CMFGEN} for the hypergiant Sk\,$-$69$^{\circ}$~279.  In comparison to our effective temperature
and mass-loss rate, theirs are larger by $700$~K and $0.4$~dex, respectively. The difference in mass-loss is due to the difference in wind clumping, as they estimate $f_{\rm vol}$ at $0.5$ versus 
$0.1$ in our analysis and $\beta$-law, which they fix at a value of $3$, whereas we estimate a value of $\beta=2.7$. The different observational data used in \citet{gvaramadze2018} and the 
variable nature of this object, which ressembles a dormant Luminous Blue Variable \citep[LBV, ][]{weis1997}, could also be a contributing 
factor to this discrepancy. \citet{gvaramadze2018} do not provide the effective surface gravity used in their models therefore we cannot compare our results.

Comparing our results to \citet{Bestenlehner2025}, we find that there is a good agreement in the $T_{\rm eff}$ and $\log{g}$. The exception to this are the  late-O hypergiants 
Sk\,$-$68$^{\circ}$~135 and Sk\,$-$69$^{\circ}$~279, where we find large differences in temperature and gravity. The discrepancy in $T_{\rm eff}$ is likely due the weak $\ion{He}{II}~\lambda4686$ 
line, the region of which is clearly offset from the continuum (see Fig.~\ref{overall_SK-69D279}). Additionally, all Balmer lines utilised by the pipeline in \citet{Bestenlehner2025} present 
heavy wind contamination (fully or partially in emission), making $\log{g}$ hard to constrain. Subsequently, all the parameters obtained from fitting those lines are not reliable. Unlike other 
\textsc{FASTWIND} based studies, we do not notice a clear offset in $T_{\rm eff}$ and $\log{g}$. We find no major discrepancies in the luminosities. Despite the constant $\beta = 1$ and 
$f_{\rm vol} = 0.1$ in the grid that was utilised by the pipeline of \citet{Bestenlehner2025}, their derived mass-loss rates are consistent with ours within the uncertainties.
    
\subsection{Comparison to numerical mass-loss rate recipes}
\label{theor_comp}
\begin{figure}
    \centering
	\includegraphics[width=\hsize]{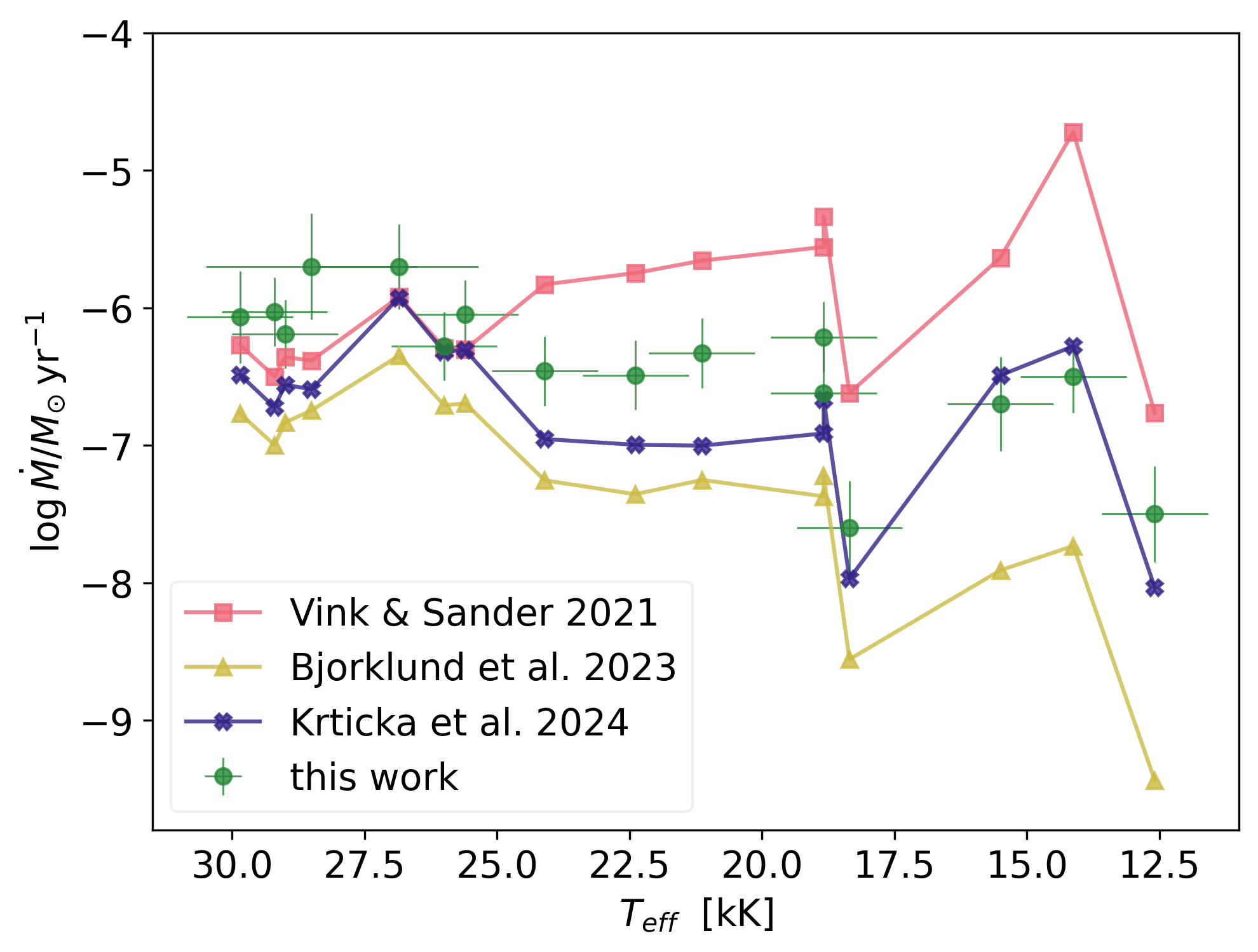}
        \caption{Mass-loss rates as a function of $T_{\rm eff}$, comparing our derived $\dot{M}$ (green circles) to mass-loss rates which we calculated via various numerical recipes. 
        Pink squares: $\dot{M}$ from \citet{vink2001} recipe.  Yellow triangles: $\dot{M}$ from \citet{bjorklund2023}. Violet crosses: $\dot{M}$ from \citet{krticka2024}.}
        \label{mass_loss_compare}
   \end{figure}

Here, we compare our derived mass-loss rates to those predicted by numerical recipes presented in \citet{vink2001}, \citet{bjorklund2023}, and \citet{krticka2024}. The mass-loss rate recipe 
produced by \citet{vink2001} is based on predictions from unified Monte-Carlo models with consistency between radiative acceleration and mechanical acceleration in the 
overall structure of the wind. The prediction of the bi-stability jump and the mass-loss rates around $T_{\rm eff} = 25-21~{\rm kK}$ in those simulations is due to the recombination of 
$\ion{Fe}{IV}$ into $\ion{Fe}{III}$ which has significantly more spectral lines resulting in a ``boost" in line-driving. In the bottommost row of Fig.~\ref{ion_struct}, it is clear 
the dominant iron ions at the connection point marking the photosphere in Sk\,$-$66$^{\circ}$~171, Sk\,$-$68$^{\circ}$~140, and Sk\,$-$68$^{\circ}$~8 are $\ion{Fe}{V}$, $\ion{Fe}{IV}$, 
and $\ion{Fe}{III}$, respectively. This means that the shift from $\ion{Fe}{IV}$ to $\ion{Fe}{III}$ is indeed reflected in our results, with $\ion{Fe}{III}$ providing the majority 
of line-driving \citep{petrov2016}, but as we discuss later, this is not accompanied by an increase in empirical mass-loss rates. \citet{petrov2016} explore another bi-stability jump at 
$T_{\rm eff}\approx10{\rm -}8$~kK, which was discussed by \citet{lamers1995} and \citet{vink1999}. This bi-stability jump is due to the change in ionization balance between $\ion{Fe}{III}$ and 
$\ion{Fe}{II}$, with $\ion{Fe}{II}$ becoming the source of most of the line acceleration.

An important element in the calculation routine is the ratio of terminal velocity to escape velocity ($\varv_{\infty}/\varv_{{\rm esc}}$). For this, we adopt the $\varv_{\infty}$ introduced in 
later study \citep{vinksander2021}. In \citet{Bjorklund2021, bjorklund2023}, a locally self-consistent wind model \textsc{FASTWIND} that solves the radiative transfer equation in the co-moving 
frame is used. In this case, no sudden increase in the mass-loss rate is produced with decreasing $T_{\rm eff}$. In this recipe, the mass-loss rates  decrease as a smooth function of stellar 
parameters.  Finally, in \citet{krticka2024} the mass-loss rates are derived using global stellar atmosphere code \textsc{METUJE} that was introduced in \citet{krtickakubat2017}, which calculates 
the velocity stratification with the radiative acceleration. This recipe predicts a much more moderate increase in the mass-loss rates compared to what is predicted by \citet{vink2001} in the 
regime bellow $T_{\rm eff}=20~{\rm kK}$. 

Fig.~\ref{mass_loss_compare} compares our derived mass-loss rates with the predicted mass-loss rates calculated by numerical recipes assuming $Z/Z_{\odot} = 0.5$. We notice that for stars with 
temperatures above $25~{\rm kK}$, the \citet{vink2001} and \citet{krticka2024} mass-loss rate predictions are relatively consistent with our derived mass-loss rates. For stars cooler that 
$25~{\rm kK}$ our derived mass-loss rates are situated between \citet{vink2001} and \citet{krticka2024} favouring the latter in values and overall trend. Predictions from \citet{bjorklund2023} have 
a similar trend to those from \citet{krticka2024}, but are consistently lower by $\approx 1$~dex. The dip in the mass-loss rates in our results and in all predictions between 
$T_{\rm eff}= 18-15~{\rm kK}$ is due to the object Sk\,$-$70$^{\circ}$~16 not being a B4 blue supergiant (Ib), but rather a lower luminosity bright giant.

\subsection{Metallicity effect}
\label{met_eff}
\begin{figure}
    \centering
	\includegraphics[width=\hsize]{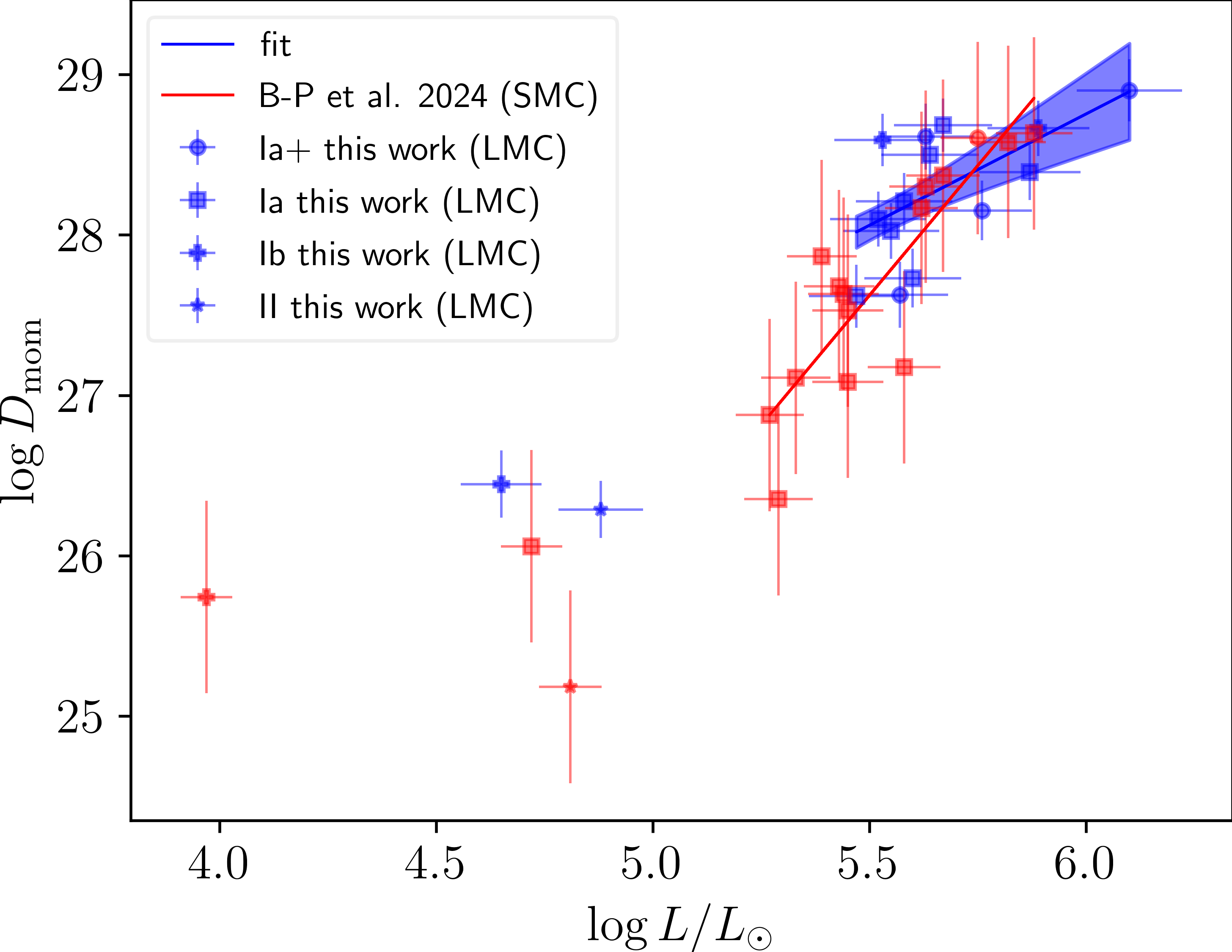}
        \caption{Modified wind momentum-luminosity distribution of LMC (blue symbols) supergiants from this work and SMC (red symbols) supergiants from \citet{bernini2024}. 
        Objects with $\log{(L_{\rm bol}/L_{\odot})}<5$ are excluded from the linear fits for both the LMC and SMC samples. The symbols for the SMC sample (red) follow the same encoding of the LMC 
        sample (blue), where each luminosity class is given a unique symbol.}
        \label{momentum}
   \end{figure}

To explore the metallicity dependence of wind properties, we derive the modified wind momentum $D_{\rm mom}$-$L_{\rm bol}$ relation ($D_{\rm mom} = \dot{M}\,\varv_{\infty}\,R_{\ast}\big/R_{\odot}$) \citep{kudritzki1995, kudritzkiandpuls2000}, and we limit our fit to 
stars with $\log{(L_{\rm bol}/L_{\odot})} > 5$. For lower luminosity stars the mass-loss rates we obtain are less reliable and can be considered upper limits due to H$\alpha$ being fully in absorption and due to the lack
of reliable P Cygni profiles for wind velocity measurements. We fit our derived values using a $D_{\rm mom}$-$L_{\rm bol}$ relation of the form:
\begin{equation}
    \label{eq:Dmom} 
      \log{D_{\rm mom}^{\rm LMC}} = x\log{(L_{\rm bol}/L_{\odot})} + \log{D_{0}}\\
\end{equation}
\begin{table}
  \caption{Slopes $x$, and offsets, $\log{D_{0}}$, of the linear fits to modified wind momentum equation (Equation~\ref{eq:Dmom}) of this study, \citet{mokiem2007}, and \citet{backs2024}}              
  \def\arraystretch{1.5}
  \label{table:dmom}      
  \centering      
  \small                               
  \addtolength{\tabcolsep}{+1.0em}
  \begin{tabular}{c c c}        
      \hline\hline{\smallskip}
          LMC          &$x$         &$\log{D_{0}}$\\
\hline 
This study            &$1.39\pm 0.54$       &$20.4\pm 3.0$    \\                          
\citet{mokiem2007}    &$1.49\pm 0.18$       &$20.4\pm 1.0$    \\ 
\citet{backs2024}     &$1.87\pm 0.11$       &$18.03\pm1.2$  \\                                              
    \noalign{\smallskip}
      \hline
  \end{tabular}
\end{table}
From Table~\ref{table:dmom}, we find that our derive relation in good alignment with the clumping-scaled relationship derived by \citet{mokiem2007} for LMC OB-stars. Our $D_{\rm mom}$-$L_{\rm bol}$ relation is 
consistent with the findings of \citet{backs2024} within the uncertainties, however they obtain a steeper relation from an LMC sample that consists mostly of O giants and bright giants, and incorporates
results from \citet{brands2025}.

We also compare our results to those of \citet{bernini2024}, the latest detailed ULLYSES/XshootU spectroscopic analysis of an SMC B-supergiant sample using CMFGEN. The main differences between our 
results and those of \citet{bernini2024} is that, whereas we include late O-stars, their sample is composed from B-stars exclusively. Their method differs from ours in that X-rays are included and 
$\varv_{\rm mic}$ is a free parameter that is determined in their fitting procedure, which could lead to different temperature determination due to the change in opacity, which effects the depth 
of silicon lines. Higher $\varv_{\rm mic}$ at larger radii in the wind could potentially also lead to underestimating $\varv_{\infty}$. 

In Fig.~\ref{momentum}, we compare our derived $D_{\rm mom}$-$L_{\rm bol}$ relation to that of \citet{bernini2024}. When fitting the results of their SMC sample we obtain:
\[\log{D_{\rm mom}^{\rm SMC}} = 3.24 (\pm 0.50) \log{(L_{\rm bol}/L_{\odot})} + 9.8 (\pm 2.8).\]
There is a clear difference in the slope and vertical offset between the linear fit of the two samples, but unlike the findings of \citet{mokiem2007}, the two fits greatly vary, which is not 
something predicted by previous studies exploring the metallicity dependence of wind properties in the LMC and SMC for neither O or B-stars (e.g. \citet{Crowther2006, mokiem2007, backs2024}).
Excluding low luminosity objects, the sample of \citet{bernini2024} spans a range of $\log{L_{\rm bol}/L_{\odot}}\approx5.2-5.9$, compared to the range of our sample 
$\log{L_{\rm bol}/L_{\odot}}\approx5.5-6.1$. Also, the sample of \citet{bernini2024} is composed of almost exclusively Ia supergiants, inferring a selection bias to strong winds, whereas 
our sample contains supergiants spanning Ib to Ia$^{+}$. Those differences in the samples could be a major contributing factor to the difference in the dependence of the slope on metallicity compared 
to previous studies.

\section{Summary and conclusions}
\label{summary}
In this paper, we have applied a quantatative analysis using \textsc{CMFGEN} models to LMC late-O- and B-supergiants, which we will also use in future analysis of SMC (ULLYSES/XshootU) and 
Milky Way samples. We have also compared our results to previous works, and we found that when taking into account the differences in analysis methods and quality of data, our parameters are 
mostly consistent with those previously derived within expected systematic variance, which is discuseed in \citet{sander2024}.

We found that our sample consists of evolved stars. The majority of the sample are main-sequence stars with a few post-main sequence stars, according to the rotating single-star evolutionary models 
of \citet{brott2011}. We find that some of the stars in our sample apear to be evolved and in the post main sequence stage. Our derived spectroscopic masses are broadly consistent with the 
evolutionary masses obtained from {\sc Bonnsai} \citep{bonnsai2014} within the uncertainties. Our results did not show the large mass discrepancy that was reported in previous studies 
\citep[e.g.,][]{herrero1992, trundle2004I, Crowther2006} \citep{schneider2018}.

We obtained a velocity-temperature relation: \[\varv_{\infty}/{\rm km\,s^{-1}}=0.076(\pm 0.011)T_{\rm eff}/{\rm K}-884(\pm260),\] which is in agreement with the relation obtinaed by \citet{xshootu3}.
We also derived a wind momentum-luminosity relation: \[\log{D_{\rm mom}^{\rm LMC}}=1.39(\pm 0.54)\log{(L_{\rm bol}/L_{\odot})}+20.4(\pm 3.0),\] which is very similar to the relation 
obtained for OB-stars in the LMC by \citet{mokiem2007}. 

We investigated the existence of the bi-stability jump, and we found that there is a lack of observational evidence in our sample that would support such a boost in line-driving, hence mass-loss 
and momentum, despite the shift from $\ion{Fe}{IV}$ to $\ion{Fe}{III}$. This is in agreement with the analysis of \citet{verhamme2024}, who conducted an analysis of a ULLUSES/XShootU LMC 
sample. This also agrees with the results of \citet{deBurgos2024}, who did not find an increase in the mass-loss rates over the bi-stability region in a 
large sample of Galactic blue supergiants. Our findings also support the results of \citet{bernini2024}, who did not detect a bi-stability jump in their sample of SMC blue supergiants.

Our results showed a strong correlation between $\varv_{\infty}/\varv_{{\rm esc}}$ and $T_{\rm eff}$, and we obtained the relation: \[\varv_{\infty}/\varv_{\rm esc}=4.1(\pm 0.8)\log{(T_{\rm eff}/{\rm K})}-16.3(\pm 3.5).\]
We did not find evidence of a sharp drop in $\varv_{\infty}/\varv_{{\rm esc}}$, which could be attributed to the shift from the low density and high velocity wind regime to the high density and low velocity regime
associated with the bi-stability jump \citep{pauldrachandpuls1990, lamers1995, vink2001, petrov2016} 

By comparing our results to a similar CMFGEN-based analysis of SMC BSGs \citep{bernini2024}, we found strong evidence of a $Z$-dependence of wind momentum. This supports previous findings and line-driving 
wind theory predictions. In this paper, we did not establish metallicity dependent mass-loss or momentum recipes. We elected to produce those empirical recipes in a future paper where we will conduct our 
own analysis of SMC and MW OB-supergiants. 

In future studies we will include stars of the same spectral range and analysis methods. This is to ensure that we are comparing like-for-like across different environments and that the results 
for each sample can be compared to the others without inherent differences in methodology. Our goal is to produce an empirical recipe for the mass-loss rates for blue supergiants, which connects 
the mass-loss rates to other environmental (metallicity) and intrinsic (temperature, mass, etc,..) parameters of the star. This, in turn, can be added as input for evolutionary models \citep{yoon2006} 
and population synthesis models \citep{leitherer1999} that currently rely on numerical mass-loss rate recipes such as the one proposed in \citet{vink2001}. Such an empirical mass-loss 
recipe would provide an important counterpart gauged by observations. Moreover, since the recipe would be produced by applying the same ethods of analysis to samples in different 
enviroments, the obtained $Z$-dependence would be more robust. 
 
\begin{acknowledgements}
TA would like to thank Science and Technology Facilities Council (STFC) for financial support through STFC scholarship ST/X508743/1. We would also like to thank Nidia Morrell for collecting and reducing 
MIKE data. JMB and PAC acknowledge financial support from the Science and Technology Facilities Council via research grant ST/V000853/1 (P.I. Vik Dhillon). DP acknowledges financial support from the FWO junior postdoctoral fellowship No.~1256225N. 
RK acknowledges financial support via the Heisenberg Research Grant funded by the Deutsche Forschungsgemeinschaft (DFG, German Research Foundation) under grant no.~KU 2849/9, project no.~445783058. 
F.N., acknowledges support by PID2022-137779OB-C41 funded by MCIN/AEI/10.13039/501100011033 by "ERDF A way of making Europe". This research has made use of the SIMBAD database, operated at CDS, 
Strasbourg, France, CMFGEN stellar atmosphere code developed by D. John Hillier, and BONNSAI software developed by Fabian Schneider. We would like to thank the referee for their critical and insightful input. 
\end{acknowledgements}

\bibliographystyle{aa}
\bibliography{bib}

\begin{thebibliography}{145}
\expandafter\ifx\csname natexlab\endcsname\relax\def\natexlab#1{#1}\fi

\bibitem[{{Alves}(2004)}]{alves2004}
{Alves}, D.~R. 2004, \nar, 48, 659

\bibitem[{{Anderson}(1985)}]{anderson1985}
{Anderson}, L.~S. 1985, \apj, 298, 848

\bibitem[{{Anderson}(1989)}]{anderson1989}
{Anderson}, L.~S. 1989, \apj, 339, 558

\bibitem[{{Ardeberg} {et~al.}(1972){Ardeberg}, {Brunet}, {Maurice}, \&
  {Prevot}}]{ardeberg1972}
{Ardeberg}, A., {Brunet}, J.~P., {Maurice}, E., \& {Prevot}, L. 1972, \aaps, 6,
  249

\bibitem[{{Asplund} {et~al.}(2005){Asplund}, {Grevesse}, \&
  {Sauval}}]{Asplund2005}
{Asplund}, M., {Grevesse}, N., \& {Sauval}, A.~J. 2005, in Astronomical Society
  of the Pacific Conference Series, Vol. 336, Cosmic Abundances as Records of
  Stellar Evolution and Nucleosynthesis, ed. T.~G. {Barnes}, III \& F.~N.
  {Bash}, 25

\bibitem[{{Backs} {et~al.}(2024){Backs}, {Brands}, {de Koter}, {Kaper}, {Vink},
  {Puls}, {Sundqvist}, {Tramper}, {Sana}, {Bernini-Peron}, {Bestenlehner},
  {Crowther}, {Hawcroft}, {Ignace}, {Kuiper}, {van Loon}, {Mahy}, {Marcolino},
  {Najarro}, {Oskinova}, {Pauli}, {Ramachandran}, {Sander}, \&
  {Verhamme}}]{backs2024}
{Backs}, F., {Brands}, S.~A., {de Koter}, A., {et~al.} 2024, \aap, 692, A88

\bibitem[{{Baum} {et~al.}(1992){Baum}, {Hamann}, {Koesterke}, \&
  {Wessolowski}}]{baum1992}
{Baum}, E., {Hamann}, W.~R., {Koesterke}, L., \& {Wessolowski}, U. 1992, \aap,
  266, 402

\bibitem[{{Beckman} \& {Crivellari}(1985)}]{abbot1985}
{Beckman}, J.~E. \& {Crivellari}, L. 1985, Science, 230, 835

\bibitem[{{Benaglia} {et~al.}(2007){Benaglia}, {Vink}, {Mart{\'\i}}, {Ma{\'\i}z
  Apell{\'a}niz}, {Koribalski}, \& {Crowther}}]{benaglia2007}
{Benaglia}, P., {Vink}, J.~S., {Mart{\'\i}}, J., {et~al.} 2007, \aap, 467, 1265

\bibitem[{{Bernini-Peron} {et~al.}(2023){Bernini-Peron}, {Marcolino}, {Sander},
  {Bouret}, {Ramachandran}, {Saling}, {Schneider}, {Oskinova}, \&
  {Najarro}}]{bernini2023}
{Bernini-Peron}, M., {Marcolino}, W.~L.~F., {Sander}, A.~A.~C., {et~al.} 2023,
  \aap, 677, A50

\bibitem[{{Bernini-Peron} {et~al.}(2024){Bernini-Peron}, {Sander},
  {Ramachandran}, {Oskinova}, {Vink}, {Verhamme}, {Najarro}, {Josiek},
  {Brands}, {Crowther}, {G{\'o}mez-Gonz{\'a}lez}, {Gormaz-Matamala},
  {Hawcroft}, {Kuiper}, {Mahy}, {Marcolino}, {Martins}, {Mehner}, {Parsons},
  {Pauli}, {Shenar}, {Schootemeijer}, {Todt}, {van Loon}, \& {XShootU
  Collaboration}}]{bernini2024}
{Bernini-Peron}, M., {Sander}, A.~A.~C., {Ramachandran}, V., {et~al.} 2024,
  \aap, 692, A89

\bibitem[{{Bestenlehner} {et~al.}(2020){Bestenlehner}, {Crowther},
  {Caballero-Nieves}, {Schneider}, {Sim{\'o}n-D{\'\i}az}, {Brands}, {de Koter},
  {Gr{\"a}fener}, {Herrero}, {Langer}, {Lennon}, {Ma{\'\i}z Apell{\'a}niz},
  {Puls}, \& {Vink}}]{bestenlehner2020}
{Bestenlehner}, J.~M., {Crowther}, P.~A., {Caballero-Nieves}, S.~M., {et~al.}
  2020, \mnras, 499, 1918

\bibitem[{{Bestenlehner} {et~al.}(2025){Bestenlehner}, {Crowther}, {Hawcroft},
  {Sana}, {Tramper}, {Vink}, {Brands}, {Sander}, \& {XShootU
  Collaboration}}]{Bestenlehner2025}
{Bestenlehner}, J.~M., {Crowther}, P.~A., {Hawcroft}, C., {et~al.} 2025, \aap,
  695, A198

\bibitem[{{Bestenlehner} {et~al.}(2024){Bestenlehner}, {En{\ss}lin},
  {Bergemann}, {Crowther}, {Greiner}, \& {Selig}}]{bestenlehner2022}
{Bestenlehner}, J.~M., {En{\ss}lin}, T., {Bergemann}, M., {et~al.} 2024,
  \mnras, 528, 6735

\bibitem[{{Bestenlehner} {et~al.}(2014){Bestenlehner}, {Gr{\"a}fener}, {Vink},
  {Najarro}, {de Koter}, {Sana}, {Evans}, {Crowther}, {H{\'e}nault-Brunet},
  {Herrero}, {Langer}, {Schneider}, {Sim{\'o}n-D{\'\i}az}, {Taylor}, \&
  {Walborn}}]{bestenlehner2014}
{Bestenlehner}, J.~M., {Gr{\"a}fener}, G., {Vink}, J.~S., {et~al.} 2014, \aap,
  570, A38

\bibitem[{{Bj{\"o}rklund} {et~al.}(2021){Bj{\"o}rklund}, {Sundqvist}, {Puls},
  \& {Najarro}}]{Bjorklund2021}
{Bj{\"o}rklund}, R., {Sundqvist}, J.~O., {Puls}, J., \& {Najarro}, F. 2021,
  \aap, 648, A36

\bibitem[{{Bj{\"o}rklund} {et~al.}(2023){Bj{\"o}rklund}, {Sundqvist}, {Singh},
  {Puls}, \& {Najarro}}]{bjorklund2023}
{Bj{\"o}rklund}, R., {Sundqvist}, J.~O., {Singh}, S.~M., {Puls}, J., \&
  {Najarro}, F. 2023, \aap, 676, A109

\bibitem[{{Brands} {et~al.}(2025){Brands}, {Backs}, {de Koter}, {Puls},
  {Crowther}, {Sana}, {Tramper}, {Kaper}, {Sundqvist}, {Bestenlehner},
  {Driessen}, {Erba}, {Hawcroft}, {Herrero}, {John Hillier}, {Ignace},
  {Lefever}, {Dylan Kee}, {Kub{\'a}tov{\'a}}, {Mahy}, {Moffat}, {Najarro},
  {Prinja}, {Ramachandran}, {Sander}, {Vink}, \& {XShootU
  Collaboration}}]{brands2025}
{Brands}, S.~A., {Backs}, F., {de Koter}, A., {et~al.} 2025, \aap, 697, A54

\bibitem[{{Brands} {et~al.}(2022){Brands}, {de Koter}, {Bestenlehner},
  {Crowther}, {Sundqvist}, {Puls}, {Caballero-Nieves}, {Abdul-Masih},
  {Driessen}, {Garc{\'\i}a}, {Geen}, {Gr{\"a}fener}, {Hawcroft}, {Kaper},
  {Keszthelyi}, {Langer}, {Sana}, {Schneider}, {Shenar}, \&
  {Vink}}]{Brands2022}
{Brands}, S.~A., {de Koter}, A., {Bestenlehner}, J.~M., {et~al.} 2022, \aap,
  663, A36

\bibitem[{{Bresolin} {et~al.}(2001){Bresolin}, {Kudritzki}, {Mendez}, \&
  {Przybilla}}]{bresolin2001}
{Bresolin}, F., {Kudritzki}, R.-P., {Mendez}, R.~H., \& {Przybilla}, N. 2001,
  \apjl, 548, L159

\bibitem[{{Brott} {et~al.}(2011){Brott}, {de Mink}, {Cantiello}, {Langer}, {de
  Koter}, {Evans}, {Hunter}, {Trundle}, \& {Vink}}]{brott2011}
{Brott}, I., {de Mink}, S.~E., {Cantiello}, M., {et~al.} 2011, \aap, 530, A115

\bibitem[{Butler \& Giddings(1985)}]{butler1985newsletter}
Butler, K. \& Giddings, J. 1985, University of London

\bibitem[{{Choudhury} {et~al.}(2016){Choudhury}, {Subramaniam}, \&
  {Cole}}]{choudhury2016}
{Choudhury}, S., {Subramaniam}, A., \& {Cole}, A.~A. 2016, \mnras, 455, 1855

\bibitem[{{Cioni} {et~al.}(2011){Cioni}, {Clementini}, {Girardi}, {Guandalini},
  {Gullieuszik}, {Miszalski}, {Moretti}, {Ripepi}, {Rubele}, {Bagheri},
  {Bekki}, {Cross}, {de Blok}, {de Grijs}, {Emerson}, {Evans}, {Gibson},
  {Gonzales-Solares}, {Groenewegen}, {Irwin}, {Ivanov}, {Lewis}, {Marconi},
  {Marquette}, {Mastropietro}, {Moore}, {Napiwotzki}, {Naylor}, {Oliveira},
  {Read}, {Sutorius}, {van Loon}, {Wilkinson}, \& {Wood}}]{VMC}
{Cioni}, M.~R., {Clementini}, G., {Girardi}, L., {et~al.} 2011, The Messenger,
  144, 25

\bibitem[{{Crowther}(2019)}]{crowther2019}
{Crowther}, P.~A. 2019, Galaxies, 7, 88

\bibitem[{{Crowther}(2024)}]{mike}
{Crowther}, P.~A. 2024, in IAU Symposium, Vol. 361, IAU Symposium, ed.
  J.~{Mackey}, J.~S. {Vink}, \& N.~{St-Louis}, 15--25

\bibitem[{{Crowther} {et~al.}(2002){Crowther}, {Hillier}, {Evans}, {Fullerton},
  {De Marco}, \& {Willis}}]{crowther2002}
{Crowther}, P.~A., {Hillier}, D.~J., {Evans}, C.~J., {et~al.} 2002, \apj, 579,
  774

\bibitem[{{Crowther} {et~al.}(2006){Crowther}, {Lennon}, \&
  {Walborn}}]{Crowther2006}
{Crowther}, P.~A., {Lennon}, D.~J., \& {Walborn}, N.~R. 2006, \aap, 446, 279

\bibitem[{{Cutri} {et~al.}(2003){Cutri}, {Skrutskie}, {van Dyk}, {Beichman},
  {Carpenter}, {Chester}, {Cambresy}, {Evans}, {Fowler}, {Gizis}, {Howard},
  {Huchra}, {Jarrett}, {Kopan}, {Kirkpatrick}, {Light}, {Marsh}, {McCallon},
  {Schneider}, {Stiening}, {Sykes}, {Weinberg}, {Wheaton}, {Wheelock}, \&
  {Zacarias}}]{cutri2003}
{Cutri}, R.~M., {Skrutskie}, M.~F., {van Dyk}, S., {et~al.} 2003, {VizieR
  Online Data Catalog: 2MASS All-Sky Catalog of Point Sources (Cutri+ 2003)},
  VizieR On-line Data Catalog: II/246. Originally published in: University of
  Massachusetts and Infrared Processing and Analysis Center, (IPAC/California
  Institute of Technology) (2003)

\bibitem[{{de Burgos} {et~al.}(2024{\natexlab{a}}){de Burgos}, {Keszthelyi},
  {Sim{\'o}n-D{\'\i}az}, \& {Urbaneja}}]{deBurgos2024}
{de Burgos}, A., {Keszthelyi}, Z., {Sim{\'o}n-D{\'\i}az}, S., \& {Urbaneja},
  M.~A. 2024{\natexlab{a}}, \aap, 687, L16

\bibitem[{{de Burgos} {et~al.}(2023){de Burgos}, {Sim{\'o}n-D{\'\i}az},
  {Urbaneja}, \& {Negueruela}}]{deBurgos2023}
{de Burgos}, A., {Sim{\'o}n-D{\'\i}az}, S., {Urbaneja}, M.~A., \& {Negueruela},
  I. 2023, \aap, 674, A212

\bibitem[{{de Burgos} {et~al.}(2024{\natexlab{b}}){de Burgos},
  {Sim{\'o}n-D{\'\i}az}, {Urbaneja}, \& {Puls}}]{deBurgos2024b}
{de Burgos}, A., {Sim{\'o}n-D{\'\i}az}, S., {Urbaneja}, M.~A., \& {Puls}, J.
  2024{\natexlab{b}}, \aap, 687, A228

\bibitem[{{Debnath} {et~al.}(2024){Debnath}, {Sundqvist}, {Moens}, {Van der
  Sijpt}, {Verhamme}, \& {Poniatowski}}]{debnath2024}
{Debnath}, D., {Sundqvist}, J.~O., {Moens}, N., {et~al.} 2024, \aap, 684, A177

\bibitem[{{Dessart} \& {Hillier}(2010)}]{Dessart2010}
{Dessart}, L. \& {Hillier}, D.~J. 2010, \mnras, 405, 2141

\bibitem[{{Dopita} {et~al.}(2019){Dopita}, {Seitenzahl}, {Sutherland},
  {Nicholls}, {Vogt}, {Ghavamian}, \& {Ruiter}}]{dopita2019}
{Dopita}, M.~A., {Seitenzahl}, I.~R., {Sutherland}, R.~S., {et~al.} 2019, \aj,
  157, 50

\bibitem[{{Evans} {et~al.}(2004){Evans}, {Crowther}, {Fullerton}, \&
  {Hillier}}]{evans2004}
{Evans}, C.~J., {Crowther}, P.~A., {Fullerton}, A.~W., \& {Hillier}, D.~J.
  2004, \apj, 610, 1021

\bibitem[{{Fitzpatrick}(1985)}]{fitzpatrick1985}
{Fitzpatrick}, E.~L. 1985, \apj, 299, 219

\bibitem[{{Fitzpatrick} \& {Massa}(1990)}]{FM1990}
{Fitzpatrick}, E.~L. \& {Massa}, D. 1990, \apjs, 72, 163

\bibitem[{Fouesneau(2025)}]{zenodopyphot}
Fouesneau, M. 2025, pyphot

\bibitem[{{Geen} {et~al.}(2023){Geen}, {Agrawal}, {Crowther}, {Keller}, {de
  Koter}, {Keszthelyi}, {van de Voort}, {Ali}, {Backs}, {Bonne}, {Brugaletta},
  {Derkink}, {Ekstr{\"o}m}, {Fichtner}, {Grassitelli}, {G{\"o}tberg},
  {Higgins}, {Laplace}, {You Liow}, {Lorenzo}, {McLeod}, {Meynet}, {Newsome},
  {Oliva}, {Ramachandran}, {Rey}, {Rieder}, {Romano-D{\'\i}az}, {Sabhahit},
  {Sander}, {Sarwar}, {Stinshoff}, {Stoop}, {Sz{\'e}csi}, {Trebitsch}, {Vink},
  \& {Winch}}]{geen2023}
{Geen}, S., {Agrawal}, P., {Crowther}, P.~A., {et~al.} 2023, \pasp, 135, 021001

\bibitem[{{Gordon} {et~al.}(2003){Gordon}, {Clayton}, {Misselt}, {Landolt}, \&
  {Wolff}}]{Gordon2003}
{Gordon}, K.~D., {Clayton}, G.~C., {Misselt}, K.~A., {Landolt}, A.~U., \&
  {Wolff}, M.~J. 2003, \apj, 594, 279

\bibitem[{{Gr{\"a}fener} {et~al.}(2002{\natexlab{a}}){Gr{\"a}fener},
  {Koesterke}, \& {Hamann}}]{grafener2002}
{Gr{\"a}fener}, G., {Koesterke}, L., \& {Hamann}, W.~R. 2002{\natexlab{a}},
  \aap, 387, 244

\bibitem[{{Gr{\"a}fener} {et~al.}(2002{\natexlab{b}}){Gr{\"a}fener},
  {Koesterke}, \& {Hamann}}]{PoWR2002}
{Gr{\"a}fener}, G., {Koesterke}, L., \& {Hamann}, W.~R. 2002{\natexlab{b}},
  \aap, 387, 244

\bibitem[{{Green} {et~al.}(2012){Green}, {Froning}, {Osterman}, {Ebbets},
  {Heap}, {Leitherer}, {Linsky}, {Savage}, {Sembach}, {Shull}, {Siegmund},
  {Snow}, {Spencer}, {Stern}, {Stocke}, {Welsh}, {B{\'e}land}, {Burgh},
  {Danforth}, {France}, {Keeney}, {McPhate}, {Penton}, {Andrews},
  {Brownsberger}, {Morse}, \& {Wilkinson}}]{COS}
{Green}, J.~C., {Froning}, C.~S., {Osterman}, S., {et~al.} 2012, \apj, 744, 60

\bibitem[{{Gvaramadze} {et~al.}(2018){Gvaramadze}, {Kniazev}, {Maryeva}, \&
  {Berdnikov}}]{gvaramadze2018}
{Gvaramadze}, V.~V., {Kniazev}, A.~Y., {Maryeva}, O.~V., \& {Berdnikov}, L.~N.
  2018, \mnras, 474, 1412

\bibitem[{{Haiman} \& {Loeb}(1997)}]{1997ApJ...483...21H}
{Haiman}, Z. \& {Loeb}, A. 1997, \apj, 483, 21

\bibitem[{{Hamann} \& {Gr{\"a}fener}(2003{\natexlab{a}})}]{hamann2003}
{Hamann}, W.~R. \& {Gr{\"a}fener}, G. 2003{\natexlab{a}}, \aap, 410, 993

\bibitem[{{Hamann} \& {Gr{\"a}fener}(2003{\natexlab{b}})}]{PoWR2003}
{Hamann}, W.~R. \& {Gr{\"a}fener}, G. 2003{\natexlab{b}}, \aap, 410, 993

\bibitem[{{Haucke} {et~al.}(2018){Haucke}, {Cidale}, {Venero}, {Cur{\'e}},
  {Kraus}, {Kanaan}, \& {Arcos}}]{haucke2018}
{Haucke}, M., {Cidale}, L.~S., {Venero}, R.~O.~J., {et~al.} 2018, \aap, 614,
  A91

\bibitem[{{Hawcroft} {et~al.}(2024){Hawcroft}, {Sana}, {Mahy}, {Sundqvist}, {de
  Koter}, {Crowther}, {Bestenlehner}, {Brands}, {David-Uraz}, {Decin}, {Erba},
  {Garcia}, {Hamann}, {Herrero}, {Ignace}, {Kee}, {Kub{\'a}tov{\'a}},
  {Lefever}, {Moffat}, {Najarro}, {Oskinova}, {Pauli}, {Prinja}, {Puls},
  {Sander}, {Shenar}, {St-Louis}, {ud-Doula}, \& {Vink}}]{xshootu3}
{Hawcroft}, C., {Sana}, H., {Mahy}, L., {et~al.} 2024, \aap, 688, A105

\bibitem[{{Herrero} {et~al.}(1992){Herrero}, {Kudritzki}, {Vilchez}, {Kunze},
  {Butler}, \& {Haser}}]{herrero1992}
{Herrero}, A., {Kudritzki}, R.~P., {Vilchez}, J.~M., {et~al.} 1992, \aap, 261,
  209

\bibitem[{{Herrero} {et~al.}(2002){Herrero}, {Puls}, \&
  {Najarro}}]{herrero2002}
{Herrero}, A., {Puls}, J., \& {Najarro}, F. 2002, \aap, 396, 949

\bibitem[{{Hillier}(1990)}]{Hillier1990}
{Hillier}, D.~J. 1990, \aap, 231, 116

\bibitem[{{Hillier}(1996)}]{hillier1996}
{Hillier}, D.~J. 1996, in Liege International Astrophysical Colloquia, Vol.~33,
  Liege International Astrophysical Colloquia, ed. J.~M. {Vreux}, A.~{Detal},
  D.~{Fraipont-Caro}, E.~{Gosset}, \& G.~{Rauw}, 509

\bibitem[{{Hillier}(1997)}]{hillier1997}
{Hillier}, D.~J. 1997, in IAU Symposium, Vol. 189, IAU Symposium, ed. T.~R.
  {Bedding}, A.~J. {Booth}, \& J.~{Davis}, 209--216

\bibitem[{{Hillier} {et~al.}(2003){Hillier}, {Lanz}, {Heap}, {Hubeny}, {Smith},
  {Evans}, {Lennon}, \& {Bouret}}]{Hillier2003}
{Hillier}, D.~J., {Lanz}, T., {Heap}, S.~R., {et~al.} 2003, \apj, 588, 1039

\bibitem[{{Hillier} \& {Miller}(1998)}]{hillier1998}
{Hillier}, D.~J. \& {Miller}, D.~L. 1998, \apj, 496, 407

\bibitem[{{Hillier} \& {Miller}(1999)}]{HillierandMiller1999}
{Hillier}, D.~J. \& {Miller}, D.~L. 1999, \apj, 519, 354

\bibitem[{{Hubeny} \& {Lanz}(1995{\natexlab{a}})}]{hubeny1995}
{Hubeny}, I. \& {Lanz}, T. 1995{\natexlab{a}}, \apj, 439, 875

\bibitem[{{Hubeny} \& {Lanz}(1995{\natexlab{b}})}]{TLUSTY1995}
{Hubeny}, I. \& {Lanz}, T. 1995{\natexlab{b}}, \apj, 439, 875

\bibitem[{{Hunter} {et~al.}(2008){Hunter}, {Brott}, {Lennon}, {Langer},
  {Dufton}, {Trundle}, {Smartt}, {de Koter}, {Evans}, \& {Ryans}}]{hunter2008}
{Hunter}, I., {Brott}, I., {Lennon}, D.~J., {et~al.} 2008, \apjl, 676, L29

\bibitem[{{Hunter} {et~al.}(2007){Hunter}, {Dufton}, {Smartt}, {Ryans},
  {Evans}, {Lennon}, {Trundle}, {Hubeny}, \& {Lanz}}]{hunter2007}
{Hunter}, I., {Dufton}, P.~L., {Smartt}, S.~J., {et~al.} 2007, \aap, 466, 277

\bibitem[{{Korn} {et~al.}(2005){Korn}, {Nieva}, {Daflon}, \&
  {Cunha}}]{korn2005}
{Korn}, A.~J., {Nieva}, M.~F., {Daflon}, S., \& {Cunha}, K. 2005, \apj, 633,
  899

\bibitem[{{Krti{\v{c}}ka} \& {Kub{\'a}t}(2017)}]{krtickakubat2017}
{Krti{\v{c}}ka}, J. \& {Kub{\'a}t}, J. 2017, \aap, 606, A31

\bibitem[{{Krti{\v{c}}ka} {et~al.}(2021){Krti{\v{c}}ka}, {Kub{\'a}t}, \&
  {Krti{\v{c}}kov{\'a}}}]{krticka2021}
{Krti{\v{c}}ka}, J., {Kub{\'a}t}, J., \& {Krti{\v{c}}kov{\'a}}, I. 2021, \aap,
  647, A28

\bibitem[{{Krti{\v{c}}ka} {et~al.}(2024){Krti{\v{c}}ka}, {Kub{\'a}t}, \&
  {Krti{\v{c}}kov{\'a}}}]{krticka2024}
{Krti{\v{c}}ka}, J., {Kub{\'a}t}, J., \& {Krti{\v{c}}kov{\'a}}, I. 2024, \aap,
  681, A29

\bibitem[{{Kudritzki} {et~al.}(2003){Kudritzki}, {Bresolin}, \&
  {Przybilla}}]{kudritzki2003}
{Kudritzki}, R.~P., {Bresolin}, F., \& {Przybilla}, N. 2003, \apjl, 582, L83

\bibitem[{{Kudritzki} {et~al.}(1995){Kudritzki}, {Lennon}, \&
  {Puls}}]{kudritzki1995}
{Kudritzki}, R.~P., {Lennon}, D.~J., \& {Puls}, J. 1995, in Science with the
  VLT, ed. J.~R. {Walsh} \& I.~J. {Danziger}, 246

\bibitem[{{Kudritzki} \& {Puls}(2000)}]{kudritzkiandpuls2000}
{Kudritzki}, R.-P. \& {Puls}, J. 2000, \araa, 38, 613

\bibitem[{{Kudritzki} {et~al.}(2024){Kudritzki}, {Urbaneja}, {Bresolin},
  {Macri}, {Yuan}, {Li}, {Anand}, \& {Riess}}]{kudritzki2024}
{Kudritzki}, R.-P., {Urbaneja}, M.~A., {Bresolin}, F., {et~al.} 2024, \apj,
  977, 217

\bibitem[{{Kudritzki} {et~al.}(2008){Kudritzki}, {Urbaneja}, {Bresolin},
  {Przybilla}, {Gieren}, \& {Pietrzy{\'n}ski}}]{kudritzki2008}
{Kudritzki}, R.-P., {Urbaneja}, M.~A., {Bresolin}, F., {et~al.} 2008, \apj,
  681, 269

\bibitem[{{Kurucz}(1979)}]{kurucz1979}
{Kurucz}, R.~L. 1979, \apjs, 40, 1

\bibitem[{{Lamers} {et~al.}(1995){Lamers}, {Snow}, \& {Lindholm}}]{lamers1995}
{Lamers}, H. J.~G.~L.~M., {Snow}, T.~P., \& {Lindholm}, D.~M. 1995, \apj, 455,
  269

\bibitem[{Langer(2012)}]{langer2012}
Langer, N. 2012, Annual Review of Astronomy and Astrophysics, 50, 107

\bibitem[{{Langer} \& {Kudritzki}(2014)}]{langerandkudritzki2014}
{Langer}, N. \& {Kudritzki}, R.~P. 2014, \aap, 564, A52

\bibitem[{{Lanz} \& {Hubeny}(2007)}]{Lanz2007}
{Lanz}, T. \& {Hubeny}, I. 2007, \apjs, 169, 83

\bibitem[{{Leitherer} {et~al.}(1999){Leitherer}, {Schaerer}, {Goldader},
  {Delgado}, {Robert}, {Kune}, {de Mello}, {Devost}, \&
  {Heckman}}]{leitherer1999}
{Leitherer}, C., {Schaerer}, D., {Goldader}, J.~D., {et~al.} 1999, \apjs, 123,
  3

\bibitem[{{Lorenzo} {et~al.}(2025){Lorenzo}, {Garcia}, {Castro}, {Najarro},
  {Cervi{\~n}o}, {Herrero}, \& {Sim{\'o}n-D{\'\i}az}}]{lorenzo2025}
{Lorenzo}, M., {Garcia}, M., {Castro}, N., {et~al.} 2025, \mnras, 537, 1197

\bibitem[{{Maeder} {et~al.}(2014){Maeder}, {Przybilla}, {Nieva}, {Georgy},
  {Meynet}, {Ekstr{\"o}m}, \& {Eggenberger}}]{maeder2014}
{Maeder}, A., {Przybilla}, N., {Nieva}, M.-F., {et~al.} 2014, \aap, 565, A39

\bibitem[{{Marcolino} {et~al.}(2009){Marcolino}, {Bouret}, {Martins},
  {Hillier}, {Lanz}, \& {Escolano}}]{marcolino2009}
{Marcolino}, W.~L.~F., {Bouret}, J.~C., {Martins}, F., {et~al.} 2009, \aap,
  498, 837

\bibitem[{{Markova} {et~al.}(2005){Markova}, {Puls}, {Scuderi}, \&
  {Markov}}]{markova2005}
{Markova}, N., {Puls}, J., {Scuderi}, S., \& {Markov}, H. 2005, \aap, 440, 1133

\bibitem[{{Martins}(2011)}]{martins2011}
{Martins}, F. 2011, Bulletin de la Societe Royale des Sciences de Liege, 80, 29

\bibitem[{{Martins} {et~al.}(2012){Martins}, {Mahy}, {Hillier}, \&
  {Rauw}}]{Martins2012}
{Martins}, F., {Mahy}, L., {Hillier}, D.~J., \& {Rauw}, G. 2012, \aap, 538, A39

\bibitem[{{Massa} {et~al.}(2024){Massa}, {Prinja}, \& {Oskinova}}]{massa2024}
{Massa}, D., {Prinja}, R.~K., \& {Oskinova}, L.~M. 2024, \apj, 971, 166

\bibitem[{{Massey}(2002)}]{massey2002}
{Massey}, P. 2002, \apjs, 141, 81

\bibitem[{{Massey} {et~al.}(2013){Massey}, {Neugent}, {Hillier}, \&
  {Puls}}]{massey2013}
{Massey}, P., {Neugent}, K.~F., {Hillier}, D.~J., \& {Puls}, J. 2013, \apj,
  768, 6

\bibitem[{{McErlean} {et~al.}(1998){McErlean}, {Lennon}, \&
  {Dufton}}]{mcerlean1998}
{McErlean}, N.~D., {Lennon}, D.~J., \& {Dufton}, P.~L. 1998, \aap, 329, 613

\bibitem[{{McErlean} {et~al.}(1999){McErlean}, {Lennon}, \&
  {Dufton}}]{mcerlean1999}
{McErlean}, N.~D., {Lennon}, D.~J., \& {Dufton}, P.~L. 1999, \aap, 349, 553

\bibitem[{{McEvoy} {et~al.}(2015){McEvoy}, {Dufton}, {Evans}, {Kalari},
  {Markova}, {Sim{\'o}n-D{\'\i}az}, {Vink}, {Walborn}, {Crowther}, {de Koter},
  {de Mink}, {Dunstall}, {H{\'e}nault-Brunet}, {Herrero}, {Langer}, {Lennon},
  {Ma{\'\i}z Apell{\'a}niz}, {Najarro}, {Puls}, {Sana}, {Schneider}, \&
  {Taylor}}]{McEvoy2015}
{McEvoy}, C.~M., {Dufton}, P.~L., {Evans}, C.~J., {et~al.} 2015, \aap, 575, A70

\bibitem[{{Menon} {et~al.}(2024){Menon}, {Ercolino}, {Urbaneja}, {Lennon},
  {Herrero}, {Hirai}, {Langer}, {Schootemeijer}, {Chatzopoulos}, {Frank}, \&
  {Shiber}}]{menon2024}
{Menon}, A., {Ercolino}, A., {Urbaneja}, M.~A., {et~al.} 2024, \apjl, 963, L42

\bibitem[{{Mokiem} {et~al.}(2007){Mokiem}, {de Koter}, {Vink}, {Puls}, {Evans},
  {Smartt}, {Crowther}, {Herrero}, {Langer}, {Lennon}, {Najarro}, \&
  {Villamariz}}]{mokiem2007}
{Mokiem}, M.~R., {de Koter}, A., {Vink}, J.~S., {et~al.} 2007, \aap, 473, 603

\bibitem[{{Moos} {et~al.}(2000){Moos}, {Cash}, {Cowie}, {Davidsen}, {Dupree},
  {Feldman}, {Friedman}, {Green}, {Green}, {Gry}, {Hutchings}, {Jenkins},
  {Linsky}, {Malina}, {Michalitsianos}, {Savage}, {Shull}, {Siegmund}, {Snow},
  {Sonneborn}, {Vidal-Madjar}, {Willis}, {Woodgate}, {York}, {Ake},
  {Andersson}, {Andrews}, {Barkhouser}, {Bianchi}, {Blair}, {Brownsberger},
  {Cha}, {Chayer}, {Conard}, {Fullerton}, {Gaines}, {Grange}, {Gummin},
  {Hebrard}, {Kriss}, {Kruk}, {Mark}, {McCarthy}, {Morbey}, {Murowinski},
  {Murphy}, {Oegerle}, {Ohl}, {Oliveira}, {Osterman}, {Sahnow}, {Saisse},
  {Sembach}, {Weaver}, {Welsh}, {Wilkinson}, \& {Zheng}}]{FUSE}
{Moos}, H.~W., {Cash}, W.~C., {Cowie}, L.~L., {et~al.} 2000, \apjl, 538, L1

\bibitem[{{Muijres} {et~al.}(2011){Muijres}, {de Koter}, {Vink},
  {Krti{\v{c}}ka}, {Kub{\'a}t}, \& {Langer}}]{muijres2011}
{Muijres}, L.~E., {de Koter}, A., {Vink}, J.~S., {et~al.} 2011, \aap, 526, A32

\bibitem[{{M{\"u}ller} \& {Vink}(2008)}]{mullerandvink2008}
{M{\"u}ller}, P.~E. \& {Vink}, J.~S. 2008, \aap, 492, 493

\bibitem[{{Negueruela} {et~al.}(2024){Negueruela}, {Sim{\'o}n-D{\'\i}az}, {de
  Burgos}, {Casasbuenas}, \& {Beck}}]{negueruela2024}
{Negueruela}, I., {Sim{\'o}n-D{\'\i}az}, S., {de Burgos}, A., {Casasbuenas},
  A., \& {Beck}, P.~G. 2024, \aap, 690, A176

\bibitem[{{Pauldrach} {et~al.}(1994){Pauldrach}, {Kudritzki}, {Puls}, {Butler},
  \& {Hunsinger}}]{pauldrach1994}
{Pauldrach}, A.~W.~A., {Kudritzki}, R.~P., {Puls}, J., {Butler}, K., \&
  {Hunsinger}, J. 1994, \aap, 283, 525

\bibitem[{{Pauldrach} \& {Puls}(1990)}]{pauldrachandpuls1990}
{Pauldrach}, A.~W.~A. \& {Puls}, J. 1990, \aap, 237, 409

\bibitem[{{Petrov} {et~al.}(2016){Petrov}, {Vink}, \&
  {Gr{\"a}fener}}]{petrov2016}
{Petrov}, B., {Vink}, J.~S., \& {Gr{\"a}fener}, G. 2016, \mnras, 458, 1999

\bibitem[{{Prinja} {et~al.}(1990){Prinja}, {Barlow}, \& {Howarth}}]{prinja1990}
{Prinja}, R.~K., {Barlow}, M.~J., \& {Howarth}, I.~D. 1990, \apj, 361, 607

\bibitem[{{Przybilla} {et~al.}(2006){Przybilla}, {Butler}, {Becker}, \&
  {Kudritzki}}]{przybilla2006}
{Przybilla}, N., {Butler}, K., {Becker}, S.~R., \& {Kudritzki}, R.~P. 2006,
  \aap, 445, 1099

\bibitem[{{Przybilla} {et~al.}(2008){Przybilla}, {Nieva}, \&
  {Butler}}]{przybilla2008}
{Przybilla}, N., {Nieva}, M.-F., \& {Butler}, K. 2008, \apjl, 688, L103

\bibitem[{{Puebla} {et~al.}(2016){Puebla}, {Hillier}, {Zsarg{\'o}}, {Cohen}, \&
  {Leutenegger}}]{puebla2016}
{Puebla}, R.~E., {Hillier}, D.~J., {Zsarg{\'o}}, J., {Cohen}, D.~H., \&
  {Leutenegger}, M.~A. 2016, \mnras, 456, 2907

\bibitem[{{Puls} {et~al.}(1996){Puls}, {Kudritzki}, {Herrero}, {Pauldrach},
  {Haser}, {Lennon}, {Gabler}, {Voels}, {Vilchez}, {Wachter}, \&
  {Feldmeier}}]{puls1996}
{Puls}, J., {Kudritzki}, R.~P., {Herrero}, A., {et~al.} 1996, \aap, 305, 171

\bibitem[{{Puls} {et~al.}(2005){Puls}, {Urbaneja}, {Venero}, {Repolust},
  {Springmann}, {Jokuthy}, \& {Mokiem}}]{Puls2005}
{Puls}, J., {Urbaneja}, M.~A., {Venero}, R., {et~al.} 2005, \aap, 435, 669

\bibitem[{{Puls} {et~al.}(2008){Puls}, {Vink}, \& {Najarro}}]{puls2008}
{Puls}, J., {Vink}, J.~S., \& {Najarro}, F. 2008, \aapr, 16, 209

\bibitem[{{Ramachandran} {et~al.}(2018){Ramachandran}, {Hamann}, {Hainich},
  {Oskinova}, {Shenar}, {Sander}, {Todt}, \& {Gallagher}}]{ramachandran2018}
{Ramachandran}, V., {Hamann}, W.~R., {Hainich}, R., {et~al.} 2018, \aap, 615,
  A40

\bibitem[{{Ram{\'\i}rez-Agudelo} {et~al.}(2013){Ram{\'\i}rez-Agudelo},
  {Sim{\'o}n-D{\'\i}az}, {Sana}, {de Koter}, {Sab{\'\i}n-Sanjul{\'\i}an}, {de
  Mink}, {Dufton}, {Gr{\"a}fener}, {Evans}, {Herrero}, {Langer}, {Lennon},
  {Ma{\'\i}z Apell{\'a}niz}, {Markova}, {Najarro}, {Puls}, {Taylor}, \&
  {Vink}}]{ramirez2013}
{Ram{\'\i}rez-Agudelo}, O.~H., {Sim{\'o}n-D{\'\i}az}, S., {Sana}, H., {et~al.}
  2013, \aap, 560, A29

\bibitem[{{Repolust} {et~al.}(2004){Repolust}, {Puls}, \&
  {Herrero}}]{repolust2004}
{Repolust}, T., {Puls}, J., \& {Herrero}, A. 2004, \aap, 415, 349

\bibitem[{{Rivero Gonz{\'a}lez} {et~al.}(2011){Rivero Gonz{\'a}lez}, {Puls}, \&
  {Najarro}}]{rivero2011}
{Rivero Gonz{\'a}lez}, J.~G., {Puls}, J., \& {Najarro}, F. 2011, \aap, 536, A58

\bibitem[{{Rivero Gonz{\'a}lez} {et~al.}(2012){Rivero Gonz{\'a}lez}, {Puls},
  {Najarro}, \& {Brott}}]{rivero2012}
{Rivero Gonz{\'a}lez}, J.~G., {Puls}, J., {Najarro}, F., \& {Brott}, I. 2012,
  \aap, 537, A79

\bibitem[{{Roman-Duval} {et~al.}(2025){Roman-Duval}, {Fischer}, {Fullerton},
  {Taylor}, {Plesha}, {Proffitt}, {Monroe}, {Fischer}, {Aloisi}, {Bouret},
  {Britt}, {Calvet}, {Carlberg}, {Crowther}, {De Rosa}, {Dixon}, {Espaillat},
  {Evans}, {Fox}, {France}, {Garcia}, {Fleming}, {Frazer}, {G{\'o}mez De
  Castro}, {Herczeg}, {Hernandez}, {Hirschauer}, {James}, {Johns-Krull},
  {Leitherer}, {Lockwood}, {Najita}, {Oey}, {Oliveira}, {Pauly}, {Reid},
  {Riedel}, {Rodriguez}, {Sahnow}, {Sankrit}, {Sembach}, {Shaw}, {Smith},
  {Sohn}, {Som}, {{\'U}beda}, \& {Welty}}]{roman-duval2025}
{Roman-Duval}, J., {Fischer}, W.~J., {Fullerton}, A.~W., {et~al.} 2025, arXiv
  e-prints, arXiv:2504.05446

\bibitem[{{Sana} {et~al.}(2024){Sana}, {Tramper}, {Abdul-Masih}, {Blomme},
  {Dsilva}, {Maravelias}, {Martins}, {Mehner}, {Wofford}, {Banyard}, {Barbosa},
  {Bestenlehner}, {Hawcroft}, {John Hillier}, {Todt}, {Larkin}, {Mahy},
  {Najarro}, {Ramachandran}, {Ram{\'\i}rez-Tannus}, {Rubio-D{\'\i}ez},
  {Sander}, {Shenar}, {Vink}, {Backs}, {Brands}, {Crowther}, {Decin}, {de
  Koter}, {Hamann}, {Kehrig}, {Kuiper}, {Oskinova}, {Pauli}, {Sundqvist},
  {Verhamme}, \& {Xshoot-U Collaboration}}]{Sana2024}
{Sana}, H., {Tramper}, F., {Abdul-Masih}, M., {et~al.} 2024, \aap, 688, A104

\bibitem[{{Sander} {et~al.}(2015){Sander}, {Shenar}, {Hainich},
  {G{\'\i}menez-Garc{\'\i}a}, {Todt}, \& {Hamann}}]{PoWR2015}
{Sander}, A., {Shenar}, T., {Hainich}, R., {et~al.} 2015, \aap, 577, A13

\bibitem[{{Sander} {et~al.}(2024){Sander}, {Bouret}, {Bernini-Peron}, {Puls},
  {Backs}, {Berlanas}, {Bestenlehner}, {Brands}, {Herrero}, {Martins},
  {Maryeva}, {Pauli}, {Ramachandran}, {Crowther}, {G{\'o}mez-Gonz{\'a}lez},
  {Gormaz-Matamala}, {Hamann}, {Hillier}, {Kuiper}, {Larkin}, {Lefever},
  {Mehner}, {Najarro}, {Oskinova}, {Sch{\"o}sser}, {Shenar}, {Todt},
  {ud-Doula}, \& {Vink}}]{sander2024}
{Sander}, A.~A.~C., {Bouret}, J.~C., {Bernini-Peron}, M., {et~al.} 2024, \aap,
  689, A30

\bibitem[{{Schmidt-Kaler} {et~al.}(1999){Schmidt-Kaler}, {Gochermann},
  {Oestreicher}, {Grothues}, {Tappert}, {Zaum}, {Bergh{\"o}fer}, \&
  {Brugger}}]{Schmidt-Kaler1999}
{Schmidt-Kaler}, T., {Gochermann}, J., {Oestreicher}, M.~O., {et~al.} 1999,
  \mnras, 306, 279

\bibitem[{{Schmutz} {et~al.}(1989){Schmutz}, {Hamann}, \&
  {Wessolowski}}]{schmutz1989}
{Schmutz}, W., {Hamann}, W.~R., \& {Wessolowski}, U. 1989, \aap, 210, 236

\bibitem[{{Schneider} {et~al.}(2014){Schneider}, {Langer}, {de Koter}, {Brott},
  {Izzard}, \& {Lau}}]{bonnsai2014}
{Schneider}, F.~R.~N., {Langer}, N., {de Koter}, A., {et~al.} 2014, \aap, 570,
  A66

\bibitem[{{Schneider} {et~al.}(2018){Schneider}, {Ram{\'\i}rez-Agudelo},
  {Tramper}, {Bestenlehner}, {Castro}, {Sana}, {Evans},
  {Sab{\'\i}n-Sanjuli{\'a}n}, {Sim{\'o}n-D{\'\i}az}, {Langer}, {Fossati},
  {Gr{\"a}fener}, {Crowther}, {de Mink}, {de Koter}, {Gieles}, {Herrero},
  {Izzard}, {Kalari}, {Klessen}, {Lennon}, {Mahy}, {Ma{\'\i}z Apell{\'a}niz},
  {Markova}, {van Loon}, {Vink}, \& {Walborn}}]{schneider2018}
{Schneider}, F.~R.~N., {Ram{\'\i}rez-Agudelo}, O.~H., {Tramper}, F., {et~al.}
  2018, \aap, 618, A73

\bibitem[{{Searle} {et~al.}(2008){Searle}, {Prinja}, {Massa}, \&
  {Ryans}}]{searle2008}
{Searle}, S.~C., {Prinja}, R.~K., {Massa}, D., \& {Ryans}, R. 2008, \aap, 481,
  777

\bibitem[{{Sim{\'o}n-D{\'\i}az}(2020)}]{diaz2020}
{Sim{\'o}n-D{\'\i}az}, S. 2020, in Reviews in Frontiers of Modern Astrophysics;
  From Space Debris to Cosmology, 155--187

\bibitem[{{Sim{\'o}n-D{\'\i}az} \& {Herrero}(2014)}]{simondiaz2014}
{Sim{\'o}n-D{\'\i}az}, S. \& {Herrero}, A. 2014, \aap, 562, A135

\bibitem[{{Skrutskie} {et~al.}(2006){Skrutskie}, {Cutri}, {Stiening},
  {Weinberg}, {Schneider}, {Carpenter}, {Beichman}, {Capps}, {Chester},
  {Elias}, {Huchra}, {Liebert}, {Lonsdale}, {Monet}, {Price}, {Seitzer},
  {Jarrett}, {Kirkpatrick}, {Gizis}, {Howard}, {Evans}, {Fowler}, {Fullmer},
  {Hurt}, {Light}, {Kopan}, {Marsh}, {McCallon}, {Tam}, {Van Dyk}, \&
  {Wheelock}}]{skrutskie2006}
{Skrutskie}, M.~F., {Cutri}, R.~M., {Stiening}, R., {et~al.} 2006, \aj, 131,
  1163

\bibitem[{{Smartt}(2009)}]{smartt2009}
{Smartt}, S.~J. 2009, \araa, 47, 63

\bibitem[{{Smith}(2014)}]{smith2014}
{Smith}, N. 2014, \araa, 52, 487

\bibitem[{{Smith} {et~al.}(2001){Smith}, {Brickhouse}, {Liedahl}, \&
  {Raymond}}]{smith2001}
{Smith}, R.~K., {Brickhouse}, N.~S., {Liedahl}, D.~A., \& {Raymond}, J.~C.
  2001, \apjl, 556, L91

\bibitem[{{Sota} {et~al.}(2011){Sota}, {Ma{\'\i}z Apell{\'a}niz}, {Walborn},
  {Alfaro}, {Barb{\'a}}, {Morrell}, {Gamen}, \& {Arias}}]{sota2011}
{Sota}, A., {Ma{\'\i}z Apell{\'a}niz}, J., {Walborn}, N.~R., {et~al.} 2011,
  \apjs, 193, 24

\bibitem[{{Stanway} \& {Eldridge}(2018)}]{BPASS2018}
{Stanway}, E.~R. \& {Eldridge}, J.~J. 2018, \mnras, 479, 75

\bibitem[{{Trundle} \& {Lennon}(2005)}]{trundle2004II}
{Trundle}, C. \& {Lennon}, D.~J. 2005, \aap, 434, 677

\bibitem[{{Trundle} {et~al.}(2004){Trundle}, {Lennon}, {Puls}, \&
  {Dufton}}]{trundle2004I}
{Trundle}, C., {Lennon}, D.~J., {Puls}, J., \& {Dufton}, P.~L. 2004, \aap, 417,
  217

\bibitem[{{Urbaneja} {et~al.}(2005{\natexlab{a}}){Urbaneja}, {Herrero},
  {Bresolin}, {Kudritzki}, {Gieren}, {Puls}, {Przybilla}, {Najarro}, \&
  {Pietrzy{\'n}ski}}]{urbaneja2005}
{Urbaneja}, M.~A., {Herrero}, A., {Bresolin}, F., {et~al.} 2005{\natexlab{a}},
  \apj, 622, 862

\bibitem[{{Urbaneja} {et~al.}(2005{\natexlab{b}}){Urbaneja}, {Herrero},
  {Kudritzki}, {Najarro}, {Smartt}, {Puls}, {Lennon}, \&
  {Corral}}]{urbaneja2005b}
{Urbaneja}, M.~A., {Herrero}, A., {Kudritzki}, R.~P., {et~al.}
  2005{\natexlab{b}}, \apj, 635, 311

\bibitem[{{Urbaneja} {et~al.}(2017){Urbaneja}, {Kudritzki}, {Gieren},
  {Pietrzy{\'n}ski}, {Bresolin}, \& {Przybilla}}]{urbaneja2017}
{Urbaneja}, M.~A., {Kudritzki}, R.~P., {Gieren}, W., {et~al.} 2017, \aj, 154,
  102

\bibitem[{{Verhamme} {et~al.}(2024){Verhamme}, {Sundqvist}, {de Koter}, {Sana},
  {Backs}, {Brands}, {Najarro}, {Puls}, {Vink}, {Crowther}, {Kub{\'a}tov{\'a}},
  {Sander}, {Bernini-Peron}, {Kuiper}, {Prinja}, {Schillemans}, {Shenar}, {van
  Loon}, \& {XShootu collaboration}}]{verhamme2024}
{Verhamme}, O., {Sundqvist}, J., {de Koter}, A., {et~al.} 2024, \aap, 692, A91

\bibitem[{{Vernet} {et~al.}(2011){Vernet}, {Dekker}, {D'Odorico}, {Kaper},
  {Kjaergaard}, {Hammer}, {Randich}, {Zerbi}, {Groot}, {Hjorth}, {Guinouard},
  {Navarro}, {Adolfse}, {Albers}, {Amans}, {Andersen}, {Andersen}, {Binetruy},
  {Bristow}, {Castillo}, {Chemla}, {Christensen}, {Conconi}, {Conzelmann},
  {Dam}, {de Caprio}, {de Ugarte Postigo}, {Delabre}, {di Marcantonio},
  {Downing}, {Elswijk}, {Finger}, {Fischer}, {Flores}, {Fran{\c{c}}ois},
  {Goldoni}, {Guglielmi}, {Haigron}, {Hanenburg}, {Hendriks}, {Horrobin},
  {Horville}, {Jessen}, {Kerber}, {Kern}, {Kiekebusch}, {Kleszcz}, {Klougart},
  {Kragt}, {Larsen}, {Lizon}, {Lucuix}, {Mainieri}, {Manuputy}, {Martayan},
  {Mason}, {Mazzoleni}, {Michaelsen}, {Modigliani}, {Moehler}, {M{\o}ller},
  {Norup S{\o}rensen}, {N{\o}rregaard}, {P{\'e}roux}, {Patat}, {Pena}, {Pragt},
  {Reinero}, {Rigal}, {Riva}, {Roelfsema}, {Royer}, {Sacco}, {Santin},
  {Schoenmaker}, {Spano}, {Sweers}, {Ter Horst}, {Tintori}, {Tromp}, {van
  Dael}, {van der Vliet}, {Venema}, {Vidali}, {Vinther}, {Vola}, {Winters},
  {Wistisen}, {Wulterkens}, \& {Zacchei}}]{vernet2011}
{Vernet}, J., {Dekker}, H., {D'Odorico}, S., {et~al.} 2011, \aap, 536, A105

\bibitem[{{Vink} {et~al.}(1999){Vink}, {de Koter}, \& {Lamers}}]{vink1999}
{Vink}, J.~S., {de Koter}, A., \& {Lamers}, H.~J.~G.~L.~M. 1999, \aap, 350, 181

\bibitem[{{Vink} {et~al.}(2000){Vink}, {de Koter}, \& {Lamers}}]{vink2000}
{Vink}, J.~S., {de Koter}, A., \& {Lamers}, H.~J.~G.~L.~M. 2000, \aap, 362, 295

\bibitem[{{Vink} {et~al.}(2001){Vink}, {de Koter}, \& {Lamers}}]{vink2001}
{Vink}, J.~S., {de Koter}, A., \& {Lamers}, H.~J.~G.~L.~M. 2001, \aap, 369, 574

\bibitem[{{Vink} {et~al.}(2023){Vink}, {Mehner}, {Crowther}, {Fullerton},
  {Garcia}, {Martins}, {Morrell}, {Oskinova}, {St-Louis}, {ud-Doula}, {Sander},
  {Sana}, {Bouret}, {Kub{\'a}tov{\'a}}, {Marchant}, {Martins}, {Wofford}, {van
  Loon}, {Grace Telford}, {G{\"o}tberg}, {Bowman}, {Erba}, {Kalari},
  {Abdul-Masih}, {Alkousa}, {Backs}, {Barbosa}, {Berlanas}, {Bernini-Peron},
  {Bestenlehner}, {Blomme}, {Bodensteiner}, {Brands}, {Evans}, {David-Uraz},
  {Driessen}, {Dsilva}, {Geen}, {G{\'o}mez-Gonz{\'a}lez}, {Grassitelli},
  {Hamann}, {Hawcroft}, {Herrero}, {Higgins}, {John Hillier}, {Ignace},
  {Istrate}, {Kaper}, {Kee}, {Kehrig}, {Keszthelyi}, {Klencki}, {de Koter},
  {Kuiper}, {Laplace}, {Larkin}, {Lefever}, {Leitherer}, {Lennon}, {Mahy},
  {Ma{\'\i}z Apell{\'a}niz}, {Maravelias}, {Marcolino}, {McLeod}, {de Mink},
  {Najarro}, {Oey}, {Parsons}, {Pauli}, {Pedersen}, {Prinja}, {Ramachandran},
  {Ram{\'\i}rez-Tannus}, {Sabhahit}, {Schootemeijer}, {Reyero Serantes},
  {Shenar}, {Stringfellow}, {Sudnik}, {Tramper}, \& {Wang}}]{xshootU1}
{Vink}, J.~S., {Mehner}, A., {Crowther}, P.~A., {et~al.} 2023, \aap, 675, A154

\bibitem[{{Vink} {et~al.}(2011){Vink}, {Muijres}, {Anthonisse}, {de Koter},
  {Gr{\"a}fener}, \& {Langer}}]{vink2011}
{Vink}, J.~S., {Muijres}, L.~E., {Anthonisse}, B., {et~al.} 2011, \aap, 531,
  A132

\bibitem[{{Vink} \& {Sander}(2021)}]{vinksander2021}
{Vink}, J.~S. \& {Sander}, A. A.~C. 2021, \mnras, 504, 2051

\bibitem[{{Weis} {et~al.}(1997){Weis}, {Chu}, {Duschl}, \& {Bomans}}]{weis1997}
{Weis}, K., {Chu}, Y.~H., {Duschl}, W.~J., \& {Bomans}, D.~J. 1997, \aap, 325,
  1157

\bibitem[{{We{\ss}mayer} {et~al.}(2022){We{\ss}mayer}, {Przybilla}, \&
  {Butler}}]{wessmayer2022}
{We{\ss}mayer}, D., {Przybilla}, N., \& {Butler}, K. 2022, \aap, 668, A92

\bibitem[{{We{\ss}mayer} {et~al.}(2023){We{\ss}mayer}, {Przybilla},
  {Ebenbichler}, {Aschenbrenner}, \& {Butler}}]{wessmayer2023}
{We{\ss}mayer}, D., {Przybilla}, N., {Ebenbichler}, A., {Aschenbrenner}, P., \&
  {Butler}, K. 2023, \aap, 677, A175

\bibitem[{{Woodgate} {et~al.}(1998){Woodgate}, {Kimble}, {Bowers}, {Kraemer},
  {Kaiser}, {Danks}, {Grady}, {Loiacono}, {Brumfield}, {Feinberg}, {Gull},
  {Heap}, {Maran}, {Lindler}, {Hood}, {Meyer}, {Vanhouten}, {Argabright},
  {Franka}, {Bybee}, {Dorn}, {Bottema}, {Woodruff}, {Michika}, {Sullivan},
  {Hetlinger}, {Ludtke}, {Stocker}, {Delamere}, {Rose}, {Becker}, {Garner},
  {Timothy}, {Blouke}, {Joseph}, {Hartig}, {Green}, {Jenkins}, {Linsky},
  {Hutchings}, {Moos}, {Boggess}, {Roesler}, \& {Weistrop}}]{STIS}
{Woodgate}, B.~E., {Kimble}, R.~A., {Bowers}, C.~W., {et~al.} 1998, \pasp, 110,
  1183

\bibitem[{{Yoon} {et~al.}(2006){Yoon}, {Langer}, \& {Norman}}]{yoon2006}
{Yoon}, S.~C., {Langer}, N., \& {Norman}, C. 2006, \aap, 460, 199

\end{thebibliography}

\begin{appendix} 
\appendix

\section{Omitted objects}
\label{Omitted}
\paragraph{Sk\,$-$69$^{\circ}$~83} For this star, we find a sign of binarity in $\ion{He}{II}~\lambda 4686$, where it is blue shifted relative to all the other lines, indicating the existence of 
an emission from a primary hotter object combined with the emission of a cooler object. We also find signs of two components in $\ion{He}{II}~\lambda 5411$ and H$\gamma$

\paragraph{Sk\,$-$67$^{\circ}$~197, Sk\,$-$66$^{\circ}$~152, Sk\,$-$67$^{\circ}$~168, Sk\,$-$70$^{\circ}$~50} These objects were omitted from the analysis due to H$\alpha$ and $\ion{He}{II}~\lambda4686$
having two symmetrical emission peaks with a narrow absorption feature in the line center which is indicative of a non-spherically symmetric geometry such as a circumstellar disk.

\paragraph{Sk\,$-$66$^{\circ}$~50} The spectrum of this object showed clear binary signatures in the form of double peaked absorption in H$\beta$, H$\epsilon$ and H$\gamma$ in addition to the
odd morphology of H$\alpha$.

\section{Comments on individual stars}
\label{individ_com}
\paragraph{Sk\,$-$66$^{\circ}$~171, Fig.~\ref{SK-66D171}, Fig.~\ref{overall_SK-66D171}}  When considering helium lines other than $\ion{He}{I}~\lambda4471$ $\ion{He}{II}~\lambda4542$, one might prefer a model with a slightly higher temperature. 
The other glaring issue is with the H$\alpha$,  although the morphology of the model line is similar to the observation, we do find that the centre of the emission in the observation is slightly 
red shifted relative to all other lines. In the UV and FUV the model fits the observations very well. Fig.~\ref{overall_SK-66D171} shows the fit for the entire spectrum, where for this object 
include a fit for the absorption profiles for $Ly\alpha$ through $Ly\eta$, which were calculated using \ion{H}{I} column density $\log{N(HI)}=20.7~{\rm cm^{-2}}$ \citep{fitzpatrick1985}. 
The observed $\ion{He}{II}~\lambda4686$ emission is predicted in absorption.

\paragraph{Sk\,$-$68$^{\circ}$~155, Fig.~\ref{SK-68D155}, Fig.~\ref{overall_SK-68D155}} This star with $v\sin{i} = 80~{\rm km/s}$ is the fastest rotator in our sample, which is reflected in the relatively broad metal and helium lines. 
The small dip in the centre of the emission in H$\alpha$ could be an indication of an optically thick disk forming around the star, other than that the overall quality of the fit is very good.
The model overpredicts the $\ion{He}{II}~\lambda4686$ absoprtion.

\paragraph{Sk\,$-$69$^{\circ}$~279, Fig.~\ref{SK-69D279}, Fig.~\ref{overall_SK-69D279}} This is one of two hypergiants in our sample, which is classified as such by the extremely strong emission in H$\alpha$ and the P Cygni shape in higher Balmer 
lines and in $\ion{He}{I}~\lambda4471$ , therefore,  we use H$\eta$ for estimating $\log{g}$, as all other Balmer lines are contaminated by winds and take $\ion{He}{I}~\lambda4026$ and 
$\ion{He}{II}~\lambda5411$ as primary diagnostics to obtain the temperature. The overall fit is quite good considering all the challenges that come with fitting the spectra of emission line 
objects, with the exception of $\ion{He}{II}~\lambda4686$ line, where the predicted emission is weaker than the observed.

\paragraph{Sk\,$-$71$^{\circ}$~41, Fig.~\ref{SK-71D41}, Fig.~\ref{overall_SK-71D41}}  For this object, we were not able to reproduce the observed unsaturated silicon lines $\ion{Si}{IV}~\lambda\lambda1393-1403$ even with models with extremely 
clumped winds $f_{\rm vol} = 0.03-0.02$. We note two other peculiarities, first of which is the large discrepancy between $\varv_{\infty}$ obtained from the $\ion{C}{IV}~\lambda\lambda1548-1551$ $(\approx 1500~{\rm km\,s^{-1}})$and 
$\ion{Si}{IV}~\lambda\lambda1394-1403$ ($\approx 1300~{\rm km\,s^{-1}}$). The other peculiarity is the unexplicably broad $\ion{O}{III}~\lambda5592$ line.

\paragraph{Sk\,$-$68$^{\circ}$~135, Fig.~\ref{SK-68D135}, Fig.~\ref{overall_SK-68D135}}  This is second hypergiant in our sample. Just as for Sk\,$-$69$^{\circ}$~279, we estimate $\log{g}$ from H$\eta$. But unlike the other hypergiant, we obtain a more 
reliable value for $T_{\rm eff}$ because the helium lines do are less contaminated by the extreme winds. We obtain decent fits for most lines, with the exception of the strong $\ion{He}{II}~\lambda4686$ emission, which is predicted in absorption in the model.

\paragraph{Sk\,$-$67$^{\circ}$~5, Fig.~\ref{SK-67D5}, Fig.~\ref{overall_SK-67D5}}   For this object, we were not able to replicate the morphology of H$\alpha$. The most glaring issue in this fit is the helium abundance, and therefore the hydrogen 
abundance as well.
\paragraph{Sk\,$-$68$^{\circ}$~52, Fig.~\ref{SK-68D52}, Fig.~\ref{overall_SK-68D52}}  This is one of the early B0 supergiants that shows weak $\ion{He}{II}$ lines. We chose not to fit $\ion{He}{II}$ lines, as increasing the temperature of the model 
by a mere $500~K$ does indeed give a better fit for $\ion{He}{II}$ lines but drastically weakens $\ion{S}{III}$ lines. Overall, we obtain an excellent fit for most lines.

\paragraph{Sk\,$-$69$^{\circ}$~43, Fig.~\ref{SK-69D43}, Fig.~\ref{overall_SK-69D43}} In the case of this object, as for Sk\,$-$67$^{\circ}$~2, we were able to accurately reproduce the emission in H$\alpha$ but not the braod, blueshifted absorption. 

\paragraph{Sk\,$-$68$^{\circ}$~140, Fig.~\ref{SK-68D140}, Fig.~\ref{overall_SK-68D140}} The unsaturated $\ion{Si}{IV}~\lambda\lambda1393-1403$ lines, which suggests clumpier winds, were not reproduced by our models even with much smaller volum 
filling factors. The fit for this stars serves as a good example of the consequences of excluding X-rays from our models, as seen on Fig.~\ref{overall_SK-68D140}, the strong P Cygni $\ion{C}{IV}~\lambda1550$ in the 
observations is much weaker in the model, this is because in this temeperature range $\ion{C}{III}$ dominates the ionization structure, but due to X-rays generated by shocks, we observe the enhanced
$\ion{C}{IV}$ line.

\paragraph{Sk\,$-$67$^{\circ}$~2, Fig.~\ref{SK-67D2}, Fig.~\ref{overall_SK-67D2}} The overall fit for this object is very good. In this case, we obtain a good fit for H$\alpha$ emission, but we were unable to fit the blueshifted absorption.  

\paragraph{Sk\,$-$67$^{\circ}$~14, Fig.~\ref{SK-67D14}, Fig.~\ref{overall_SK-67D14}} We notice on this object that $\ion{Si}{IV}~\lambda\lambda1393-1403$ would suggest less emissive winds (higher volume $f_{\rm vol}$), whereas 
$\ion{Al}{III}~\lambda\lambda1856-1860$ would suggest more emissive winds (lower volume $f_{\rm vol}$).  

\paragraph{Sk\,$-$69$^{\circ}$~52, Fig.~\ref{SK-69D52}, Fig.~\ref{overall_SK-69D52}} For this star, we were not able to reproduce the odd morphology of H$\alpha$, which seem to have a sudden cut-off and the peak of the emission.

\paragraph{Sk\,$-$67$^{\circ}$~78, Fig.~\ref{SK-67D78}, Fig.~\ref{overall_SK-67D78}} The most noticeable issue with the fit of this object is the extended wings and the blue shifted absorption of the model H$\alpha$, which 
we could not eliminate. Although, we were able to obtain a good match to the emission in H$\alpha$.

\paragraph{Sk\,$-$70$^{\circ}$~16, Fig.~\ref{SK-70D16}, Fig.~\ref{overall_SK-70D16}} For this object we obtain a very good fit, except for H$\alpha$, where we could not reproduce the weak, but broad absorption.

\paragraph{Sk\,$-$68$^{\circ}$~8, Fig.~\ref{SK-68D8}, Fig.~\ref{overall_SK-68D8}} For this object, we were able to obtain a good fit for H$\alpha$ emission and to the general morphology. 
We do note the infilling of the H$\gamma$ line core, which could be due to nebular contamination. The $\ion{Si}{II}~\lambda\lambda6347-6371$ lines in the model match the observations quite well.
This can be used as a sanity check for determining the effective temperature for mid and late B-supergiants.

\paragraph{Sk\,$-$67$^{\circ}$~195, Fig.~\ref{SK-67D195}, Fig.~\ref{overall_SK-67D195}} For this low luminosity late B-supergiant we do not find any vaiable oxygen lines, therefore we do not change 
the oxygen mass fraction when applying our pipline to this star. We consider the mass-loss rate of this object to be an upper limit due to H$\alpha$ being fully in absorption. We note that the predicted 
$\ion{Si}{II}~\lambda\lambda6347-6371$ lines excellently match the observed, albeit we did not use them as diagnostics. 

\clearpage
\section{$\varv_{\rm black}$ and $\varv_{\rm edge}$}
In Table~\ref{table_app_1} we present the measured $\varv_{\rm black}$ from each line for all our sample. Similarly, in Table~\ref{table_app_1} we present the measured $\varv_{\rm edge}$.                  

    \begin{table*}
        \caption{$\varv_{\rm black}$ for the stars in our sample from all the available saturated P Cygni lines in the UV. Single velocity measurement from each line is subject to an 
        uncertainty $\Delta\varv_{\rm black} = 30~{\rm km\,s^{-1}}$ which take into acount the velocity resolution of the UV spectra and the uncertainty in the radial velocity correction which 
        were added in quadrature.}            
        \label{table_app_1}      
        \centering                                    
        \begin{tabular}{c c c c c c c c c}        
            \hline{\smallskip}
            Target &$\ion{Si}{iv}~1393$ &$\ion{Si}{iv}~1403$ &$\ion{C}{iv}~1548$ &$\ion{Al}{III}~1855$ &$\ion{Al}{III}~1863$ &$\ion{Mg}{II}~2796$ &$\ion{Mg}{II}~2803$ &$\overline{\varv_{\rm black}}$ \\
            \hline                                    
            Sk\,$-$66$^{\circ}$~171 &1704 &- &1846 &- &- &- &-	&1775$\pm 63$	\\      
            Sk\,$-$68$^{\circ}$~155 &- &1485 &1482 &- &- &- &-	&1520$\pm 61$\\       
            Sk\,$-$69$^{\circ}$~279 &613 &640 &- &- &- &- &-	  &630$\pm 62$	\\      
            Sk\,$-$71$^{\circ}$~41  &- &- &- &- &- &- &-	&-	\\                      
            Sk\,$-$68$^{\circ}$~135 &855 &909 &- &- &- &- &-	  &880$\pm 62$	\\      
            Sk\,$-$67$^{\circ}$~5	&1231 &- &1237 &- &- &- &-    &1230$\pm 61$	\\      
            Sk\,$-$68$^{\circ}$~52  &1050 &1097 &- &- &- &- &-  &1150$\pm 61$\\       
            Sk\,$-$69$^{\circ}$~43  &790 &- &914 &- &- &- &-	  &825$\pm 64$    \\    
            Sk\,$-$68$^{\circ}$~140 &1005 &984 &1016 &- &- &-   &-	&1000$\pm 62$	\\  
            Sk\,$-$67$^{\circ}$~2   &377 &361 &- &384 &381 &-   &-	&380$\pm 61$	\\  
            Sk\,$-$67$^{\circ}$~14  &796 &796 &839 &- &- &-	&-  &810$\pm 67$	\\      
            Sk\,$-$69$^{\circ}$~52  &473 &457 &- &- &- &- &-	  &465$\pm 61$    \\    
            Sk\,$-$67$^{\circ}$~78  &- &- &- &- &- &- &-	&-	\\                      
            Sk\,$-$70$^{\circ}$~16  &- &- &- &- &- &- &-	&-	\\                      
            Sk\,$-$68$^{\circ}$~8   &- &- &- &- &- &- &-	&-	\\                      
            Sk\,$-$67$^{\circ}$~195 &- &- &- &- &- &218 &211	  &215$\pm 62$	\\      
            \noalign{\smallskip}
            \hline
        \end{tabular}
    \end{table*}

    \begin{table*}
        \caption{$\varv_{\rm edge}$ for the stars in our sample from all the available P Cygni line profiles in the UV. Single velocity measurement from each line is subject to an uncertainty 
        $\Delta\varv_{\rm edge} = 30~{\rm km\,s^{-1}}$ which take into acount the velocity resolution of the UV spectra and the uncertainty in the radial velocity correction which were added in 
        quadrature.}              
        \label{table_app_2}     
        \centering                                     
        \begin{tabular}{c c c c c c c c c}         
            \hline{\smallskip}
            Target &$\ion{Si}{iv}~1393$ &$\ion{Si}{iv}~1403$ &$\ion{C}{iv}~1548$ &$\ion{Al}{III}~1855$ &$\ion{Al}{III}~1863$ &$\ion{Mg}{II}~2796$ &$\ion{Mg}{II}~2803$ &$\overline{\varv_{\rm black}}/\overline{\varv_{\rm edge}}$ \\
            \hline
            Sk\,$-$66$^{\circ}$~171 &2135 &- &2119 &- &- &- &-	&0.83	\\
            Sk\,$-$68$^{\circ}$~155 &1860 &- &1799 &- &- &- &-	&0.81	\\
            Sk\,$-$69$^{\circ}$~279 &866 &- &- &- &- &- &-	&0.72	\\
            Sk\,$-$71$^{\circ}$~41 &1729 &- &1737 &- &- &- &-	&-	\\
            Sk\,$-$68$^{\circ}$~135 &1247 & &1226 &- &- &- &-	&0.71	\\
            Sk\,$-$67$^{\circ}$~5	 &1651 &- &1629 &- &- &- &-   &0.75	\\
            Sk\,$-$68$^{\circ}$~52 &1583 &1441 &1581 &- &- &- &- &0.75\\
            Sk\,$-$69$^{\circ}$~43 &1059 &1075 &- &- &- &- &-	&0.80    \\
            Sk\,$-$68$^{\circ}$~140 &1333 &1419 &1328 &- &- &- &-	&0.74	\\
            Sk\,$-$67$^{\circ}$~2 &- &505 &- &465 &464 &- &-	&0.79	\\ 
            Sk\,$-$67$^{\circ}$~14 &1005 &1059 &1081 &- &- &- &- &0.77	\\
            Sk\,$-$69$^{\circ}$~52 &608 &570 &613 &- &- &- &- &0.78    \\
            Sk\,$-$67$^{\circ}$~78 &473 &- &441 &- &- &- &- &-	\\
            Sk\,$-$70$^{\circ}$~16 &304 &- &275 &- &- &- &-	&-	\\
            Sk\,$-$68$^{\circ}$~8 &256 &258 &262 &255 &257 &-	&-  &-	\\
            Sk\,$-$67$^{\circ}$~195 &- &- &- &- &- &263 &261	&0.82	\\ 
            \noalign{\smallskip}
            \hline
        \end{tabular}
    \end{table*}

\clearpage
\section{Atomic data}
In Table~\ref{table_atomic} we list the ions, number of important levels, super-levels, full levels, and transitions considered per ion, which constitutes the atomic model used in our 
\textsc{CMFGEN} model grid to synthesize the spectrum. The grid is split into two groups, with the boundry between the two is set at $\approx25~{\rm kK}$. 

         \begin{table*}
        \caption{Ions and number of important levels, super-levels, full levels, and transitions considered per ion, which constitutes the atomic model used by the code to synthesize the spectrum.
         We divide our model grid into a 'cool' ($<~25~kK$) and a 'hot' ($>~25~kK$) sections which correspond to different ions. ``all" in the models column means that this ion has was included in both the hot and cool sections of our grid}      
 	\def\arraystretch{0.75}
        \label{table_atomic}    
        \centering                                     
        \begin{tabular}{c c c c c c}          
            \hline{\smallskip}           
    Ion             &important levels &super-levels &full levels  &Transitions    &models   \\
\hline{\smallskip}
		$\ion{H}{I}$    &20               &20           &30           &435            &all      \\   
\\
    $\ion{He}{I}$   &45               &45           &69           &905            &all      \\   
		$\ion{He}{II}$  &22               &22           &30           &435            &all      \\
\\
    $\ion{C}{I}$    &52               &52           &100          &10204          &$<25~kK$ \\      
		$\ion{C}{II}$   &40               &40           &92           &8017           &all      \\
		$\ion{C}{III}$  &51               &51           &84           &5528           &all      \\
		$\ion{C}{IV}$   &64               &64           &64           &1446           &all      \\
		$\ion{C}{V}$    &0                &35           &67           &2196           &$>25~kK$ \\
\\
    $\ion{N}{I}$    &44               &44           &104          &855            &$<25~kK$ \\
		$\ion{N}{II}$   &45               &45           &85           &7879           &all      \\
		$\ion{N}{III}$  &104              &104          &164          &6710           &all      \\  
		$\ion{N}{IV}$   &107              &107          &202          &6943           &all      \\
		$\ion{N}{V}$    &45               &45           &67           &1664           &$>25~kK$ \\
\\  
    $\ion{O}{I}$    &53               &53           &133          &2138           &$<25~kK$ \\
		$\ion{O}{II}$   &54               &54           &123          &8937           &all      \\
		$\ion{O}{III}$  &788              &88           &170          &6561           &all      \\
		$\ion{O}{IV}$   &78               &78           &154          &7599           &all      \\
		$\ion{O}{V}$    &32               &32           &56           &2324           &all      \\
		$\ion{O}{VI}$   &25               &25           &31           &1475           &$>25~kK$ \\
\\
		$\ion{Mg}{I}$   &36               &36           &56           &3511           &$<25~kK$ \\
		$\ion{Mg}{II}$  &27               &27           &60           &1993           &all      \\
		$\ion{Mg}{III}$ &0                &39           &175          &3052           &all      \\
		$\ion{Mg}{IV}$  &0                &36           &169          &5706           &$>25~kK$ \\
\\
		$\ion{Ca}{I}$   &0                &66           &88           &3976           &$<25~kK$ \\
		$\ion{Ca}{II}$  &0                &39           &46           &1736           &$<25~kK$ \\
		$\ion{Ca}{III}$ &0                &29           &88           &3497           &$<25~kK$ \\
		$\ion{Ca}{IV}$  &0                &30           &123          &8532           &$<25~kK$ \\
\\
		$\ion{Al}{I}$   &54               &54           &82           &4985           &$<25~kK$ \\
		$\ion{Al}{II}$  &0                &40           &67           &3490           &$<25~kK$ \\
		$\ion{Al}{III}$ &0                &37           &60           &2011           &all      \\
		$\ion{Al}{IV}$  &0                &56           &163          &3052           &$>25~kK$ \\
\\
		$\ion{Si}{II}$  &22               &22           &43           &2294           &all      \\
		$\ion{Si}{III}$ &33               &33           &33           &1639           &all      \\
		$\ion{Si}{IV}$  &22               &22           &33           &1090           &all      \\
\\
		$\ion{P}{III}$  &0                &59           &128          &5576           &$>25~kK$ \\
		$\ion{P}{IV}$   &0                &30           &90           &2537           &$>25~kK$ \\
		$\ion{P}{V}$    &0                &16           &62           &561            &$>25~kK$ \\
\\
		$\ion{S}{II}$   &0                &21           &65           &8208           &all      \\
		$\ion{S}{III}$  &0                &24           &44           &6193           &all      \\
		$\ion{S}{IV}$   &0                &51           &142          &3598           &all      \\
		$\ion{S}{V}$    &0                &31           &98           &3462           &all      \\
\\
		$\ion{Fe}{I}$   &100              &100          &297          &141821         &$<25~kK$ \\
		$\ion{Fe}{II}$  &0                &318          &2430         &21544          &$<25~kK$ \\
		$\ion{Fe}{III}$ &104              &104          &1433         &136060         &all      \\
		$\ion{Fe}{IV}$  &74               &74           &540          &72223          &all      \\
		$\ion{Fe}{V}$   &50               &50           &220          &71983          &$>25~kK$ \\
		$\ion{Fe}{VI}$  &44               &44           &433          &185392         &$>25~kK$ \\
		$\ion{Fe}{VII}$ &29               &29           &153          &86504          &$>25~kK$ \\
\\
		$\ion{Ni}{II}$  &0                &42           &433          &51707          &$<25~kK$ \\
		$\ion{Ni}{III}$ &0                &46           &849          &66486          &all      \\
		$\ion{Ni}{IV}$  &0                &36           &200          &72898          &all      \\
		$\ion{Ni}{V}$   &0                &69           &305          &75541          &$>25~kK$ \\
		$\ion{Ni}{VI}$  &0                &65           &314          &79169          &$>25~kK$ \\
            \noalign{\smallskip}
            \hline
        \end{tabular}
    \end{table*}

\clearpage
\section{Spectral energy distribution (SED)}
\label{SED_app}
In Fig.~\ref{SED_1} we present the SED fits for the entire sample.
\begin{figure*}
\captionsetup[subfloat]{labelformat=empty}
\centering
\subfloat[]{
  \includegraphics[width=95mm]{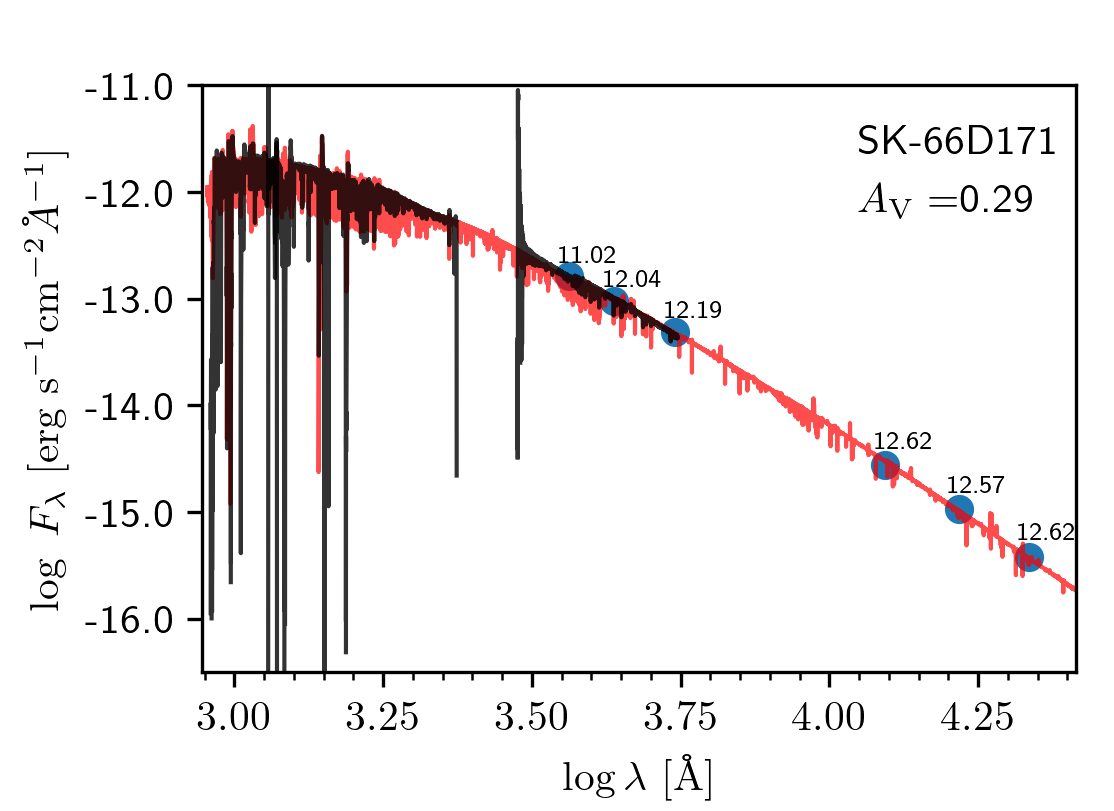}
}
\subfloat[]{
  \includegraphics[width=95mm]{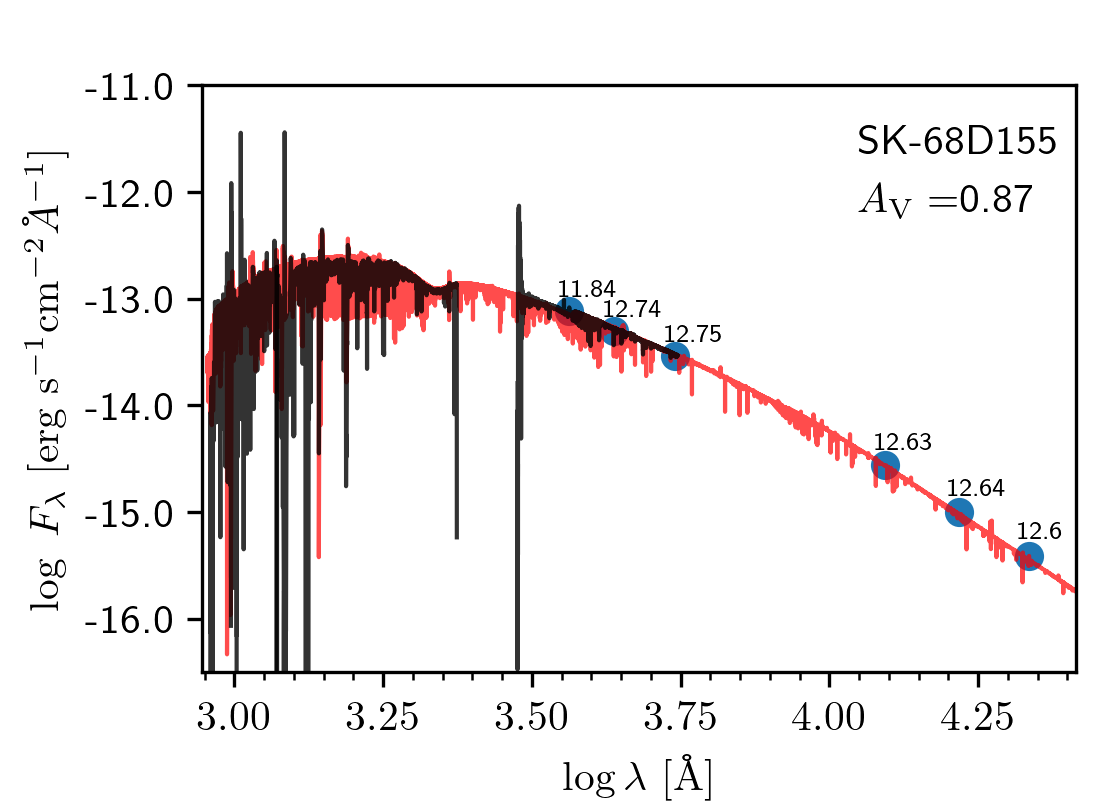}
}
\hspace{0mm}
\subfloat[]{
  \includegraphics[width=95mm]{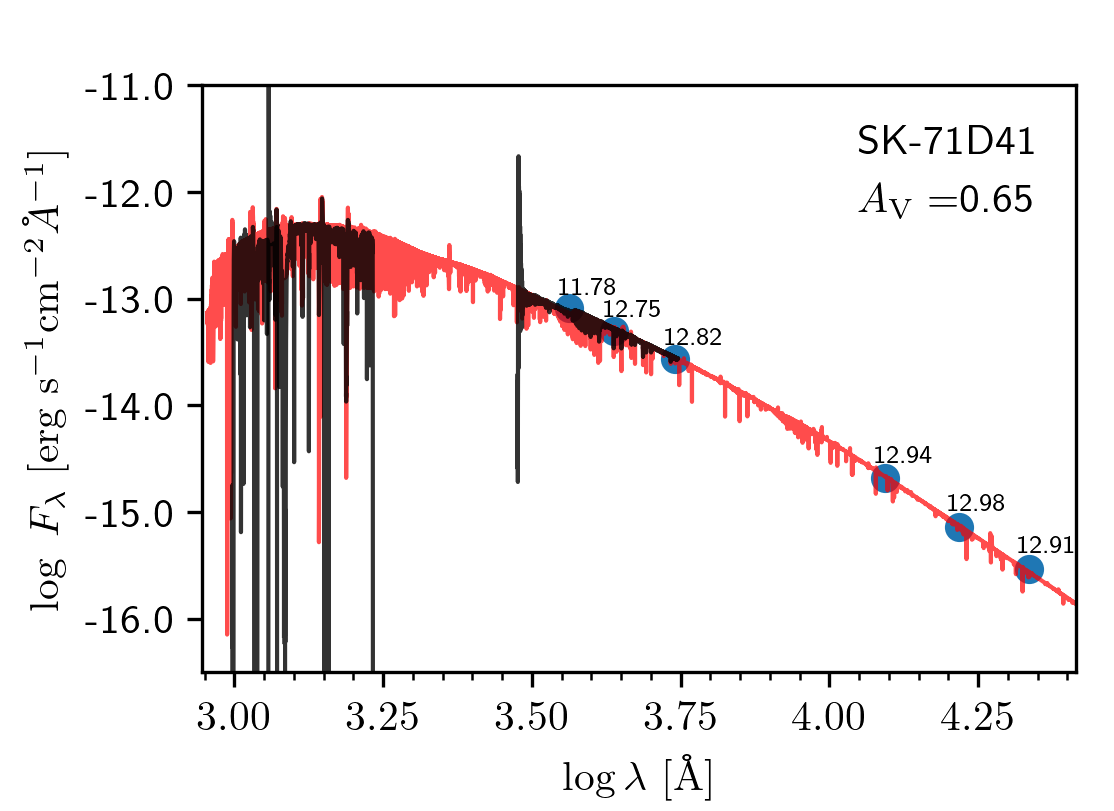}
}
\subfloat[]{  
  \includegraphics[width=95mm]{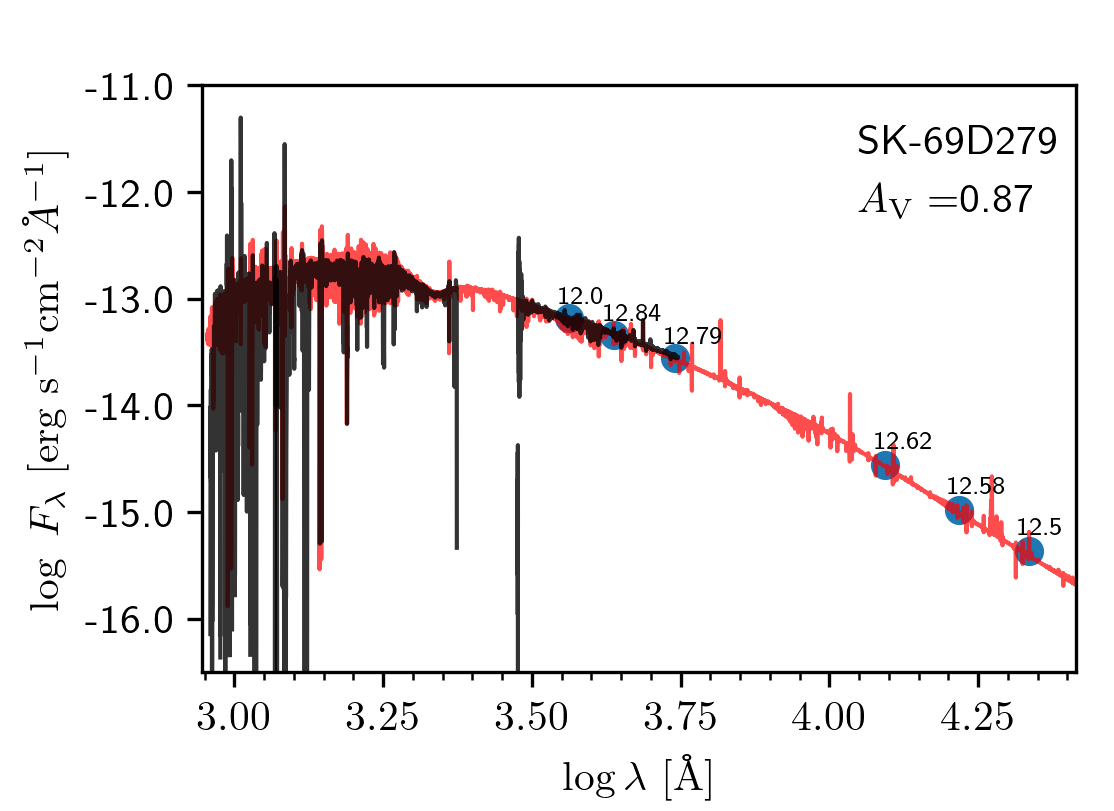}
}
\hspace{0mm}
\subfloat[]{
  \includegraphics[width=95mm]{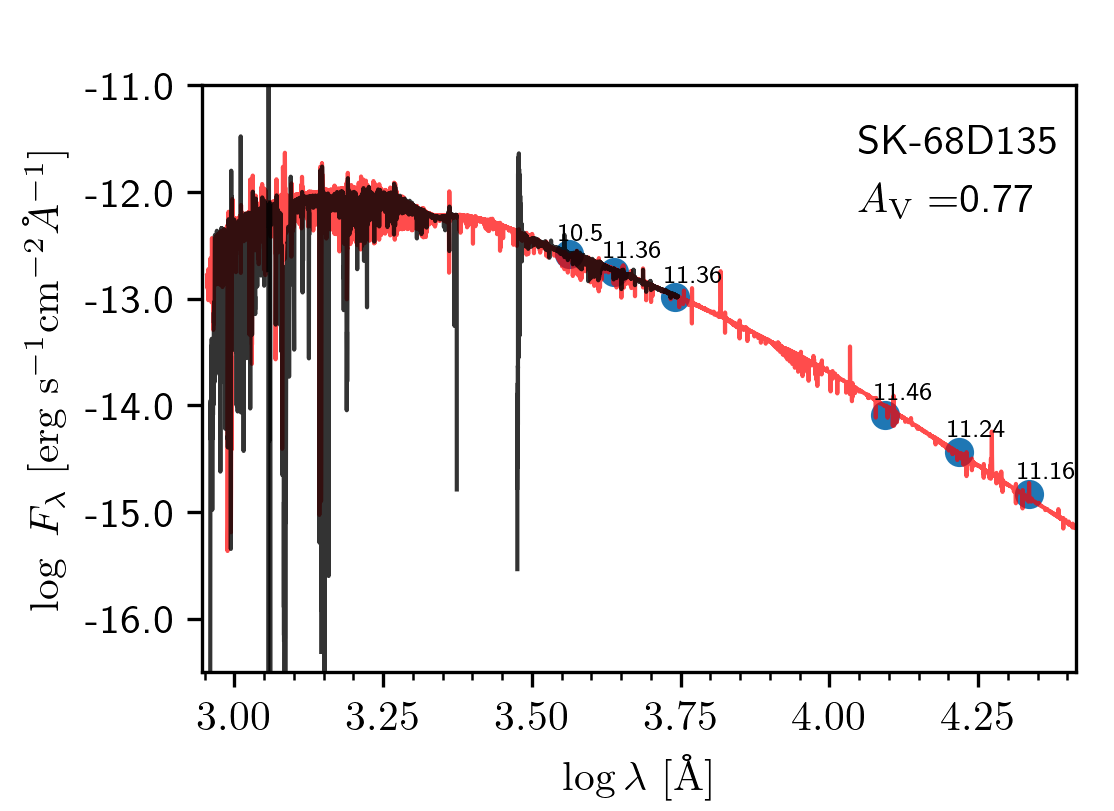}
}
\subfloat[]{  
  \includegraphics[width=95mm]{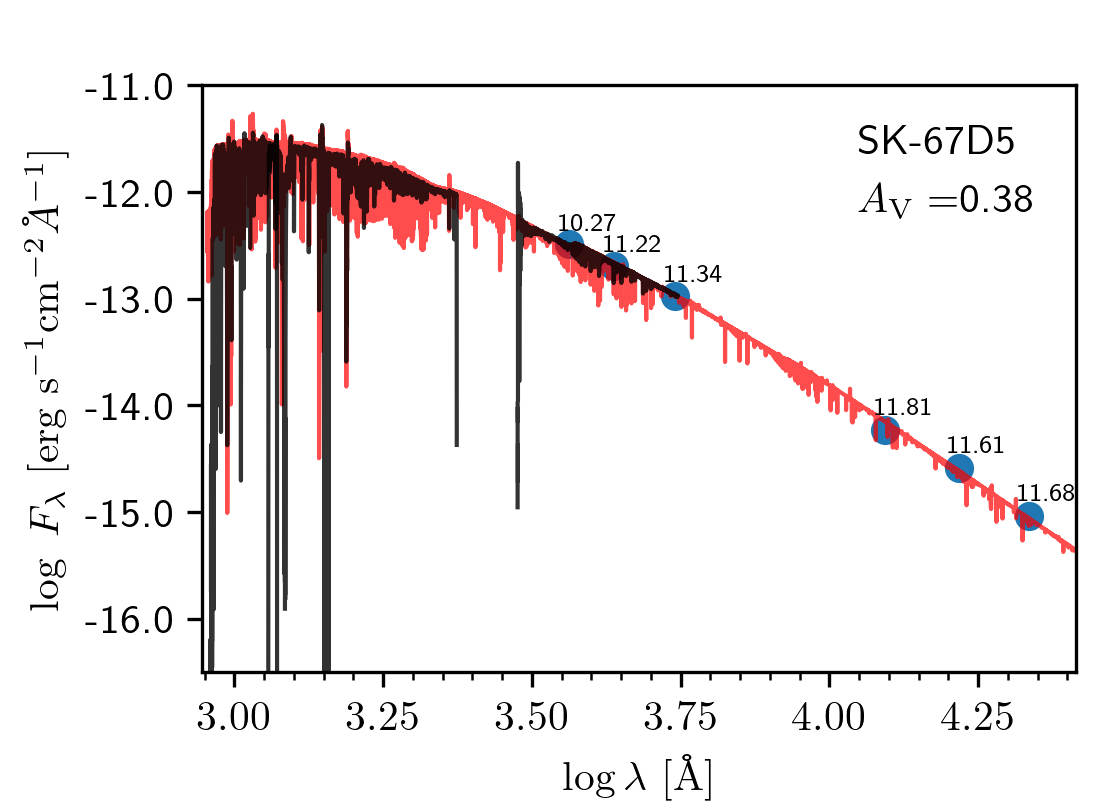}
}
\caption{Spectral energy distribution for all the stars in our sample. Red solid line: model SED, Black solid line: observed SED. The photometric points are indicated as blue circles.}
\label{SED_1}
\end{figure*}

\begin{figure*}
\captionsetup[subfloat]{labelformat=empty}
\centering
\ContinuedFloat
\subfloat[]{
  \includegraphics[width=95mm]{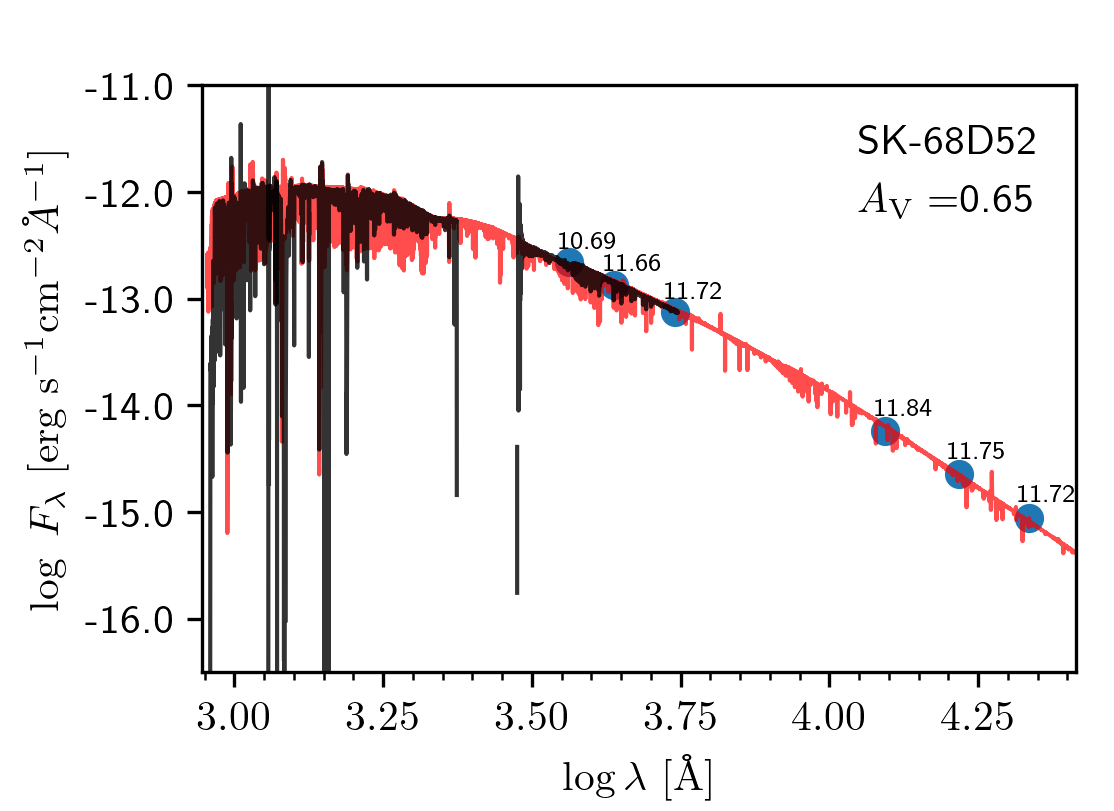}
}
\subfloat[]{
  \includegraphics[width=95mm]{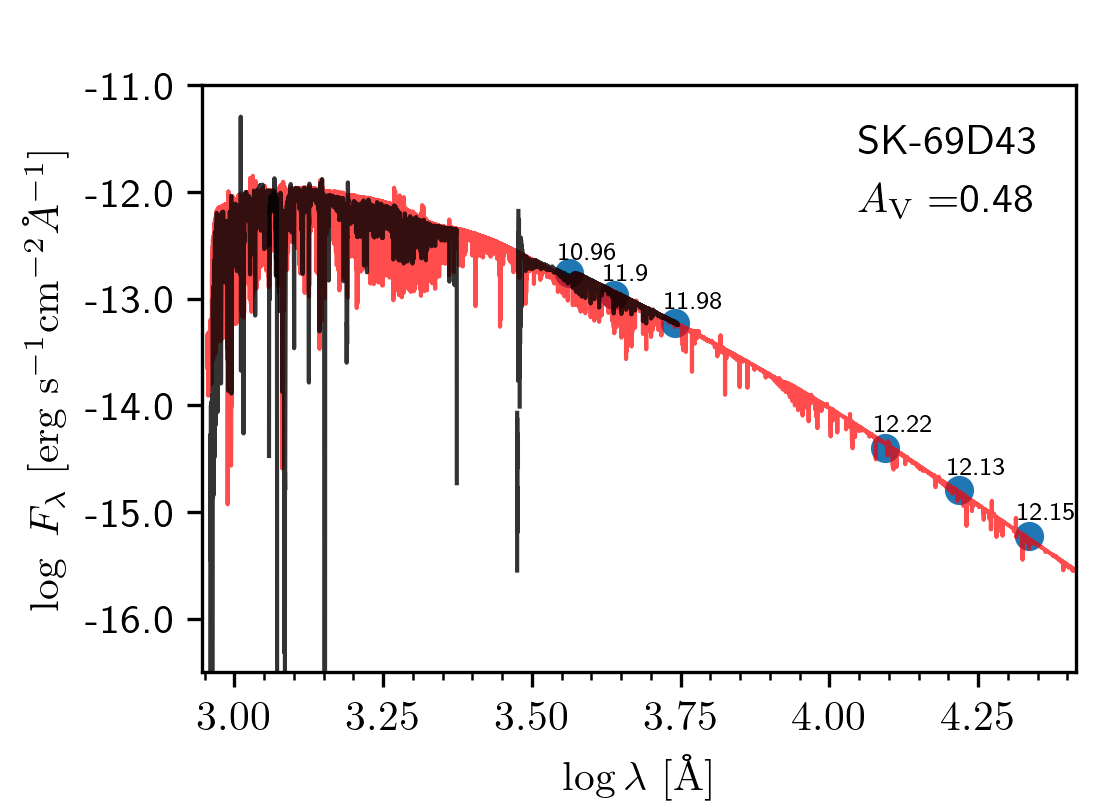}
}
\hspace{0mm}
\subfloat[]{
  \includegraphics[width=95mm]{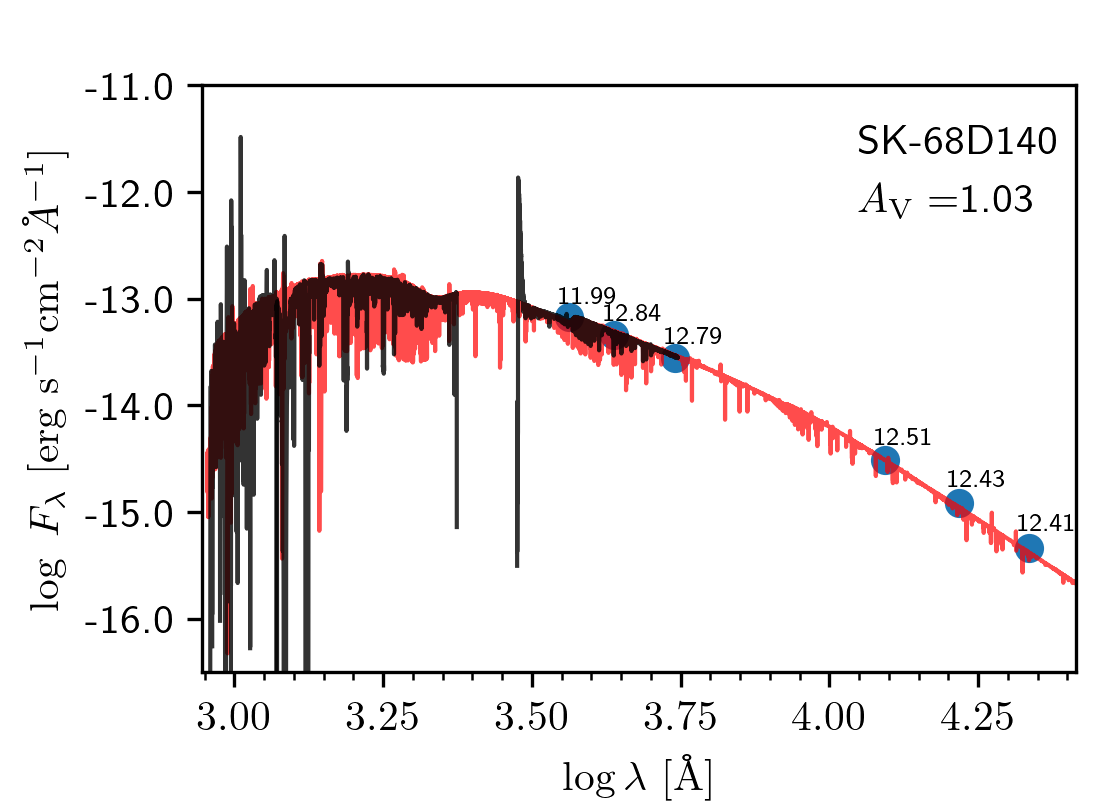}
}
\subfloat[]{
  \includegraphics[width=95mm]{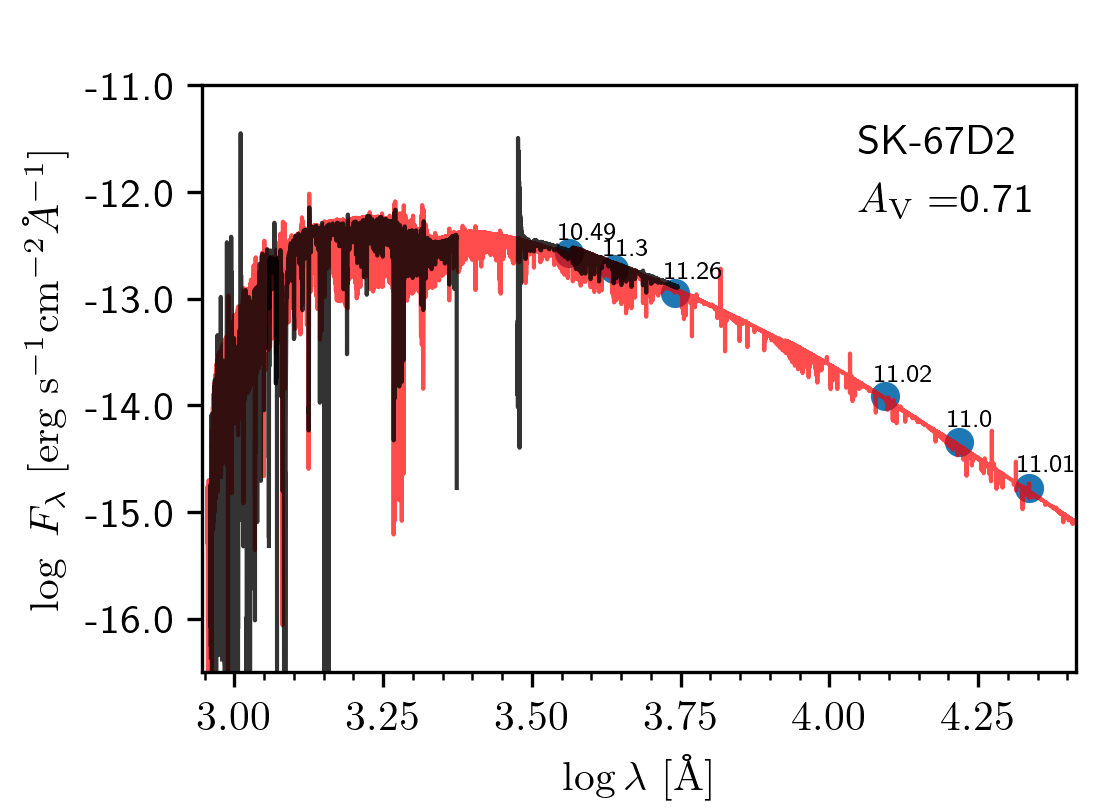}
}
\hspace{0mm}
\subfloat[]{
  \includegraphics[width=95mm]{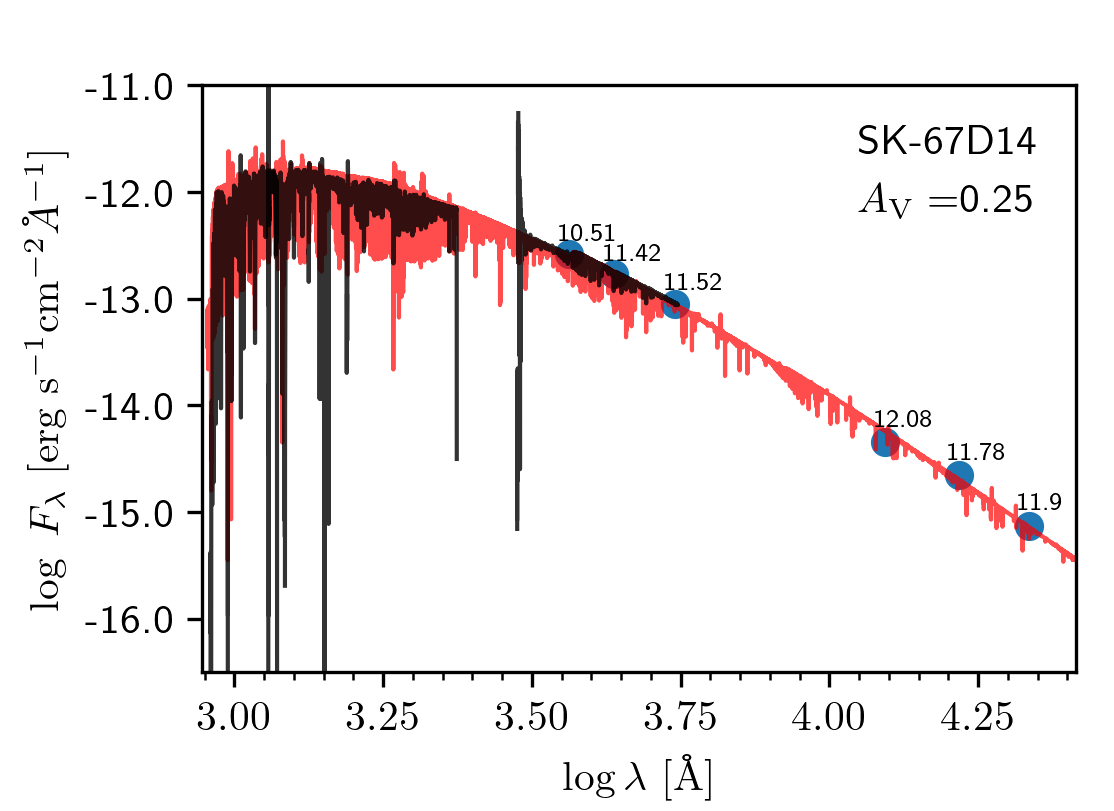}
}
\subfloat[]{   
  \includegraphics[width=95mm]{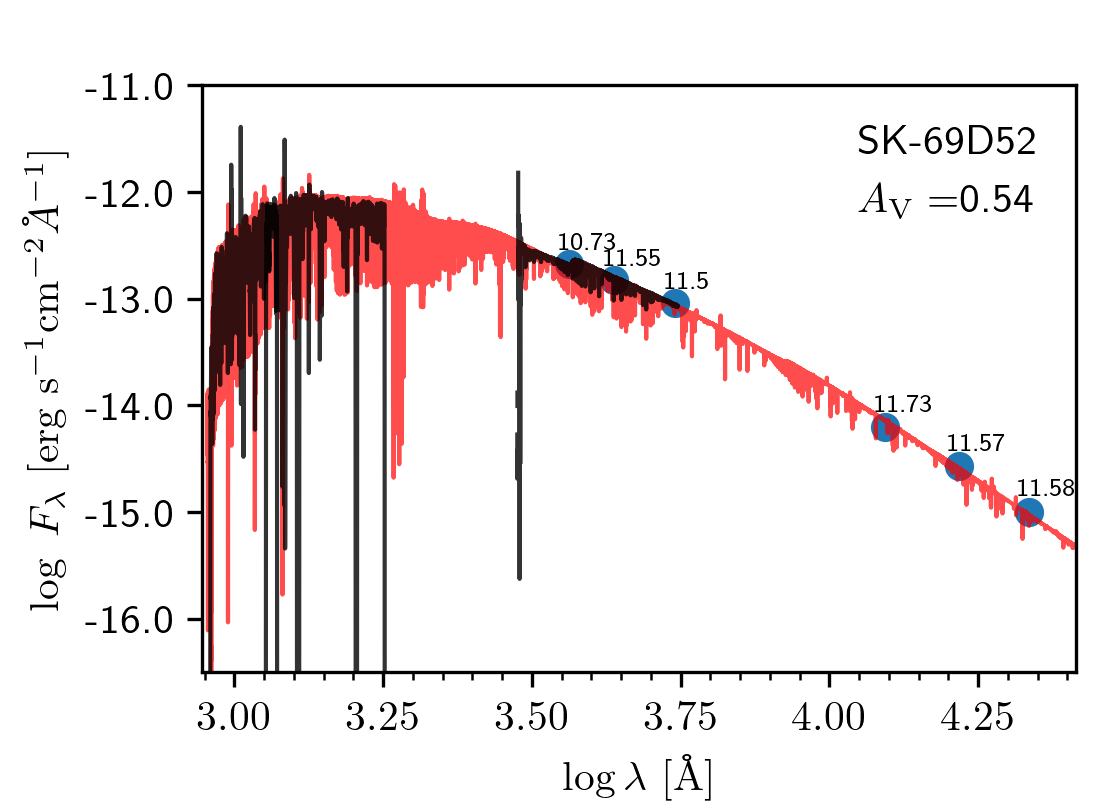}
}
\caption{continued}
\end{figure*}

\begin{figure*}
\captionsetup[subfloat]{labelformat=empty}
\centering
\ContinuedFloat
\subfloat[]{
  \includegraphics[width=95mm]{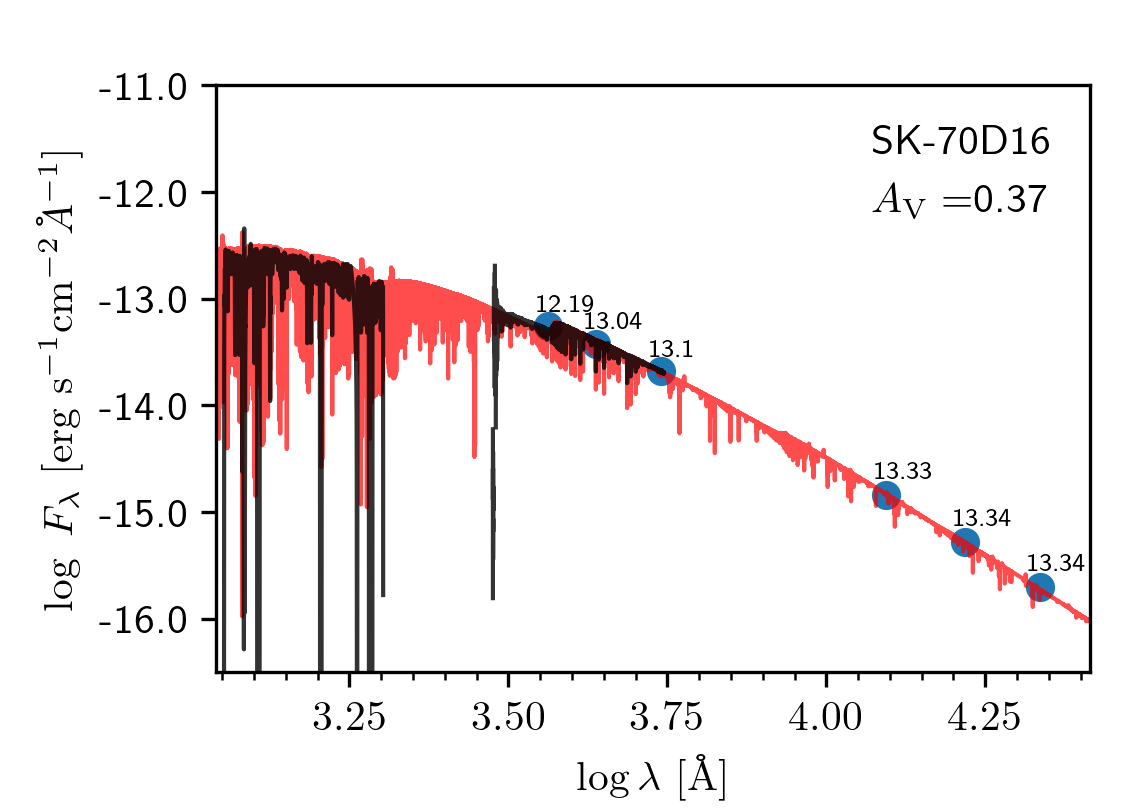}
}
\subfloat[]{
  \includegraphics[width=95mm]{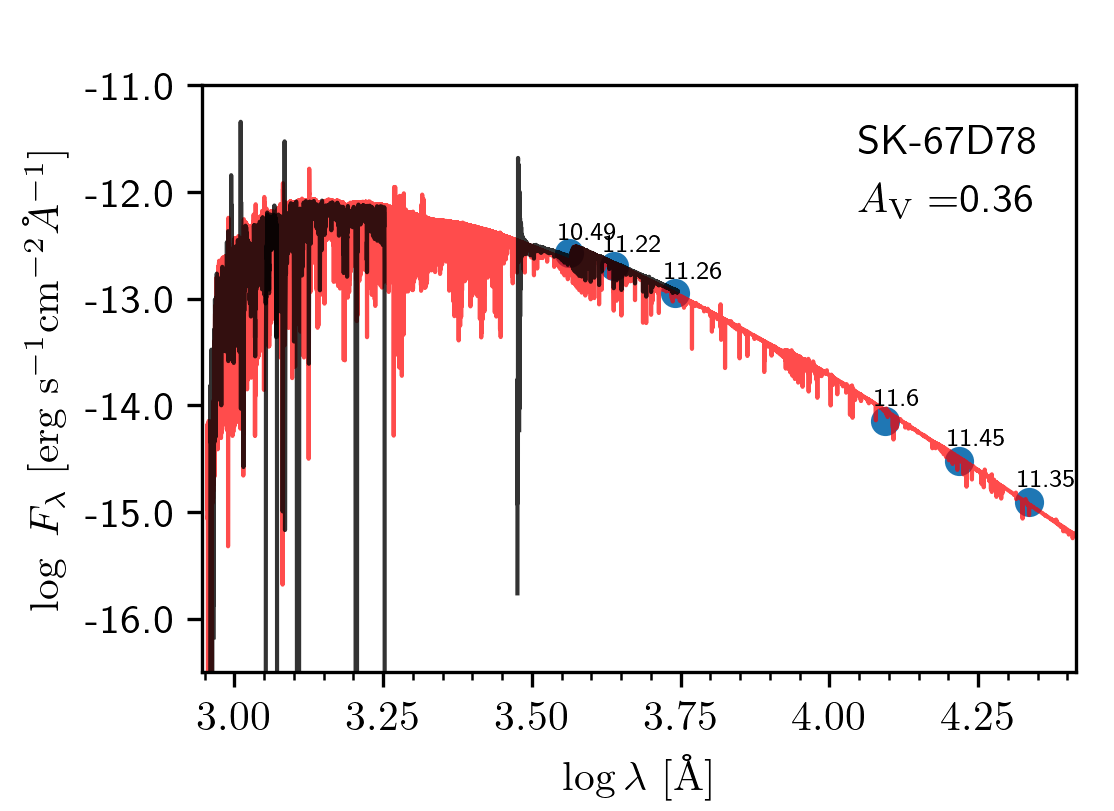}
}
\hspace{0mm}
\subfloat[]{
  \includegraphics[width=95mm]{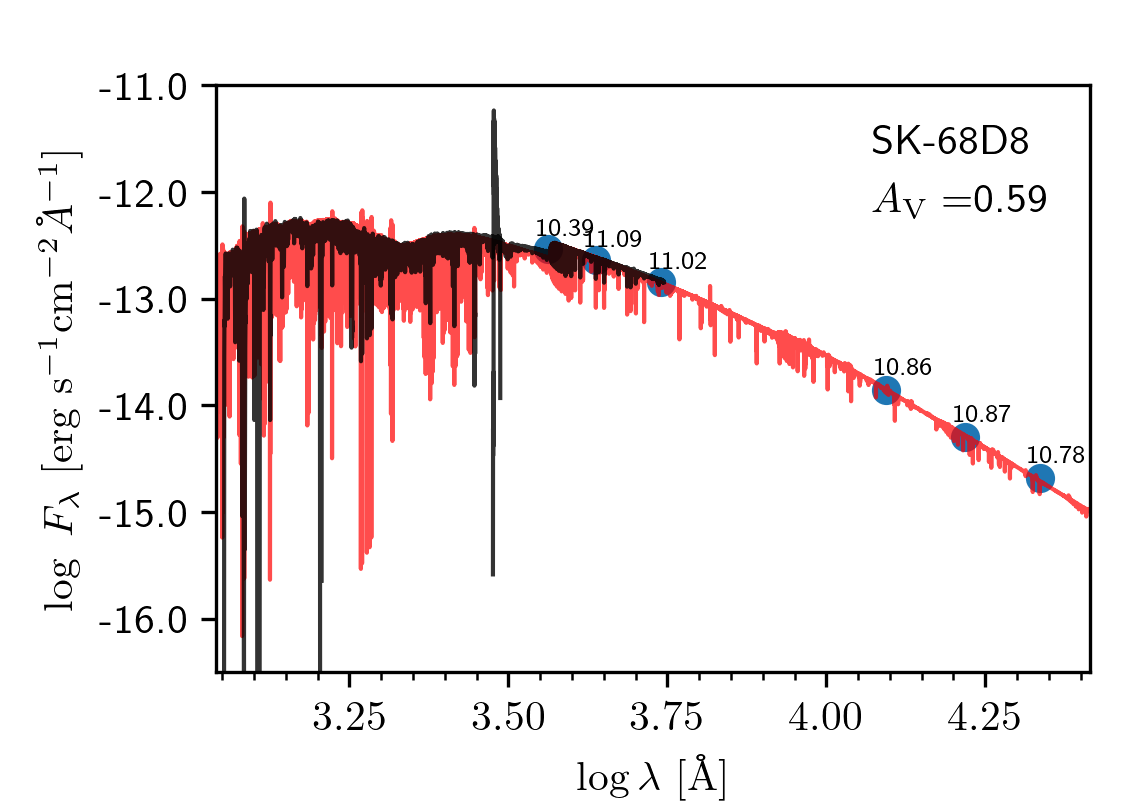}
}
\subfloat[]{
  \includegraphics[width=95mm]{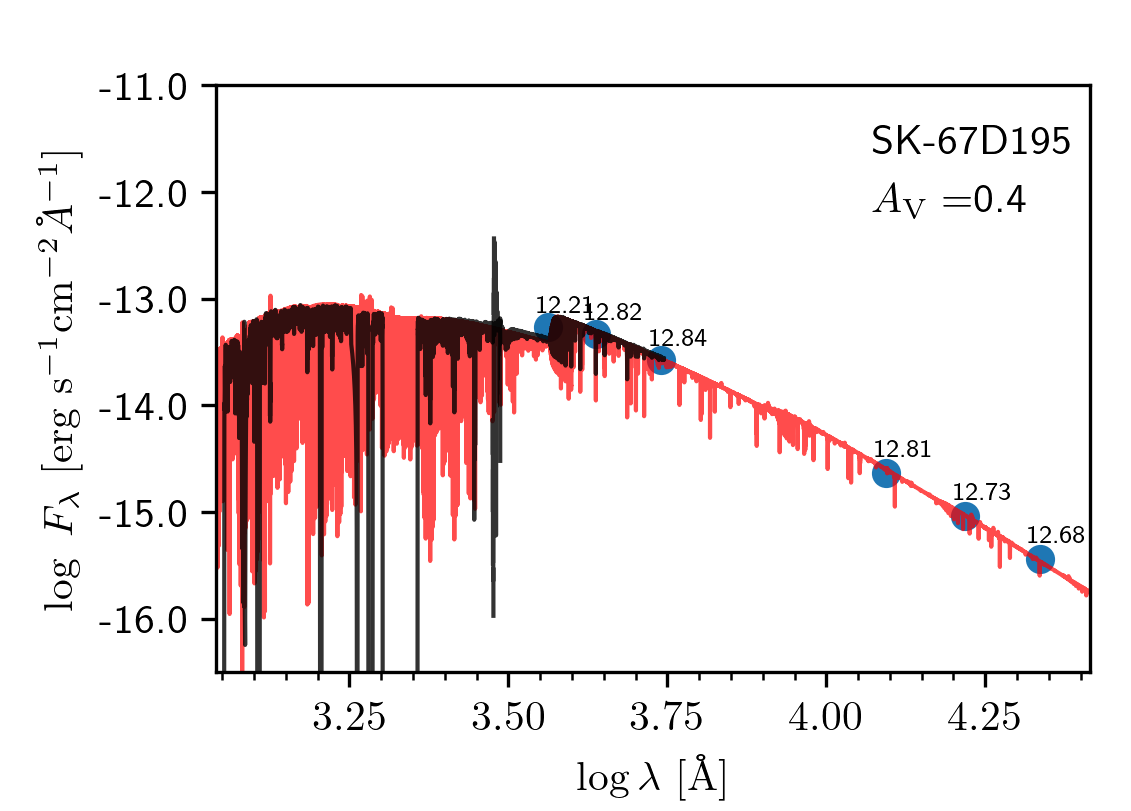}
}
\caption{continued}
\end{figure*}

\clearpage
\section{Comparison to literature}
In this section, we include Table~\ref{comparison_table}, in which we present our obtained values for $T_{\rm eff}$, $\log{g}$, $\log{\dot{M}}$, $f_{\rm vol}$, and $\beta$ in comparison to values 
presented in other previous studies. The volume filling factor for studies that utilise the optically-thick clumping is calculated from the equation introduced in \citet{sander2024}: 
\begin{equation}
    \label{eq:BC} 
    f_{\rm vol} = \frac{(1-f_{\rm ic})^2}{f_{\rm cl}-2f_{\rm ic}+f_{\rm ic}^2},
\end{equation}
where $f_{\rm cl} = \langle\rho^2\rangle/\langle\rho\rangle^2$ describes the contrast between the density in the clump and the mean density, and $f_{\rm ic}$ describes the 
contrast between the clump density and the density of the inter-clump medium.

\label{app:comp}
\begin{table*}
  \centering   
\begin{threeparttable}[b]
    \caption{Comparison of derived parameters to previous studies. Obtained using F: \textsc{FASTWIND}, P: \textsc{PoWR}, C: \textsc{CMFGEN}, T: \textsc{TLUSTY}. The mass-loss rates are not corrected for clumping.}     
        \def\arraystretch{1.0}
        \label{comparison_table}                           
        \small                              
        \addtolength{\tabcolsep}{-0.0em}
        \begin{tabular}{c c c c c c c c c c c}         
            \hline
            Target&         &$T_{\rm eff}$          &$\log{g}$                  &$\log{L}$                &$\log{\dot{M}}$            &$f_{\rm vol}$  &$\beta$  &Ref.       &wavelength &Code \\[1.50 pt]
            
            Sk\,$-$&        &kK                     &${\rm cm\,s^{-2}}$         &$L_{\odot}$              &$M_{\odot}\,{\rm yr}^{-1}$ &               &         &           &           &\\     
            \hline
            \\[0.35 pt]
	$66^{\circ}$ 171&         &$29.9_{-1.2}^{+1.2}$   &$3.12_{-0.19}^{+0.14}$     &$5.66_{-0.08}^{+0.08}$  &$-5.99_{-0.23}^{+0.85}$     &0.1            &1.00     &JB25       &Opt        &F    \\[1.50 pt]
                  &	        &$32.3_{-1.75}^{+0.70}$	&$3.20_{-0.12}^{+0.47}$	    &$5.80_{-0.07}^{+0.03}$   &$-6.18_{-0.07}^{+0.55}$	  &0.0207         &1.60     &SB25       &UV+Opt     &F    \\[1.50 pt]                  
				          &         &$29.9_{-1.0}^{+1.0}$	  &$3.11_{-0.20}^{+0.20}$	    &$5.67_{-0.11}^{+0.11}$   &$-6.07_{-0.33}^{+0.33}$	  &0.03           &1.70     &This study &UV+Opt     &C    \\[1.50 pt]
                                                                                                                                                                          \\[0.35 pt]
	$68^{\circ}$ 155&         &$29.9_{-1.2}^{+1.2}$   &$3.12_{-0.19}^{+0.14}$     &$5.70_{-0.08}^{+0.08}$   &$-6.19_{-0.23}^{+0.23}$    &0.1            &1.00     &JB25       &Opt        &F    \\[1.50 pt]
                  &         &$31.5_{-0.25}^{+0.75}$	&$3.20_{-0.05}^{+0.20}$	    &$5.81_{-0.02}^{+0.03}$   &$-5.85_{-0.10}^{+0.12}$	  &0.0366         &1.65     &SB25       &UV+Opt     &F    \\[1.50 pt]
                  &         &$32.5_{-2.00}^{+0.25}$ &$3.25_{-0.05}^{+0.10}$     &$5.77_{-0.04}^{+0.04}$   &$-6.00_{-0.15}^{+0.05}$    &0.0016         &1.35     &V24        &UV+Opt     &F    \\[1.50 pt]
					        &         &$29.0_{-1.0}^{+1.0}$   &$3.06_{-0.2}^{+0.2}$       &$5.64_{-0.11}^{+0.11}$   &$-6.19_{-0.25}^{+0.25}$	  &0.03           &1.50     &This study &UV+Opt     &C    \\[1.50 pt]
                                                                                                                                                                          \\[0.35 pt]  
	$69^{\circ}$ 279&         &$(15.1_{-0.8}^{+0.4})$ &$(1.50_{-0.00}^{+0.10})$   &$(2.37_{-0.10}^{+0.07})$ &$(-7.75_{-0.22}^{+0.30})$  &0.1            &1.00     &JB25       &Opt        &F    \\[1.50 pt]
                  &	        &$29.5_{-0.5}^{+0.5}$   &-		                      &$5.54_{-0.06}^{+0.06}$   &$-5.26_{-0.04}^{+0.04}$	  &0.5            &3.00     &G18        &UV+Opt     &C    \\[1.50 pt]
					        &         &$28.5_{-2.0}^{+2.0}$   &$2.95_{-0.22}^{+0.22}$     &$5.63_{-0.11}^{+0.11}$   &$-5.70_{-0.39}^{+0.39}$	  &0.10           &2.70     &This study &UV+Opt     &C    \\[1.50 pt]
                                                                                                                                                                          \\[0.35 pt]
	$71^{\circ}$ 41 &         &$29.9_{-1.2}^{+1.2}$   &$3.31_{-0.14}^{+0.14}$     &$5.60_{-0.08}^{+0.08}$   &$-6.22_{-0.23}^{+0.23}$    &0.1            &1.00     &JB25       &Opt        &F    \\[1.50 pt]
                  &	        &$31.0_{-1.25}^{+1.00}$	&$3.12_{-0.07}^{+0.23}$	    &$5.65_{-0.05}^{+0.04}$   &$-6.55_{-0.05}^{+0.10}$	  &0.0190         &2.90     &SB25       &UV+Opt     &F    \\[1.50 pt]
                  &         &$30.0_{-2.0}^{+2.0}$   &$3.40_{-0.20}^{+0.20}$	    &$5.50_{-0.10}^{+0.10}$   &$-6.09_{-0.10}^{+0.10}$	  &0.10           &1.00     &R18        &UV+Opt     &P    \\[1.50 pt]
					        &         &$29.2_{-1.0}^{+1.0}$	  &$3.12_{-0.20}^{+0.20}$     &$5.53_{-0.11}^{+0.11}$   &$-6.03_{-0.25}^{+0.25}$ 	  &0.10           &1.20     &This work  &UV+Opt     &C    \\[1.50 pt]
                                                                                                                                                                          \\[0.35 pt]
	$68^{\circ}$ 135&         &$(22.3_{-1.2}^{+0.8})$ &$(2.31_{-0.14}^{+0.14})$   &$(5.09_{-0.10}^{+0.08})$ &($-5.94_{-0.23}^{+0.30})$  &0.1            &1.00     &JB25       &Opt        &F    \\[1.50 pt]
                  &         &$27.5$                 &$2.70$	                    &$6.97$                   &$-5.12$	                  &-              &3.50     &E04        &UV+Opt     &C    \\[1.50 pt]
					        &         &$26.9_{-1.5}^{+1.5}$	  &$2.82_{-0.21}^{+0.21}$     &$6.10_{-0.12}^{+0.12}$   &$-5.70_{-0.31}^{+0.31}$	  &0.10           &2.30     &This study &UV+Opt     &C    \\[1.50 pt]
                                                                                                                                                                          \\[0.35 pt]
	$67^{\circ}$ 5	&         &$26.8_{-1.2}^{+1.2}$   &$3.22_{-0.33}^{+0.14}$     &$5.96_{-0.15}^{+0.09}$   &$-5.95_{-0.23}^{+0.23}$    &0.1            &1.00     &JB25       &Opt        &F    \\[1.50 pt]
                  &	        &$27.5_{-1.00}^{+1.00}$	&$2.98_{-0.15}^{+0.10}$	    &$5.40_{-0.04}^{+0.04}$   &$-6.64_{-0.12}^{+0.15}$	  &0.0200         &2.25     &SB25       &UV+Opt     &F    \\[1.50 pt]
					        &         &$25.6_{-1.0}^{+1.0}$	  &$2.81_{-0.20}^{+0.20}$     &$5.89_{-0.12}^{+0.12}$   &$-6.05_{-0.14}^{+0.14}$	  &0.10           &1.30     &This study &UV+Opt     &C    \\[1.50 pt]
                                                                                                                                                                          \\[0.35 pt]
	$68^{\circ}$ 52	&         &$25.2_{-1.2}^{+1.2}$   &$2.88_{-0.14}^{+0.19}$     &$5.86_{-0.09}^{+0.09}$   &$-6.03_{-0.23}^{+0.23}$    &0.1            &1.00     &JB25       &Opt        &F    \\[1.50 pt]
                  &         &$28.3_{-0.75}^{+0.75}$ &$3.10_{-0.10}^{+0.25}$     &$5.96_{-0.02}^{+0.02}$   &$-6.28_{-0.25}^{+0.25}$	  &0.0836         &1.95     &V24        &UV+Opt     &F    \\[1.50 pt]
                  &         &$24.5$	                &$2.70$	                    &$5.76$                   &$-5.49$	                  &-              &2.75     &E04        &UV+Opt     &C    \\[1.50 pt]
					        &         &$26.0_{-1.0}^{+1.0}$   &$2.85_{-0.20}^{+0.20}$     &$5.87_{-0.12}^{+0.12}$   &$-6.28_{-0.25}^{+0.25}$	  &0.03           &2.00     &This study &UV+Opt     &C    \\[1.50 pt]
                                                                                                                                                                          \\[0.35 pt]
	$69^{\circ}$ 43	&         &$23.7_{-3.5}^{+1.2}$   &$2.69_{-0.33}^{+0.14}$     &$5.62_{-0.23}^{+0.10}$   &$-6.34_{-0.23}^{+0.23}$    &0.1            &1.00     &JB25       &Opt        &F    \\[1.50 pt]
                  &         &$25.0_{-2.00}^{+0.75}$	&$2.85_{-0.25}^{+0.05}$	    &$5.62_{-0.09}^{+0.02}$   &$-6.30_{-0.25}^{+0.05}$	  &0.2350         &2.45     &V24        &UV+Opt     &F    \\[1.50 pt]
					        &         &$22.8_{-0.22}^{+0.25}$ &$2.60_{-0.02}^{+0.02}$	    &$5.48$                    &-		                      &-              &-        &U17        &Opt        &F    \\[1.50 pt]	
					        &         &$22.4_{-1.0}^{+1.0}$	  &$2.71_{-0.21}^{+0.21}$     &$5.55_{-0.11}^{+0.11}$   &$-6.49_{-0.25}^{+0.25}$	  &0.10           &2.00     &This study &UV+Opt     &C    \\[1.50 pt]
                                                                                                                                                                          \\[0.35 pt]
	$68^{\circ}$ 140&         &$23.7_{-2.0}^{+1.2}$   &$3.12_{-0.29}^{+0.19}$     &$5.29_{-0.27}^{+0.10}$   &$-6.18_{-0.52}^{+0.23}$    &0.1            &1.00     &JB25       &Opt        &F    \\[1.50 pt]
                  &	        &$23.5_{-1.00}^{+1.00}$ &$2.75_{-0.10}^{+0.10}$     &$5.64_{-0.10}^{+0.10}$   &-		                      &-              &-        &M15        &Opt        &T    \\[1.50 pt]
					        &         &$24.1_{-1.0}^{+1.0}$   &$2.81_{-0.21}^{+0.21}$     &$5.52_{-0.11}^{+0.11}$   &$-6.46_{-0.25}^{+0.25}$	  &0.10           &2.20     &This study &UV+Opt     &C    \\[1.50 pt]
                                                                                                                                                                          \\[0.35 pt]
  $67^{\circ}$ 2	&         &$21.3_{-1.2}^{+0.8}$   &$2.50_{-0.14}^{+0.14}$     &$5.96_{-0.09}^{+0.06}$   &$-6.03_{-0.23}^{+0.85}$    &0.1            &1.00     &JB25       &Opt        &F    \\[1.50 pt]
                  &	        &$19.9_{-0.39}^{+0.36}$	&$2.30_{-0.02}^{+0.03}$	    &$5.81$                   &-		                      &-              &-        &U17        &Opt        &F    \\[1.50 pt]
                  &         &$18.8_{-1.0}^{+1.0}$   &$2.31_{-0.21}^{+0.21}$     &$5.76_{-0.12}^{+0.12}$   &$-6.21_{-0.26}^{+0.26}$	  &0.14           &3.00     &This study &UV+Opt     &C    \\[1.50 pt]
                                                                                                                                                                          \\[0.35 pt]
                  \hline
\end{tabular}

\begin{tablenotes}
  \item U17: \citep{urbaneja2017}, G18: \citep{gvaramadze2018}, R18: \citep{ramachandran2018}, E04: \citep{evans2004}, 
  M15: \citep{McEvoy2015}, V24: \citep{verhamme2024}, SB25: \citep{brands2025}, JB25: \citep{Bestenlehner2025}.
\end{tablenotes}
\end{threeparttable}
\end{table*}

\begin{table*}
  \ContinuedFloat
  \centering   
\begin{threeparttable}[b]
    \caption{Continued}          
        \def\arraystretch{1.1}                       
        \small                                
        \addtolength{\tabcolsep}{-0.0em}
        \begin{tabular}{c c c c c c c c c c c}          
            \hline
            Target&         &$T_{\rm eff}$          &$\log{g}$              &$\log{L}$                  &$\log{\dot{M}}$                &$f_{\rm vol}$  &$\beta$  &Ref.       &wavelength &Code \\[1.50 pt]
            
            Sk\,$-$&        &kK                     &${\rm cm\,s^{-2}}$     &$L_{\odot}$                &$M_{\odot}\,{\rm yr}^{-1}$     &               &         &           &           &\\     
            \hline
            \\[0.35 pt]
	$67^{\circ}$ 14	&         &$22.5_{-1.2}^{+0.8}$   &$2.50_{-0.14}^{+0.14}$ &$5.60_{-0.10}^{+0.08}$     &$-6.24_{-0.24}^{+0.23}$        &0.1            &1.00     &JB25       &Opt        &F    \\[1.50 pt]
                  &         &$23.3_{-0.75}^{+1.50}$ &$2.55_{-0.50}^{+0.35}$	&$5.53_{-0.03}^{+0.06}$     &$-6.25_{-0.30}^{+0.05}$	      &0.0028         &1.65     &V24        &UV+Opt     &F    \\[1.50 pt]
                  &         &$22.9_{-0.13}^{+0.12}$	&$2.70_{-0.01}^{+0.01}$ &$5.62$                     &-		                          &-              &-        &U17        &Opt        &F    \\[1.50 pt]
					        &         &$21.1_{-1.0}^{+1.0}$   &$2.51_{-0.21}^{+0.21}$ &$5.58_{-0.11}^{+0.11}$     &$-6.33_{-0.29}^{+0.29}$	      &0.10           &2.00     &This study &UV+Opt     &C    \\[1.50 pt]
                                                                                                                                                                          \\[0.35 pt]
	$69^{\circ}$ 52	&         &$20.1_{-1.2}^{+3.5}$   &$2.31_{-0.48}^{+0.38}$ &$5.58_{-0.11}^{+0.26}$     &$-6.89_{-0.30}^{+0.34}$        &0.1            &1.00     &JB25       &Opt        &F    \\[1.50 pt]
                  &         &$22.5_{1.75}^{+0.25}$	&$2.60_{-0.10}^{+0.05}$	&$5.73_{-0.08}^{+0.00}$     &$-6.70_{-0.05}^{+0.10}$	      &0.0002         &2.60     &V24        &UV+Opt     &F    \\[1.50 pt]
		 			        &         &$18.8_{-1.0}^{+1.0}$	  &$2.31_{-0.21}^{+0.21}$ &$5.60_{-0.11}^{+0.11}$     &$-6.62_{-0.38}^{+0.38}$	      &0.14           &2.50     &This study &UV+Opt     &C    \\[1.50 pt]
                                                                                                                                                                          \\[0.35 pt]
	$67^{\circ}$ 78	&         &$16.6_{-1.2}^{+0.8}$   &$2.31_{-0.33}^{+0.14}$ &$5.45_{-0.13}^{+0.09}$     &$-6.72_{-0.51}^{+0.23}$        &0.1            &1.00     &JB25       &Opt        &F    \\[1.50 pt]
                  &         &$17.8_{-0.25}^{+1.50}$	&$2.35_{-0.25}^{+0.05}$	&$5.62_{-0.08}^{+0.06}$     &$-6.82_{-0.03}^{+0.05}$	      &0.0362         &3.45     &V24        &UV+Opt     &F    \\[1.50 pt]
					        &         &$16.2_{-0.18}^{+0.98}$ &$2.20_{-0.02}^{+0.01}$	&$5.36$                     &-		                          &-              &-        &U17        &Opt        &F    \\[1.50 pt]
					        &         &$15.5_{-1.0}^{+1.0}$	  &$2.10_{-0.21}^{+0.21}$	&$5.47_{-0.11}^{+0.11}$     &$-6.70_{-0.40}^{+0.40}$	      &0.20           &3.00     &This study &UV+Opt     &C    \\[1.50 pt]
                                                                                                                                                                          \\[0.35 pt]
	$70^{\circ}$ 16	&         &$20.1_{-1.6}^{+0.5}$   &$2.69_{-0.24}^{+0.14}$ &$4.93_{-0.14}^{+0.06}$     &$-8.15_{-0.73}^{+1.06}$        &0.1            &1.00     &JB25       &Opt        &F    \\[1.50 pt]
                  &         &$20.9_{+0.70}^{+2.60}$	&$2.85_{-0.20}^{+0.25}$	&$4.92_{-0.04}^{+0.14}$     &$-7.70_{-1.150}^{+0.3}$	      &0.0305         &2.90     &V24        &UV+Opt     &F    \\[1.50 pt]
					        &         &$18.4_{-1.0}^{+1.0}$	  &$2.61_{-0.21}^{+0.21}$ &$4.88_{-0.10}^{+0.10}$     &$-7.60_{-0.34}^{+0.34}$	      &0.10           &1.00     &This study &UV+Opt     &C    \\[1.50 pt]
                                                                                                                                                                          \\[0.35 pt]	
	$68^{\circ}$ 8	&         &$13.5_{-0.8}^{+1.2}$   &$1.50_{-0.00}^{+0.29}$ &$5.23_{-0.10}^{+0.13}$     &$-6.85_{-0.37}^{+0.24}$        &0.1            &1.00     &JB25       &Opt        &F    \\[1.50 pt]
                  &         &$14.8_{-0.75}^{+2.00}$	&$1.95_{-0.25}^{+0.25}$	&$5.56_{-0.11}^{+0.00}$     &$-6.90_{-0.05}^{+0.60}$	      &0.0055         &2.00     &V24        &UV+Opt     &F    \\[1.50 pt]
	 				        &         &$14.1_{-1.0}^{+1.0}$   &$1.81_{-0.21}^{+0.21}$ &$5.57_{-0.11}^{+0.11}$     &$-6.50_{-0.26}^{+0.26}$	      &0.10           &1.00     &This study &UV+Opt     &C    \\[1.50 pt]
                                                                                                                                                                          \\[0.35 pt]
	$67^{\circ}$ 195&         &$12.0_{-0.4}^{+0.4}$   &$1.88_{-0.14}^{+0.19}$ &$4.55_{-0.08}^{+0.08}$     &$-7.58_{-0.29}^{+0.36}$        &0.1            &1.00     &JB25       &Opt        &F    \\[1.50 pt]
                  &	        &$14.2_{-0.50}^{+0.90}$	&$2.45_{-0.20}^{+0.20}$	&$4.72_{-0.03}^{+0.06}$     &$-7.45_{-1.50}^{+0.60}$	      &0.0267         &0.65     &V24        &UV+Opt     &F    \\[1.50 pt]
					        &         &$12.6_{-1.0}^{+1.0}$	  &$2.11_{-0.22}^{+0.22}$	&$4.65_{-0.10}^{+0.10}$     &$-7.50_{-0.35}^{+0.35}$	      &0.10           &1.00     &This study &UV+Opt     &C    \\[1.50 pt]
                                                                                                                                                                          \\[0.35 pt]
                  \hline
\end{tabular}
\begin{tablenotes}
  \item U17: \citep{urbaneja2017}, G18: \citep{gvaramadze2018}, R18: \citep{ramachandran2018}, E04: \citep{evans2004}, 
  M15: \citep{McEvoy2015}, V24: \citep{verhamme2024}, SB25: \citep{brands2025}, JB25: \citep{Bestenlehner2025}.
\end{tablenotes}
\end{threeparttable}
\end{table*}

\clearpage
\section{Line fitting for individual stars}
\label{individ_app}
\begin{figure*}
	\includegraphics[scale=0.30]{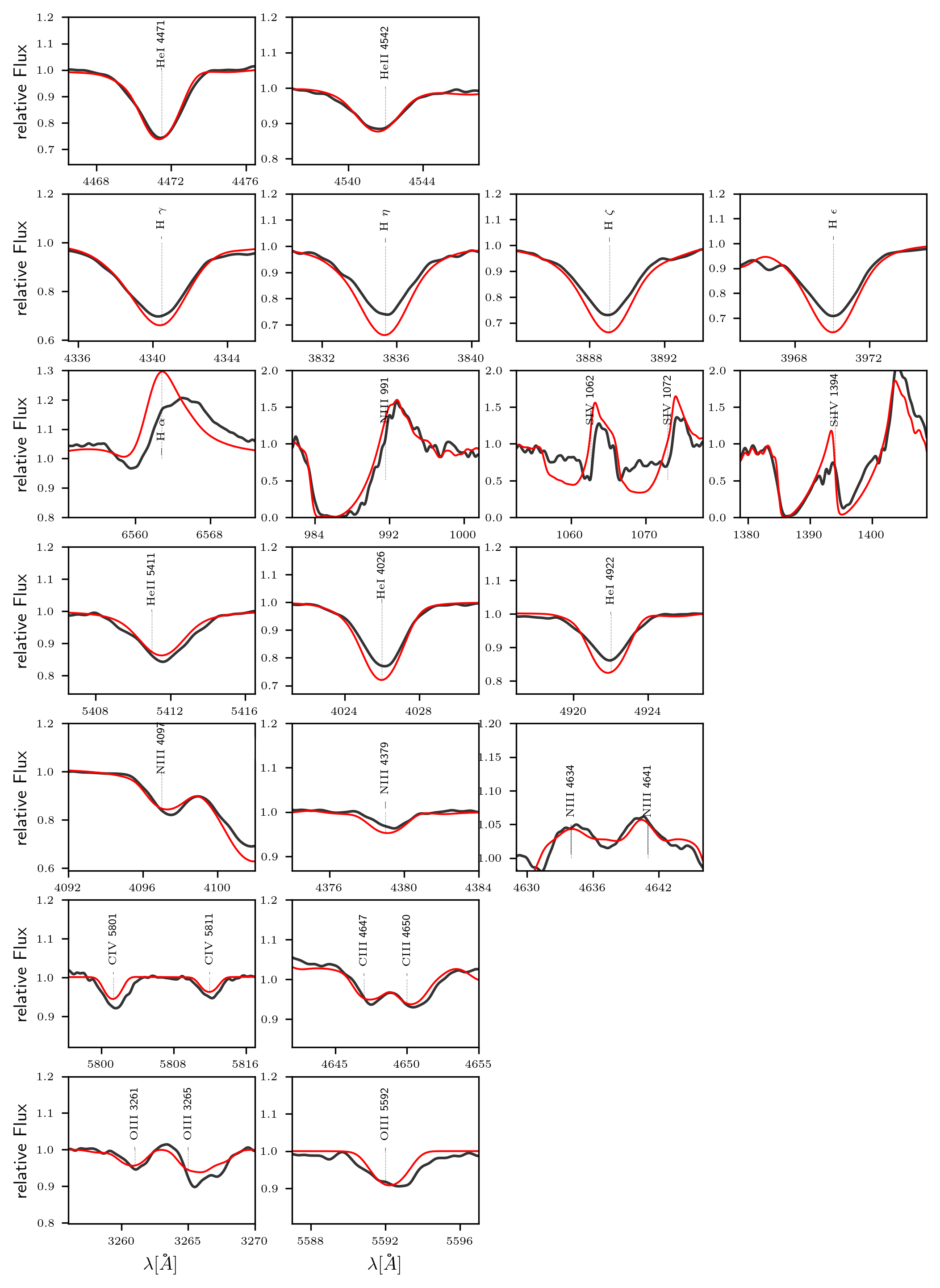}
    \caption{Individual line fits for Sk\,$-$66$^{\circ}$~171. Black solid line: observed spectrum. Red solid line: spectrum of the best fitting model. The observed spectrum is corrected for $\varv_{\rm rad}=410 {\rm km\,s^{-1}}$}
\label{SK-66D171}
\end{figure*}

\begin{figure*}
	\includegraphics[scale=0.95]{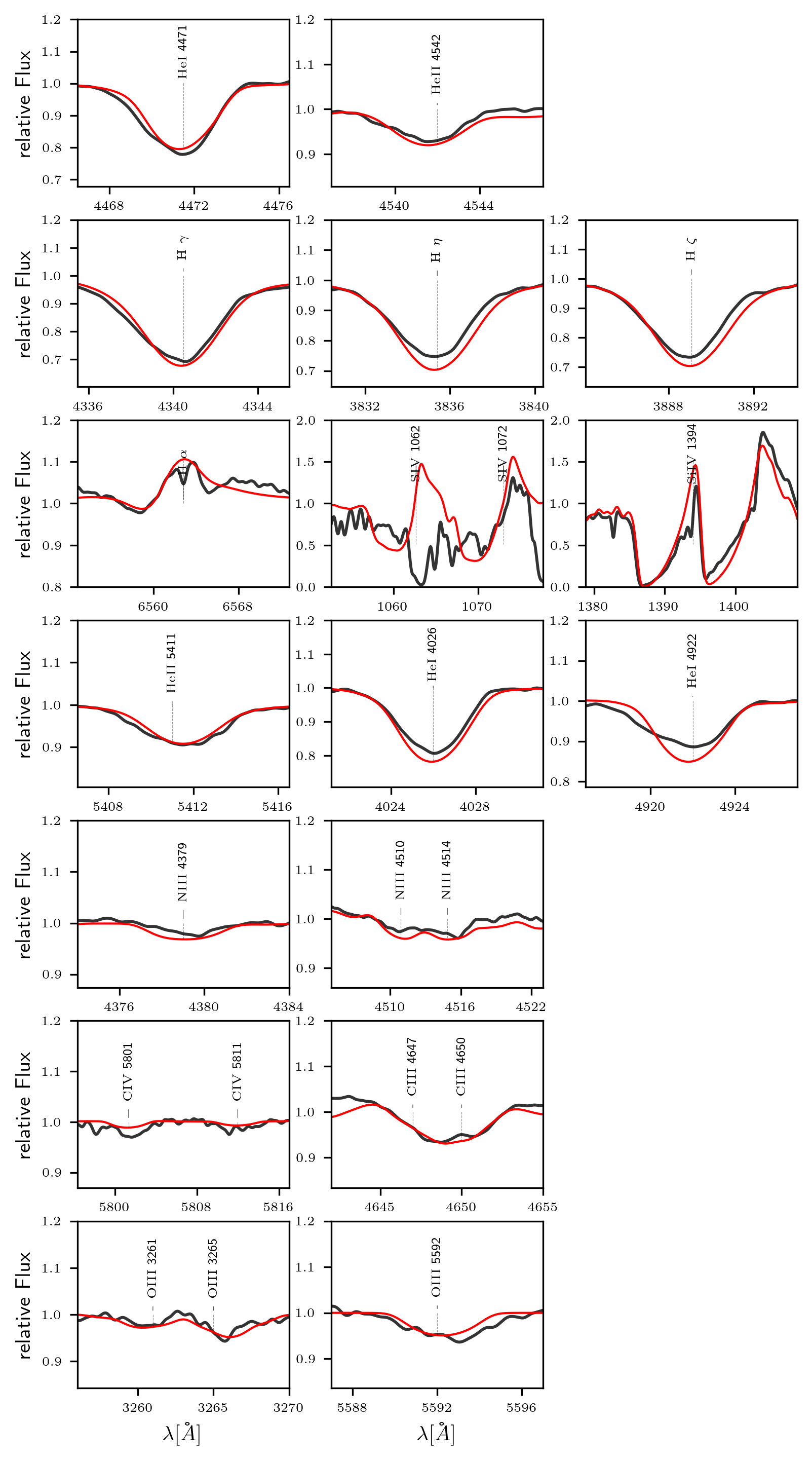}
    \caption{Individual line fits for Sk\,$-$68$^{\circ}$~155. Black solid line: observed spectrum. Red solid line: spectrum of the best fitting model. The observed spectrum is corrected for $\varv_{\rm rad}=240 {\rm km\,s^{-1}}$}
\label{SK-68D155}
\end{figure*}

\begin{figure*}
	\includegraphics[scale=0.95]{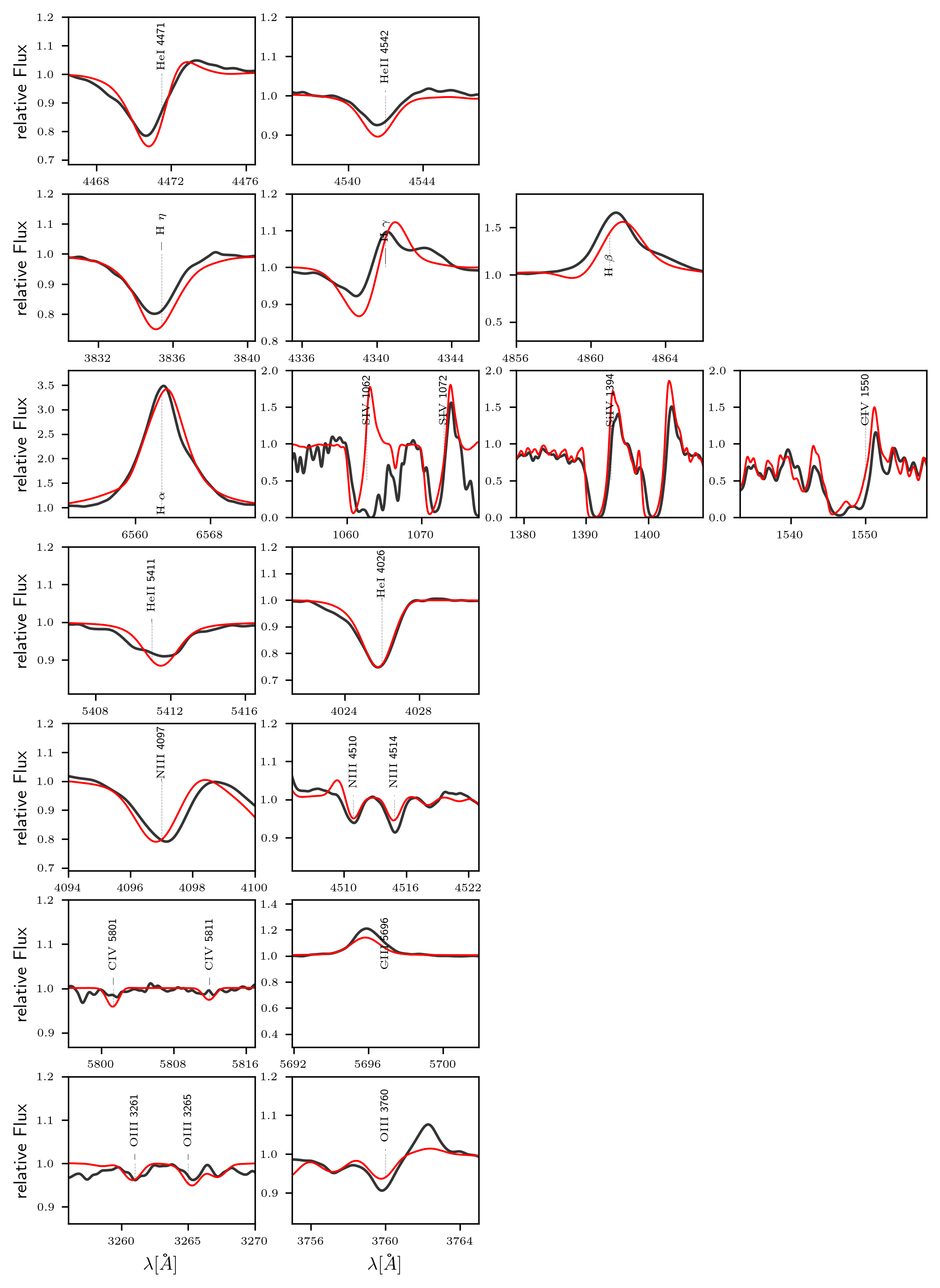}
    \caption{Individual line fits for Sk\,$-$69$^{\circ}$~279. Black solid line: observed spectrum. Red solid line: spectrum of the best fitting model. The observed spectrum is corrected for $\varv_{\rm rad}=230 {\rm km\,s^{-1}}$}
\label{SK-69D279}
\end{figure*}

\begin{figure*}
	\includegraphics[scale=0.95]{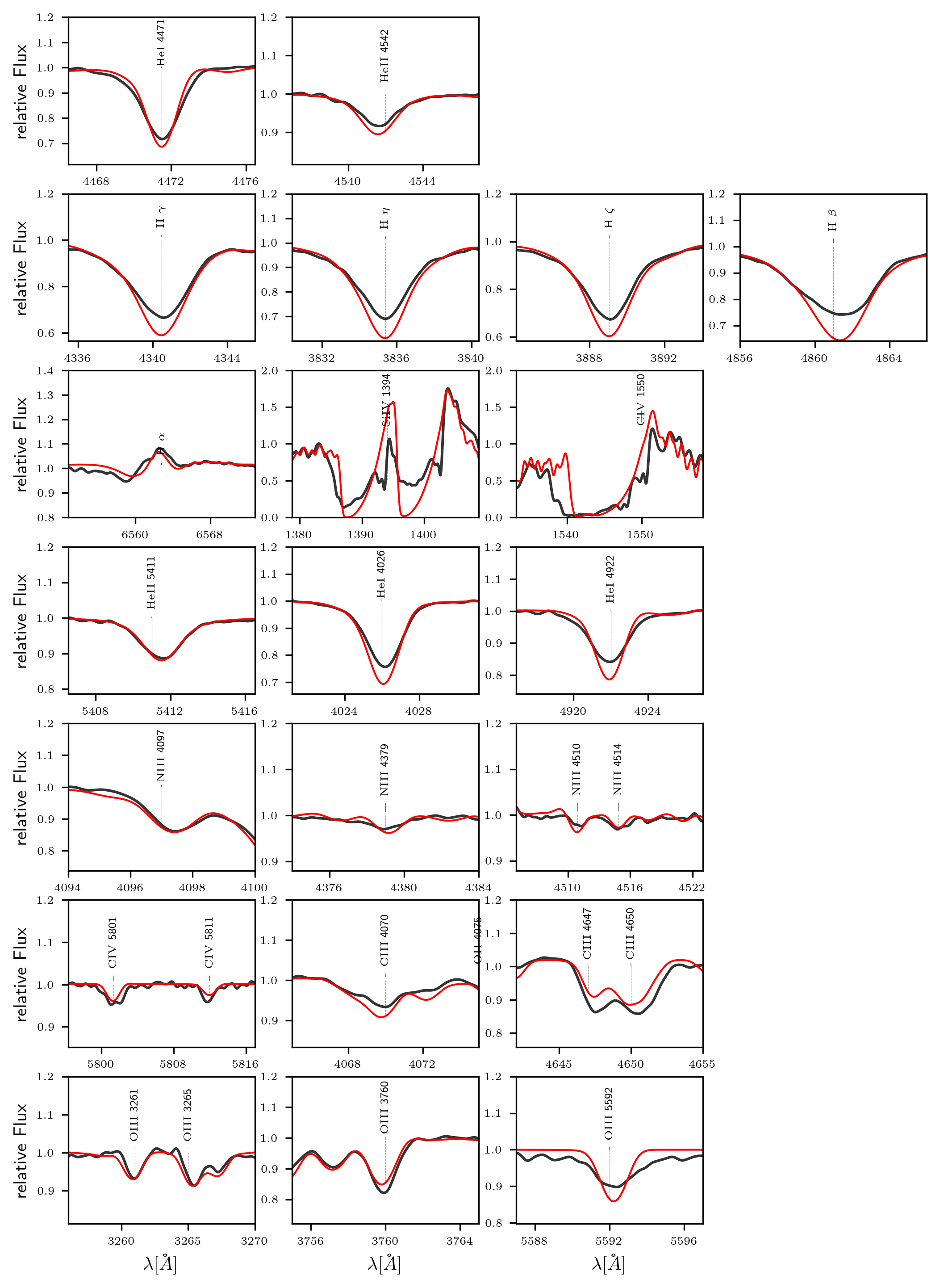}
    \caption{Individual line fits for Sk\,$-$71$^{\circ}$~41. Black solid line: observed spectrum. Red solid line: spectrum of the best fitting model. The observed spectrum is corrected for $\varv_{\rm rad}=260 {\rm km\,s^{-1}}$}
\label{SK-71D41}
\end{figure*}

\begin{figure*}
	\includegraphics[scale=0.95]{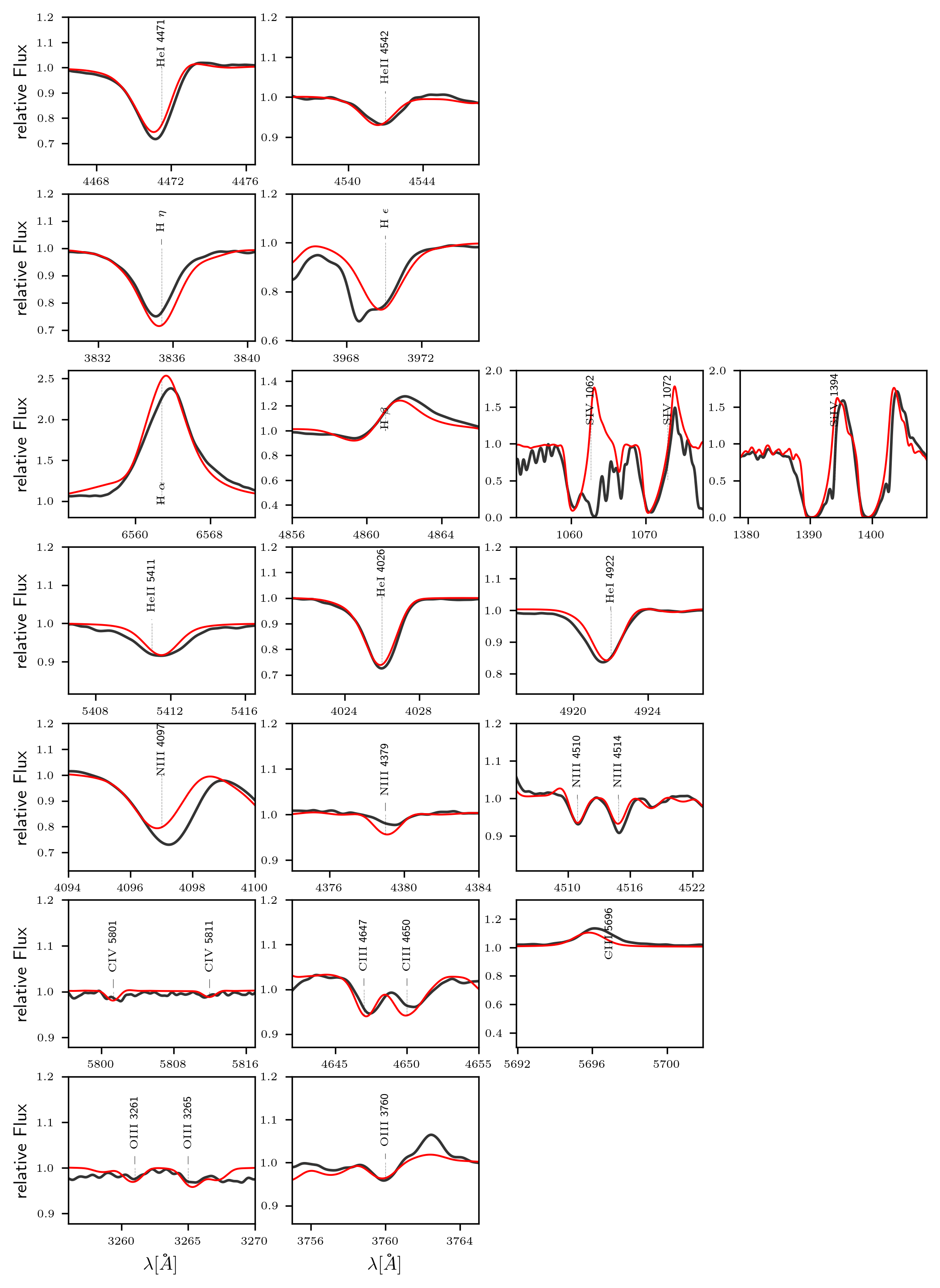}
    \caption{Individual line fits for Sk\,$-$68$^{\circ}$~135. Black solid line: observed spectrum. Red solid line: spectrum of the best fitting model. The observed spectrum is corrected for $\varv_{\rm rad}=270 {\rm km\,s^{-1}}$}
\label{SK-68D135}
\end{figure*}

\begin{figure*}
	\includegraphics[scale=0.95]{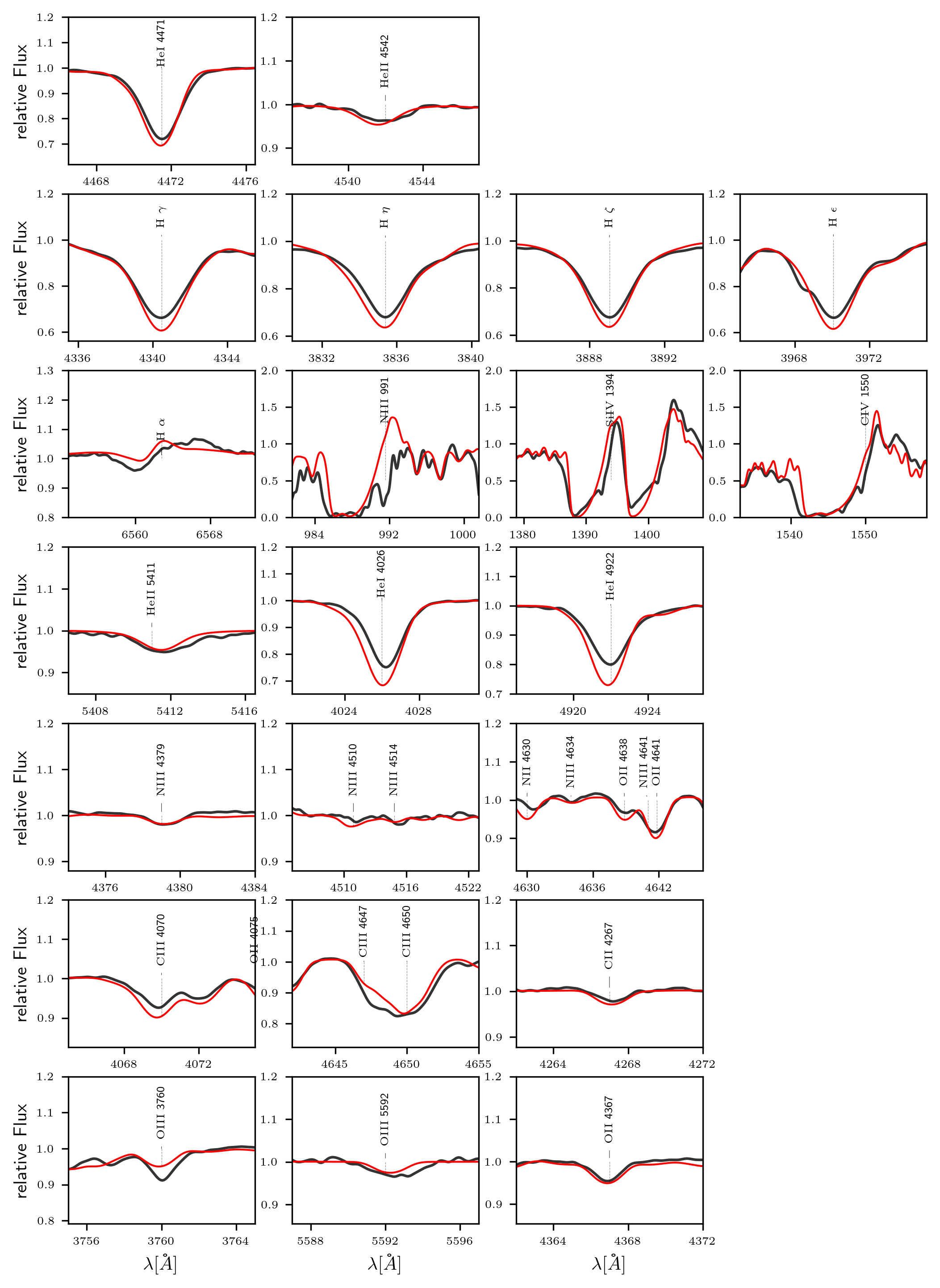}
    \caption{Individual line fits for Sk\,$-$67$^{\circ}$~5. Black solid line: observed spectrum. Red solid line: spectrum of the best fitting model. The observed spectrum is corrected for $\varv_{\rm rad}=295 {\rm km\,s^{-1}}$}
\label{SK-67D5}
\end{figure*}

\begin{figure*}
	\includegraphics[scale=0.30]{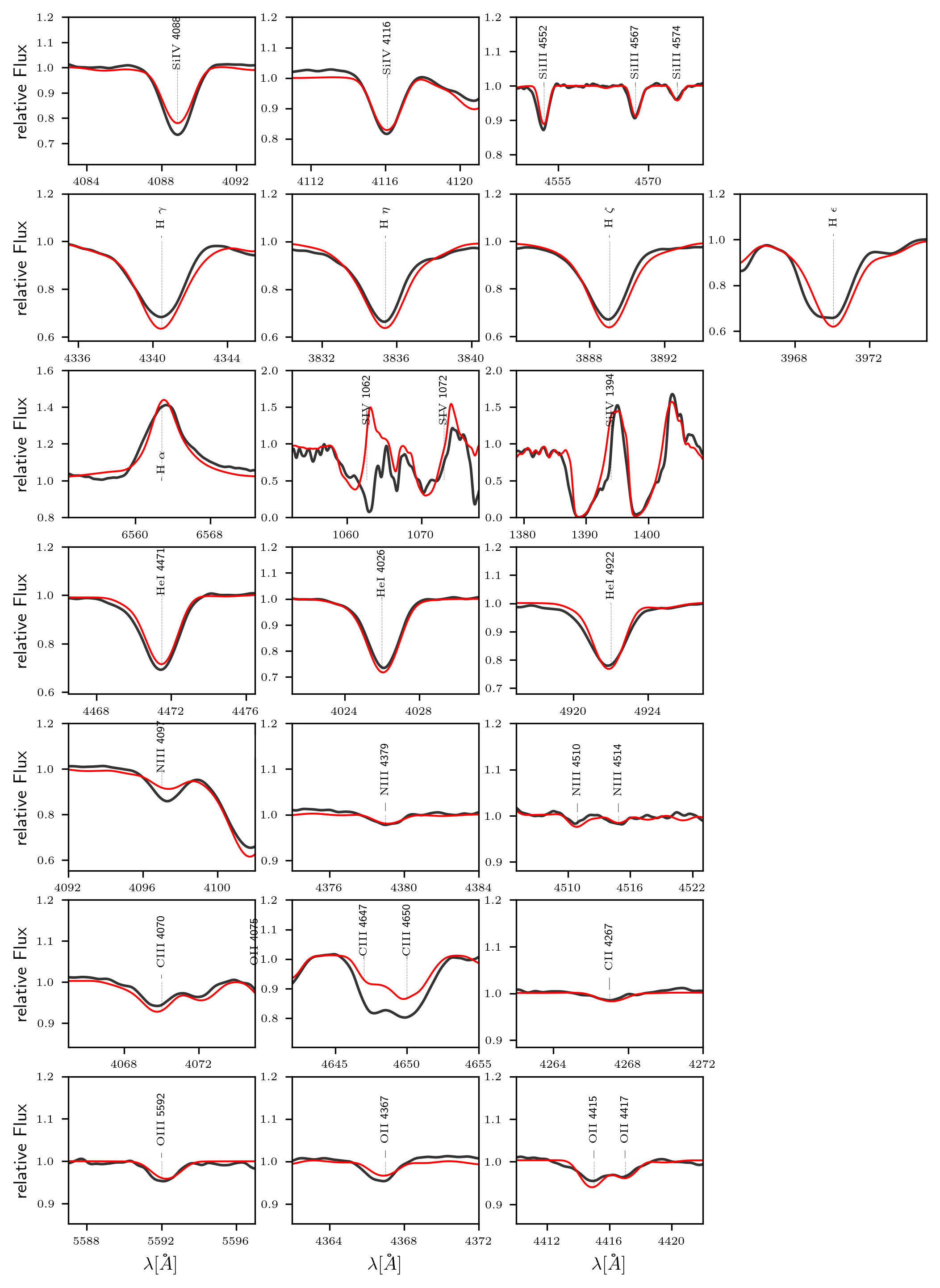}
    \caption{Individual line fits for Sk\,$-$68$^{\circ}$~52. Black solid line: observed spectrum. Red solid line: spectrum of the best fitting model. The observed spectrum is corrected for $\varv_{\rm rad}=255 {\rm km\,s^{-1}}$}
\label{SK-68D52}
\end{figure*}

\begin{figure*}
	\includegraphics[scale=0.30]{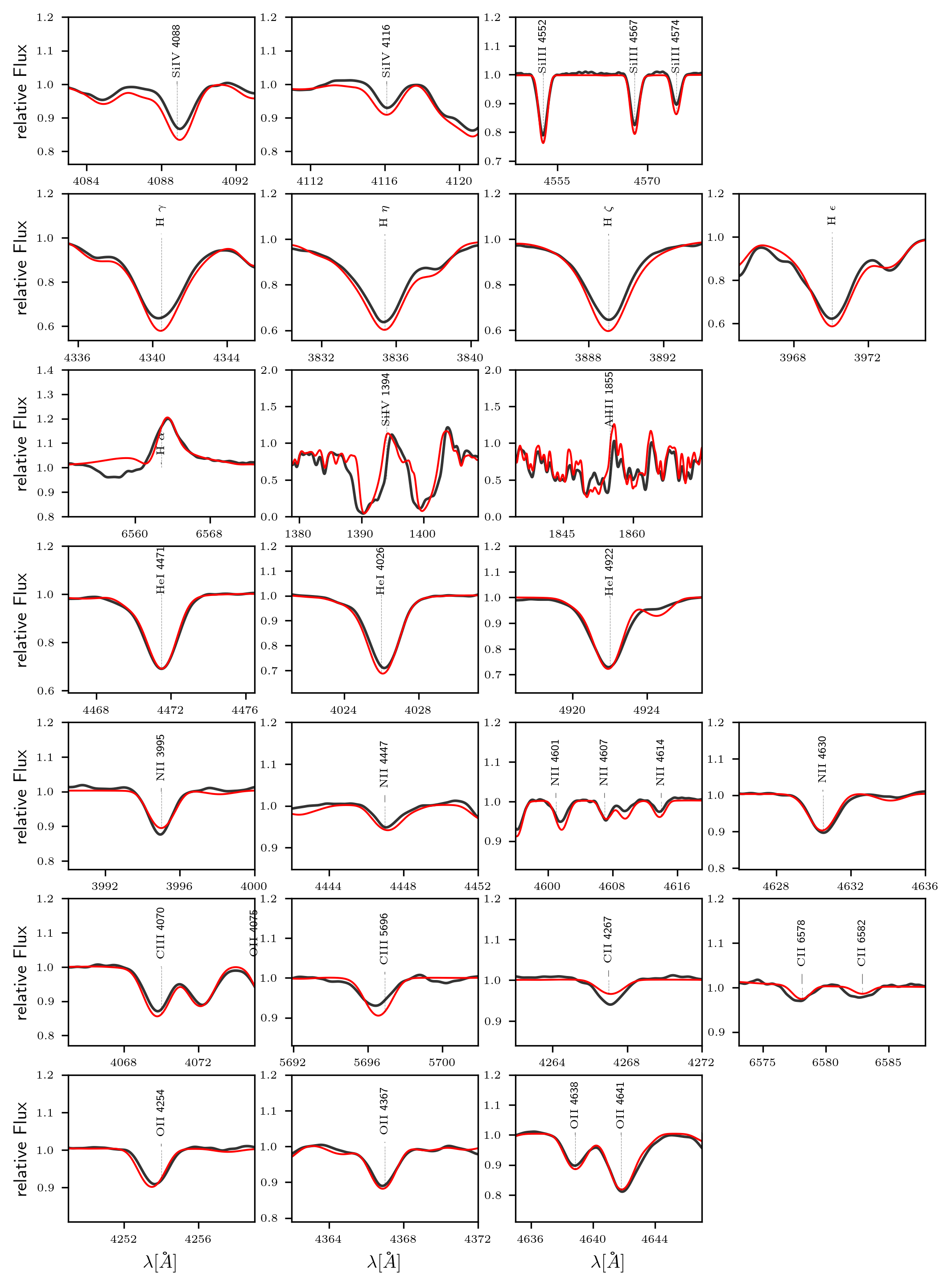}
    \caption{Individual line fits for Sk\,$-$69$^{\circ}$~43. Black solid line: observed spectrum. Red solid line: spectrum of the best fitting model. The observed spectrum is corrected for $\varv_{\rm rad}=255 {\rm km\,s^{-1}}$}
\label{SK-69D43}
\end{figure*}

\begin{figure*}
	\includegraphics[scale=0.30]{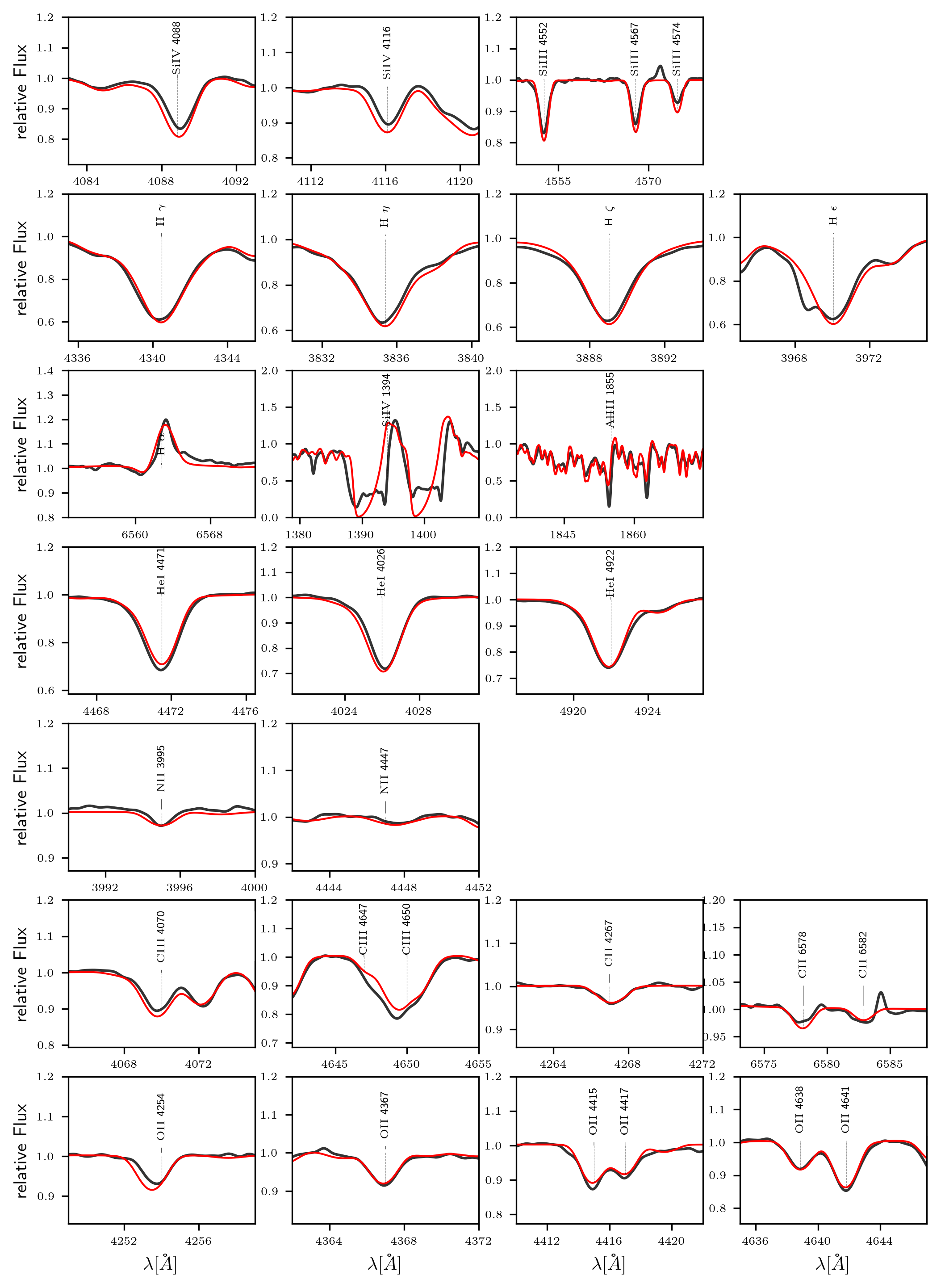}
    \caption{Individual line fits for Sk\,$-$68$^{\circ}$~140. Black solid line: observed spectrum. Red solid line: spectrum of the best fitting model. The observed spectrum is corrected for $\varv_{\rm rad}=260 {\rm km\,s^{-1}}$}
\label{SK-68D140}
\end{figure*}

\begin{figure*}
	\includegraphics[scale=0.95]{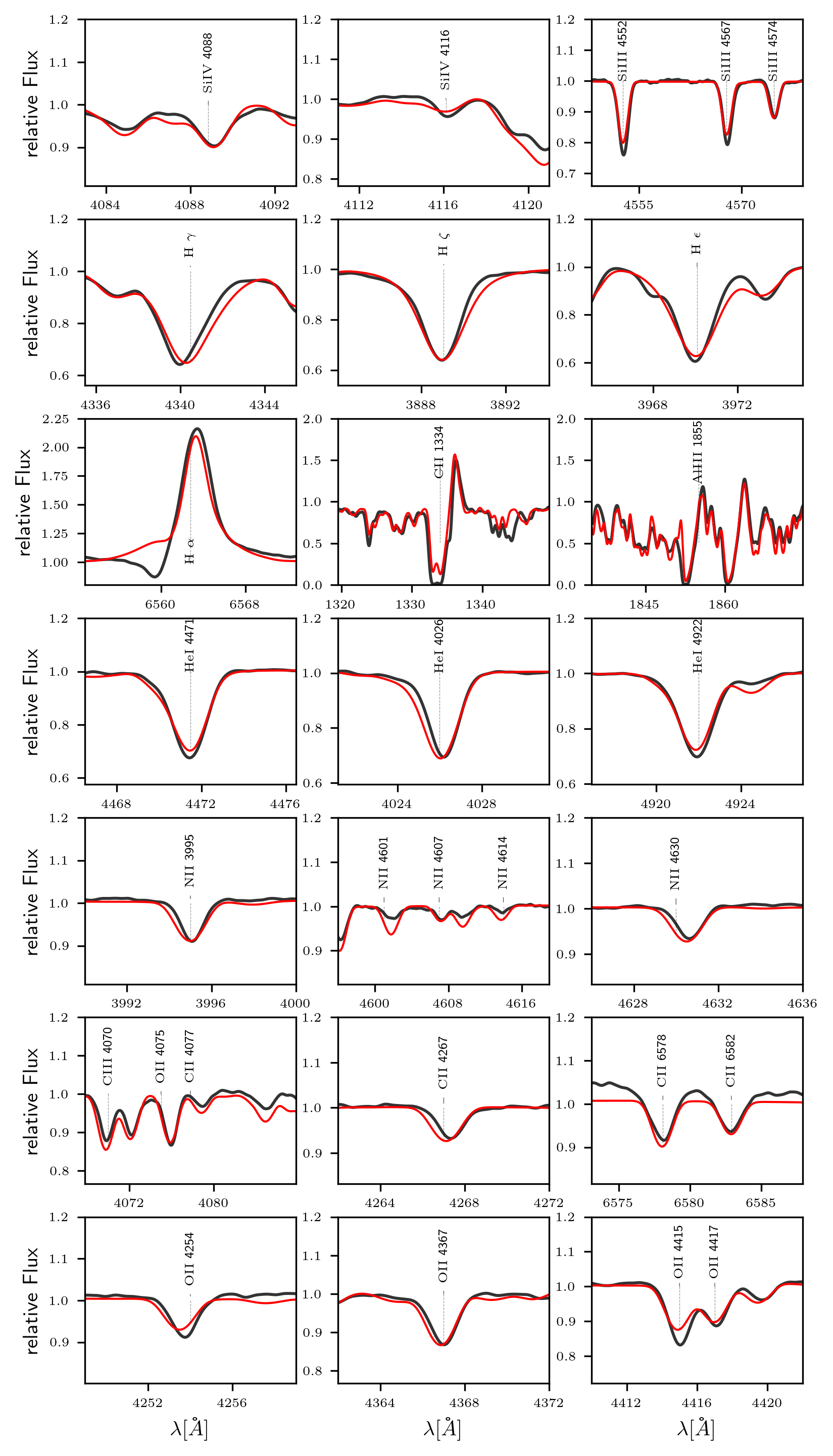}
    \caption{Individual line fits for Sk\,$-$67$^{\circ}$~2. Black solid line: observed spectrum. Red solid line: spectrum of the best fitting model. The observed spectrum is corrected for $\varv_{\rm rad}=320 {\rm km\,s^{-1}}$}
\label{SK-67D2}
\end{figure*}

\begin{figure*}
	\includegraphics[scale=0.30]{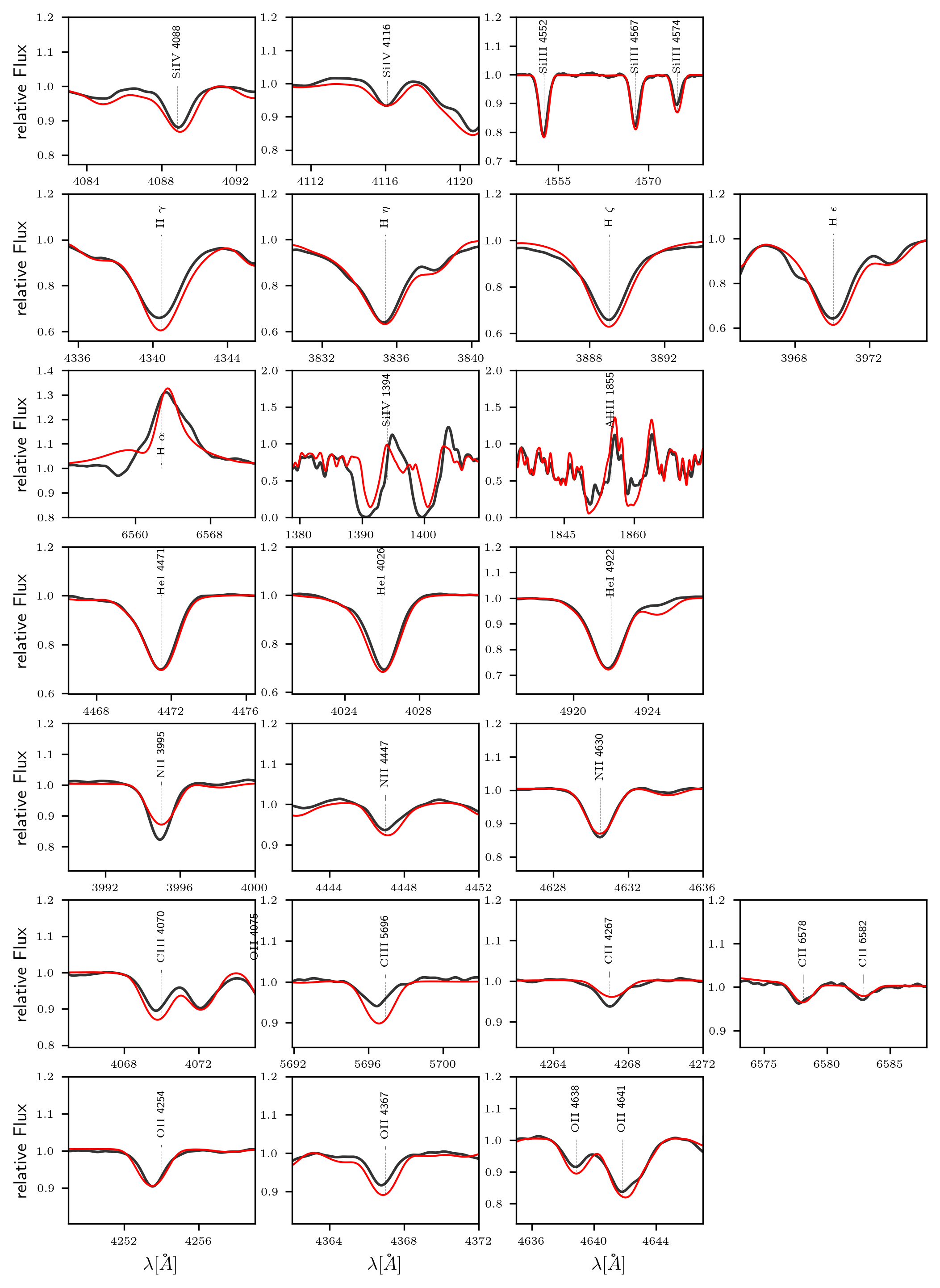}
    \caption{Individual line fits for Sk\,$-$67$^{\circ}$~14. Black solid line: observed spectrum. Red solid line: spectrum of the best fitting model. The observed spectrum is corrected for $\varv_{\rm rad}=300 {\rm km\,s^{-1}}$}
\label{SK-67D14}
\end{figure*}

\begin{figure*}
	\includegraphics[scale=0.95]{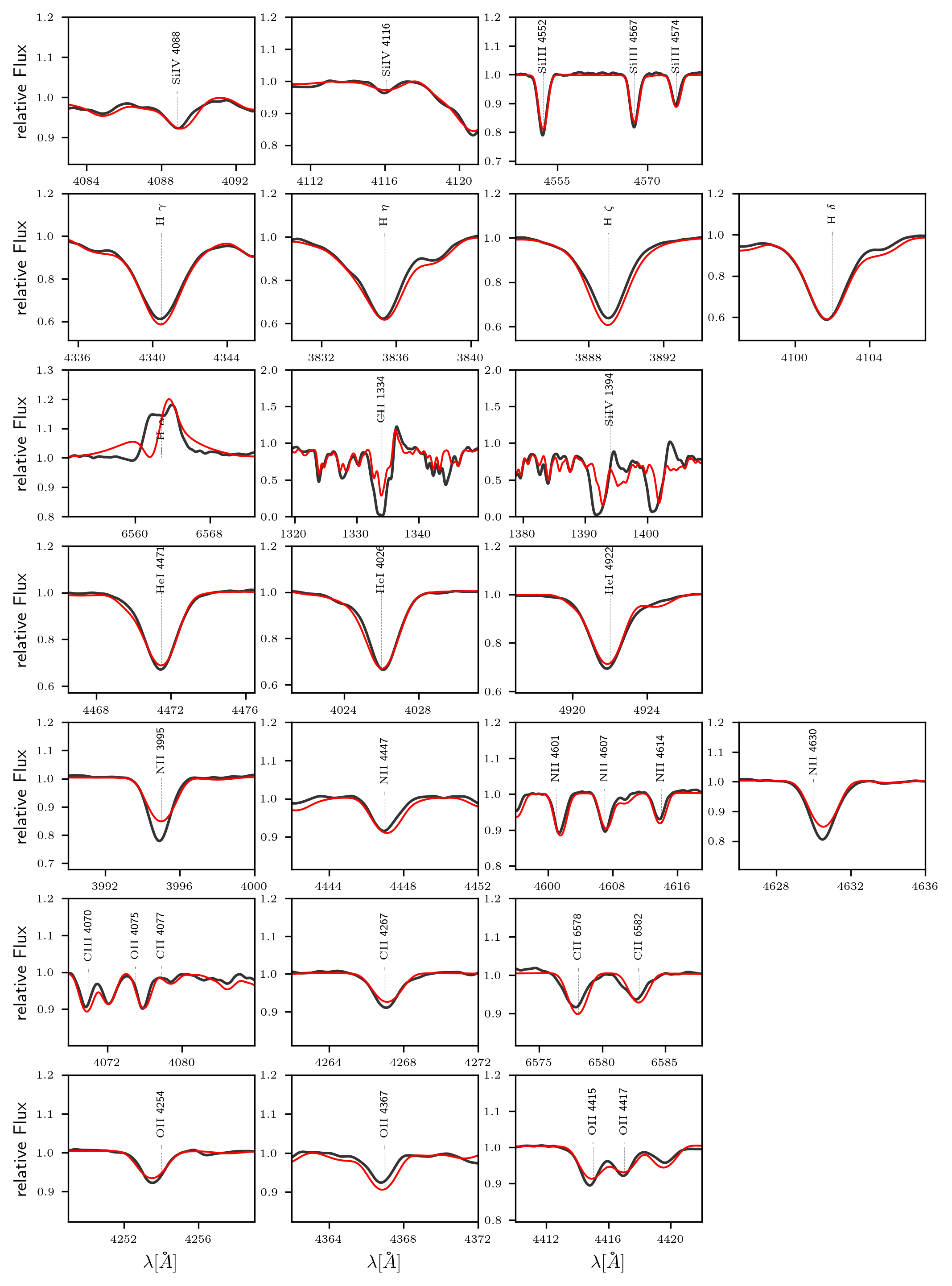}
    \caption{Individual line fits for Sk\,$-$69$^{\circ}$~52. Black solid line: observed spectrum. Red solid line: spectrum of the best fitting model. The observed spectrum is corrected for $\varv_{\rm rad}=260 {\rm km\,s^{-1}}$}
\label{SK-69D52}
\end{figure*}

\begin{figure*}
	\includegraphics[scale=0.95]{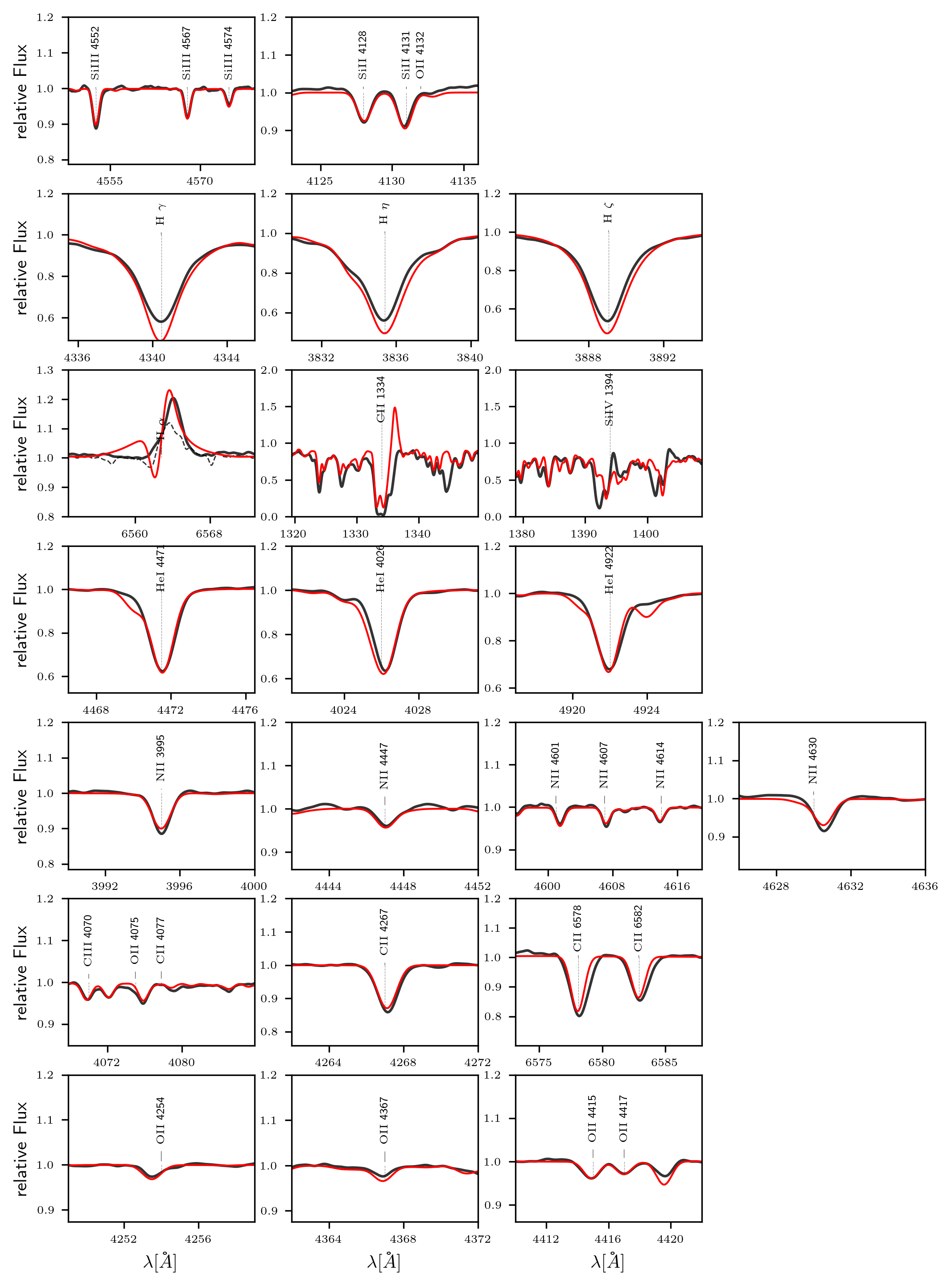}
    \caption{Individual line fits for Sk\,$-$67$^{\circ}$~78. Black solid line: observed spectrum. Red solid line: spectrum of the best fitting model. The observed spectrum is corrected for $\varv_{\rm rad}=310 {\rm km\,s^{-1}}$}
\label{SK-67D78}
\end{figure*}

\begin{figure*}
	\includegraphics[scale=0.30]{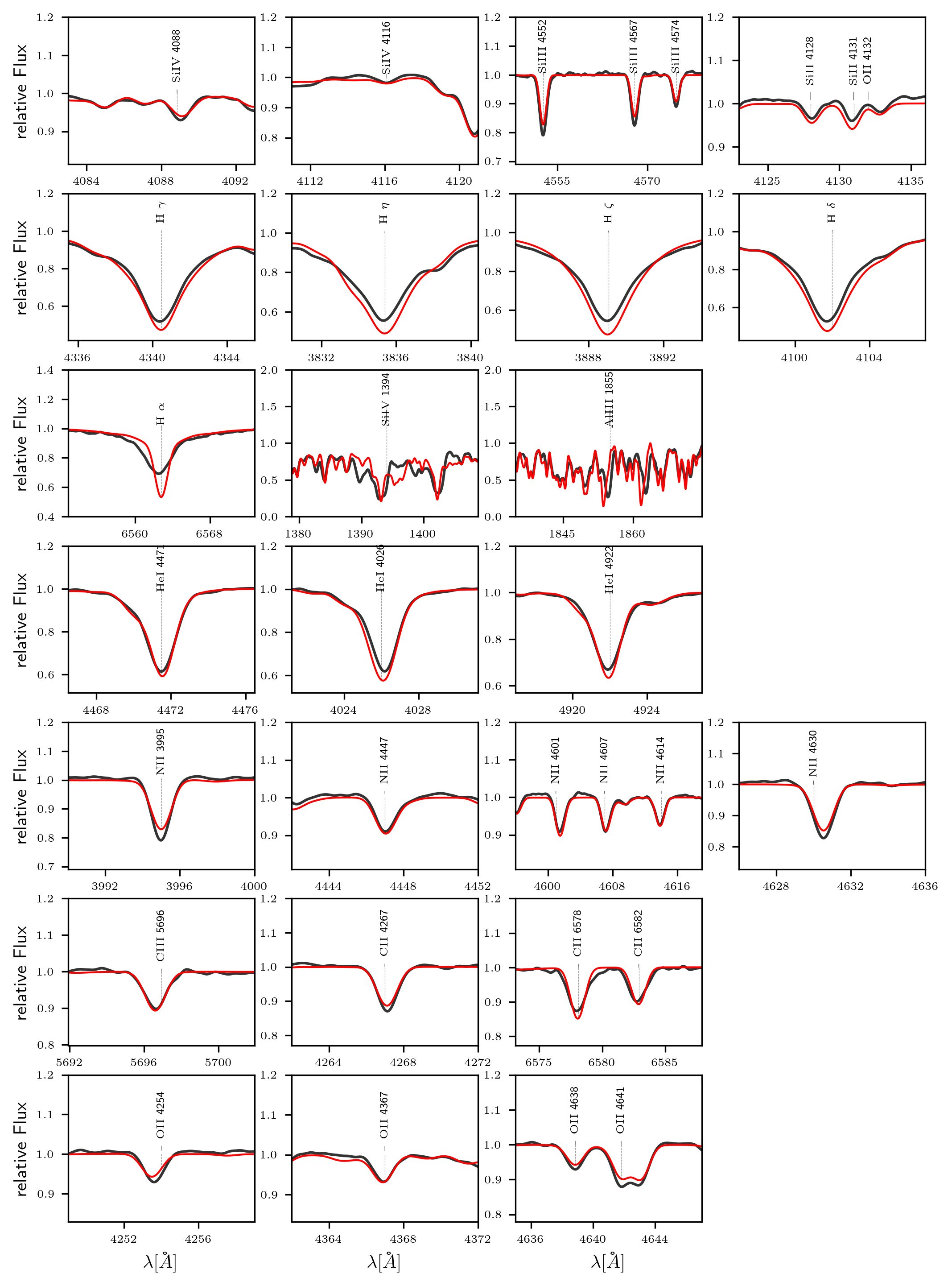}
    \caption{Individual line fits for Sk\,$-$70$^{\circ}$~16. Black solid line: observed spectrum. Red solid line: spectrum of the best fitting model. The observed spectrum is corrected for $\varv_{\rm rad}=265 {\rm km\,s^{-1}}$}
\label{SK-70D16}
\end{figure*}

\begin{figure*}
	\includegraphics[scale=0.95]{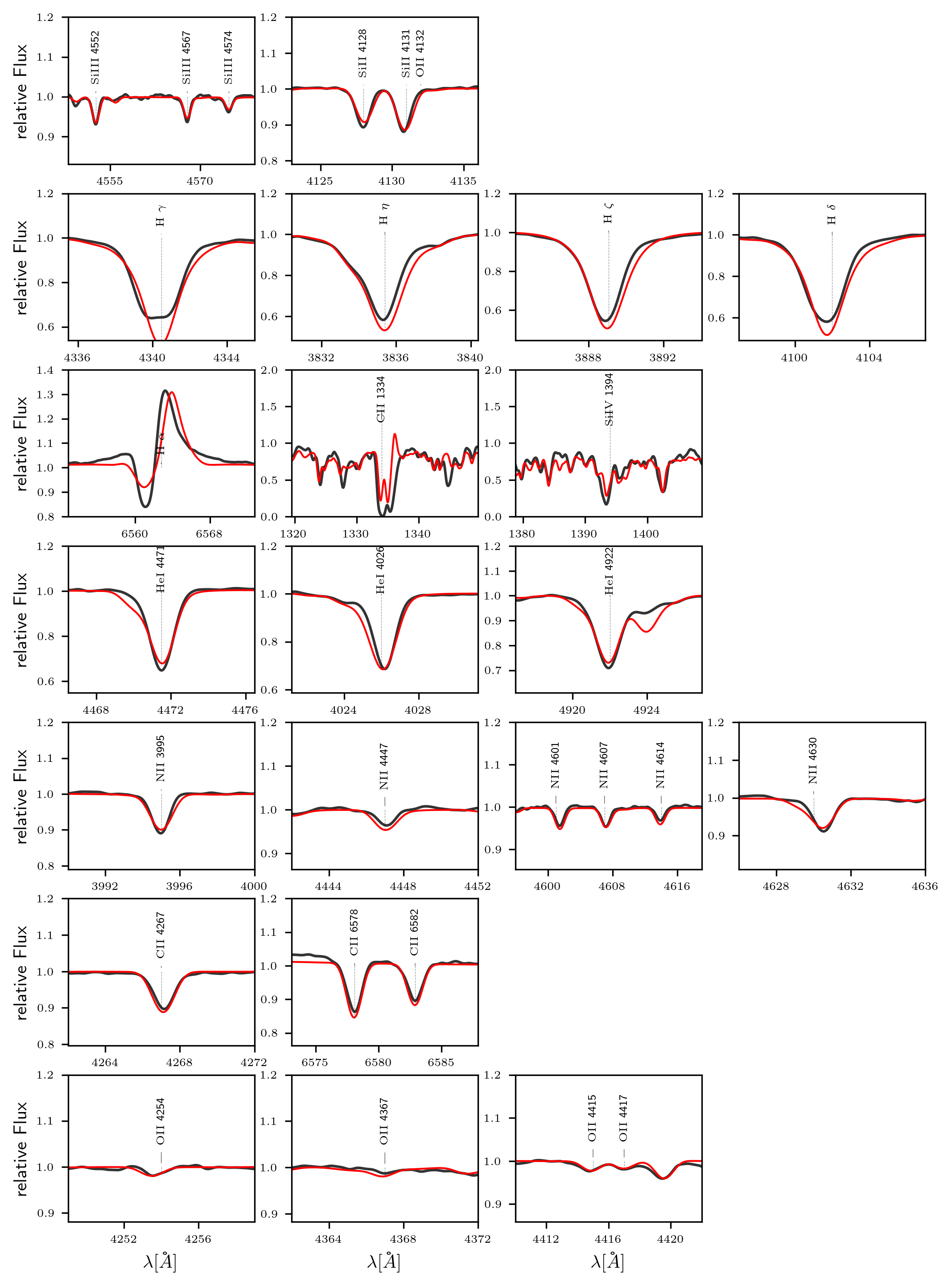}
    \caption{Individual line fits for Sk\,$-$68$^{\circ}$~8. Black solid line: observed spectrum. Red solid line: spectrum of the best fitting model. The observed spectrum is corrected for $\varv_{\rm rad}=250 {\rm km\,s^{-1}}$}
\label{SK-68D8}
\end{figure*}

\begin{figure*}
	\includegraphics[scale=0.95]{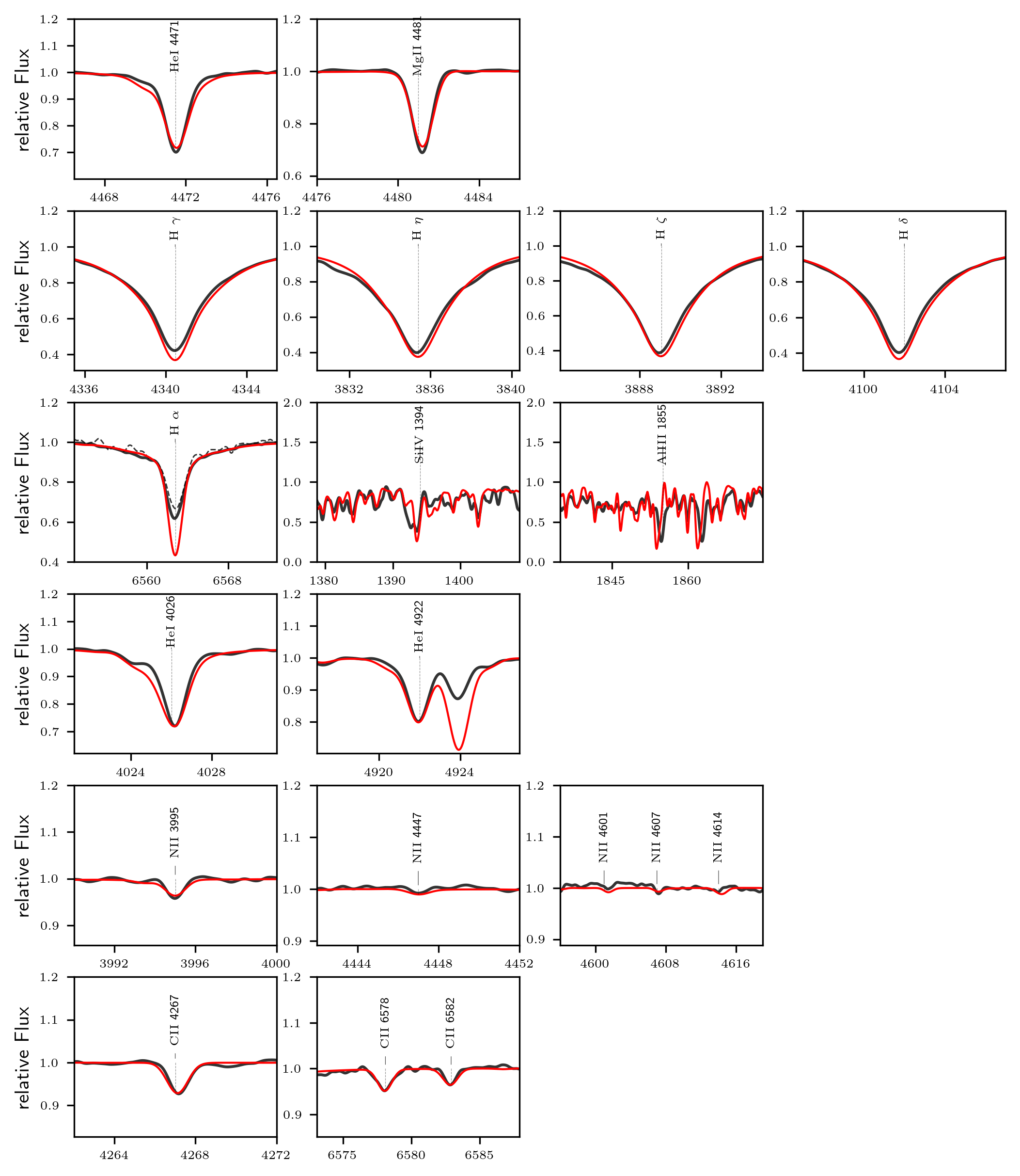}
    \caption{Individual line fits for Sk\,$-$67$^{\circ}$~195. Black solid line: observed spectrum. Red solid line: spectrum of the best fitting model. The observed spectrum is corrected for $\varv_{\rm rad}=300 {\rm km\,s^{-1}}$}
\label{SK-67D195}
\end{figure*}

Fig.~\ref{SK-66D171}-\ref{SK-67D195} are the individual diagnostic line fits for each star in our sample. On each row, from top to bottom, are the diagnostic lines used to determine the effective 
temperature, the effective surface gravity, the wind density paramaters (mass-loss rate, $\beta$, clumping paramters), the helium abundance, the nitrogen abundace, the carbon abundance, and 
the oxygen abundace, respectively.

\clearpage
\section{Overall fitting for individual stars}
\label{overall_app}
\begin{figure*}
  \centering
	\includegraphics[scale=0.73]{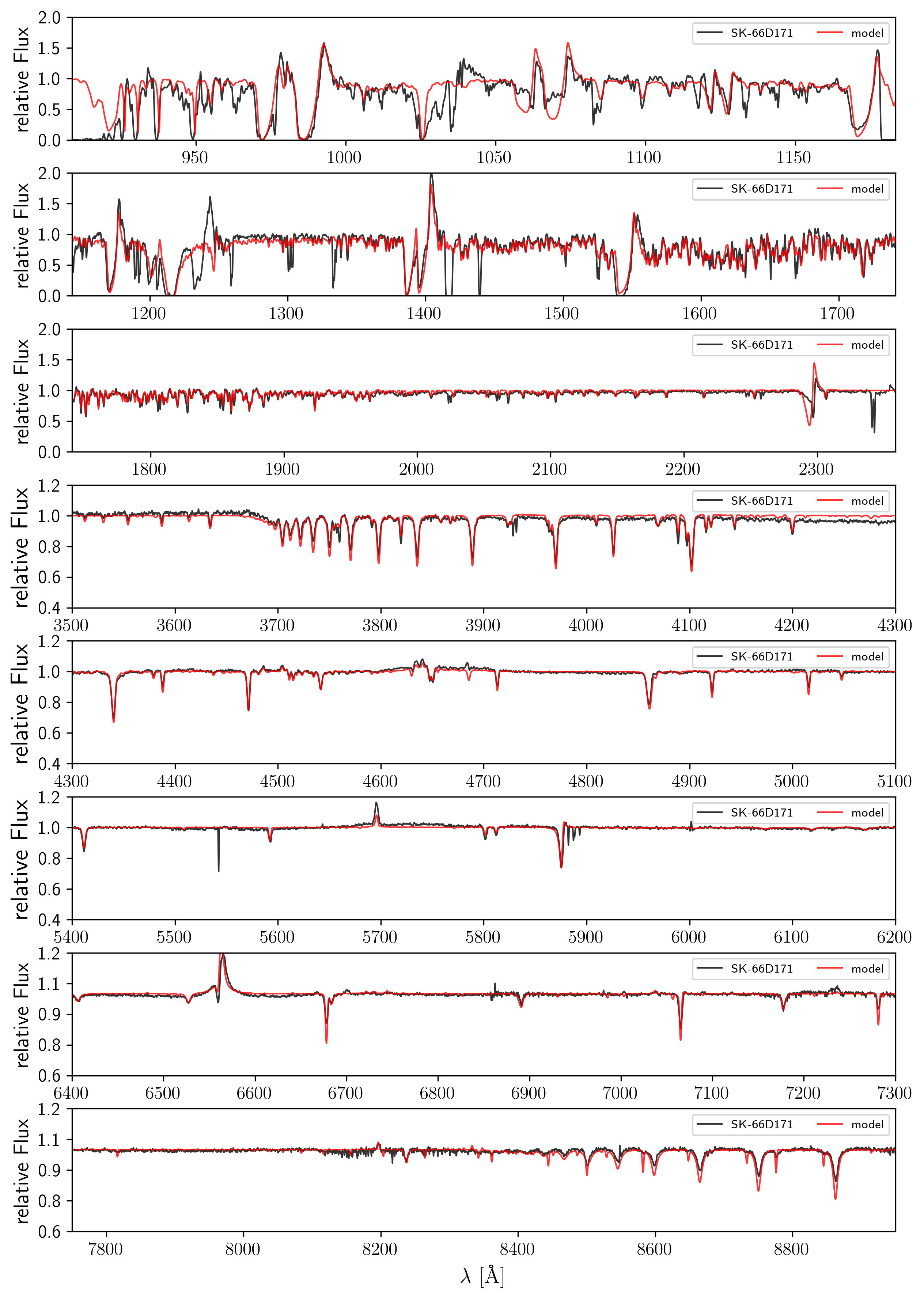}
    \caption{An overall view of the best fit (red solid line) to key regions of the observed spectrum (black solid line) of Sk\,$-$66$^{\circ}$~171. From the top, the first panel is the FUV 
    FUSE range, containing interstellar features such as $Ly\delta~\lambda950$, $Ly\gamma~\lambda973$, and $Ly\beta~\lambda1026$. Second and third panels are STIS E140M FUV and STIS E230M NUV
    spectral ranges, respectively, and we note some interstellar features: $Ly\alpha~\lambda1216$, $\ion{O}{I}+\ion{P}{II}~\lambda1302$, $\ion{C}{II}~\lambda1335$, $\ion{Si}{II}~\lambda1527$, 
    and $\ion{Al}{II}\lambda1671$. The rest of the panels are the UBV and VIS XShooter spectra, with the break between the two arms at $\approx5600~\AA$. The lines
    $\ion{Ca}{II}~{\rm H+K}$ and $\ion{Na}{I}~{\rm D}$ are interstellar features. For this star, we fit the absorption profiles for $Ly\alpha$ through $Ly\eta$, which were calculated using 
    \ion{H}{I} column density $\log{N(HI)}=20.7~{\rm cm^{-2}}$ \citep{fitzpatrick1985}.}
    \label{overall_SK-66D171}
\end{figure*}

\begin{figure*}
  \centering
	\includegraphics[scale=0.75]{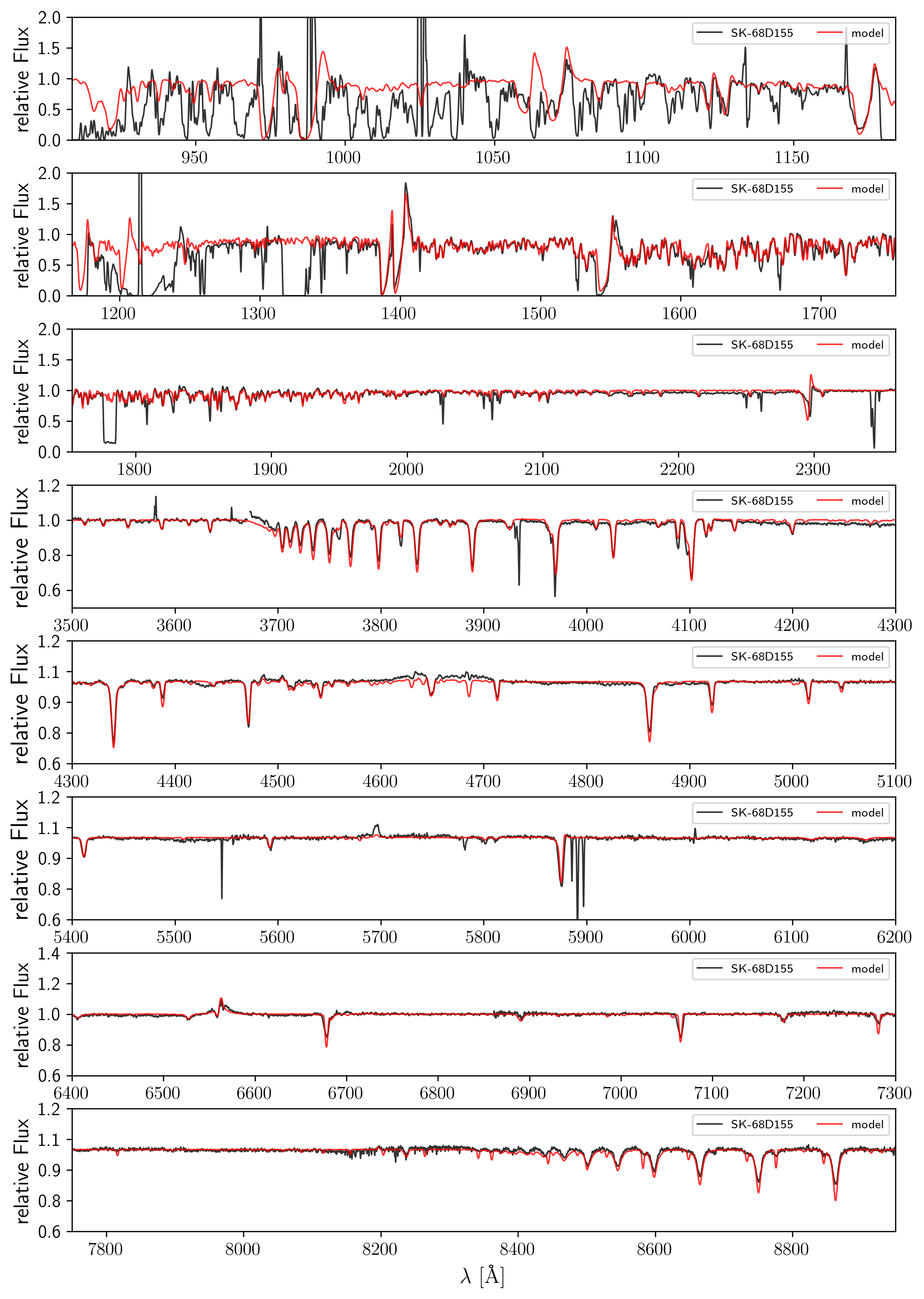}
    \caption{An overall view of the best fit (red solid line) to key regions of the observed spectrum (black solid line) of Sk\,$-$68$^{\circ}$~155. From the top, the first panel is the FUV 
    FUSE range, containing interstellar features such as $Ly\delta~\lambda950$, $Ly\gamma~\lambda973$, and $Ly\beta~\lambda1026$. The second and third panels are the COS G130M+G160M FUV and 
    STIS E230M NUV spectral ranges, respectively, and we note some interstellar features: $Ly\alpha~\lambda1216$, $\ion{O}{I}+\ion{P}{II}~\lambda1302$, $\ion{C}{II}~\lambda1335$, 
    $\ion{Si}{II}~\lambda1527$, and $\ion{Al}{II}\lambda1671$. The rest of the panels are the UBV and VIS XShooter spectra, with the break between the two arms at $\approx5600~\AA$. The lines 
    $\ion{Ca}{II}~{\rm H+K}$ and $\ion{Na}{I}~{\rm D}$ are interstellar features.}
    \label{overall_SK-68D155}
\end{figure*}

\begin{figure*}
  \centering
	\includegraphics[scale=0.75]{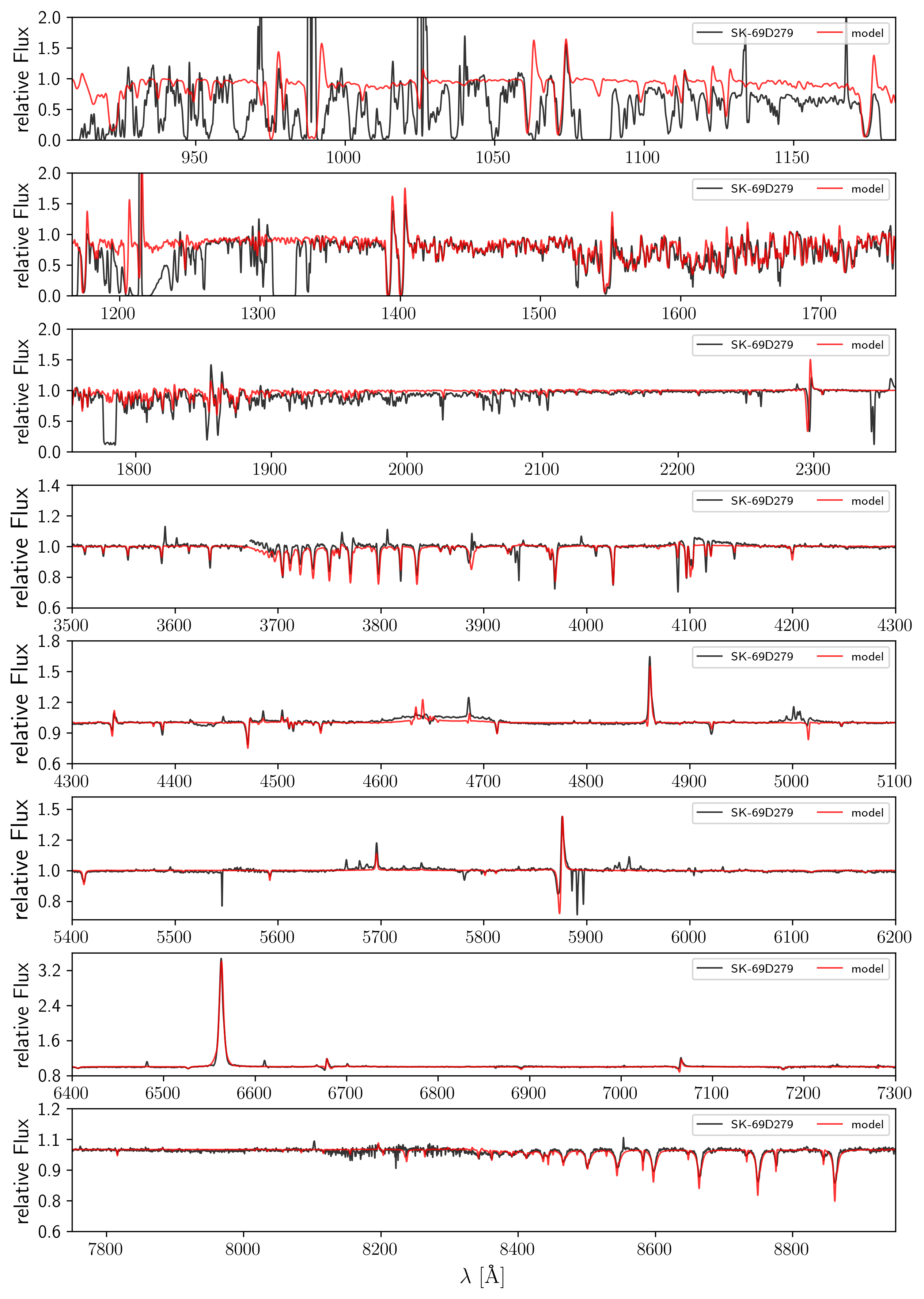}
    \caption{An overall view of the best fit (red solid line) to key regions of the observed spectrum (black solid line) of Sk\,$-$69$^{\circ}$~279. From the top, the first panel is the FUV 
    FUSE range, containing interstellar features such as $Ly\delta~\lambda950$, $Ly\gamma~\lambda973$, and $Ly\beta~\lambda1026$. Second and third panels are STIS E140M FUV and STIS E230M NUV
    spectral ranges, respectively, and we note some interstellar features: $Ly\alpha~\lambda1216$, $\ion{O}{I}+\ion{P}{II}~\lambda1302$, $\ion{C}{II}~\lambda1335$, $\ion{Si}{II}~\lambda1527$, 
    and $\ion{Al}{II}\lambda1671$. The rest of the panels are the UBV and VIS XShooter spectra, with the break between the two arms at $\approx5600~\AA$. The lines
    $\ion{Ca}{II}~{\rm H+K}$ and $\ion{Na}{I}~{\rm D}$ are interstellar features.}
    \label{overall_SK-69D279}
\end{figure*}

\begin{figure*}
  \centering
	\includegraphics[scale=0.75]{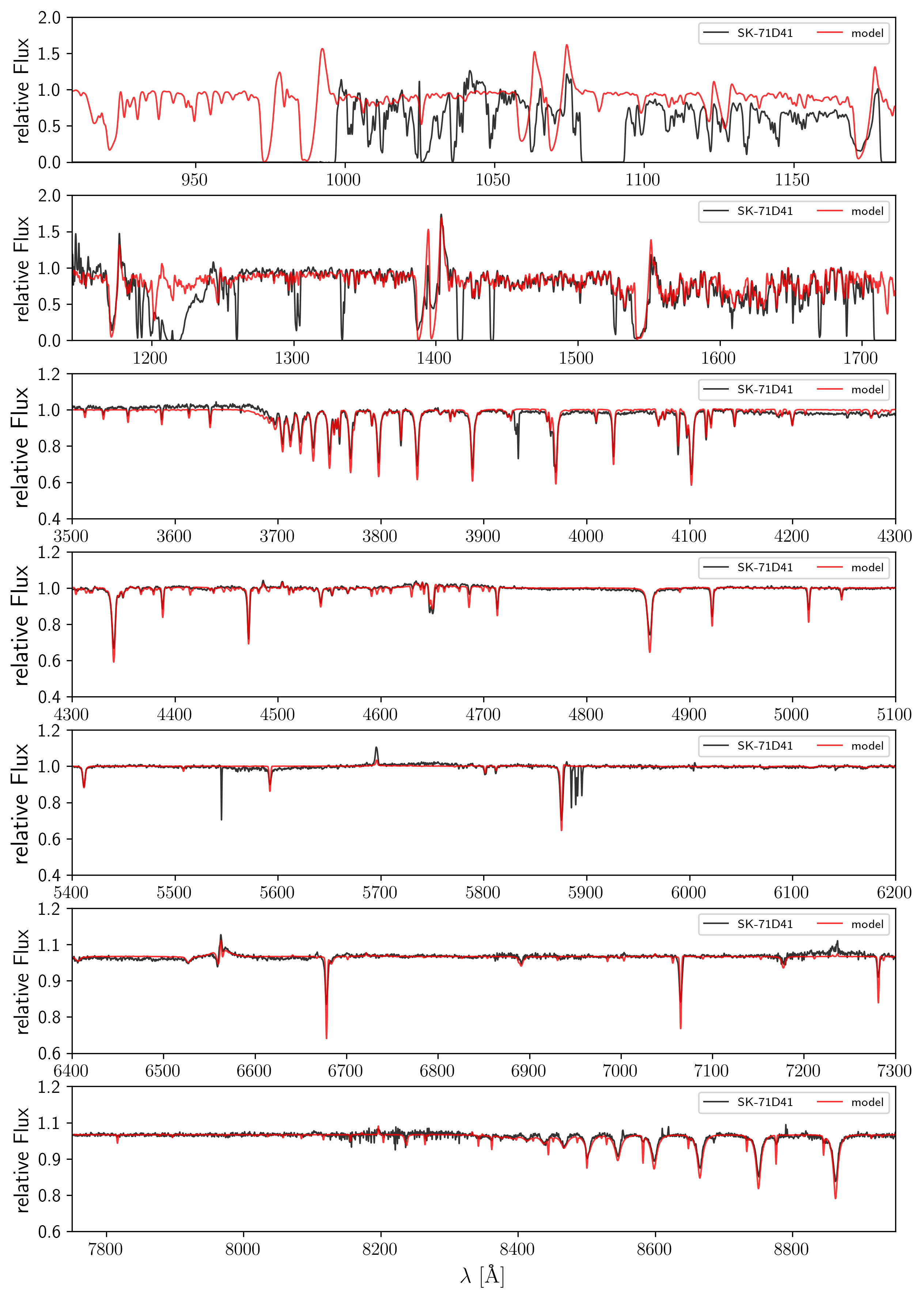}
    \caption{An overall view of the best fit (red solid line) to key regions of the observed spectrum (black solid line) of Sk\,$-$71$^{\circ}$~41. From the top, the first panel is the FUV 
    FUSE range, containing interstellar features such as $Ly\delta~\lambda950$, $Ly\gamma~\lambda973$, and $Ly\beta~\lambda1026$. Second panel is STIS E140M FUV, and we note some 
    interstellar features: $Ly\alpha~\lambda1216$, $\ion{O}{I}+\ion{P}{II}~\lambda1302$, $\ion{C}{II}~\lambda1335$, $\ion{Si}{II}~\lambda1527$, 
    and $\ion{Al}{II}\lambda1671$. The rest of the panels are the UBV and VIS XShooter spectra, with the break between the two arms at $\approx5600~\AA$. The lines
    $\ion{Ca}{II}~{\rm H+K}$ and $\ion{Na}{I}~{\rm D}$ are interstellar features.}
    \label{overall_SK-71D41}
\end{figure*}

\begin{figure*}
  \centering
	\includegraphics[scale=0.75]{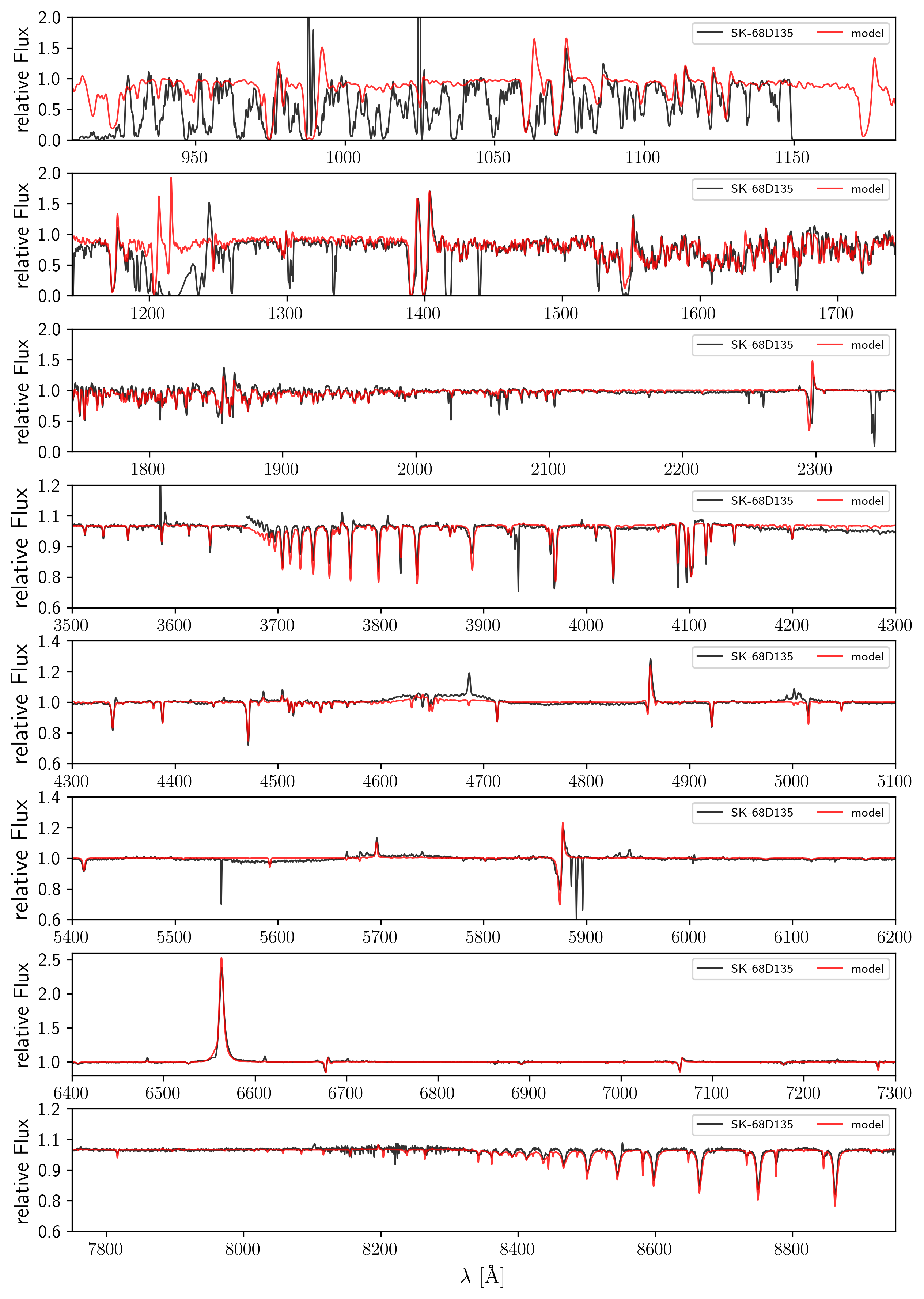}
    \caption{An overall view of the best fit (red solid line) to key regions of the observed spectrum (black solid line) of Sk\,$-$68$^{\circ}$~135. From the top, the first panel is the FUV 
    FUSE range, containing interstellar features such as $Ly\delta~\lambda950$, $Ly\gamma~\lambda973$, and $Ly\beta~\lambda1026$. Second and third panels are STIS E140M FUV and STIS E230M NUV
    spectral ranges, respectively, and we note some interstellar features: $Ly\alpha~\lambda1216$, $\ion{O}{I}+\ion{P}{II}~\lambda1302$, $\ion{C}{II}~\lambda1335$, $\ion{Si}{II}~\lambda1527$, 
    and $\ion{Al}{II}\lambda1671$. The rest of the panels are the UBV and VIS XShooter spectra, with the break between the two arms at $\approx5600~\AA$. The lines
    $\ion{Ca}{II}~{\rm H+K}$ and $\ion{Na}{I}~{\rm D}$ are interstellar features.}
    \label{overall_SK-68D135}
\end{figure*}

\begin{figure*}
  \centering
	\includegraphics[scale=0.75]{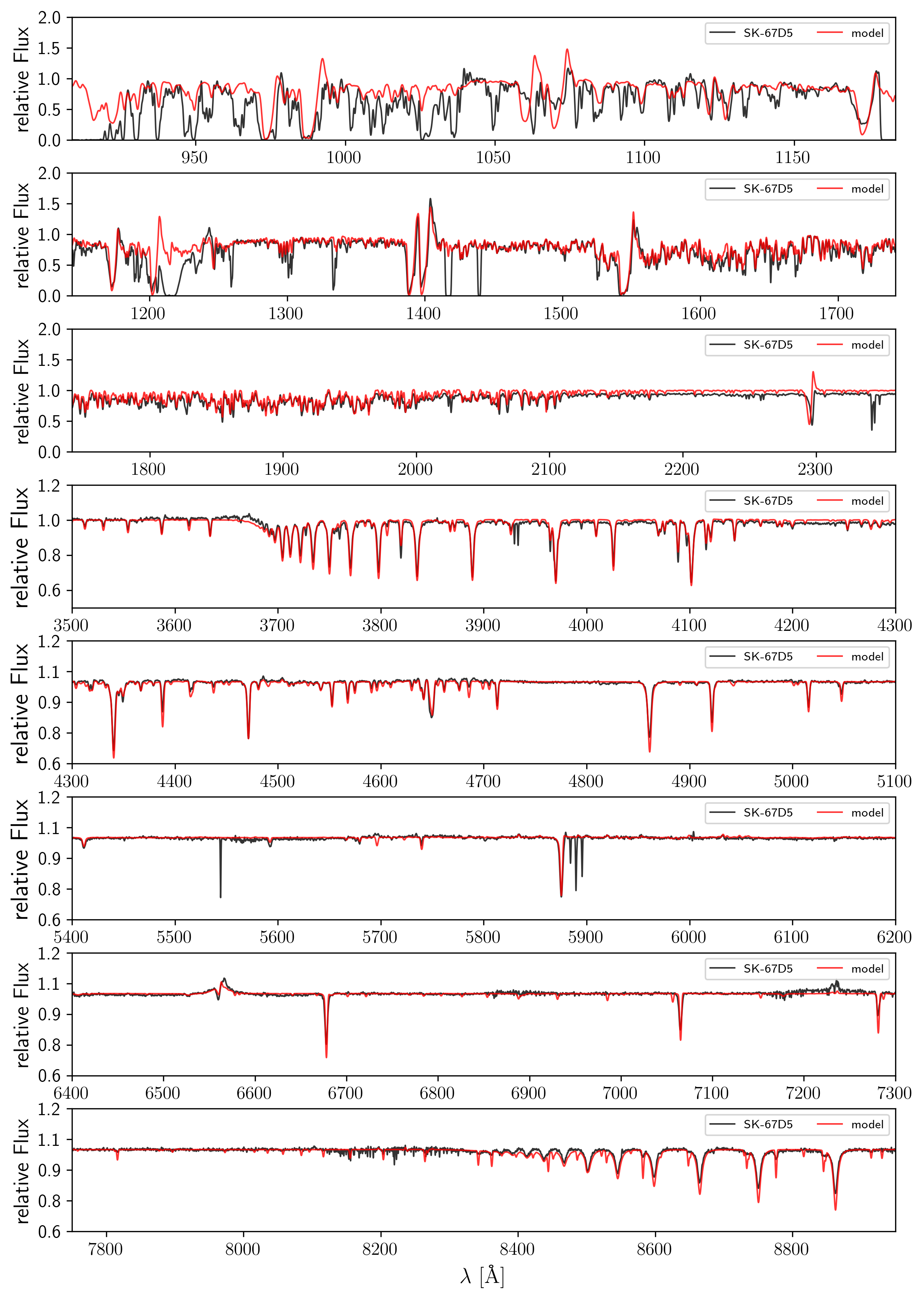}
    \caption{An overall view of the best fit (red solid line) to key regions of the observed spectrum (black solid line) of Sk\,$-$67$^{\circ}$~5. From the top, the first panel is the FUV 
    FUSE range, containing interstellar features such as $Ly\delta~\lambda950$, $Ly\gamma~\lambda973$, and $Ly\beta~\lambda1026$. Second and third panels are STIS E140M FUV and STIS E230M NUV
    spectral ranges, respectively, and we note some interstellar features: $Ly\alpha~\lambda1216$, $\ion{O}{I}+\ion{P}{II}~\lambda1302$, $\ion{C}{II}~\lambda1335$, $\ion{Si}{II}~\lambda1527$, 
    and $\ion{Al}{II}\lambda1671$. The rest of the panels are the UBV and VIS XShooter spectra, with the break between the two arms at $\approx5600~\AA$. The lines
    $\ion{Ca}{II}~{\rm H+K}$ and $\ion{Na}{I}~{\rm D}$ are interstellar features.}
    \label{overall_SK-67D5}
\end{figure*}

\begin{figure*}
  \centering
	\includegraphics[scale=0.75]{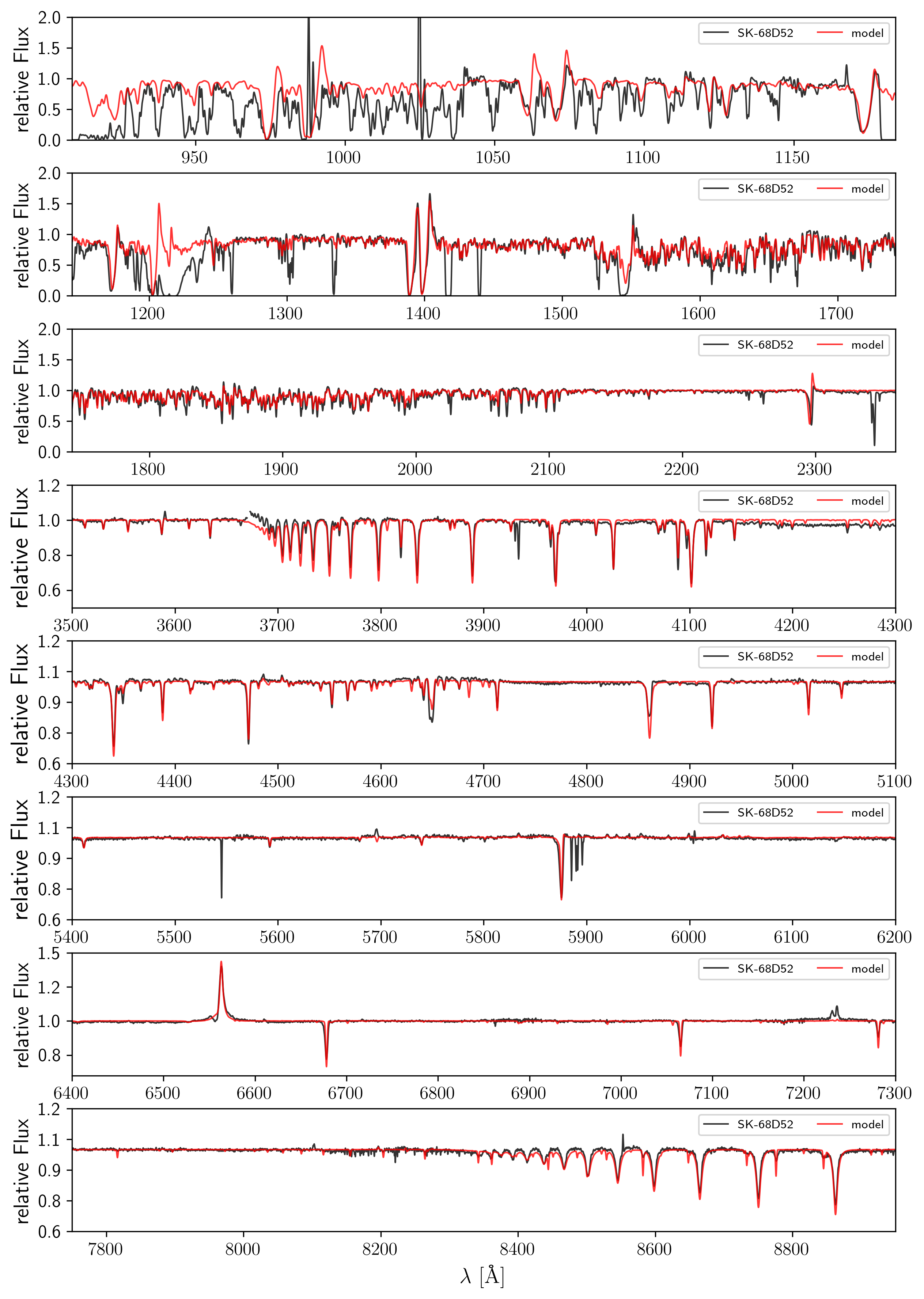}
    \caption{An overall view of the best fit (red solid line) to key regions of the observed spectrum (black solid line) of Sk\,$-$68$^{\circ}$~52. From the top, the first panel is the FUV 
    FUSE range, containing interstellar features such as $Ly\delta~\lambda950$, $Ly\gamma~\lambda973$, and $Ly\beta~\lambda1026$. Second and third panels are STIS E140M FUV and STIS E230M NUV
    spectral ranges, respectively, and we note some interstellar features: $Ly\alpha~\lambda1216$, $\ion{O}{I}+\ion{P}{II}~\lambda1302$, $\ion{C}{II}~\lambda1335$, $\ion{Si}{II}~\lambda1527$, 
    and $\ion{Al}{II}\lambda1671$. The rest of the panels are the UBV and VIS XShooter spectra, with the break between the two arms at $\approx5600~\AA$. The lines
    $\ion{Ca}{II}~{\rm H+K}$ and $\ion{Na}{I}~{\rm D}$ are interstellar features.}
    \label{overall_SK-68D52}
\end{figure*}

\begin{figure*}
  \centering
	\includegraphics[scale=0.75]{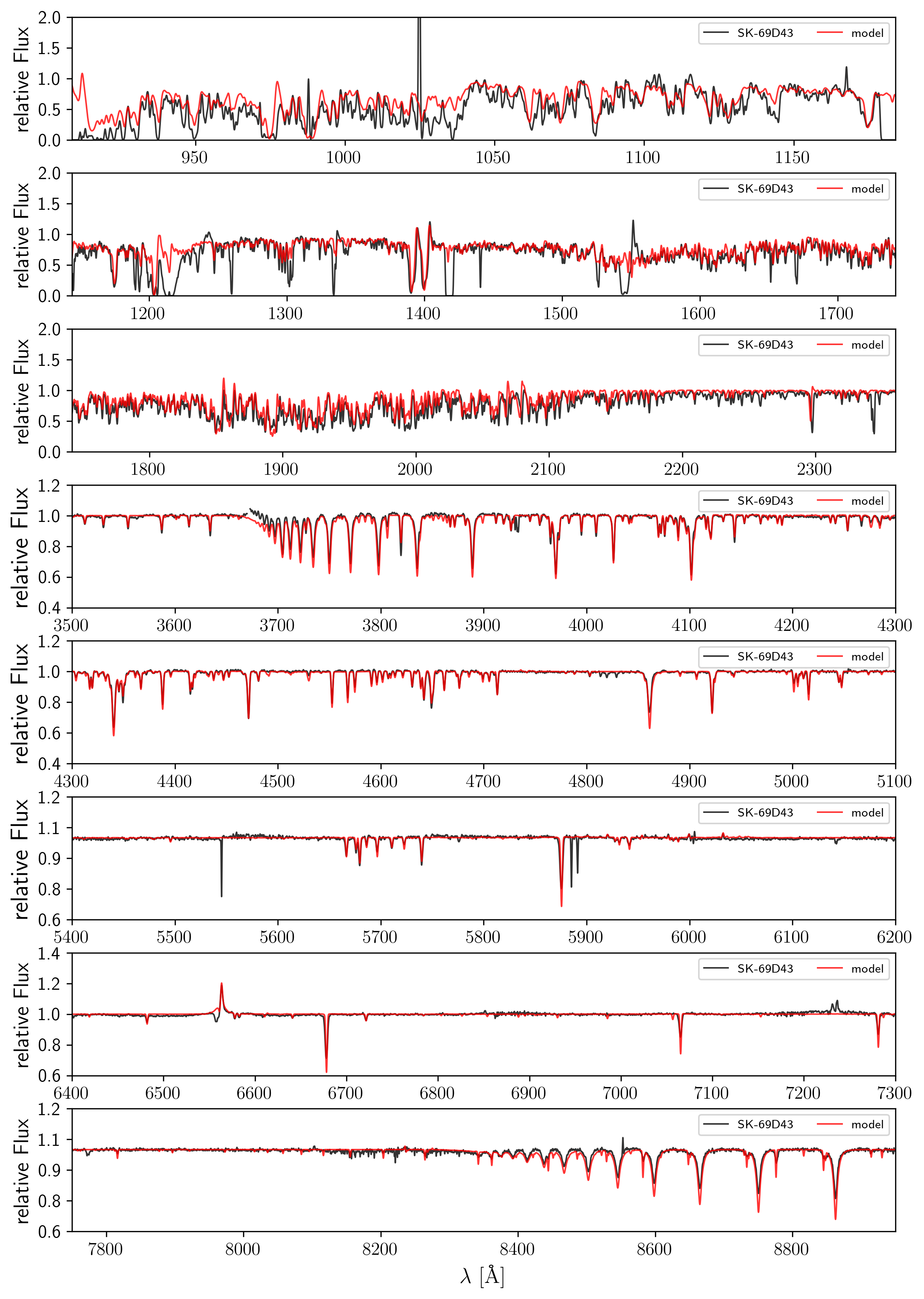}
    \caption{An overall view of the best fit (red solid line) to key regions of the observed spectrum (black solid line) of Sk\,$-$69$^{\circ}$~43. From the top, the first panel is the FUV 
    FUSE range, containing interstellar features such as $Ly\delta~\lambda950$, $Ly\gamma~\lambda973$, and $Ly\beta~\lambda1026$. Second and third panels are STIS E140M FUV and STIS E230M NUV
    spectral ranges, respectively, and we note some interstellar features: $Ly\alpha~\lambda1216$, $\ion{O}{I}+\ion{P}{II}~\lambda1302$, $\ion{C}{II}~\lambda1335$, $\ion{Si}{II}~\lambda1527$, 
    and $\ion{Al}{II}\lambda1671$. The rest of the panels are the UBV and VIS XShooter spectra, with the break between the two arms at $\approx5600~\AA$. The lines
    $\ion{Ca}{II}~{\rm H+K}$ and $\ion{Na}{I}~{\rm D}$ are interstellar features.}
    \label{overall_SK-69D43}
\end{figure*}

\begin{figure*}
  \centering
	\includegraphics[scale=0.75]{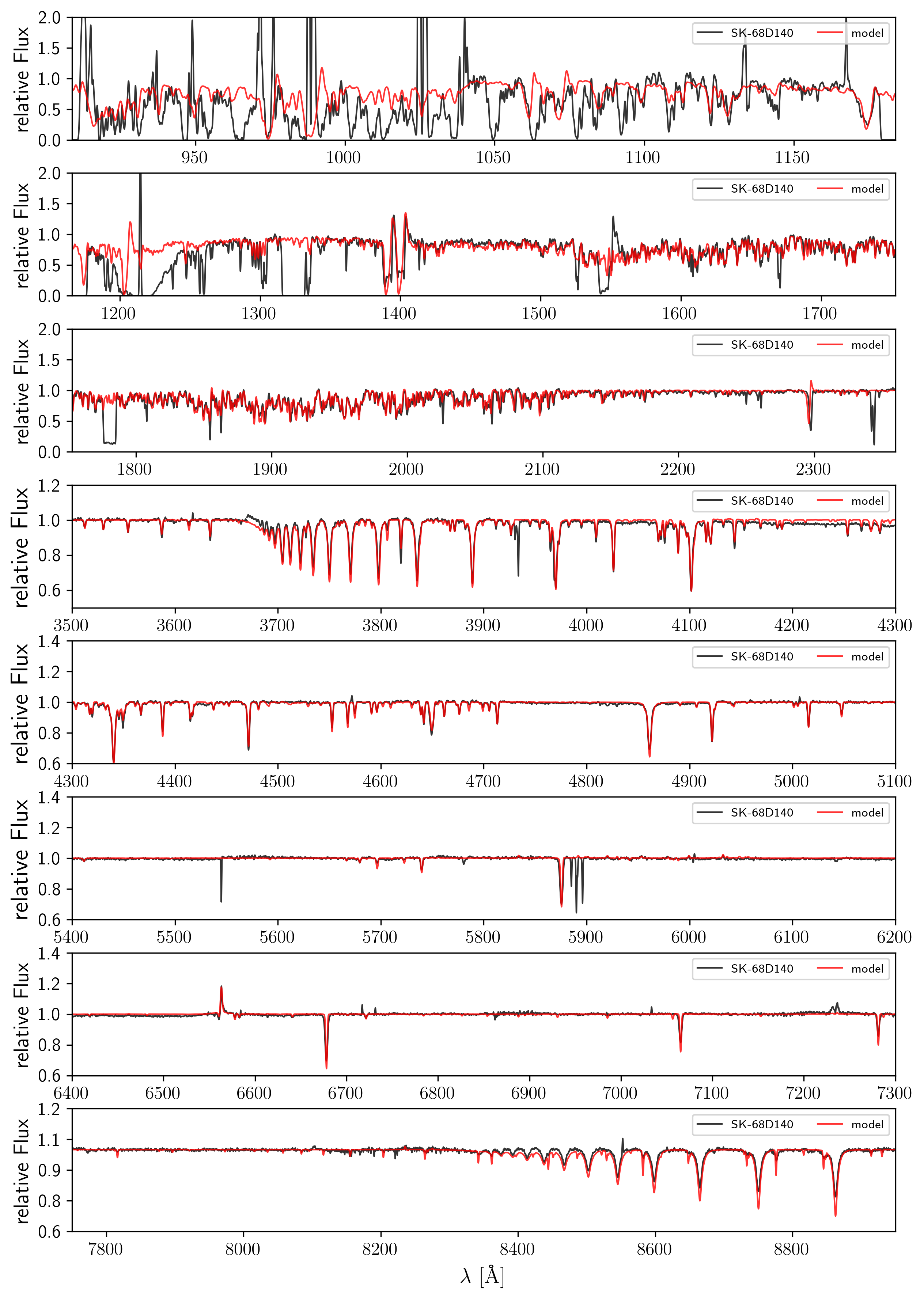}
    \caption{An overall view of the best fit (red solid line) to key regions of the observed spectrum (black solid line) of Sk\,$-$68$^{\circ}$~140. From the top, the first panel is the FUV 
    FUSE range, containing interstellar features such as $Ly\delta~\lambda950$, $Ly\gamma~\lambda973$, and $Ly\beta~\lambda1026$. Second and third panels are COS G130M+G160M FUV and STIS E230M NUV
    spectral ranges, respectively, and we note some interstellar features: $Ly\alpha~\lambda1216$, $\ion{O}{I}+\ion{P}{II}~\lambda1302$, $\ion{C}{II}~\lambda1335$, $\ion{Si}{II}~\lambda1527$, 
    and $\ion{Al}{II}\lambda1671$. The rest of the panels are the UBV and VIS XShooter spectra, with the break between the two arms at $\approx5600~\AA$. The lines
    $\ion{Ca}{II}~{\rm H+K}$ and $\ion{Na}{I}~{\rm D}$ are interstellar features.}
    \label{overall_SK-68D140}
\end{figure*}

\begin{figure*}
  \centering
	\includegraphics[scale=0.75]{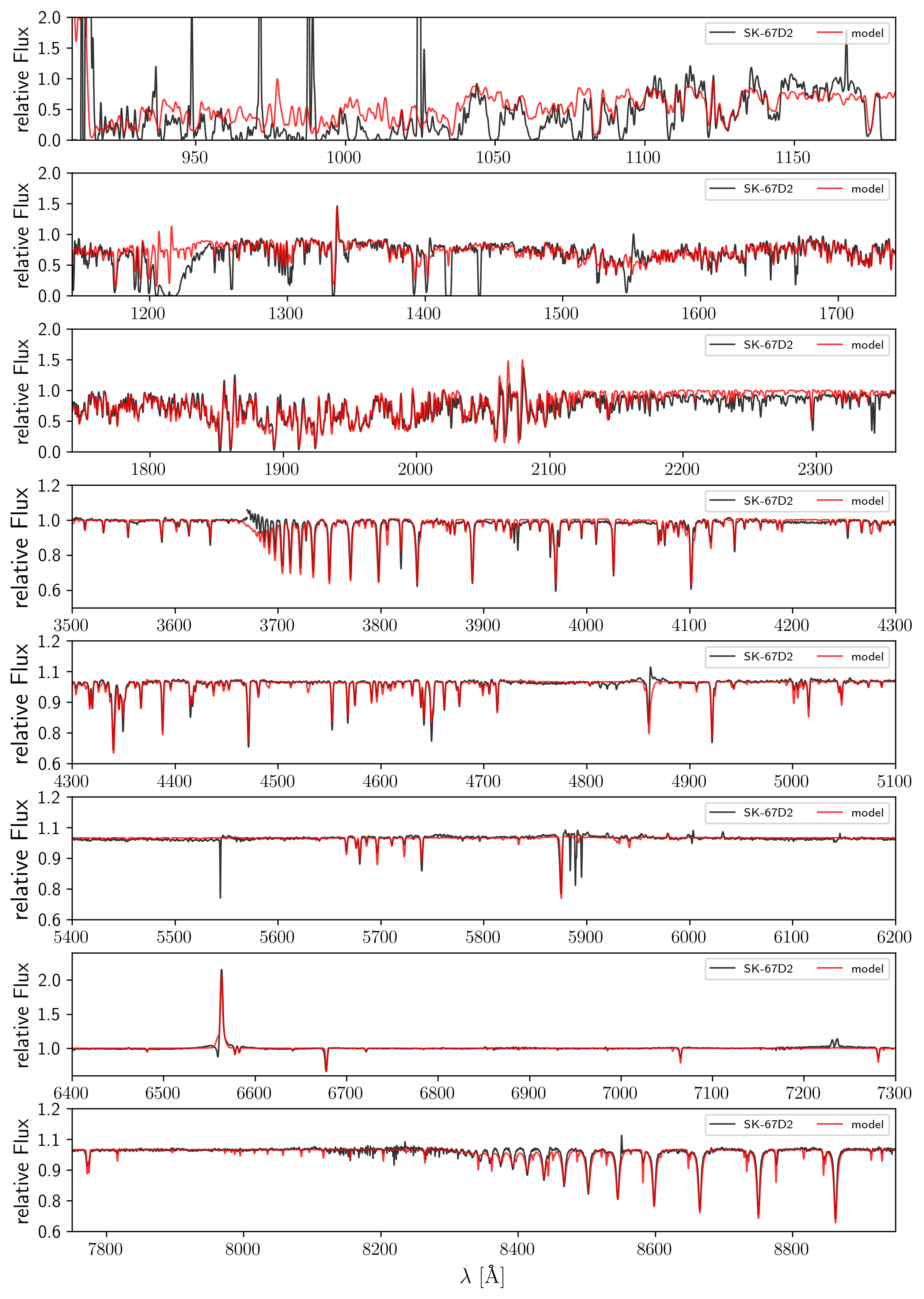}
    \caption{An overall view of the best fit (red solid line) to key regions of the observed spectrum (black solid line) of Sk\,$-$67$^{\circ}$~2. From the top, the first panel is the FUV 
    FUSE range, containing interstellar features such as $Ly\delta~\lambda950$, $Ly\gamma~\lambda973$, and $Ly\beta~\lambda1026$. Second and third panels are STIS E140M FUV and STIS E230M NUV
    spectral ranges, respectively, and we note some interstellar features: $Ly\alpha~\lambda1216$, $\ion{O}{I}+\ion{P}{II}~\lambda1302$, $\ion{C}{II}~\lambda1335$, $\ion{Si}{II}~\lambda1527$, 
    and $\ion{Al}{II}\lambda1671$. The rest of the panels are the UBV and VIS XShooter spectra, with the break between the two arms at $\approx5600~\AA$. The lines
    $\ion{Ca}{II}~{\rm H+K}$ and $\ion{Na}{I}~{\rm D}$ are interstellar features.}
    \label{overall_SK-67D2}
\end{figure*}

\begin{figure*}
  \centering
	\includegraphics[scale=0.75]{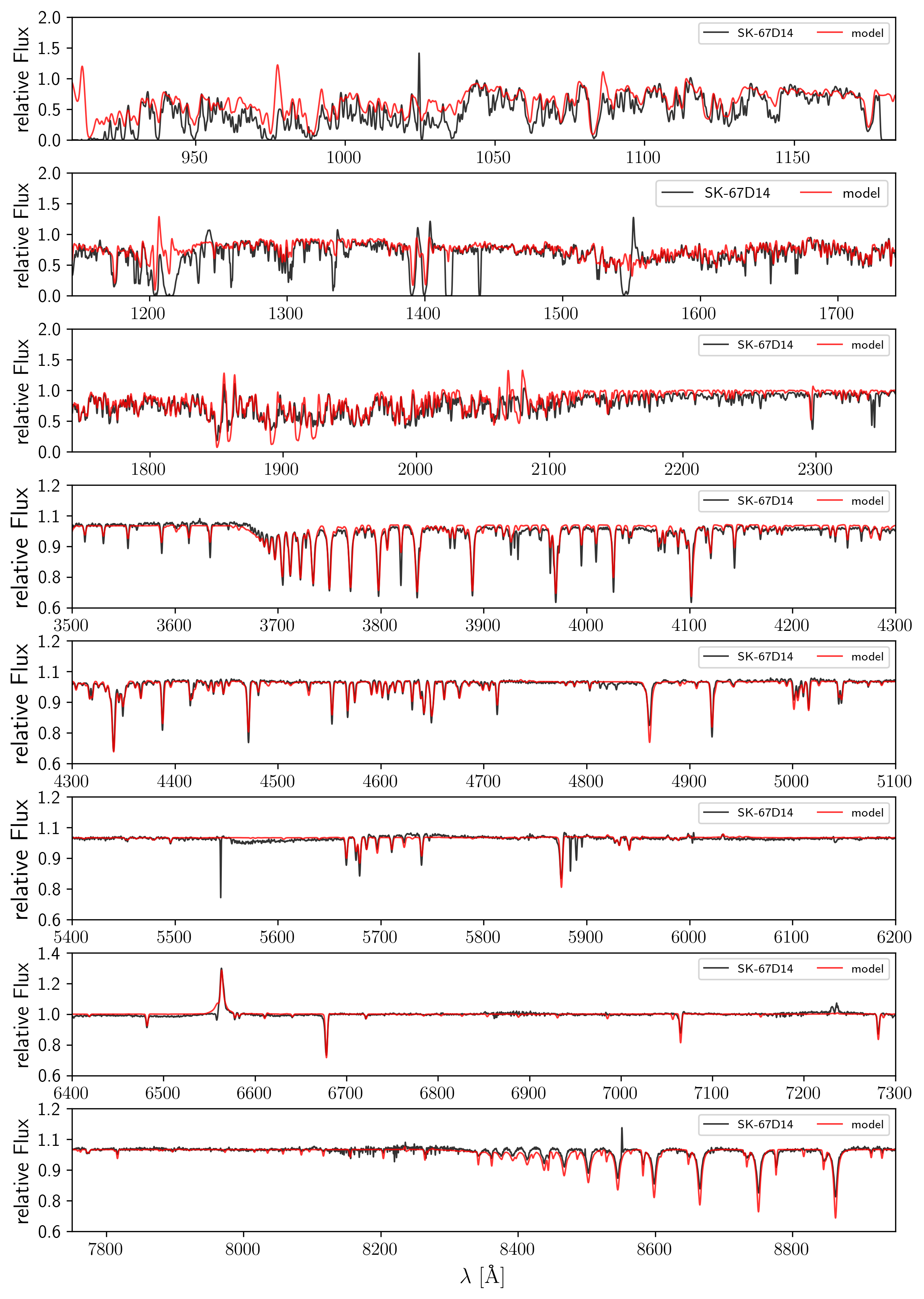}
    \caption{An overall view of the best fit (red solid line) to key regions of the observed spectrum (black solid line) of Sk\,$-$67$^{\circ}$~14. From the top, the first panel is the FUV 
    FUSE range, containing interstellar features such as $Ly\delta~\lambda950$, $Ly\gamma~\lambda973$, and $Ly\beta~\lambda1026$. Second and third panels are COS G130M+G160M FUV and STIS E230M NUV
    spectral ranges, respectively, and we note some interstellar features: $Ly\alpha~\lambda1216$, $\ion{O}{I}+\ion{P}{II}~\lambda1302$, $\ion{C}{II}~\lambda1335$, $\ion{Si}{II}~\lambda1527$, 
    and $\ion{Al}{II}\lambda1671$. The rest of the panels are the UBV and VIS XShooter spectra, with the break between the two arms at $\approx5600~\AA$. The lines
    $\ion{Ca}{II}~{\rm H+K}$ and $\ion{Na}{I}~{\rm D}$ are interstellar features.}
    \label{overall_SK-67D14}
\end{figure*}

\begin{figure*}
  \centering
	\includegraphics[scale=0.75]{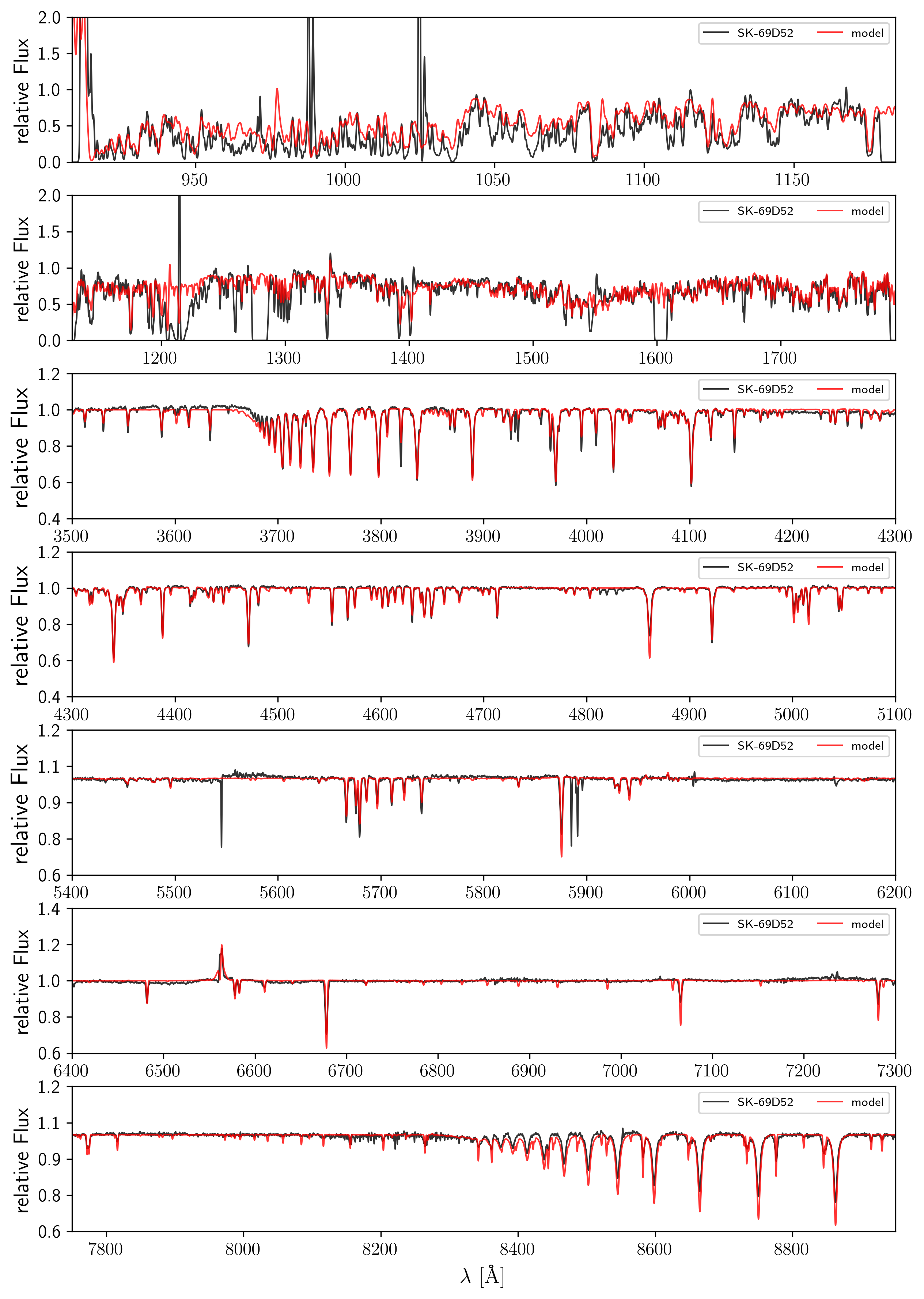}
    \caption{An overall view of the best fit (red solid line) to key regions of the observed spectrum (black solid line) of Sk\,$-$69$^{\circ}$~52. From the top, the first panel is the FUV 
    FUSE range, containing interstellar features such as $Ly\delta~\lambda950$, $Ly\gamma~\lambda973$, and $Ly\beta~\lambda1026$. Second and third panels are COS G130M+G160M FUV and STIS E230M NUV
    spectral ranges, respectively, and we note some interstellar features: $Ly\alpha~\lambda1216$, $\ion{O}{I}+\ion{P}{II}~\lambda1302$, $\ion{C}{II}~\lambda1335$, $\ion{Si}{II}~\lambda1527$, 
    and $\ion{Al}{II}\lambda1671$. The rest of the panels are the UBV and VIS XShooter spectra, with the break between the two arms at $\approx5600~\AA$. The lines
    $\ion{Ca}{II}~{\rm H+K}$ and $\ion{Na}{I}~{\rm D}$ are interstellar features.}
    \label{overall_SK-69D52}
\end{figure*}

\begin{figure*}
  \centering
	\includegraphics[scale=0.75]{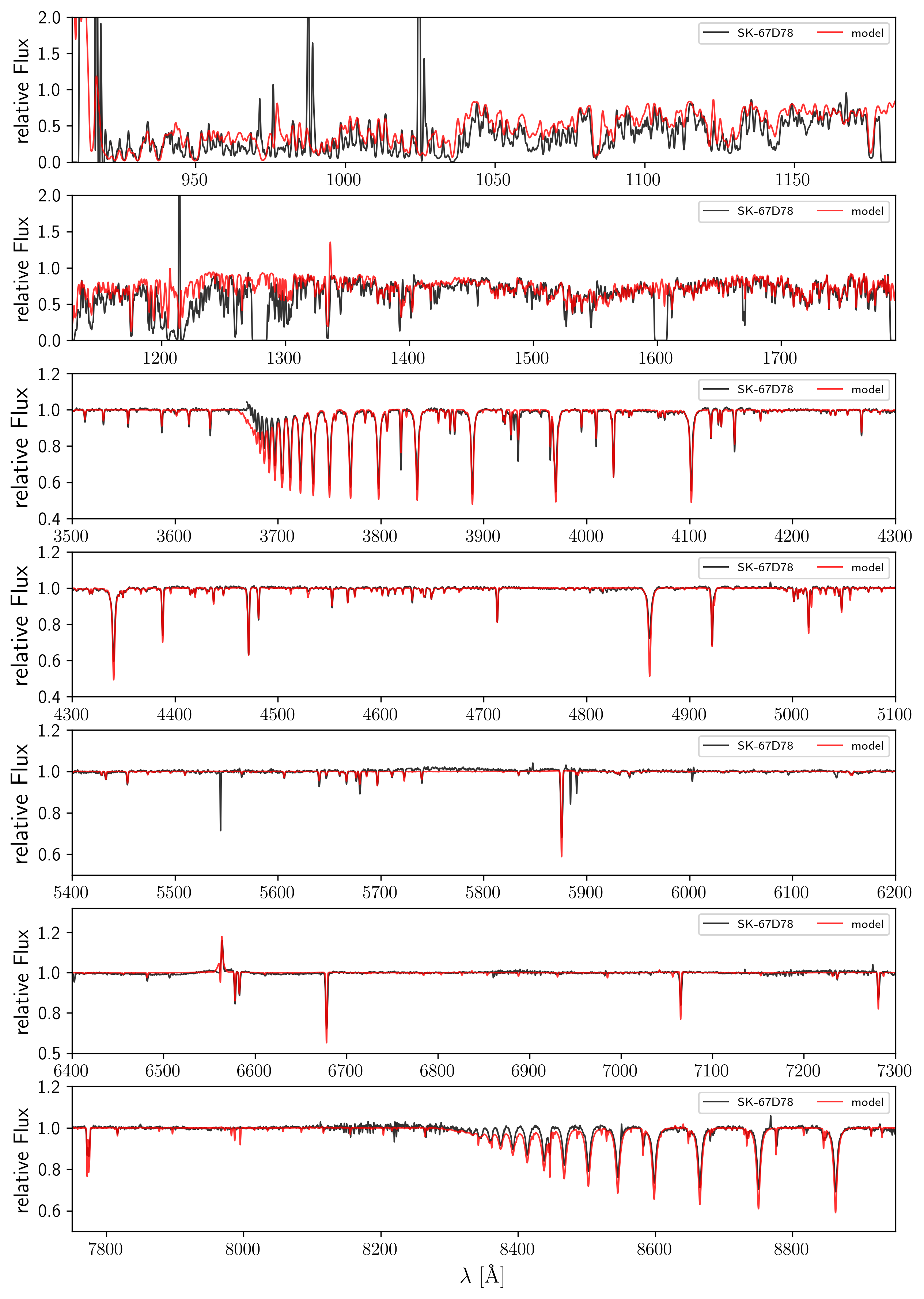}
    \caption{An overall view of the best fit (red solid line) to key regions of the observed spectrum (black solid line) of Sk\,$-$67$^{\circ}$~78. From the top, the first panel is the FUV 
    FUSE range, containing interstellar features such as $Ly\delta~\lambda950$, $Ly\gamma~\lambda973$, and $Ly\beta~\lambda1026$. Second and third panels are COS G130M+G160M FUV and STIS E230M NUV
    spectral ranges, respectively, and we note some interstellar features: $Ly\alpha~\lambda1216$, $\ion{O}{I}+\ion{P}{II}~\lambda1302$, $\ion{C}{II}~\lambda1335$, $\ion{Si}{II}~\lambda1527$, 
    and $\ion{Al}{II}\lambda1671$. The rest of the panels are the UBV and VIS XShooter spectra, with the break between the two arms at $\approx5600~\AA$. The lines
    $\ion{Ca}{II}~{\rm H+K}$ and $\ion{Na}{I}~{\rm D}$ are interstellar features.}
    \label{overall_SK-67D78}
\end{figure*}

\begin{figure*}
  \centering
	\includegraphics[scale=0.75]{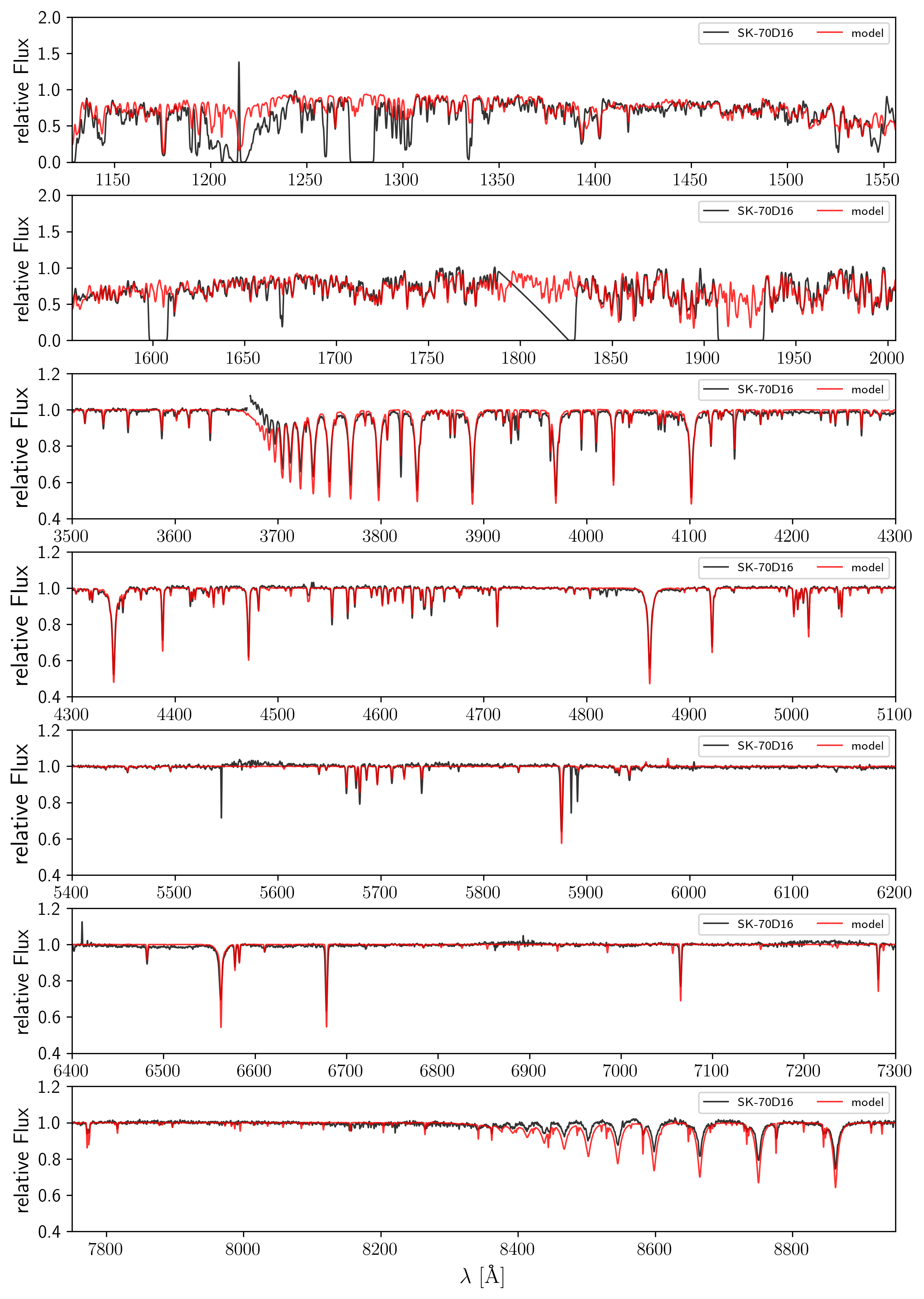}
    \caption{An overall view of the best fit (red solid line) to key regions of the observed spectrum (black solid line) of Sk\,$-$70$^{\circ}$~16. From the top, the first and second panels are 
    COS G130M+G160M FUV and COS G160M+G185M NUV spectral ranges, respectively, and we note some interstellar features: $Ly\alpha~\lambda1216$, $\ion{O}{I}+\ion{P}{II}~\lambda1302$, 
    $\ion{C}{II}~\lambda1335$, $\ion{Si}{II}~\lambda1527$, and $\ion{Al}{II}\lambda1671$. The rest of the panels are the UBV and VIS XShooter spectra, with the break between the two 
    arms at $\approx5600~\AA$. The lines $\ion{Ca}{II}~{\rm H+K}$ and $\ion{Na}{I}~{\rm D}$ are interstellar features.}
    \label{overall_SK-70D16}
\end{figure*}

\begin{figure*}
  \centering
	\includegraphics[scale=0.75]{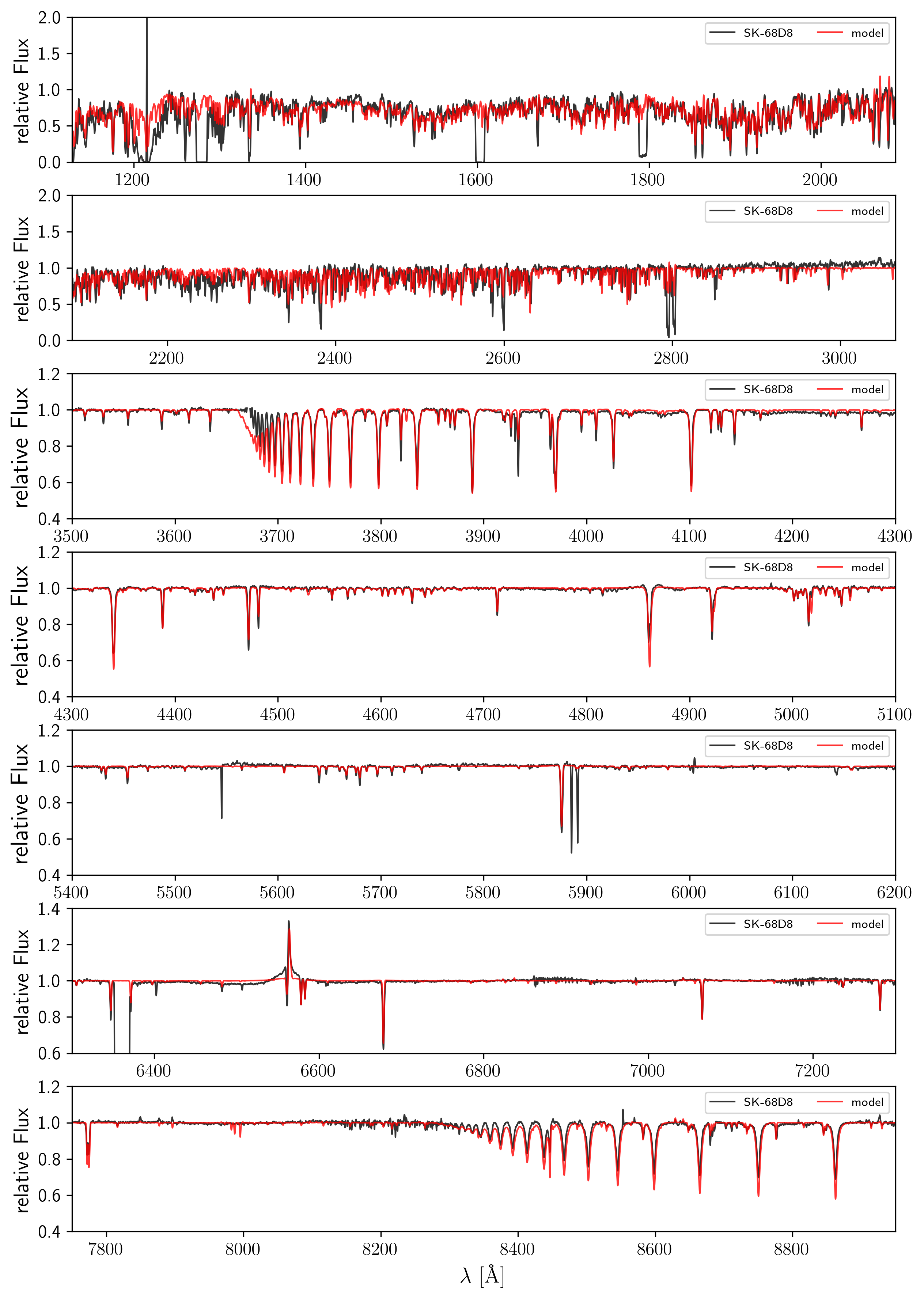}
    \caption{An overall view of the best fit (red solid line) to key regions of the observed spectrum (black solid line) of Sk\,$-$68$^{\circ}$~8. From the top, the first and second panels are 
    COS G130M+G160M FUV and STIS E230M NUV spectral ranges, respectively, and we note some interstellar features: $Ly\alpha~\lambda1216$, $\ion{O}{I}+\ion{P}{II}~\lambda1302$, 
    $\ion{C}{II}~\lambda1335$, $\ion{Si}{II}~\lambda1527$, and $\ion{Al}{II}\lambda1671$. The rest of the panels are the UBV and VIS XShooter spectra, with the break between the two 
    arms at $\approx5600~\AA$. The lines $\ion{Ca}{II}~{\rm H+K}$ and $\ion{Na}{I}~{\rm D}$ are interstellar features.}
    \label{overall_SK-68D8}
\end{figure*}

\begin{figure*}
  \centering
	\includegraphics[scale=0.75]{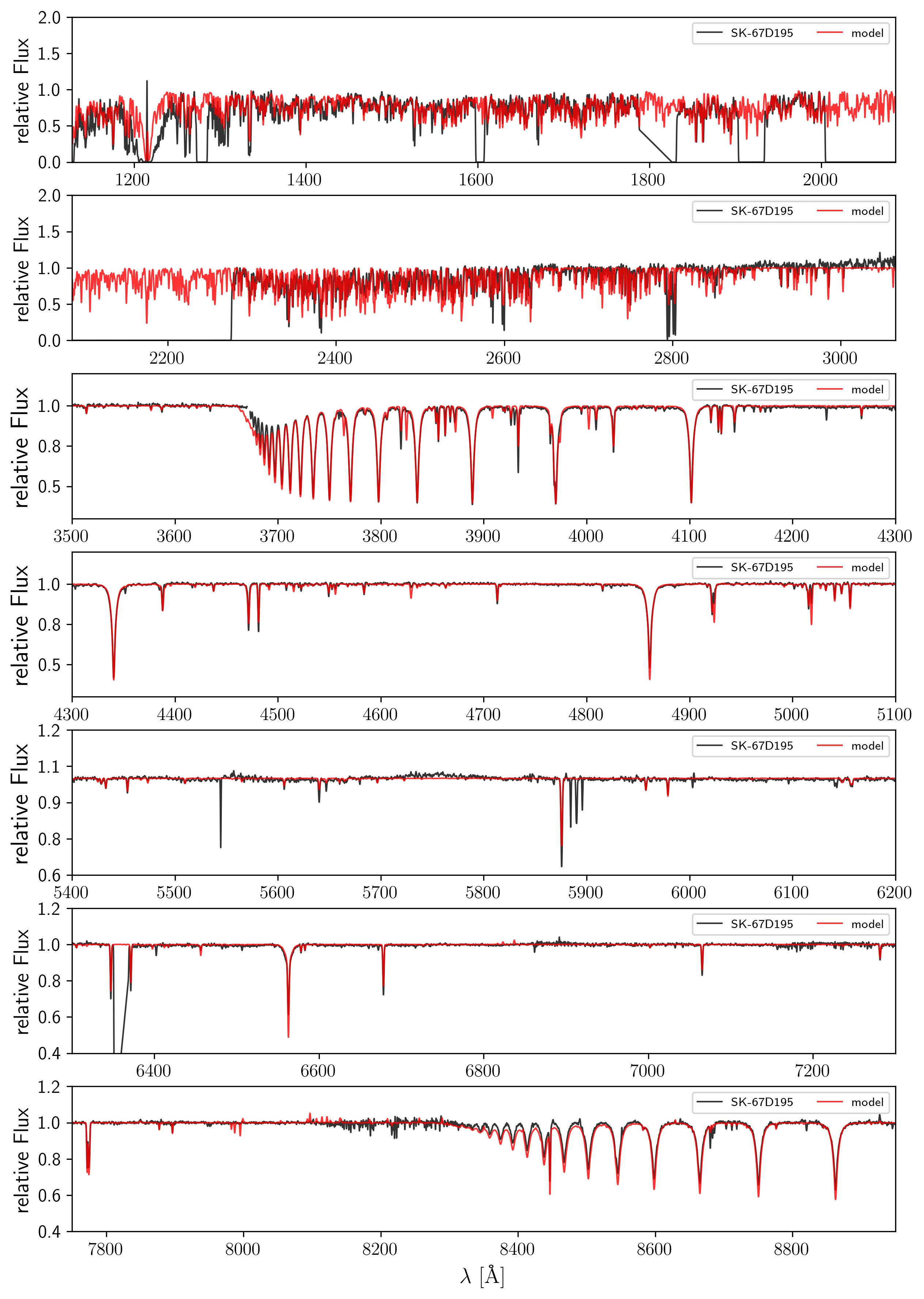}
    \caption{An overall view of the best fit (red solid line) to key regions of the observed spectrum (black solid line) of Sk\,$-$67$^{\circ}$~195. From the top, the first and second panels are 
    COS G130M+G160M FUV and STIS E230M NUV spectral ranges, respectively, and we note some interstellar features: $Ly\alpha~\lambda1216$, $\ion{O}{I}+\ion{P}{II}~\lambda1302$, 
    $\ion{C}{II}~\lambda1335$, $\ion{Si}{II}~\lambda1527$, and $\ion{Al}{II}\lambda1671$. The rest of the panels are the UBV and VIS XShooter spectra, with the break between the two 
    arms at $\approx5600~\AA$. The lines $\ion{Ca}{II}~{\rm H+K}$ and $\ion{Na}{I}~{\rm D}$ are interstellar features.}
    \label{overall_SK-67D195}
\end{figure*}

Fig.~\ref{overall_SK-66D171}-\ref{overall_SK-67D195} are the fits for the overall spectrum of each star in our sample.

\end{appendix}
\end{document}